\documentclass[rmp,reprint,longbibliography]{revtex4-1}

\usepackage[utf8]{inputenc}
\usepackage[english]{babel}
\usepackage{amssymb,amsthm,amsmath,amstext,amsbsy,amsopn}
\usepackage{bbm}
\usepackage{nicefrac}
\usepackage{slashed}
\usepackage{pstricks}
\usepackage{graphicx}
\usepackage{hyperref}
\usepackage{leftidx}
\usepackage{environ}
\usepackage{suffix}
\usepackage{mathtools}
\usepackage{xspace}
\usepackage{overpic}

\makeatletter
\def\NAT@sort{\z@}
\makeatother

\providecommand{\BibitemShut}[1]{}

\newcommand{\ie}{\textit{i.e.}\xspace}
\newcommand{\eg}{\textit{e.g.}\xspace}
\newcommand{\cf}{\textit{cf.}\xspace}
\newcommand{\etc}{\textit{etc.}\xspace}
\newcommand{\apriori}{\textit{a priori}\xspace}

\newcommand{\abinitio}{\textit{ab initio}\xspace}
\newcommand{\Abinitio}{\textit{Ab initio}\xspace}
\newcommand{\viceversa}{\textit{vice versa}\xspace}
\newcommand{\via}{\textit{via}\xspace}

\newcommand{\mathspace}{\ \ }
\newcommand{\mathtext}[1]{\mathspace\text{#1}\mathspace}

\newcommand{\vecp}{\mathbf{p}}
\newcommand{\veck}{\mathbf{k}}
\newcommand{\vecq}{\mathbf{q}}

\newcommand{\vNabla}{\boldsymbol{\nabla}}

\newcommand*\rvec[1]{%
\ensuremath{\overset{\smash{\raisebox{-1.5pt}{\tiny$\rightarrow$}}}{#1}}%
}
\newcommand*\lvec[1]{%
\ensuremath{\overset{\smash{\raisebox{-1.5pt}{\tiny$\leftarrow$}}}{#1}}%
}
\newcommand*\lrvec[1]{%
\ensuremath{\overset{\smash{\raisebox{-1.5pt}{\tiny$\leftrightarrow$}}}{#1}}%
}
\newcommand{\vNablaL}{\lvec{\vNabla}}
\newcommand{\vNablaR}{\rvec{\vNabla}}
\newcommand{\vNablaLR}{(\vNablaL - \vNablaR)}
\WithSuffix\newcommand\vNablaLR*{\lrvec{\vNabla}}

\newcommand{\dd}{\mathrm{d}}

\newcommand{\dq}[1]{\!\!\frac{\mathrm{d}^3#1}{(2\pi)^3}}

\newcommand{\ddq}{\dq{q}}

\newcommand{\gsim}{\gtrsim}

\newcommand{\ii}{\mathrm{i}}
\newcommand{\eex}{\mathrm{e}}

\newcommand{\hc}{\mathrm{H.c.}}
\newcommand{\OO}{\mathcal{O}}

\newcommand{\NN}{\mathcal{N}}

\newcommand{\eps}{\varepsilon}

\newcommand{\leviciv}{\epsilon}

\newcommand{\Laplace}{\vNabla^2}

\newcommand{\braket}[2]{\langle #1|#2\rangle}

\newcommand{\mbraket}[3]{\langle #1|#2|#3\rangle}

\newcommand{\abs}[1]{|#1|}

\newcommand{\CG}[6]{\ensuremath{\braket{#1\,#2\,#3\,#4}{#5\,#6}}}
\WithSuffix\newcommand\CG*[6]{\ensuremath{\braket{#1\,#2}{#3\,#4\,#5\,#6}}}

\NewEnviron{subalign}[1][]{%
\begin{subequations}\begin{align}
  \BODY
\end{align}\label{#1}\end{subequations}
}

\NewEnviron{spliteq}{%
\begin{equation}\begin{split}
  \BODY
\end{split}\end{equation}
}

\newcommand*{\vcenteredhbox}[1]
{\begingroup\setbox0=\hbox{#1}\parbox{\wd0}{\box0}\endgroup}

\newcommand{\GeV}{\ensuremath{\mathrm{GeV}}}
\newcommand{\MeV}{\ensuremath{\mathrm{MeV}}}
\newcommand{\keV}{\ensuremath{\mathrm{keV}}}
\newcommand{\fm}{\ensuremath{\mathrm{fm}}}

\newcommand{\LL}{\mathcal{L}}

\newcommand{\MN}{\ensuremath{m_N}}

\newcommand{\mpi}{\ensuremath{m_\pi}}
\newcommand{\Mpi}{\mpi}

\newcommand{\gamd}{\ensuremath{\gamma_d}}
\newcommand{\rhod}{\ensuremath{\rho_d}}

\newcommand{\Bd}{\ensuremath{B_d}}

\newcommand{\ThreeSOne}{\ensuremath{{}^3S_1}\xspace}
\newcommand{\OneSNot}{\ensuremath{{}^1S_0}\xspace}
\newcommand{\ThreePNot}{\ensuremath{{}^3P_0}\xspace}

\newcommand{\LamNoPi}{M_{\text{hi}}}
\renewcommand{\aleph}{M_{\text{lo}}}

\newcommand{\ann}{a_{nn}}

\newcommand{\Triton}{\ensuremath{{}^3\mathrm{H}}\xspace}
\newcommand{\ThreeH}{\Triton}
\newcommand{\ThreeHe}{\ensuremath{{}^3\mathrm{He}}\xspace}
\newcommand{\FourHe}{\ensuremath{{}^4\mathrm{He}}\xspace}
\newcommand{\SixHe}{\ensuremath{{}^6\mathrm{He}}\xspace}
\newcommand{\SixLi}{\ensuremath{{}^6\mathrm{Li}}\xspace}
\newcommand{\SixteenO}{\ensuremath{{}^{16}\mathrm{O}}\xspace}

\newcommand{\LO}{\text{LO}\xspace}
\newcommand{\NLO}{\text{NLO}\xspace}
\newcommand{\NNLO}{\text{N$^2$LO}\xspace}
\newcommand{\NNNLO}{\text{N$^3$LO}\xspace}
\newcommand{\NNNNLO}{\text{N$^4$LO}\xspace}

\newcommand{\pPt}{\ensuremath{P_{\mathrm{t}}}}
\newcommand{\pPs}{\ensuremath{P_{\mathrm{s}}}}

\newcommand{\bqa}{\begin{eqnarray}}
\newcommand{\eqa}{\end{eqnarray}}
\newcommand{\be}{\begin{equation}}
\newcommand{\ee}{\end{equation}}
\newcommand{\ba}{\begin{eqnarray}}
\newcommand{\ea}{\end{eqnarray}}
\newcommand{\beq}{\begin{equation}}
\newcommand{\eeq}{\end{equation}}
\newcommand{\beqa}{\begin{eqnarray}}
\newcommand{\eeqa}{\end{eqnarray}}

\newcommand{\galnab}{\tensor{\nabla}}

\newcommand{\simle}{\stackrel{<}{{}_\sim}}
\newcommand{\simge}{\stackrel{>}{{}_\sim}}

\newcommand{\boldS}{\mbox{\boldmath $S$}}

\newcommand{\boldD}{\mbox{\boldmath $D$}}
\newcommand{\boldcalD}{\mbox{\boldmath ${\cal D}$}}

\newcommand{\boldq}{\mbox{\boldmath $q$}}
\newcommand{\boldp}{\mbox{\boldmath $p$}}

\newcommand{\boldl}{\mbox{\boldmath $l$}}
\newcommand{\boldsigma}{\mbox{\boldmath $\sigma$}}

\newcommand{\MQCD}{M_{\text{QCD}}}
\newcommand{\Mhi}{M_{\text{hi}}}
\newcommand{\Mlo}{M_{\text{lo}}}
\newcommand{\MNN}{M_{N\! N}}

\begin{document}

\title{Nuclear effective field theory: status and perspectives}

\author{H.-W.~Hammer}
\email{Hans-Werner.Hammer@physik.tu-darmstadt.de}
\affiliation{Institut f\"ur Kernphysik, Technische Universit\"at Darmstadt,\\
64289 Darmstadt,\\
Germany}
\affiliation{ExtreMe Matter Institute EMMI, GSI Helmholtzzentrum 
f\"ur Schwerionenforschung GmbH,\\
64291 Darmstadt,\\
Germany}

\author{Sebastian K\"onig}
\email{sekoenig@theorie.ikp.physik.tu-darmstadt.de}
\affiliation{Institut f\"ur Kernphysik, Technische Universit\"at Darmstadt,\\
64289 Darmstadt,\\
Germany}
\affiliation{ExtreMe Matter Institute EMMI, GSI Helmholtzzentrum 
f\"ur Schwerionenforschung GmbH,\\
64291 Darmstadt,\\
Germany}
\affiliation{Department of Physics, The Ohio State University,\\
Columbus, Ohio 43210,\\
USA}
\affiliation{Department of Physics, North Carolina State University,\\
Raleigh, NC 27695,\\
USA}

\author{U.~van Kolck}
\email{vankolck@ipno.in2p3.fr}
\affiliation{Universit\'e Paris-Saclay, CNRS/IN2P3, IJCLab,\\
91405 Orsay,\\
France}
\affiliation{Department of Physics, University of Arizona,\\
Tucson, AZ 85721,\\
USA}

\date{\today}

\begin{abstract}
{The nuclear physics landscape has been redesigned as}
a sequence of effective field theories (EFTs) connected to the Standard Model 
through symmetries and lattice simulations of Quantum Chromodynamics (QCD).
EFTs in this sequence are expansions around different low-energy limits of QCD,
each with its own characteristics, scales, and ranges of applicability regarding
energy and number of nucleons.  We review each of the three main nuclear
EFTs---Chiral, Pionless, Halo/Cluster---highlighting their similarities, 
differences, and connections.  In doing so, we survey the structural properties
and reactions of nuclei that have been derived from the \abinitio solution of 
the few- and many-body problem built upon EFT input.
\end{abstract}

\maketitle

\tableofcontents

\section{Introduction} \label{sec:introduction}
The problem of obtaining the properties of atomic nuclei from the interactions
among the constituent nucleons has been central to nuclear physics since its 
inception.  Attempts to derive nuclear forces and currents from the exchange of 
mesons---in particular the lightest meson, the pion---were derailed in the 
1950s by a lack of renormalizability, that is, by an uncontrolled sensitivity 
to physics at short distances.\footnote{Not long after the successful 
renormalization of QED, it was understood that the only relativistic
pion-nucleon coupling that is renormalizable in the same sense is 
pseudoscalar~\cite{Matthews:1951sk}. However, pseudoscalar coupling differs 
from pseudovector coupling by a large nucleon-pair term, which was found
to be in conflict with pion phenomenology~\cite{Marshak:1952NN}.  The favored 
pseudovector coupling required the introduction of short-distance cutoffs, on 
which description of two-nucleon data depended sensitively (see, for example, 
\cite{Gartenhaus:1955zz}).  Subsequent work increasingly emphasized 
the phenomenology of short-range interactions.
{A brief history of nuclear potential models is given by 
\textcite{Machleidt:2011zz} and \textcite{Machleidt:2017vls}.}}  
The rise in the 1970s of a
renormalizable theory of the strong interactions, Quantum Chromodynamics (QCD),
did not immediately offer a path forward: because QCD---formulated in terms of
quarks and gluons---is nonperturbative for processes characterized by external
momenta $Q\simle \MQCD\sim 1~\GeV$, it is very difficult to calculate the
properties of hadrons and nuclei, a problem that becomes more severe as the
number $A$ of nucleons increases.  {Yet, a precise and accurate description 
of nuclei is crucial for the transition from the perturbative regime of the 
Standard Model to the atomic domain and beyond, governed by Quantum 
Electrodynamics (QED) and its small fine-structure constant.} 
{Examples} {relying on input from nuclear physics include
tests of fundamental symmetries (such as neutrinoless double-beta decay to probe
lepton number violation) and reactions in astrophysical environments.}

About a quarter of a century ago effective field theories (EFTs) entered 
nuclear physics~\cite{Weinberg:1990rz,Rho:1990cf,Weinberg:1991um,%
Ordonez:1992xp,Weinberg:1992yk,VanKolck:1993ee}.  EFTs had been developed in 
particle and condensed-matter physics to deal with systems containing multiple 
momentum scales.  An EFT captures the most general dynamics among low-energy 
degrees of freedom that is consistent with some assumed symmetries.  In nuclear 
physics, where the symmetries of QCD are known, an EFT provides a realization 
of QCD in terms of hadrons instead of quarks and gluons.  All the details of 
the QCD dynamics at short distances are encoded in the EFT interaction strengths
(``Wilson coefficients'' or ``low-energy constants'').  Scattering amplitudes
(and their poles representing bound states and resonances) are calculated as
expansions in $Q/\Mhi$ and $\Mlo/\Mhi$, {with $\Mhi$ the momentum 
scale where the EFT breaks down and $\Mlo$ standing for low-energy scales 
of} physics we want to capture.  An EFT is renormalizable in the sense that at 
each order in the expansion the sensitivity to unaccounted short-distance
physics is small, that is, of relative $\OO(Q/\Mhi,\Mlo/\Mhi)$.  EFTs opened
the door to a description of nuclear phenomena with systematic error estimates.

EFTs have,
{in fact, revolutionized} nuclear physics.  Most of the ``\abinitio'' 
studies of nuclear structure---based on the explicit solution of the 
Schr\"odinger equation or its equivalents---are now carried out with potentials
inspired by EFT.  A host of nuclear properties has been predicted or postdicted 
from two- and three-nucleon forces, and one- and two-nucleon currents {with}
low-energy constants were determined from $A\le 3$ experimental data.
In parallel, starting with~\textcite{Beane:2006mx}, fully dynamical simulations
of QCD on a discretized and boxed spacetime have been able to access some 
$A\le 4$ properties.  Matching an EFT {to results from} {lattice QCD 
(LQCD)}, {and not just to experiment,} allows for a determination of the
low-energy constants.  EFTs thus build a bridge between QCD and nuclear
structure and reactions.

Historically the first nuclear EFT was \textbf{Chiral (or Pionful) 
EFT}~\cite{Weinberg:1990rz,Rho:1990cf,%
Weinberg:1991um,Ordonez:1992xp,Weinberg:1992yk,VanKolck:1993ee},
which is designed for momenta of the order of the pion mass.  In addition to
nucleons, it includes explicit pions, whose interactions are 
constrained by an approximate global symmetry of QCD, the chiral symmetry of 
independent flavor rotations of left- and right-handed quarks.  Chiral EFT 
generalizes a popular hadronic EFT, Chiral Perturbation 
Theory~\cite{Weinberg:1978kz,Gasser:1983yg,Gasser:1984gg}, to $A\ge 2$.  Despite
its phenomenological successes, Chiral EFT has proven to be extremely 
challenging to renormalize due to the singularity of the dominant interactions, 
which have to be treated nonperturbatively in order to produce bound states and 
resonances, \ie, nuclei.  Originally conceived as a renormalization playground,
a simpler EFT---\textbf{Pionless (or Contact) EFT}---focuses on momenta below
the pion mass~{\cite{Bedaque:1997qi,vanKolck:1997ut,Kaplan:1998tg,%
Bedaque:1998mb,Kaplan:1998we,vanKolck:1998bw,Birse:1998dk}}.  This EFT, whose 
renormalization is relatively well understood, is constrained only by QCD
spacetime symmetries.  It exhibits a high degree of universality, and except for
the degrees of freedom it is formally identical to other EFTs where all
interactions are of short range.  Light nuclei are well described within the
same framework that has been successful for atomic systems with large scattering
lengths (for example, near a Feshbach resonance)~\cite{Braaten:2004rn}.  A
variant of this EFT, \textbf{Halo/Cluster EFT}, has been
applied~\cite{Bertulani:2002sz,Bedaque:2003wa} to bound states and reactions 
involving halo and cluster nuclei, characterized by such low energies
that one or more tight clusters of nucleons can be treated as elementary 
degrees of freedom.

{No doubt there are ``more effective'' EFTs to be discovered for larger 
nuclei.  In fact, a description of rotational and vibrational bands in heavy 
nuclei has been initiated by \textcite{Papenbrock:2010yg}, with 
successful applications to different nuclei and 
processes~\cite{Papenbrock:2013cra,Perez:2015ufa,%
Perez:2016nwc,Perez:2017ksf,Perez:2018cly,Chen:2018jtd,Chen:2019ugo}.  These 
recent developments extend the EFT paradigm to generalized degrees of freedom 
that capture the low-energy physics of deformed nuclei.}

{Here we present a summary of the main ideas, achievements, and prospects
for the nuclear EFTs formulated in terms of nucleons and clusters thereof.}  
These theories can be thought of as a tower of EFTs at successively lower
$\Mhi$, starting at $\MQCD$.  Our emphasis is not on phenomenology, but on the
conceptual similarities and differences among  Chiral, Pionless, and
Halo/Cluster EFTs.  Our hope is that a focused approach will stimulate a
reformulation of our  understanding of heavy nuclei, just as these EFTs have
shed new light on the structure and reactions of light nuclei.

In the remainder of this section some common aspects of nuclear EFTs are 
presented.  Each of the three following sections deal with one nuclear EFT.
In Sec.~\ref{sec:other} we address the connection between these EFTs and QCD, 
as well as broader applications.  We conclude in Sec.~\ref{sec:conclusion}.
We use throughout units such that $\hbar=1$ and $c=1$.

\subsection{Nuclei from the perspective of QCD}
\label{subsec:QCDperspective}

As an $SU(3)_{\rm c}$ gauge theory of colored quarks and gluons, QCD is 
characterized by a coupling constant $g_s$ that becomes strong at low momenta.
The fact that most hadron masses are about $1~\GeV$ or higher---for example, 
the nucleon mass $m_N\simeq 940$ MeV---reveals that nonperturbative QCD
phenomena are associated with a mass scale $\MQCD\sim 1~\GeV$.  The EFT at the 
scale of a few GeV includes not only strong interactions, but also electroweak
and even weaker interactions.  In contrast to most textbooks, for convenience we
refer to this EFT, which is our starting point, simply as QCD.  Focusing on the
two lightest (up and down) quarks most relevant to nuclear physics, the QCD
Lagrangian is written in terms of quark $q=( u \; d)^T$, gluon $G_{\mu}$ and
photon $A_{\mu}$ fields as
\begin{multline}
 \LL_{\text{QCD}} = 
 \bar{q} \left[\ii\gamma^\mu \left(\partial_\mu + \ii g_s G_\mu + \ii e Q 
 A_\mu\right) 
 + \bar{m} \left(1 - \varepsilon \tau_3\right)\right]q \\
 \null - \frac{1}{2} {\rm Tr} \; G_{\mu\nu}G^{\mu\nu}
 - \frac{1}{4} F_{\mu\nu}F^{\mu\nu} + {\cdots},
\label{QCDL}
\end{multline}
where $\gamma_\mu$ and $\tau_i$ are the Dirac and Pauli matrices, and 
$G_{\mu\nu}$ and $F_{\mu\nu}$ are the gauge and photon field strengths.
{Neglecting the ``$\cdots$'', which include for example weak interactions,}
the only parameters in QCD are the quark 
masses and electromagnetic charges.  We can express quark masses in terms of the
common mass $\bar{m} = (m_u+m_d)/2 \sim 5~\MeV$ and of the relative mass 
splitting $\varepsilon=(m_d-m_u)/(m_u+m_d)\sim 1/3$.  The quark charges are
fractions $Q={\text{diag}}(2/3, -1/3)$ of the proton charge 
$e=\sqrt{4\pi\alpha}\sim 1/3$.

Below $\MQCD$, QCD is best represented as a theory of colorless hadrons, where 
$SU(3)_{\mathrm{c}}$ is realized trivially.  An important role is played by
pions, which arise as pseudo-Goldstone bosons from the spontaneous breaking of 
approximate $SU(2)_{\mathrm{L}}\times SU(2)_{\mathrm{R}}$ chiral symmetry down 
to its vector subgroup $SU(2)_{\mathrm{V}}$ of isospin.  In the chiral limit
$\bar m=0$, $\varepsilon=0$ and $e=0$, chiral symmetry is exact, and pions are
massless and interact purely derivatively.  Away from the chiral limit, the
common quark mass breaks chiral symmetry explicitly and leads to a nonzero
common pion mass $m_\pi \simeq 140~\MeV$ and nonderivative pion interactions.
The QCD interactions associated with $\varepsilon$ and $e$ further break
isospin, and appear in relatively small quantities such as the pion mass
splitting $\delta m_\pi^2=m_{\pi^\pm}^2-m_{\pi^0}^2\simeq (36~\MeV)^2$ and the
neutron-proton mass difference $\delta m_N=m_n - m_p\simeq 1.3~\MeV$.

A more complete understanding of the low-energy consequences of QCD can be
achieved if we consider alternative realities where $\bar m$, $\varepsilon$,
and $e$ are varied from their real-world values.  So far,
{LQCD}
simulations of nuclear quantities have been carried out in the isospin-symmetry
limit, where $\varepsilon=0$ and $e=0$.  The only remaining QCD parameter, $\bar
m$, can be traded for the pion mass $m_\pi$.  Because the signal-to-noise ratio
for $A$-nucleon correlation functions at large time $t$ scales as
$\exp{[-A(m_N-3m_\pi/2)t]}$~\cite{Lepage:1989hd,Beane:2010em}, current
simulations are limited to unphysically large $m_\pi$ and to small $A$.
Although one can expect future simulations at smaller pion masses and more
nucleons, it is more efficient to switch to an EFT description suited to the
large distances involved in nuclear physics.

\begin{figure}[tb]
\centering
\includegraphics[clip,width=0.90\columnwidth]{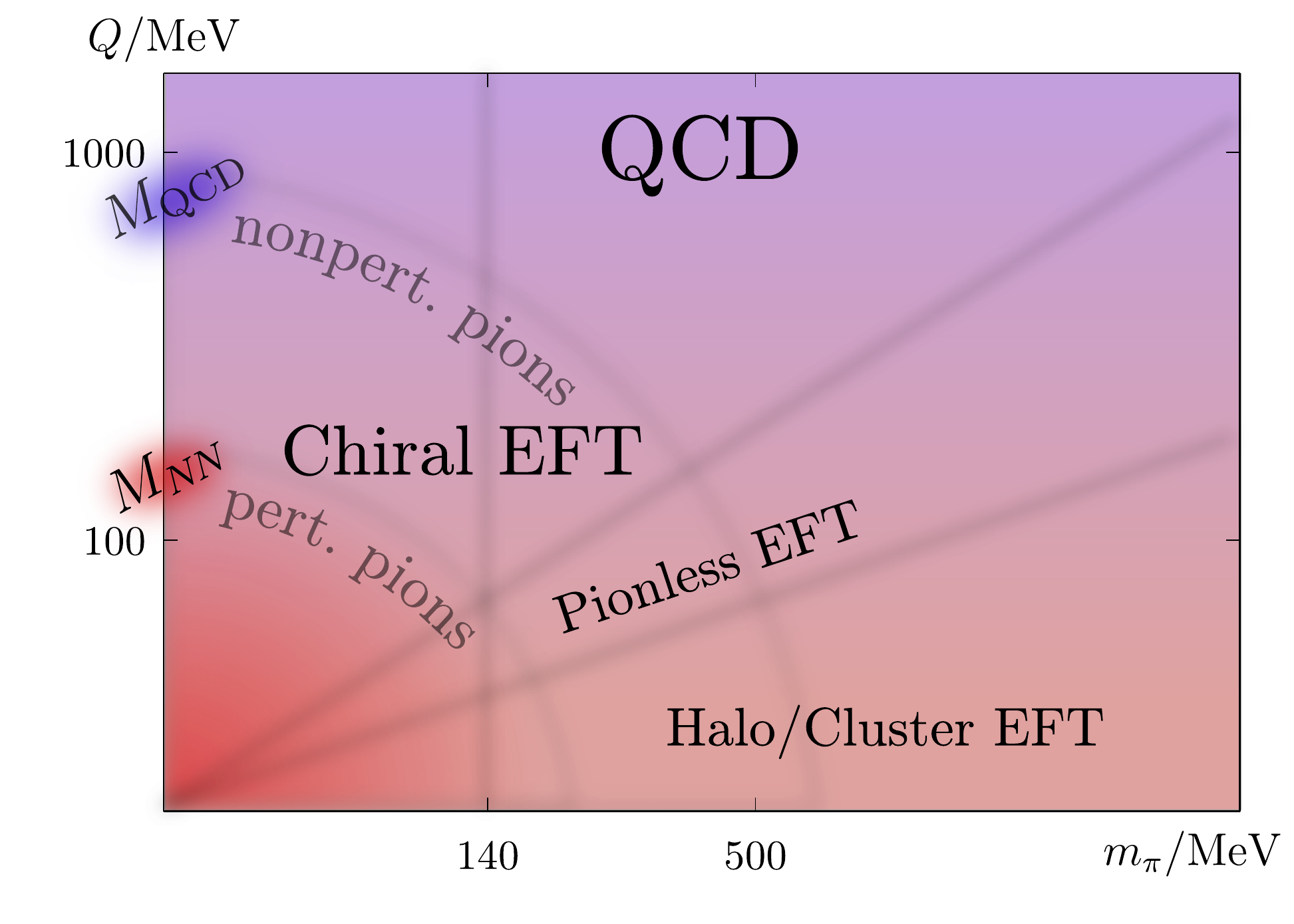}
\caption{Landscape of nuclear EFTs in the plane of typical momentum $Q$ and pion
 mass $m_\pi$.}
\label{fig:EFTscape}
\end{figure}

Years of experience suggest that nuclei can be seen as bound states or 
resonances made out of nucleons, or perhaps clusters of nucleons.  The 
choice of degrees of freedom determines the range of validity $\Mhi$ of the 
respective EFT.  Because isospin violation is a relatively small effect for most
nuclear dynamics (more so for light nuclei), we can classify nuclear EFTs by 
their regions of applicability according to typical momentum
and pion mass,
see Fig.~\ref{fig:EFTscape}.  A possible estimate of the typical binding
momentum, where each nucleon contributes equally to the binding energy $B_A$, is
$Q_A\sim \sqrt{2m_N B_A/A}$.  {Nuclear saturation for large $A$ leads, at 
physical pion mass,
to a constant $B_A/A \sim 10$ MeV and nuclear radii $R_A\sim R_0 A^{1/3}$, 
where $R_0\sim 1.2~\fm$.}  Numerically, $Q_A \sim
R_0^{-1}$ is not very different from $m_\pi$, and it has been assumed
that Chiral EFT is best suited for typical nuclei.  (In fact, we will see in
Sec.~\ref {sec:chiral} how $B_A/A \sim 10$ MeV arises naturally within Chiral
EFT.)  At sufficiently small $Q$ and $m_\pi$ (\ie, below a scale 
{$\MNN\sim270~\MeV$ at the physical pion mass, see Eq.~\eqref{OPEscale} for a
precise definition}), one expects pions to be perturbative.  As $Q$
increases  at fixed $m_\pi$, chiral-symmetric pion interactions become
nonperturbative (for $A\ge 2$), and as $Q$ increases further the EFT  eventually
ceases to converge.  As $m_\pi$ increases at fixed $Q$, chiral-symmetry breaking
becomes more important and again the Chiral EFT expansion eventually fails.
We expect that $\Mhi \sim \MQCD$, but the exact breakdown values of $Q$ and
$m_\pi$ are not well known.  It seems that for $A=0$, for example, Chiral EFT
(in the form of ChPT) has $\Mhi\le 500~\MeV$~\cite{Durr:2014oba}.

Light nuclei are weakly bound and radii scale differently than in the 
saturation regime.  Pions can be treated as short-range interactions
and in Pionless EFT we expect $\Mhi \sim m_\pi$ at all $m_\pi$, including values
beyond the breakdown of Chiral EFT such as in LQCD simulations {to date}.  
For $Q$ smaller
than the inverse radius of a nucleus, the nucleus itself can be treated as an
elementary particle in more complex systems where it appears as a sub-unit.
In the Halo/Cluster EFT relevant for clusterized nuclei, $\Mhi \sim R_c^{-1}$, 
the {inverse} cluster radius.  
Pionless and Halo/Cluster EFTs carry the information of 
QCD to the large distances of nuclear dynamics near the driplines.

\subsection{The way of EFT}
\label{subsec:WayofEFT}

How does one ensure that a nuclear EFT reproduces QCD in the appropriate energy
domain?  Once degrees of freedom have been selected according to the energies
of interest, one constructs the most general Lagrangian $\mathcal{L}$ 
involving the corresponding set of fields $\{\psi\}$, which is constrained only
by the 
QCD symmetries,
\begin{equation}
 \LL = \sum_i c_i(\Mlo,\Mhi,\Lambda) \, O_i(\{\psi\}),
\label{Lexp}
\end{equation}
where the $O_i(\{\psi\})$ are operators that involve fields at the same 
spacetime point but contain an arbitrary number of derivatives, and 
$c_i(\Mlo,\Mhi,\Lambda)$ are the low-energy constants (LECs).  Here $\Lambda$
denotes an arbitrary regulator parameter with dimension of mass.  With $\LL$, or
the corresponding Hamiltonian, the propagation and interaction of the low-energy
degrees of freedom can be calculated.  The procedure might be entirely
perturbative, as represented by Feynman diagrams with a finite number 
of loops, or partially nonperturbative, as obtained by an infinite sum
of Feynman diagrams or the solution of an equivalent integral or differential
equation such as, respectively, the Lippmann-Schwinger or the Schr\"odinger 
equation.  In either case, the interactions are singular, which requires
regularization.  When the calculation can be reduced to a finite number of 
loops, dimensional regularization can be employed, which introduces a 
renormalization scale $\mu$.  However, in nuclear physics we are most often 
faced with summing an infinite number of loops with overlapping momenta
which, with present techniques, can only be made finite by the introduction,
at either interaction vertices or propagators, of a momentum-regulator function
$f(p/\Lambda)$ such that $f(0)=1$ and
{$f(x\to \infty) =0$}.  Here $p$ refers to the
momentum of a nucleon, in which case the regulator is separable, or the
transferred momentum, when the regulator is nonseparable.  We can alternatively
look at position space, where the nonseparable regulator is local (\ie, a 
function of position only) whereas the separable regulator is nonlocal.

The goal is to construct the $T$ matrix for a low-energy process
as an expansion in $Q/\Mhi<1$, schematically,
{
\begin{multline}
T(Q) = \NN \sum_{\nu= 0}^\infty \, 
 \left(\frac{Q}{\Mhi}\right)^{\!\nu} \\
 \null \times 
 F^{(\nu)}\!\left(\frac{Q}{\Mlo}, \frac{Q}{\Lambda}; 
 \gamma^{(\nu)}\!\left(\frac\Mlo\Mhi,\frac\Lambda\Mhi\right)
 \right) \,,
\label{Texp}
\end{multline}
}
where $\NN$ is a normalization factor, the $F^{(\nu)}$ are functions generated 
by the dynamics of the $\{\psi\}$, the $\gamma^{(\nu)}=\OO(1)$ are 
dimensionless combinations of the $c_i$, and $\nu$ is a counting index.
``Power counting'' is the relation between $\nu$ and the interaction label $i$ 
in Eq.~\eqref{Lexp}.  While the form of the $O_i$ in the Lagrangian~\eqref{Lexp}
depends on the choice of fields, the expansion~\eqref{Texp} must not
{\cite{Chisholm:1961tha,Kamefuchi:1961sb}}.  Likewise,
observables obtained from Eq.~(\ref{Texp}) must not depend on the arbitrary
regularization procedure---renormalization-group (RG) invariance.

Once the expansion~(\ref{Texp}) has been achieved, one can truncate the sum at 
a given $\nu=\mathcal{V}$ with a small error,
\begin{equation}
 T(Q\sim \Mlo)=T^{(\mathcal{V})}(Q,\Lambda)\left[1
 + \OO\left(\frac{Q^{{\mathcal{V}}+1}}{\Mhi^{{\mathcal{V}}+1}}\right)\right] \,.
\label{Ttrunc}
\end{equation}
Before renormalization, non-negative powers of $\Lambda$ can appear, which 
originate in the short-distance part of loops.  The uncertainty principle 
ensures that such contributions cannot be separated from that of LECs.
Renormalization is the procedure that fixes the cutoff dependence of the LECs
so that the truncated amplitude $T^{(\mathcal{V})}(Q,\Lambda)$ satisfies 
approximate RG invariance,
\begin{equation}
 \frac{\Lambda}{T^{(\mathcal{V})}(Q,\Lambda)}
 \frac{\dd T^{(\mathcal{V})}(Q,\Lambda)}{\dd\Lambda}
 =  \OO\left(\frac{Q^{{\mathcal{V}}+1}}{\Mhi^{\mathcal{V}}\Lambda}\right) \,.
\label{TRGtrunc}
\end{equation}
This condition ensures the error introduced by the arbitrary regularization 
procedure is no larger than the $Q/\Mhi$ error stemming from the neglect of
higher-order terms in Eq.~\eqref{Ttrunc}, as long as $\Lambda \simge \Mhi$.
In this ``modern view'' of renormalization, there is no need to take the 
$\Lambda \to \infty$ limit~\cite{Lepage:1989hf}.  However, while in analytical 
calculations Eq.~\eqref{TRGtrunc} can be verified explicitly, in numerical 
calculations varying the regulator parameter widely above the breakdown scale
is usually the only tool available to check RG invariance.  In contrast, 
$\Lambda < \Mhi$ generates relatively large errors from the regularization
procedure.  Failure to satisfy Eq.~\eqref{TRGtrunc} altogether means
uncontrolled sensitivity to short-distance physics: results depend on the value
of $\Lambda$ and on the choice of the regulator function $f(x)$, which acquires
the status of a physical, model-dependent ``form factor.''

After renormalization, when the contribution from momenta of the order
of the large cutoff have been removed, the dominant terms in loop integrals come
from momenta of $\OO(Q)$.  Counting powers of $Q$ in individual contributions
to Eq.~\eqref{Texp} is similar to determining the superficial degree of
divergence of diagrams.  There is, in general, also residual $\Lambda$
dependence (Eq.~\eqref{TRGtrunc}) which can be absorbed in the LECs of
higher-derivative interactions.
{Since shuffling short-range physics between loops and LECs does
not change observables, the finite part of {an} LEC is expected to be set
by the replacement $\Lambda\to \Mhi$ (see, for example,
\cite{Veltman:1980mj})\footnote{%
{\textcite{Burgess:2013ara} offers a clear discussion in the specific 
context of the cosmological constant.}
}, which then places an upper bound on the order these interactions appear 
{at}.
The exception is when a symmetry suppresses the corresponding 
interaction~\cite{tHooft:1979rat}.
``Naturalness'' assumes that all terms in the effective
Lagrangian (respecting the relevant symmetries) have dimensionless 
coefficients
of $\OO(1)$ when the appropriate powers of $\Mlo$ and $\Mhi$ are factored out.
Renormalization is thus a powerful tool to estimate sizes of the 
LECs.}

This framework is a generalization of the ancient requirement of 
renormalizability by a finite set of parameters.  If 
{all interactions needed for Eq.~\eqref{TRGtrunc}
are present at each order,}
the resulting $S$ matrix incorporates the relations among QCD 
$S$-matrix elements demanded by symmetries, with no other assumption than an 
expansion in $Q/\Mhi<1$.  
Every low-energy observable depends on a finite 
number of LECs at leading order (LO), where $\mathcal{V}=0$, a few more at 
next-to-leading order (NLO), where $\mathcal{V}=1$, {\textit{etc}.}  Once 
the LECs are determined from a finite number of data, all other observables can 
be pre- or postdicted with a controlled error.  Traditionally the input data 
have been experimental, but LQCD results can now be used
instead~\cite{Barnea:2013uqa,Beane:2015yha,Kirscher:2015yda}.

One of the virtues of the model independence encoded in Eq.~\eqref{TRGtrunc}
is that it provides an {\it a priori} estimate of theoretical errors.  At the 
simplest level errors can be estimated from the higher-order terms in 
Eq.~\eqref{Ttrunc} with a guess for $\Mhi$.  A lower bound on the theoretical 
error is provided by cutoff variation from $\Mhi$ to much higher values.
The breakdown scale $\Mhi$ itself can be inferred comparing the energy 
dependence at various orders with data~\cite{Lepage:1997cs}.  Reliance on data 
can be minimized by using instead EFT results at different
cutoffs~\cite{Griesshammer:2015osb}.  Up to now both data fitting and
propagation of errors have employed standard statistical analyses previously 
used for models.  However, these methods can lead to biases because they are 
not particularly well suited to the \apriori EFT error estimates, which 
typically increase with $Q$, while experimental data are
{sometimes} more 
precise at higher $Q$.  A comprehensive theory of EFT error analysis based on 
Bayesian methods is currently being 
developed~\cite{Schindler:2008fh,Furnstahl:2014xsa,Furnstahl:2015rha,
Wesolowski:2015fqa}
with the promise of becoming the standard in the field.

\subsection{Nuclear EFTs}
\label{subsec:NuclearEFTS}

The implementation of these ideas in nuclear physics has posed some unexpected 
challenges.  They can be traced to the fact that at LO some interactions need 
to be fully iterated---or, equivalently, a dynamical equation should be solved 
exactly---in order to produce the bound states and resonances that we refer to 
as nuclei.  

Nuclear EFTs typically include fields for the nucleon or clusters of nucleons.
These particles have masses of $\OO(\MQCD)$, and the expansion~(\ref{Texp})
includes a $Q/m_N$ expansion around the nonrelativistic limit.  Creation of 
{virtual} heavy particle-antiparticle pairs takes place at small distances 
$r\simle 1/(2m_N)$ and its effects can be absorbed in the LECs.  As a 
consequence, a process involving $A$ heavy particles is not affected by 
interactions in Eq.~\eqref{Lexp} involving more than $2A$ fields associated with
these heavy particles.  The simplest way to incorporate the fact that the
(large) particle rest energy does not play any role is to employ a ``heavy
field'' from which the trivial evolution factor due to the rest energy is 
removed~\cite{Jenkins:1990jv}.  Lorentz invariance for these fields is encoded 
in ``reparametrization invariance''~\cite{Luke:1992cs}.  Kinetic terms reduce
to the standard nonrelativistic form {that respects Galilean invariance}, 
with relativistic corrections {suppressed by inverse powers of $m_N$}
appearing at higher orders.

There is a crucial difference between $A= 1$ and $A\ge 2$ 
processes.  The former processes involve also light particles (\eg, photons) in 
initial and final states with momenta $Q\sim \Mlo$.  They deposit on the 
nucleon an energy of $\OO(Q)$ which is larger than the recoil of 
$\OO(Q^2/(2m_N))$, so that the nucleon is essentially static---deviation from 
the static limit can be treated as a perturbation.  Intermediate states differ 
in energy from the initial state by an amount of $\OO(Q)$.  In contrast, there 
are Feynman diagrams for the $T$ matrix of an $A\ge 2$ process---whether it 
involves external probes or not---that include intermediate states which differ 
in energy from the initial state only by a small difference in nucleon kinetic 
energies of $\OO(Q^2/(2m_N))$.  In these ``reducible'' diagrams nucleons are 
not static, and there is an infrared (IR) enhancement relative to intermediate 
states for $A= 1$ processes~\cite{Weinberg:1991um}.  Nucleon recoil cannot be 
treated perturbatively, although relativistic corrections remain small.

The ``full'' nuclear potential $V$ is defined as the sum of irreducible diagrams
for a process involving $A$ nucleons in initial and final states.  The full $T$ 
matrix~\eqref{Texp} is obtained by sewing potential subdiagrams with nucleon 
lines representing the free $A$-body Green's function $G$.  This gives rise to
the Lippmann-Schwinger equation, schematically
\begin{equation}
 T = V + \int V G T = V + \int V G V + \cdots \,,
\label{LS}
\end{equation}
or alternatively to the Schr\"odinger equation and its many-body relatives.  The
full potential so defined involves all $A$ bodies but it includes components
with $1\le C\le A-1$ separately connected pieces.  Frequently the potential is
thought of as one of these connected pieces.  One thus defines the ``$A$-nucleon
($AN$) potential'' as the sum of diagrams with $C=1$ in the $A$-nucleon system.
For $A=2$ all diagrams in the nuclear potential are connected ($C=1$), but
starting at $A=3$ multiply connected diagrams appear, \ie, the full potential is
made up of a sum of fewer-body potentials.  Diagrams with $C=A-1$ are made out
of the $2N$ potential and $A-2$ disconnected nucleon lines.  Diagrams in the
full potential that have $1< C <A-1$ are made of combinations of lower-$A$
potentials and disconnected nucleon lines.

In contrast to phenomenological models, all mesons with masses $\simge \MQCD$
and nucleon excitations heavier than the nucleon by the same amount can be
integrated out because their effects can be captured by the LECs.  As we are
going to see in Sec.~\ref{sec:pionless}, in Pionless EFT the potential consists
purely of contact interactions, while in Chiral EFT pion exchanges are present
as well (Sec.~\ref{sec:chiral}).  In either case, the potential involves small
transfers of energy $\OO(Q^2/(2m_N))$, and the total exchanged four-momentum is
close to the total transferred three-momentum.  Dependence on the latter
translates into a function of the position in coordinate space---the potential
is local.  Meanwhile, dependence on other nucleon momenta leads to derivatives
with respect to position, \ie, the momentum operator in quantum mechanics---the
potential becomes nonlocal.  We expect to be able to expand the potential
in momentum space analogously to Eq.~\eqref{Texp},
\begin{multline}
{
 V(Q,\Lambda) = \sum_{\mu= 0}^\infty V^{(\mu)}(Q,\Lambda)
 = \tilde {\mathcal{N}} 
 \sum_{\mu= 0}^\infty \, 
 \left(\frac{Q}{\Mhi}\right)^{\!\mu}} \\
 {\null\times
 \tilde{F}^{(\mu)}\!\left(\frac{Q}{\Mlo};\frac{Q}{\Lambda};
 \tilde{\gamma}^{(\mu)}\!\left(\frac\Mlo\Mhi,\frac\Lambda\Mhi\right)  
\right) \,,}
\label{Vexp}
\end{multline}
where $\tilde{\mathcal{N}}$ is a normalization factor,\footnote{Note that, in 
the units we use, the momentum-space potential, like the $T$ matrix, has mass 
dimension $-2$.  Its Fourier transform, which involves three powers of momentum,
has mass dimension $+1$, as it should.} the $\tilde{F}^{(\mu)}$ are functions 
obtained from irreducible diagrams, the $\tilde \gamma^{(\mu)}=\OO(1)$ are 
dimensionless combinations of the $c_i$, and $\mu$ is a counting index for the 
potential.

When the nucleus is disturbed by low-momentum external probes (photons, leptons,
perhaps pions), similar considerations apply.  One can define nuclear currents
(or reaction kernels) as the sum of irreducible diagrams to which the probes are
attached.  Again, currents involve all $A$ nucleons but include disconnected
diagrams.  A subtlety is that a probe can deposit an energy $\OO(Q)$ on a
nucleon line, and thus there can be purely nucleonic intermediate states in
irreducible diagrams.  Observables come from the sandwich of currents between
wavefunctions of the initial and final states.  Currents have an expansion
similar to Eq.~\eqref{Vexp}.

The nuclear potential and associated currents can always be defined as such 
intermediate quantities between $\LL$ and $T$.  We have reduced the EFT to a
quantum-mechanical problem, but one in which the form of the potential and
currents is determined.  This is a distinct improvement over a purely
phenomenological approach, particularly in what concerns the bewildering variety
of many-body potentials and currents one can construct.
{This feature is one of the major reasons for
the dominant role nuclear EFTs play nowadays in the nuclear theory community.}

However, one should keep in mind that the potential and currents are not
directly observable.  There are important differences between Eqs.~\eqref{Vexp}
and~\eqref{Texp}:
\begin{itemize}
\item
The potential does not need to obey an equation such as~\eqref{TRGtrunc}. 
EFT potentials involve terms that are singular and often attractive, in the 
sense of diverging faster than $-1/(4m_Nr^2)$ as the relative position 
$r\to 0$.  The potential would generate strong regulator dependence in 
Eq.~\eqref{LS} integrals if it did not itself depend strongly on 
$\Lambda${, see, \eg,} {a pedagogical discussion by
\textcite{Lepage:1997cs}.}
\item
Since after renormalization $\Lambda$ disappears from $T$ (apart from 
arbitrarily small terms),
\begin{equation}
 \int V G V\sim \frac{Q^3}{4\pi} \frac{m_N}{Q^2} V^2\sim \frac{m_NQ V}{4\pi}V
\label{VGV}
\end{equation}
and the expansion~\eqref{LS} is in the dimensionless ratio $m_N Q
V/(4\pi)$~\cite{Bedaque:2002mn}.  For $Q\simge 4\pi/(m_N V^{(0)})$, $F^{(0)}$ in
Eq.~\eqref{Texp} stems from an infinite iteration of the LO potential
$V^{(0)}$.  This is good, because nuclear bound states and resonances, as poles
of $T$ matrices, can only be obtained from a nonperturbative LO.
\item
Equations~\eqref{Vexp} and~\eqref{LS} do \emph{not} imply that all terms in 
$V$ should be treated on the same footing.  One cannot immediately identify
$\mu$ with $\nu$ because a term in $V$ contributes to various orders in the $T$ 
matrix.  Higher-order $F^{(\nu>0)}$ can be obtained from $V^{(\mu>0)}$ in a
distorted-wave Born expansion: $F^{(1)}$ from a single insertion of $V^{(1)}$,
$F^{(2)}$ from a single insertion of $V^{(2)}$ or two insertions of $V^{(1)}$, 
and so on.  Treating the potential truncated at a subleading order 
exactly---\ie, treating it as a phenomenological potential---is in general 
not correct from a renormalization point of view.  In an expansion in $Q$,
the potential gets more and more singular with increasing order.  Resumming a 
partial subset of higher-order terms will in general not include all the LECs
needed for proper renormalization.\footnote{{%
An example of resummation of higher-order interactions is found in lattice
implementations of Nonrelativistic QCD (NRQCD)~\cite{Thacker:1990bm}.
In Heavy Quark Effective Theory (HQET) all $Q/m_Q$ corrections in the heavy 
quark mass $m_Q$ are treated perturbatively, and lattice simulations have
a continuum limit~\cite{Sommer:2010ic}.  For NRQCD, lattice practice is to treat
exactly not only heavy quark recoil but also the associated, subleading gluon
interactions.  Thus, only for relatively large values of the lattice spacing $a$
do observables look like they might converge, before $1/a$-type effects take 
over.  There are also situations where one can resum higher-order interactions
without introducing essential regulator dependence. An example is given by
\textcite{Lepage:1997cs}.}}
\end{itemize}

The age-long challenge in nuclear physics has been to achieve RG invariance 
when some interactions are nonperturbative and yet some others can be treated 
as small.  In an EFT, that translates into the nontrivial task of developing a 
power counting that guarantees  Eqs.~\eqref{Ttrunc} and~\eqref{TRGtrunc}.
In a purely perturbative context the cutoff dependence of loops can be obtained 
analytically.  Assuming naturalness and looking at individual loop diagrams, 
a simple rule has been devised for the size of the LECs needed for perturbative
renormalization~\cite{Manohar:1983md,Georgi:1986kr}.  This ``na\"ive dimensional
analysis'' (NDA) states that, for an operator $O_i$ in Eq.~\eqref{Lexp} with
canonical dimension $D_i$ involving $N_i$ fields $\psi$,
\begin{equation}
 c_i = \OO\left(\frac{(4\pi)^{N_i-2}}{\MQCD^{D_i-4}}\,
 c_{i {\text{red}}}\right) \,,
\label{NDA}
\end{equation}
where the dimensionless ``reduced'' LEC $c_{i \text{red}}$ is of the order 
of the combination of {reduced} QCD parameters that give rise to it.  
Examples for Chiral
Perturbation Theory are given in Sec.~\ref{sec:chiral}.  It is, however, not
immediately obvious that NDA applies to LECs of operators involving four or
more nucleon fields {subject to nonperturbative renormalization, \ie, 
which} {are renormalized once LO interactions are resummed.}
In
fact, as we are going to see below, cutoff variations in the Lippmann-Schwinger
equation~\eqref{LS} require significant departures from NDA for contact 
interactions among nucleons.  These departures were first understood within
Pionless EFT.  Its simplicity makes Pionless EFT the poster-child for nuclear 
EFT, and we {therefore} make it the start of this review.

\section{Pionless EFT} \label{sec:pionless}
\subsection{Motivation}
\label{sec:pionlessmotivation}

At very low energies---\ie, for momenta $Q\ll m_\pi$---few-nucleon systems
are not sensitive to the details associated with pion (or other meson) exchange.
This fact makes it possible to describe such systems with short-range
interactions alone (\ie interactions of finite range or falling off at least as
an exponential {in} the interparticle distance), an approach dating
back to Bethe and his effective range expansion (ERE) for nucleon-nucleon
($N\!N$) scattering~\cite{Bethe:1949yr}---see
also {related work
by~\textcite{Bethe:1935aa,Bethe:1935ab,Fermi:1936XX,Schwinger:1947zz,
Jackson:1950zz}}.
Casting this basic idea into a modern
systematic framework leads directly to what has become known as Pionless EFT.

Historically, Pionless EFT emerged out of the effort to understand the
renormalization of EFTs where a certain class of interactions need to be treated
nonperturbatively.  It had been shown by \textcite{Kaplan:1996xu},
\textcite{Phillips:1997xu}, and \textcite{Beane:1997pk} that the original
prescription~\cite{Weinberg:1990rz,Weinberg:1991um} to extend Chiral
Perturbation Theory to few-nucleon systems (discussed in Sec.~\ref{sec:chiral})
could not be implemented satisfying RG invariance.  It turned out that there is
a surprisingly rich structure of phenomena in the low-energy regime where
explicit pion exchange cannot be resolved.

\begin{figure}[tb]
\centering
\includegraphics[clip,width=0.75\columnwidth]{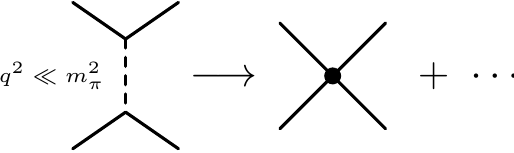}
\caption{Reduction of pion exchange (dashed line) to a series of contact
interactions between nucleons (solid lines) for $q^2\ll \Mpi^2$.}
\label{fig:PionsOut}
\end{figure}

Formally, the pion can be regarded as ``integrated out'' if all other dynamical
scales are much smaller than the pion mass.  Consider, for example, the Yukawa
potential corresponding to one-pion exchange:
\begin{equation}
 {\mbraket{\veck'}{V_{2N,\pi}}{\veck}
 \propto \frac{1}{\vecq^2 + \Mpi^2}
 \;,\;
 \vecq = \veck' - \veck \,,}
\label{eq:V-pi}
\end{equation}
where $\veck$ and $\veck'$ are incoming and outgoing momenta of two scattered
nucleons (in their center-of-mass frame).  If these are both small compared to
$\Mpi$, Eq.~\eqref{eq:V-pi} can be expanded in $\vecq^2$, with the leading term
being just a constant and the following terms coming with ever higher powers of
$\vecq^2$.  This shrinking of the original interaction to a point is illustrated
in Fig.~\ref{fig:PionsOut}.  Fourier-transforming into configuration space one
obtains a series of
{delta functions} with a growing number of derivatives.  In
Chiral EFT, which includes pions, analogous contact interactions
{represent the exchange} of heavier mesons.  Integrating out
pions to arrive at Pionless EFT means merging unresolved pion exchange with
these operators.  It should be noted, however, that Chiral EFT is based on an
expansion around a vanishing pion mass, whereas Pionless EFT treats $\Mpi$ as a
large scale.  As such, these two EFTs are very different{---in particular,
the respective LECs cannot in general be related by perturbative
matching---}but they are both well-defined low-energy limits of QCD.

In practice, one does not have to derive Pionless EFT from a more fundamental
EFT by integrating out explicit pions.  Instead, one can just follow the EFT
paradigm and write down an effective Lagrangian{, Eq.~\eqref{Lexp},} with
all contact interactions
between nucleons that are allowed by symmetry.  This restriction means that one
requires invariance under
{``small'' Lorentz boosts} (Galilean
boosts plus systematic relativistic corrections), {rotations, isospin, and
discrete symmetries like parity and time reversal}, the systematic breaking of
which can also be accounted for.  The same EFT with other particles substituted
for nucleons can describe different systems where the important dynamics takes
place at distances beyond the range of the force.  Some of these systems are
discussed in Secs.~\ref{sec:halo} and~\ref{sec:other}.  In particular{,}
Pionless EFT captures the universal aspects of Efimov physics
\cite{Braaten:2004rn}.

In this section we discuss the basic features and formalism of Pionless EFT,
first in the context of two-body systems (Sec.~\ref{sec:pionlessweaklyboundS})
and later for a larger number of particles
(Sec.~\ref{sec:pionlesslightnuclei}). Some of the outstanding issues are raised
in Sec.~\ref{sec:pionlesssummary}.

\subsection{Weakly bound $S$-wave systems}
\label{sec:pionlessweaklyboundS}

Two very-low-energy particles, represented by a field $\psi$, can be described
by an effective Lagrangian
\begin{multline}
 \LL
 = \psi^\dagger \left(\ii\partial_0 + \frac{\Laplace}{2\MN}\right)\psi
 - \frac{C_0}2 (\psi^\dagger\psi)^2 \\
 + \frac{C_2}{16} \big[(\psi\psi)^\dagger\,(\psi\,\vNablaLR*^2
   \psi)  + \hc\big]
 + \cdots \,,
\label{eq:L-NN-simple}
\end{multline}
where $\vNablaLR*=\vNablaL-\vNablaR$ is the Galilei-invariant derivative and
$\hc$ denotes the Hermitian conjugate.  The ``$\cdots$'' represent local
operators with other combinations of derivatives, including relativistic
corrections.  Here we have adopted the notation of~\textcite{Hammer:2000xg},
but various forms for the Lagrangian---differing by prefactors absorbed in the
low-energy constants ($C_0$, $C_2$, \etc) or choice of equivalent
operators---exist in the literature.  One can treat the two $N\!N$ $S$-wave
channels (\ThreeSOne or \OneSNot, in the spectroscopic notation $^{2s+1}l_j$
where $l$, $s$, and $j$ denote respectively orbital angular momentum, spin, and
total angular momentum) simultaneously using a nucleon field $N$ that is a
doublet in spin and isospin space.  We will come back to this after discussing
the general features of the two-body sector on the basis of
Eq.~\eqref{eq:L-NN-simple}.

\subsubsection{Two-body scattering amplitude}
\label{sec:pionlessscattamp}

To fill the theory described by the effective Lagrangian~\eqref{eq:L-NN-simple}
with physical meaning, we need to equip it with a power counting.  We seek an
expansion of the form~\eqref{Texp} where $\LamNoPi$ is expected to be set by
the pion mass $\Mpi$, since pion exchange has been integrated out.  In
particular, we want to reproduce the ERE \cite{Bethe:1949yr} for the on-shell
$N\!N$ scattering amplitude:
\begin{subalign}[eq:T-kcot]
  T(k,\cos\theta) &= {-}\frac{4\pi}{\MN}\sum_l\frac{(2l+1)P_l(\cos\theta)}
  {k\cot\delta_l(k)-\ii k} \,,
 \label{eq:T-kcot-a} \\
 k^{2l+1}\cot\delta_l(k) &= {-}\frac{1}{a_l} + \frac{r_l}{2}k^2 + \OO(k^4) \,,
 \label{eq:T-kcot-b}
\end{subalign}
with a Legendre polynomial $P_l$, {the scattering angle $\theta$
and energy} $E=k^2/\MN$ in the center-of-mass
frame, and where $\delta_l(k)$ is the scattering phase shift in the $l$-th
partial wave, while $a_l$ and $r_l$ denote the corresponding scattering length
and effective range, respectively.  Here, we focus on $S$ waves with $l=0$.
Higher partial waves will be discussed below.

In a ``natural'' scenario, the LECs $C_{2n}$ in Eq.~\eqref{eq:L-NN-simple} would
scale with inverse powers of their mass dimension, \eg, $C_0
\propto\LamNoPi^{-1}$.  (Note that an overall scaling with $1/\MN$ from the
nonrelativistic framework is shared by all terms in the effective Lagrangian.)
In this case, to lowest order $T$ would simply be given by the tree-level $C_0$
vertex, and we could identify $C_0=4\pi a_0/\MN$.  However, the low-energy
$N\!N$ system is {\it not} natural.  From the above relation for $C_0$ it is
immediately clear what this means here: the actual $N\!N$ scattering lengths
($a_{\OneSNot}\simeq {-}23.7~\fm$ and $a_{\ThreeSOne}\simeq 5.4~\fm$)
are large compared to the pion Compton wavelength
{$\Mpi^{-1}\simeq 1.4~\fm$},
so $C_0=4\pi a_0/\MN$ is incompatible with $C_0 \propto\LamNoPi^{-1}$
if one assumes $\LamNoPi\sim\Mpi$.  Turning the argument around, the
perturbative expansion in $C_0$ has a breakdown scale set by $1/a_0 \ll \Mpi$,
rendering it useful only for the description of extremely low-energy $N\!N$
scattering.

The physical reason for the rapid breakdown of the perturbative expansion is
that the large $N\!N$ $S$-wave scattering lengths correspond to low-energy
(``shallow'') bound states (virtual, in the case of the \OneSNot channel).  For
example, it is well known that the deuteron binding momentum
$\gamd = \sqrt{\mathstrut\MN\Bd} \simeq 45.7~\MeV$ is given to about 30\%
accuracy by $1/a_{\ThreeSOne}$.  These states directly correspond to poles of
the amplitude $T$ (located on the imaginary axis of the complex $k$ plane,
or on the negative energy axis in the first or second Riemann sheet).  It is
clear that a (Taylor) expansion of $T$ in $k^2$ will only converge up to the
nearest pole \emph{in any direction in the complex plane}.  Thus, the presence
of the shallow $N\!N$ bound states limits the range for a perturbative
description of $N\!N$ scattering.

\begin{figure}[tb]
\centering
\includegraphics[clip,width=0.9\columnwidth]{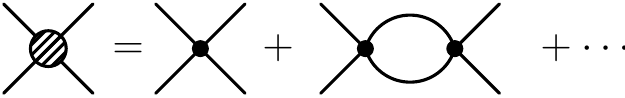}
\caption{Bubble chain for a generic $S$-wave $N\!N$ scattering amplitude
at LO from the $C_0$ interaction (solid circle).}
\label{fig:BubbleChain}
\end{figure}

{A} nonperturbative treatment is necessary to generate poles in $T$,
since a finite sum of polynomials can never have a pole.  As pointed out
by \textcite{Weinberg:1991um}, this can be achieved by ``resumming'' the $C_0$
interaction, \ie, by writing the LO amplitude as the tree-level $C_0$ diagram
plus any number of $C_0$ vertices with intermediate propagation,
as shown in Fig.~\ref{fig:BubbleChain} (see also {a related analysis
by~\textcite{Luke:1996hj}}).  The result for a single generic $N\!N$ channel is
\begin{align}
 T^{(0)} &= {C_0 + C_0\,I_0(k)\,C_0 + C_0\,I_0(k)\,C_0\,I_0(k)\,C_0 + \cdots}
 \nonumber \\
 &= {\left[C_0^{-1} - I_0(k)\right]^{-1}} \,,
\label{eq:T-C0}
\end{align}
where $I_0$ is the two-body ``bubble integral,'' discussed in more detail
below.  Having $C_0$ now in the denominator means that it can be adjusted to
give a pole at the desired position.

\subsubsection{Power counting}
\label{sec:PowerCounting}

Of course, the power counting of the theory should be such that it actually
\emph{mandates} this procedure.  The small inverse $N\!N$ scattering lengths
introduce a genuine new low-momentum scale $\Mlo$ (or large length $\Mlo^{-1}$).
Typically, this is referred to as ``fine tuning'' because the existence of this
scale---at odds with the perfectly natural assumption that pion exchange should
set the lowest energy scale---implies that different contributions from quarks
and gluons have to combine in just the right way to produce this scenario (see
Sec.~\ref{subsec:qcdconn}).

Equation \eqref{eq:T-C0} is nothing but Eq.~\eqref{LS} for a two-body potential
$C_0$, which, from the discussion above, is enhanced by a factor $\Mlo^{-1}$.
The loops connecting two insertions of the potential contain nucleon
propagators, which from Eq.~\eqref{eq:L-NN-simple} we read off to be
\begin{equation}
 \ii D_N(p_0,\vecp)
 = \ii \left(p_0-\dfrac{\vecp^2}{2\MN}+\cdots + \ii\eps\right)^{-1} \,.
\end{equation}
Here, $p_0$ and $\vecp$ are the energy and momentum associated with a nucleon
line in Fig.~\ref{fig:BubbleChain}.  If a total momentum $k \sim Q$ runs
through the diagram, we see that, after regularization effects have been
removed by renormalization, the dominant contribution in a loop integral
$\dd q_0\dd^3 q$ will come from the region where $q\sim Q$.  Hence, keeping
in mind that $q_0$ is a nonrelativistic kinetic energy $\sim q^2/\MN$, we count
\begin{subalign}[eq:PionlessPC]
 \text{nucleon propagator} &\sim m_N Q^{-2} \,,
 \label{nucprop} \\
 \text{(reducible) loop integral} &\sim (4\pi m_N)^{-1} Q^5 \,.
 \label{redloopint}
\end{subalign}
Equations \eqref{nucprop} and~\eqref{redloopint} lead directly to the
estimate~\eqref{VGV} and imply that the one-loop contribution in
Fig.~\ref{fig:BubbleChain} scales like the tree-level one times a factor
$Q/\aleph$.  In fact, each additional dressing by one loop with a $C_0$ vertex
contributes such a factor.  Hence,
{in the regime} where $Q\sim\aleph \ll \LamNoPi$ each such {diagram} is
equally
important, and they all have to be summed up to get the LO amplitude
nonperturbatively.  On the other hand, for $Q\ll\aleph$ one can still use
a perturbative approach, so the counting here is able to capture both scenarios.

Operators with derivatives in the effective Lagrangian must contain inverse
powers of $\Mhi$ in order not to introduce additional low-energy poles in the
LO $T$ matrix.  They provide corrections to the $2N$ potential,
\begin{equation}
 V_{2N}(\vecp',\vecp) = C_{0} + C_{2} \left(\vecp'^2 + \vecp^2\right) + \cdots\,.
\label{eq:V-pp}
\end{equation}
Being suppressed, higher orders can be calculated in perturbation theory and
matched to an expansion of Eq. \eqref{eq:T-kcot-a}{,
\begin{equation}
 T = \frac{4\pi}{\MN}\frac{1}{1/a_0+\ii k}
 \left(1 + \frac{r_0k^2/2}{1/a_0+\ii k} + \cdots \right).
\label{eq:EREexp}
\end{equation}
The specific scaling with
$\Mhi$ can be inferred from this} and from regulator effects considered below.

\begin{figure}[tb]
\centering
\includegraphics[clip,width=0.85\columnwidth]{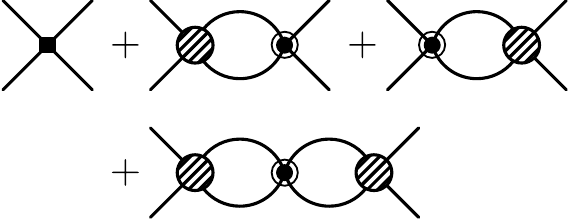}
\caption{NLO correction to the $N\!N$ scattering amplitude from the $C_2$
 interaction (circled circle).}
\label{fig:C2-corr}
\end{figure}

For example, the NLO amplitude $T^{(1)}$ is the result from inserting a single
$C_2$ vertex into each combination that can be formed with the LO amplitude
\cite{Bedaque:1997qi,vanKolck:1997ut,Kaplan:1998tg,Kaplan:1998we,%
Bedaque:1998mb,vanKolck:1998bw}, as shown in Fig.~\ref{fig:C2-corr}.  Matching
to the {$k^2$} coefficient in {Eq.~\eqref{eq:EREexp}}
shows that the $C_2$ contributions are related to the effective ranges.  Since
the values of the $N\!N$ $S$ waves
{are} $r_{\ThreeSOne}\simeq 1.75~\fm$ and $r_{\OneSNot}\simeq 2.7~\fm$,
and thus of the order $\Mpi^{-1}\sim\Mhi^{-1}$, we
conclude $C_2$ is indeed an NLO effect,
\begin{equation}
 \frac{C_2 Q^2}{C_0} \sim \frac{Q^2}{\Mlo\Mhi} \,.
\label{eq:scaling-re}
\end{equation}
For comparison, given that the $C_0$ term is a dimension-6 operator whereas the
one with $C_2$ is {dimension-8}, the na\"ive (natural) scaling is
$C_2 Q^2 / C_0 \sim (Q/\LamNoPi)^2$.  The additional low-energy enhancement
$\aleph$ also occurs in the scaling of the $C_2$ parameters.

\begin{figure}[tb]
\begin{center}
\includegraphics[width=0.9\columnwidth]{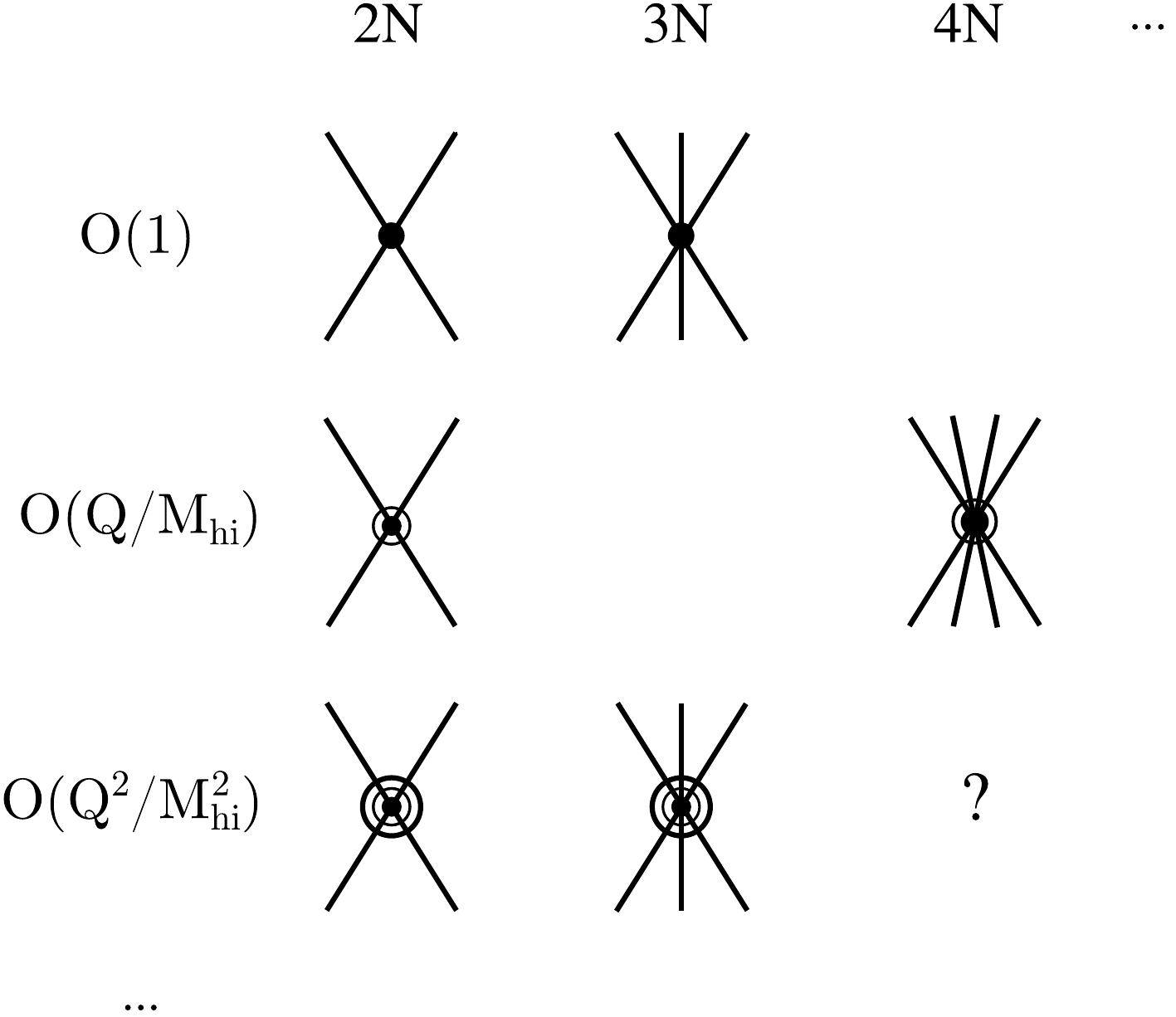}
\end{center}
\caption{Diagrams representing the $AN$ nuclear potential in Pionless EFT.  The
 order of the contributions is indicated as $\OO(Q^\mu/\Mhi^\mu)$, $\mu\geq0$,
 where $Q\sim\Mlo$ and $\Mhi\sim\mpi$, so that circles around the central
 solid circle denotes inverse powers of $\Mhi$.
}
\label{fig:pionlesspot}
\end{figure}

This procedure can be generalized to higher orders and other operators.  At
N$^2$LO we must consider two insertions of $C_2$ and one insertion of $C_4$; the
latter is determined entirely in terms of $r_0^2$, the shape parameter emerging
at N$^3$LO~\cite{Kaplan:1998tg,Kaplan:1998we,vanKolck:1998bw}.
{Generally, enhancements depend on the partial waves involved.}
The interactions contributing to such waves are operators in the ``$\cdots$'' of
Eq.~\eqref{eq:L-NN-simple} that make $T$ dependent on the scattering angle.
There is no enhancement for operators that contribute only to higher waves, as
long as there are no other low-energy poles as is the case in $N\!N$ scattering.
Thus, for example, a $P$-wave operator leading to a term $\propto
\veck'\cdot\veck$ appears first at N$^3$LO.  The enhancement is only partial for
operators that connect an $S$ wave to other waves.  The short-range tensor force
that connects $S$ and $D$ waves is present at N$^2$LO because it is enhanced by
one power of $\aleph^{-1}$~\cite{Chen:1999tn}.  The lowest orders in the
potential are shown schematically in Fig.~\ref{fig:pionlesspot}.

Summarizing, the $2N$ potential~\eqref{eq:V-pp} is a particularly simple form of
Eq.~\eqref{Vexp} where there are no non-analytic functions and
\begin{align}
 &\mu = d/2 \;\; (s_2=2)
 \mathtext{,}
 \mu = d+1-s_2 \;\; (s_2=0,1) \,,
 \label{pionlesspot}
 \\
 &\tilde{\mathcal{N}} = \OO\left((4\pi)^{A-1} \MN^{-1} \Mlo^{5-3A}\right)
 \,,
\label{pionlesspotnorm}
\end{align}
with $A=2$, $d$ the number of derivatives, and $s_2=0,1,2$ the number of $S$
waves connected by the
operator~\cite{Bedaque:1997qi,vanKolck:1997ut,Kaplan:1998tg,Kaplan:1998we,%
Bedaque:1998mb,vanKolck:1998bw}.  Using the standard graph equalities to
eliminate the number of internal lines $I$ and loops $L$, $I=\sum_i V_i+L-1$ and
$I=-A+ \sum_i f_iV_i/2$, where $V_i$ is the number of vertices with $f_i$
nucleon lines, we obtain Eq.~\eqref{Texp} for the amplitude with
\begin{equation}
 \nu = \sum_i V_i \mu_i \mathtext{,}
 \mathcal{N} = \OO\left((4\pi)^{A-1} m_N^{-1} \Mlo^{5-3A}\right) \,.
\label{pionlessamp}
\end{equation}
Assuming $\Mlo\sim \gamma_d$, a rough estimate of the expansion parameter is
$\Mlo/\Mhi\sim \gamma_d/m_\pi \sim 1/3$.

\subsubsection{Regularization and renormalization}
\label{subsubsec:regren}

Loops in a quantum field theory are often not convergent, and the same in
true in Pionless EFT. Observables are rendered finite by renormalization.
For example, if we introduce a regulator function $f(x)$,
the nucleon bubble integral becomes
\begin{multline}
 I_0(k) = \MN\int\ddq
 \frac{f(\vecq^2/\Lambda^2)}{k^2- \vecq^2 + \ii\eps} \\
 = {-}\frac{\MN}{4\pi}
 \left[\theta_1\Lambda- \sqrt{-k^2-\ii\eps}+\OO(k^2/\Lambda) \right]
 \,,
\label{eq:I0-cutoff}
\end{multline}
where $\theta_1$ is a dimensionless number that depends
on the form of $f(x)$ (for example, $\theta_1=2/\pi$ for a step function).
With Eq.~\eqref{eq:T-kcot-b} truncated at the scattering length as a
renormalization condition, the choice
\begin{equation}
 C_0(\Lambda)= \frac{4\pi}{\MN}\frac{1}{1/a_0-\theta_1\Lambda}
\label{eq:C0-Lambda}
\end{equation}
ensures, to this order, that the physical amplitude is independent of
$\Lambda$, up to corrections that vanish as $\Lambda\to\infty$.
The latter can be removed by higher-order LECs, such as $C_2(\Lambda)$.
It is the non-analytic
dependence on energy, which is regulator independent, that characterizes a
loop. The corresponding term in Eq.~\eqref{eq:I0-cutoff} is an explicit example
of the estimates~\eqref{nucprop} and~\eqref{redloopint}.

\paragraph{Schemes and power counting}

In early stages, there was much confusion about whether or not the choice of
regularization should be understood to affect the power-counting scheme.
The difference between the \emph{artificial} regulator parameter $\Lambda$ and
the breakdown scale $\LamNoPi$ of the theory has not always been appreciated.
For example, \textcite{Kaplan:1998tg} have argued that $C_0\propto\Lambda^{-1}$
would again give a theory with a very limited range of applicability.  The need
to choose $\Lambda\simge \Mhi$ in order to suppress regulator artifacts
$\sim\OO(1/\Lambda)$ does seem to invalidate the scaling
$C_0\propto\aleph^{-1}$, but there are correlations among the diagrams which are
captured by determining $C_0(\Lambda)$ {\emph{after} resummation},
reflecting the original $\aleph$ counting.

If one uses dimensional regularization to render integrals finite, the $I_0$
bubble does not have a pole in four spacetime dimensions, so in the minimal
subtraction scheme there would be no divergence at all.  {Instead of this,
\textcite{Kaplan:1998tg} {advocate explicitly subtracting} the pole in
three dimensions (corresponding to the linear divergence in the cutoff scheme),
thereby introducing a renormalization scale $\mu$, which can be chosen freely,
and giving Eq.~\eqref{eq:C0-Lambda} with $\theta_1\Lambda\to \mu$.
This procedure, called ``power divergence subtraction'' (PDS),
makes the need for resummation of the bubble diagrams more transparent.
Picking $\mu\sim Q$, the \emph{running} coupling $C_0$ scales
like $Q^{-1}$, implying again that each diagram in Fig.~\ref{fig:BubbleChain} is
of the same order.}  With this scheme, power counting is ``manifest'' in the
sense that it is reflected by the scaling of coupling constants even after
renormalization has been carried out.  \textcite{Phillips:1998uy} have shown
that if \emph{all} poles of a divergent loop integral are subtracted---like the
original PDS, one particular choice of infinitely many possibly schemes---one
recovers exactly the same result as with a simple momentum cutoff.

{Under an appropriate power counting, changing the low-energy points used
as renormalization conditions affects the running of the LECs by $1/\Lambda$
terms, and leads to the same $T$ matrix up to higher-order terms.
Taking for example the pole position $\ii\gamma$ instead of zero energy
generates the LO amplitude with $1/a_0 \to \gamma$; the relative difference is
an NLO correction $\sim r_0/a_0 = \OO(\Mlo/\Mhi)$.
While the \apriori EFT error estimate is always determined by neglected higher
orders, the freedom to choose what input parameters are used at a given order
can improve agreement of the \emph{central values} with experimental data.}
{\textcite{Gegelia:1998iu} discusses the relation of subtractive
renormalization to the other approaches mentioned above.}

It was eventually realized~{\cite{Lepage:1997cs}} that cutoff variation can
be used (and is particularly useful in numerical calculations) as a diagnostic
for missing interactions at a given order, an example of which will be given in
Sec.~\ref{sec:Triton}.  \textcite{Long:2012ve} pointed out that also the leading
\emph{residual} cutoff dependence can be used to infer the existence of
next-order operators.  Equation~\eqref{eq:I0-cutoff}, for example, indicates
that $C_2\propto \Mhi^{-1}$ in order for the residual dependence on
$\Lambda\simge \Mhi$ to be no larger than NLO.  Thus renormalization provides
guidance for the power counting.

\paragraph{Subleading resummation}

Experience with nonsingular potentials makes it almost automatic to solve the
Schr\"odinger equation exactly with a truncation of the
potential~\eqref{eq:V-pp}.  At LO this is equivalent to the
resummation~\eqref{eq:T-C0}.  Renormalization of the truncation at the level of
$C_2$, however, leads to $r_0\simle
\Lambda^{-1}$~\cite{Cohen:1996my,Phillips:1996ae,Scaldeferri:1996nx}, a version
of the so-called ``Wigner bound''~\cite{Wigner:1955zz}.  This is problematic
for $N\!N$ scattering where $r_0>0$.  At first interpreted as a failure of EFT,
this observation reveals instead the danger of resumming subleading singular
potentials~\cite{vanKolck:1998bw}.  Such a resummation includes a subset of
arbitrarily high-order contributions without all the LECs needed for
perturbative renormalization, such as $C_4$ when $C_2$ is inserted twice at
N$^2$LO.  It is still possible to work with a fixed cutoff that reproduces
$r_0$, at the cost of losing the ability to use cutoff variation $\Lambda\simge
\Mhi$ as a diagnostic for missing interactions.  Moreover, there is no
guarantee that results for other observables will be any better than those
obtained from a perturbative treatment of subleading corrections.  An example
is given by \textcite{Stetcu:2010xq}.

\subsubsection{Renormalization group}
\label{sec:RG}

\paragraph{Running coupling}
Imposing renormalizability of physical amplitudes leads to solutions of
RG equations.  Their detailed form depends on the regularization scheme.  For
example, in PDS one finds for the dimensionless coupling constant
$\hat{C_0}\equiv\MN\mu C_0/(4\pi)$~\cite{Kaplan:1998tg},
\begin{equation}
 \mu \frac{\dd}{\dd\mu} \hat{C_0} =
 \hat{C_0}(1+\hat{C_0})\,,
\label{eq:C0-running-mu}
\end{equation}
where the right-hand side is given by the beta function.  It is convenient to
consider the flow of $\hat{C_0}$ instead of $C_0$ in order to separate the
behavior of the operator from the behavior of  the coupling constant.  The RG
equation~\eqref{eq:C0-running-mu} has two fixed points: the free fixed point
$\hat{C_0}=0$ and a nontrivial fixed point
$\hat{C_0}=-1$~\cite{Weinberg:1991um}, which correspond to $a_0=0$ and to the
unitary limit $1/a_0=0$, respectively.  Similar equations can be derived for
all coupling constants in the effective Lagrangian, and the beta function will
in general change as one goes to higher orders.  Thus the expansion in Pionless
EFT can be thought of as an expansion around the unitary limit of infinite
scattering length, similar to the expansion in Chiral EFT around the chiral
limit of vanishing quark masses.  An equation similar
to~\eqref{eq:C0-running-mu} holds for a simple momentum cutoff $\Lambda$,
leading then to Eq.~\eqref{eq:C0-Lambda}.  In dimensional regularization with
minimal subtraction, on the other hand, the coupling $C_0$ is independent
of $\mu$~\cite{Kaplan:1996xu}.  In this scheme the unitary limit cannot be
reached for any finite value of the coupling.

\paragraph{Wilsonian renormalization group}
The RG is more generally useful to study the behavior of the EFT.  Extending
previous
work~\cite{Weinberg:1990rz,Weinberg:1991um,Adhikari:1995uu,Adhikari:1997dz,
Beane:1997pk,Phillips:1997xu,Kaplan:1998tg,Kaplan:1998we,Kaplan:1998sz},
\textcite{Birse:1998dk} studied the RG flow of an effective potential of the
form
\begin{equation}
 V_{2N}(\vecp',\vecp, k)= V_{2N}(\vecp',\vecp)+ C_{02} k^2 + \cdots \,,
\label{eq:V-ppk}
\end{equation}
where the additional energy-dependent terms compared to Eq.~\eqref{eq:V-pp}
come from a different choice of operators in the effective
Lagrangian~\eqref{eq:L-NN-simple}.  It is possible to trade energy dependence
for momentum dependence and \viceversa by field redefinitions or,
alternatively,
using the equation of motion.  Within a Wilsonian formulation of the
RG~\cite{Wilson-83}, demanding that the off-shell amplitude stays invariant
under a decrease in the momentum cutoff $\Lambda$ in the Lippmann-Schwinger
equation defines a ``running'' potential $V(p,p',k,\Lambda)$ which satisfies
\begin{equation}
 \frac{\partial V}{\partial \Lambda}
 = \frac{\MN}{2\pi^2} V(p',\Lambda,k,\Lambda)
 \frac{\Lambda^2}{\Lambda^2 - k^2}
 V(\Lambda,p,k,\Lambda) \,.
\end{equation}
Defining further a rescaled potential $\hat{V}$ by multiplying all quantities
with appropriate powers of $\Lambda$, \textcite{Birse:1998dk} showed that in
the limit where $\Lambda\to0$ there exist two IR fixed points satisfying
$\partial\hat{V}/\partial\Lambda=0$.
One of these, $\hat{V} = 0$, is trivial
whereas the second, nontrivial one corresponds to the unitary limit.
{Additional fixed points are accessible with further fine tuning
\cite{Birse:2015iea}.
An} extensive study including also higher waves was carried by
\textcite{Harada:2006cw} and \textcite{Harada:2007ua}.
The RG analysis captures
the results obtained from Feynman diagrams, which yield directly the solutions
of the RG equations.\footnote{{\textcite{Weinberg:1995mt} gives a
general discussion of the connection between the Wilsonian RG and the
conventional renormalization program.}}
It unifies both the natural and fine-tuned cases discussed
in Sec.~\ref{sec:PowerCounting}, and it is possible to derive the power counting
for either case by studying perturbations of the potential around the fixed
points.

\subsubsection{Dibaryon fields}
\label{sec:Dibaryons}

It is possible to efficiently capture the physics associated with the shallow
$S$-wave two-body states by introducing in the effective Lagrangian ``dimeron''
(``molecular'' or, here, ``dibaryon'') fields with their quantum numbers, an
idea first introduced in EFT by \textcite{Kaplan:1996nv}.  For {any single
channel} we can write, instead of~\eqref{eq:L-NN-simple},
\begin{multline}
 {\!\LL
 = \psi^\dagger
 \Big(\ii\partial_0 + \frac{\Laplace}{2\MN}\Big)\psi
  + g\left[d^\dagger (\psi\psi) + \hc\right]} \\
  {\null + d^\dagger \Big[\eta\Big(\ii\partial_0
  + \frac{\Laplace}{4\MN}\Big) -\Delta \Big] d
  + \cdots \,,}
\label{eq:L-NN-simple-d}
\end{multline}
where $\eta=\pm 1$ is a parameter that determines the sign of the effective
range. It will be fixed to $\eta=-1$ in the remainder of this section to ensure
$r_0>0$.  Instead of $C_0$ and $C_2$, we have the new parameters $\Delta$ (the
``residual mass'') and $g$.  With this choice, nucleons no longer couple
directly, but only through the $s$-channel exchange of the dibaryon {$d$}.
If one neglects the kinetic term for this field, it is possible to recover the
leading terms in Eq.~\eqref{eq:L-NN-simple} by using the equation of motion
for {$d$},
\begin{equation}
 {d = \frac{g}{\Delta} \psi\psi\,,}
\end{equation}
and identifying $C_0 = -g^2/\Delta$.  Because of this redundancy, without loss
of generality one may fix $g$ at LO; a convenient choice is
$g^2\equiv 4\pi/m_N$~\cite{Griesshammer:2004pe} so that $\Delta=-1/a_0$
represents the low-energy scale $\Mlo$.
The {$d$} kinetic term leads to both energy- and
momentum-dependent four-nucleon-field interactions, corresponding to a choice of
operators that differs from Eq.~\eqref{eq:L-NN-simple}, but can be shown to be
equivalent up to higher orders and field redefinitions~\cite{Bedaque:1999vb}.

\begin{figure}[tb]
\centering
\includegraphics[clip,width=0.99\columnwidth]{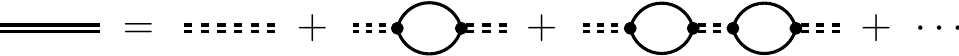}
\caption{Bubble sum for the dressed dibaryon propagator
 obtained from the bare propagator (double dashed line).
}
\label{fig:DibaryonProp}
\end{figure}

The original bubble series with $C_0$ vertices turns into a self-energy
correction for the dibaryon field: whereas the tree-level bare propagator is
just $\ii D(p_0,\vecp) = {-}\ii/\Delta$, summing up all bubble insertions as
shown in Fig.~\ref{fig:DibaryonProp} gives the full LO propagator as
\begin{equation}
 \ii D^{(0)}(p_0,\vecp)
 = {-}\ii \left[ \Delta + g^2 \, I_0\!\left(\sqrt{\MN p_0-\vecp^2/4}\right)
 \right]^{-1} \,.
\label{eq:Delta-L0}
\end{equation}
The {center-of-mass} $N\!N$ scattering amplitude is recovered by attaching
nucleon-dibaryon
vertices on both ends: $T^{(0)} = {-}g^2 D^{(0)}(p_0=k^2/\MN,\vecp=0)$.

Not only is the dibaryon formalism useful to study processes with deuterons in
the initial and/or final state (see below), where it can conveniently be used
as an interpolating field, but it also makes higher-order corrections
particularly
simple.  For example, where before we had to insert $C_2$ vertices in different
places (see Fig.~\ref{fig:C2-corr}), we now only have to insert the dibaryon
kinetic-energy operator into the LO propagator, giving
\begin{equation}
 \ii D^{(1)}(p_0,\vecp)
  = \ii \left(p_0-\frac{\vecp^2}{4m_N}\right) \left(D^{(0)}(p_0,\vecp)\right)^2
\label{eq:Delta-NLO}
\end{equation}
at NLO.  As in the case without dibaryons, renormalization is carried out
by relating $D^{(1)}$ to the NLO amplitude correction $T^{(1)}(k)$ and matching
to the effective-range term in Eq.~\eqref{eq:T-kcot-b}.  A difference is,
however, that this is now carried out with an energy-dependent operator---note
the dependence of Eq.~\eqref{eq:Delta-NLO} on the Galilei-invariant energy
$\tilde{p}_0=p_0-\vecp^2/(4\MN)$---whereas our choice of $C_2$ terms in
Eq.~\eqref{eq:L-NN-simple} only includes momentum-dependent operators.  The NLO
component of $g$ can be adjusted to reproduce $r_0$, and $g$ and $\Delta$ are
now independent.  This means that these parameters have RG runnings that differ
from those for $C_0$ and $C_2$~\cite{Birse:1998dk}.

With a dibaryon, range effects can be resummed using the propagator
\begin{equation}
 \ii D^{\text{resum}}(p_0,\vecp)
 = \frac{{-}\ii}
 {\Delta + g^2 \, I_0\!\left(\sqrt{\MN \tilde{p}_0\vphantom{2^3}}\right)
 - \tilde{p}_0} \,.
\label{eq:Prop-d-resum}
\end{equation}
The Wigner bound is automatically avoided by allowing the dibaryon to be a
ghost field.  In fact, \textcite{Beane:2000fi} proposed that the relatively
large sizes of the $N\!N$ effective ranges (about $2m_\pi^{-1}$) justify their
resummation as an LO effect.  However, this procedure leads to two $S$-matrix
poles per channel and is thus more likely to be interpreted as a resummation of
NLO interactions, which includes additional higher-order effects.

\subsubsection{Spin-isospin projection and parametrizations}

{For a fixed $N\!N$ channel it is convenient to use the effective
Lagrangian~\eqref{eq:L-NN-simple}, with $\psi$ a
nucleon field for which the combination $\psi^\dagger\psi$ has definite spin and
isospin $(S,I)$.  The Pauli principle dictates that only isospin-triplet
{$\mathrm{t}\equiv(0,1)$} and isospin-singlet {$\textrm{s}\equiv(1,0)$}
are allowed combinations.}  We use subscripts ``$\mathrm{s}$'' and
``$\mathrm{t}$'' here in reference to isospin, with a warning {that
the same subscripts are sometimes used in reference to spin}.  To go beyond the
description of an isolated two-nucleon system, it is desirable to treat both
combinations on the same footing.  To this end, it is convenient to introduce a
nucleon field $N$ that is a doublet in both spin and isospin space, along with
projection operators
\begin{equation}
 (P_{\textrm{s}})^i = \sigma^2\sigma^i\tau^2 / \sqrt8
 \mathtext{,}
 (P_{\textrm{t}})^A = \sigma^2\tau^2\tau^A/\sqrt8 \,,
\label{eq:P-t-s}
\end{equation}
where $\sigma^i$ ($\tau^A$) denotes the three Pauli matrices in spin (isospin)
space, and we have used lower- (upper-) case indices to further distinguish the
two spaces.  The Lagrangian for the $N\!N$ system can then be written as
\begin{multline}
 \LL
 = N^\dagger \left(\ii\partial_0 + \frac{\Laplace}{2\MN}\right) N
 - \frac{C_{0\textrm{s}}}2  (N^T P_{\textrm{s}} N)^\dagger(N^T P_{\textrm{s}}N)
 \\
 - \frac{C_{0\textrm{t}}}2 (N^T P_{\textrm{t}} N)^\dagger(N^T P_{\textrm{t}} N)
 + \cdots \,,
\label{eq:L-NN}
\end{multline}
where the ellipses represent analogous terms with $C_{2\textrm{s/t}}$ as well
as higher-order operators.  Fierz rearrangements can be used to generate
equivalent interactions.  Analogously, Eq.~\eqref{eq:L-NN-simple-d} is
generalized to the nuclear case by introducing two dibaryon fields---one for
each $N\!N$ $S$-wave channel--- using the same projection operators
$P_{\textrm{s},\textrm{t}}$~\cite{Bedaque:1997qi}.

The two channels are somewhat different concerning both sign and magnitude of
the scattering lengths.  However, it has been customary to treat both
$a_{\mathrm{s}}^{-1}$ and $|a_{\mathrm{t}}|^{-1}$ as $\Mlo$, although we return
to this issue in Sec.~\ref{sec:pionlessCoulomb}.  The ERE,
Eq.~\eqref{eq:T-kcot-b}, has a certain radius of convergence, set by the
nearest singularity to the expansion point $k^2=0$.
{The pion-exchange cut
on the imaginary $k$ axis starting at $\mpi$/2}
{leaves the deuteron pole within the radius of convergence of the ERE,
and indeed it is well known that the properties of this pole can be
expressed in terms of the ERE parameters~\cite{Goldberger:1967},
\cf~Sec.~\ref{sec:pionlessscattamp}.}
For example, the deuteron binding momentum is
\begin{equation}
 \gamd = \frac{1}{a_{\rm s}}
 \left(1+\frac{r_{\rm s}}{2a_{\mathrm{s}}}+\cdots\right) \,.
\label{eq:gammad}
\end{equation}
Alternatively, and this is in fact what was done first
historically~\cite{Bethe:1949yr}, one can perform the ERE directly about
this pole (\ie, about the point $\ii\gamd$ in the complex momentum plane),
\begin{equation}
 k\cot\delta_{d}(k)
 = {-}\gamd + \dfrac{\rhod}2\big(k^2+\gamd^2\big) + \cdots \,,
\label{eq:ERE-d}
\end{equation}
where $\rho_d\simeq 1.765~\fm$~\cite{deSwart:1995ui} is the deuteron effective
range.  The motivation for using Eq.~\eqref{eq:ERE-d} instead of the ERE about
zero momentum is that it captures the exact location of the pole already at LO.
\textcite{Griesshammer:2004pe} extended the procedure to the \OneSNot channel,
where it is possible to define the ERE about the virtual-state pole.

\begin{figure}[tb]
\centering
\includegraphics[clip,width=1\columnwidth]{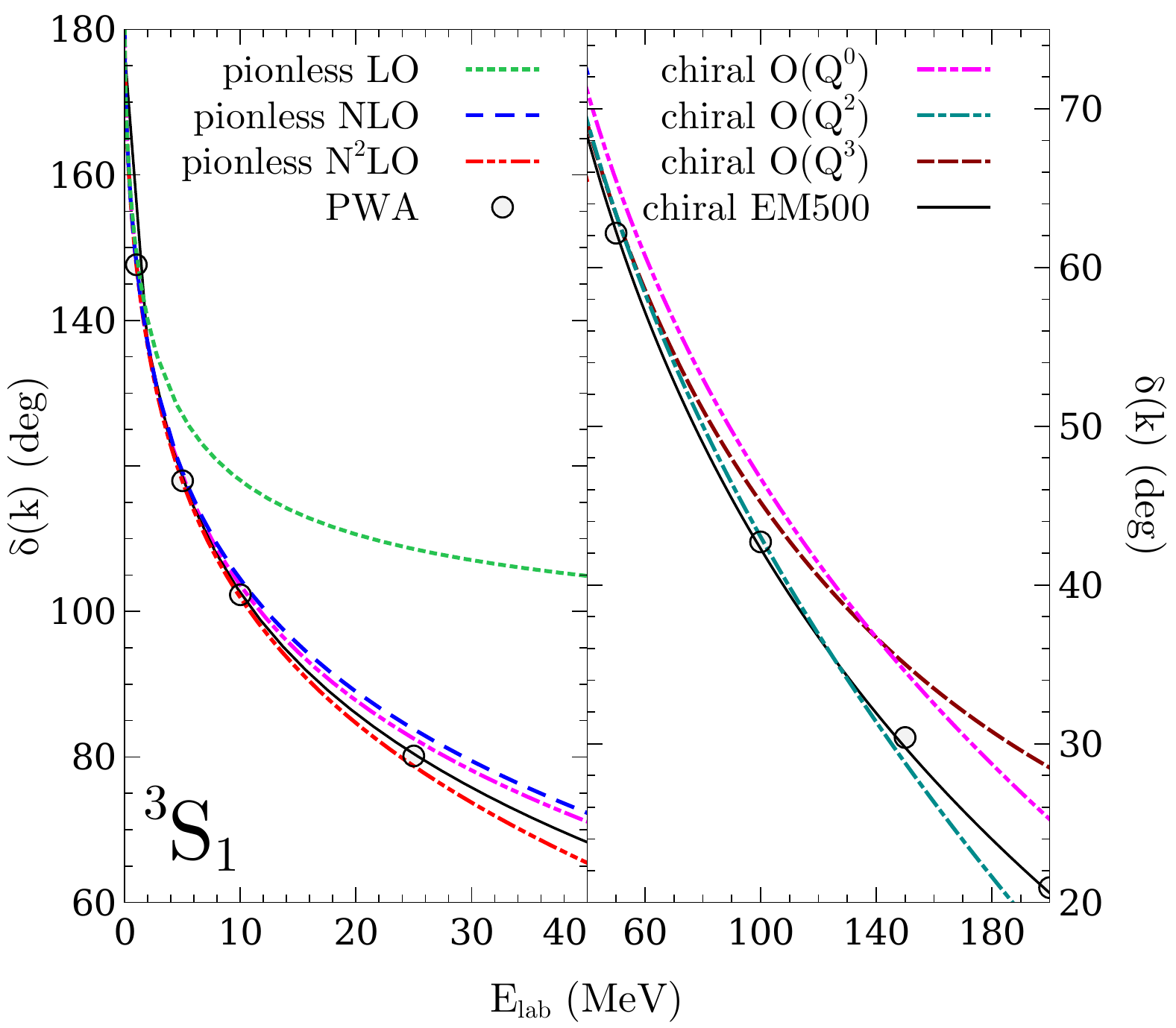}
\caption{{The $N\!N$ scattering phase shift $\delta$ as a function
 of the nucleon laboratory energy $E_{\text{lab}}$ in the \ThreeSOne partial
 wave for Pionless EFT and Chiral EFT at various
 orders~\cite{Long:2011xw} (results kindly provided by C.-J.~Yang), and
 the chiral potential ``EM500''~\cite{Entem:2003ft}.
 For comparison we show the partial-wave analysis (PWA)
 of~\textcite{Perez:2013jpa}, with error bars smaller than the symbols.}}
\label{fig:ThreeSOnePhase}
\end{figure}

The first detailed comparison of the \ThreeSOne phase shift obtained in Pionless
EFT with empirical values was carried out up to N$^2$LO by
\textcite{Chen:1999tn}, with LECs fitted to Eq.~\eqref{eq:ERE-d}.  In
Fig.~\ref{fig:ThreeSOnePhase} we show results {fitted to}
Eq.~\eqref{eq:T-kcot-b} instead, which are qualitatively similar: convergence is
seen at low energies and already at NLO a very good description is achieved.
The corresponding results for \OneSNot up to N$^2$LO were presented
by \textcite{Beane:2000fx}.

\subsubsection{Coulomb effects and other isospin breaking}
\label{sec:pionlessCoulomb}

Since almost all nuclear systems involve more than one proton, the inclusion of
electromagnetic effects is {generally} important.  In the low-energy
regime, the dominant effect is given by ``Coulomb photons'', \ie, the familiar,
static potential {($\sim\alpha/r$)} between charged particles.  It
originates from the
replacement of derivatives in the effective Lagrangian with covariant ones,
\begin{equation}
 D_\mu = \partial_\mu + \ii eA_\mu \hat{Q} \,,
\label{eq:minimal}
\end{equation}
where $\hat{Q}$ is an appropriate charge operator (\eg{,}
$\hat{Q}=(1+\tau_3)/2$ for nucleons).  The Coulomb photon-nucleon coupling $\ii
e$ comes from the
gauging of the nucleon {time derivative in Eq.~\eqref{eq:L-NN}}, while the
Coulomb-photon ``propagator'' is $\ii/(\vecq^2+\lambda^2)$, where $\lambda$ is
an IR-regulating photon mass that is eventually taken to zero.  Finer
electromagnetic effects enter through operators with more covariant derivatives
and also directly through the field strength, or alternatively the electric
{(}$E_i=\partial_0A_i-\partial_iA_0${)} and magnetic {(}$B_i =
\leviciv_{ijm} \partial^j A^m${)} fields.

\textcite{Kong:1998sx,Kong:1999sf} were the first to study proton-proton ($pp$)
scattering in Pionless EFT.  The challenge here lies in the fact that the
Coulomb interaction is important at very low energies: we see from
Eq.~\eqref{VGV} for the Coulomb potential $V\sim e^2/Q^2$ {that} Coulomb is
nonperturbative for ${Q\simle \alpha m_N / 2 \equiv k_C}$,
which is in the low-energy region of Pionless EFT.
Subtracting the pure-Coulomb amplitude $T_C$
from the full amplitude $T$, one can write
\begin{equation}
 T_{SC}
 = T - T_C
 = {-}\frac{4\pi}{\MN}\frac{\eex^{2\ii \sigma_C}}{k\cot\delta_{pp}(k) - \ii k}
\label{eq:T-SC}
\end{equation}
in terms of the ``subtracted'' $pp$ phase shift $\delta_{pp}(k)$ and the
pure-Coulomb phase shift $\sigma_C = \arg{\Gamma(1 + \ii\eta)}$.
Renormalization can be carried out by matching to the ``Coulomb-modified''
ERE~\cite{Bethe:1949yr},
\begin{multline}
 C_\eta^2 \left(k\cot\delta_{pp}(k) - \ii k\right) + \alpha\MN H(\eta) \\
 = {-}\frac{1}{a_{pp}} + \frac{r_{pp}}{2} k^2 + \cdots \,,
\label{eq:ERE-pp}
\end{multline}
where $a_{pp}\simeq -7.8$ fm
and $r_{pp}\simeq 2.8~\fm$~\cite{Bergervoet:1988zz} are the ERE parameters,
$C_\eta^2 = 2\pi\eta [\exp(2\pi\eta) - 1]^{-1}$ is the
Sommerfeld factor in terms of {$\eta = k_C/k$}, and
$H(\eta)=\Re [\psi(1+\ii\eta)]-\ln\eta+\ii C_\eta^2/(2\eta)$
in terms of the digamma function $\psi$.
{It should be emphasized that the $pp$
scattering amplitude, and thus also the effective range parameters, are
always defined in the presence of the Coulomb interaction and cannot be divided
into strong and electromagnetic parts in a model-independent
way~\cite{Kong:1998sx,Gegelia:2003ta}.}
{For particles with non-unit charges
the definition of the Coulomb momentum $k_C$ is generalized in
Sec.~\ref{sec:halo}, see Eq.~\eqref{eq:Cb-mom}.}

\begin{figure}[tb]
\centering
\includegraphics[clip,width=0.9\columnwidth]{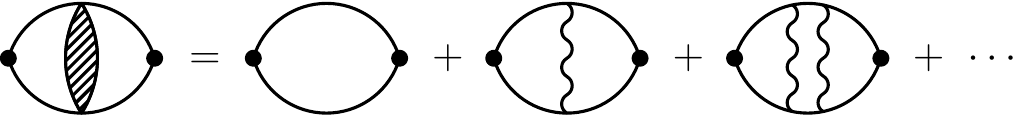}
\caption{Bubble diagram dressed with Coulomb exchange (wavy line).}
\label{fig:DressedBubble}
\end{figure}

In Pionless EFT, $T_{SC}$ is obtained by replacing all empty bubbles in
Fig.~\ref{fig:BubbleChain} with the dressed one shown in
Fig.~\ref{fig:DressedBubble}.  {The} initial and final-state Coulomb
interactions are {accounted for} by the construction in Eq.~\eqref{eq:T-SC}.
``Dressing'' here refers to resumming the Coulomb interaction to all orders
between each pair of $C_0$ vertices, which \textcite{Kong:1998sx,Kong:1999sf}
were able to do using a known analytic expression for the pure Coulomb Green's
function.  With dimensional regularization,
\begin{equation}
 \frac{4\pi}{\MN C_{0}^{pp}(\mu)} =
 \frac{1}{a_{{pp}}} - \mu
  + \alpha\MN\left(\frac{1}{\eps} +\ln\frac{\mu}{\alpha m_N}
  + \text{const.} \right) .
\label{eq:app-renorm}
\end{equation}
The term linear in the renormalization scale $\mu$ comes from the PDS
prescription, but Coulomb exchange now introduces an additional logarithmic
divergence, reflected in the pole in $\eps = d-3$, where $d$ is the number of
spatial dimensions.  Range corrections have been considered at \NLO by
\textcite{Kong:1999sf} and at \NNLO by \textcite{Ando:2007fh}.  An equivalent
formulation in terms of a $pp$ dibaryon exists~\cite{Ando:2010wq}.  The RG
analysis of \textcite{Birse:1998dk} discussed in Sec.~\ref{sec:RG} has also
been
extended to the charged-particle sector~\cite{Barford:2002je,Ando:2008jb}.

The LEC $C_{0}^{pp}=C_{0\textrm{t}}+\Delta C_{0(+)}$ in
Eq.~\eqref{eq:app-renorm} contains an isospin-dependent contribution $\Delta
C_{0(+)}$, which is a short-range (or ``indirect'') electromagnetic effect.
The EFT includes also isospin breaking from the quark
masses~\cite{VanKolck:1993ee,vanKolck:1995cb}.  While electromagnetic
interactions break isospin more generally (``charge dependence''), effects
linear in the quark masses break charge symmetry (a rotation of $\pi$ around
the second axis in isospin space) specifically.  Introducing the projectors
$P_{(\pm)}=(P_{\mathrm{t}}^{1}\mp\ii P_{\mathrm{t}}^{2})/\sqrt{2}$
onto the $pp$/$nn$ channel, the isospin-breaking Lagrangian takes the form
\begin{multline}
 \LL_{\textrm{ib}}
 = \delta m_N N^\dagger \tau_3 N
 - \frac{\Delta C_{0(+)}}{2}
   (N^T P_{(+)} N)^\dagger(N^T P_{(+)} N) \\
 - \frac{\Delta C_{0(-)}}{2}
   (N^T P_{(-)} N)^\dagger(N^T P_{(-)} N)
 + \cdots \,.
\label{eq:L-pp}
\end{multline}
NDA~\eqref{NDA} indicates that the neutron-proton mass splitting
$\delta m_N=\OO(\varepsilon {\bar m}, \alpha m_N/(4\pi))$.  It is well known
that the two types of contributions are comparable in magnitude, $\varepsilon
{\bar m}\sim \alpha m_N/4\pi$, {valid up to a (scale-dependent) factor of a
few,} but have opposite signs, the quark masses tilting the balance in favor of
the neutron.  The mass-splitting term can be
removed by a redefinition of the nucleon field~\cite{Friar:2004ca}, and
reappears as an $\OO(\delta m_N/m_N)$ effect in the nucleon kinetic term.
The most important quark-mass effects in the $N\!N$ system lie in the
short-range LECs $\Delta C_{0(\pm )}$.  The reduced quark mass is
$(\varepsilon {\bar m})_{\mathrm{red}}=\varepsilon {\bar m}/\Mhi$ and, together
with the $S$-to-$S$-wave enhancement discussed in Sec.~\ref{sec:PowerCounting},
leads to {$(a_{nn}-a_{\mathrm{t}})/a_{\mathrm{t}} \sim
\Delta C_{0(-)}/C_{0\mathrm{t}} = \OO(\varepsilon
{\bar m}\, a_{\mathrm{t}}) \simeq 0.2$, \cf~\cite{Konig:2015aka}.}
A similar contribution exists for $\Delta C_{0(+)}$ which is, however, dominated
by the electromagnetic contribution $\Delta C_{0(+)}/C_{0\textrm{t}}=\OO(\alpha
m_N a_{\mathrm{t}})$, consistent with Eq.~\eqref{eq:app-renorm}.

For most of the region of validity of Pionless EFT, $Q\simge \alpha m_N$
and \emph{all} electromagnetic interactions are expected to be perturbative.
In this region, $Q\simge 1/a_{\mathrm{t}}$ as well.  \textcite{Konig:2015aka}
developed an expansion in powers of $\alpha m_N/Q$ and $1/(Qa_{\mathrm{t}})$ in
addition to the standard $Q/\Mhi$ expansion.  For simplicity, they paired the
expansions by taking $\alpha m_N\sim a_{\mathrm{t}}^{-1} = \OO(\Mlo^2/\Mhi)$
and $\varepsilon {\bar m}=\OO(\Mlo^3/\Mhi^2)$.  In this case, LO in the
\OneSNot channel consists of the isospin-symmetric unitary amplitude, that is,
Eq.~\eqref{eq:T-kcot-a} with $k\cot \delta_{\textrm{t}}=0$.  The first
short-range and electromagnetic corrections break isospin symmetry at NLO,
reproducing $a_{pp}$ and leading to equal scattering lengths in the other two
\OneSNot isospin channels.  In addition, at NLO there is the standard,
isospin-symmetric $C_{2\mathrm{t}}$ interaction, while quark-mass effects (and
the $nn$ splitting from $np$) first enter at N$^2$LO.  This is consistent with
the observed relation $r_{pp}\simeq r_{\mathrm{t}}$.

\subsubsection{External currents}
\label{sec:Currents-TwoBody}

One of the great advantages of the EFT approach is that it is straightforward
to include external currents in addition to interactions between nucleons.
Power counting leads to a systematic expansion of current operators, which had
previously been classified only as one-body and many-body pieces (also known as
``meson-exchange currents'').

Photons are introduced in the effective Lagrangian as described above.   In
addition, weak interactions are accounted for by current-current interactions,
where the currents have the well-known vector-axial ($V-A$) form.  Power
counting is similar to that described in Sec.~\ref{sec:PowerCounting}, with
current operators subject to the same enhancement by powers of $\aleph^{-1}$
when $S$ waves are involved~\cite{Chen:1999tn}.
{Electromagnetic couplings were analyzed with the Wilsonian
RG by \textcite{Kvinikhidze:2018dnw}.}

The earliest example in the context of Pionless EFT are calculations of static
deuteron properties by \textcite{Chen:1999tn}, paralleling previous work by
{\textcite{Kaplan:1998sz} and \textcite{Savage:1998ae}} in Chiral EFT with
perturbative pions.
\textcite{Chen:1999tn} calculated several deuteron properties (charge, magnetic
dipole and electric quadrupole form factors, as well {as} electric
polarizabilities)
beyond LO, including also relativistic corrections.  Results were found to be
in
very good agreement with both experimental data and, at low orders, with those
obtained from effective-range
theory~\cite{Lucas:1968aa,Friar:1984zzb,Wong:1994sy}.  At higher orders, the
EFT goes beyond the effective-range approach (which is based on input from
elastic $N\!N$ scattering only) because new operators appear with undetermined
coefficients.  For example, there are magnetic four-nucleon-one-photon
couplings at NLO,
\begin{multline}
 \LL_{\text{mag}}^{(1)}
 = e L_1 \left(N^T P^i_{\textrm{s}} N\right)^\dagger
         \left(N^T P^3_{\textrm{t}} N\right) B_i \\
 - \ii e L_2 \, \leviciv_{ijk} \left(N^T P^i_{\textrm{s}} N\right)^\dagger
                              \left(N^T P^j_{\textrm{s}} N\right) B^k
 + \hc \,.
\label{eq:L-mag-L1L2}
\end{multline}
This is the two-nucleon analog of the single-particle ``Pauli term'' that
describes the direct $\vec{S}\cdot \vec{B}$ coupling of the nucleon spin to a
magnetic field, which accounts for the nucleon anomalous magnetic moment.
Here $L_1$ and $L_2$ are LECs that contribute to the deuteron dipole magnetic
moment as well as to the capture process $np \to d\gamma$.

Motivated by the original work of \textcite{Bethe:1949yr} and
\textcite{Bethe:1950jm}, \textcite{Phillips:1999hh} proposed a new scheme to
incorporate \NLO and higher orders in processes involving the deuteron.  Up to
higher-order corrections contained in the ellipses we can read off the residue
of the deuteron pole from Eq.~\eqref{eq:ERE-d},
\begin{equation}
 Z_d = \left(1-\gamd\rhod\right)^{-1}
 = 1 + \gamd\rhod + (\gamd\rhod)^2 + \cdots \,.
\label{eq:Z-d}
\end{equation}
This residue is directly related to the long-range tail of the deuteron
wavefunction in configuration space.  \textcite{Phillips:1999hh} argued that
convergence of deuteron observables (at least those  sensitive to the
long-range tail of the wavefunction) can be dramatically improved by fitting
to $Z_d$ exactly right at \NLO---rather than building it up perturbatively as
given in Eq.~\eqref{eq:Z-d}---while not spoiling convergence for the
\ThreeSOne phase shifts.

A deuteron dibaryon field (see Sec.~\ref{sec:Dibaryons}) is particularly
convenient for processes with external deuterons.  The dressed dibaryon can be
used directly as an interpolating field to define the $S$ matrix, provided its
wavefunction renormalization is properly taken into account. With a dibaryon,
the effects of a resummation of the effective range can be
assessed~\cite{Beane:2000fi,Ando:2004mm}.

A number of processes has been carefully addressed with these tools.  A precise
and controlled theoretical prediction of the $np \to d\gamma$ cross section is
important because {it enters as an input parameter into big-bang
nucleosynthesis calculations.}  The low-energy values required are difficult
to access experimentally, but are ideally suited for an application of Pionless
EFT.  The {pionless} analysis of this process started by
\textcite{Chen:1999tn} was refined in subsequent
papers~\cite{Chen:1999vd,Chen:1999bg}.  \textcite{Rupak:1999rk} carried out the
analysis {to \NNNNLO}, giving a prediction that is accurate to
a theoretical uncertainty below 1\%. This reaction was revisited with dibaryon
fields at NLO and a resummation of effective-range effects by
\textcite{Ando:2005cz}.

\begin{figure}[tb]
\centering
\includegraphics[clip,width=0.9\columnwidth]{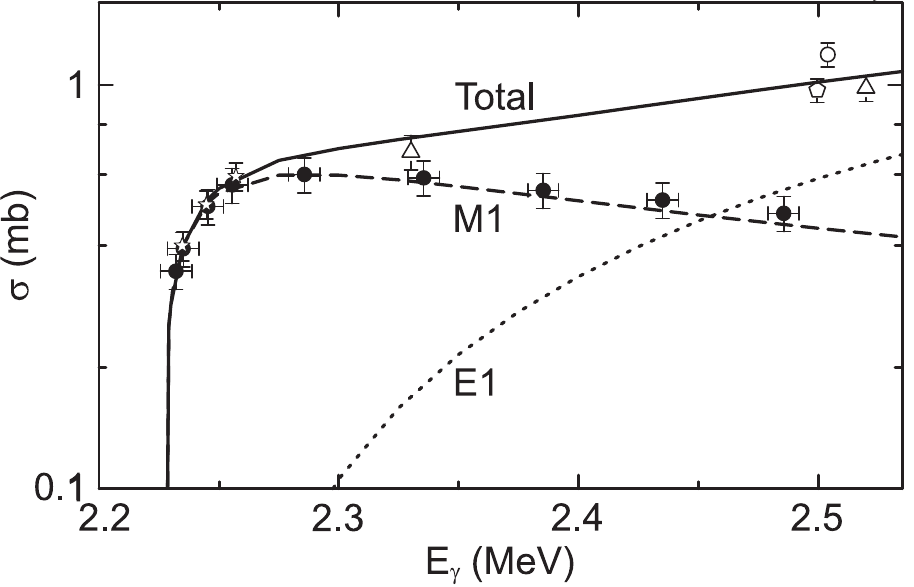}
\caption{Total cross section $\sigma$ for deuteron electrodisintegration as
 function of the photon energy $E_\gamma$.
 Good agreement is seen
 between measured data points and theoretical calculation (lines), which
 was achieved after the pionless calculation of \textcite{Christlmeier:2008ye}
 helped resolve a problem with the experimental analysis.
 Figure adapted from \textcite{Ryezayeva:2008zz} by H.~W.~Grie\ss hammer.
 Courtesy of H.~W.~Grie\ss hammer.}
\label{fig:GrieChrist}
\end{figure}

The related processes of deuteron electro- and photodisintegration are
experimentally accessible, and discrepancies between phenomenological potential
models and data have been reported.  Dibaryon fields implementing a resummation
of range effects have been used to \NNLO for $ed\to
e'pn$~\cite{Christlmeier:2008ye} and $d\gamma\to
np$~\cite{Ando:2011nv,Song:2017bkb}, with results generally supporting
phenomenological models.  For example, \textcite{Christlmeier:2008ye} concluded
that no consistent theoretical calculation could describe the data because the
EFT calculation, unlike the potential-model approach, comes with a rigorous
uncertainty estimate.  Subsequently, the
resolution of a problem with the data analysis gave agreement between
experiment and the EFT calculation~\cite{Ryezayeva:2008zz}{, see
Fig.~\ref{fig:GrieChrist}.}

The proton-proton fusion process $pp \to d e^+ \nu_e$ is of similar importance
for an understanding of the Sun.  Obviously, Coulomb effects play an important
role for this reaction at very low energies.
\textcite{Kong:1999tw,Kong:1999mp,Kong:2000px}, building upon their previous
work on $pp$ scattering {(see Sec.~\ref{sec:pionlessCoulomb})}, presented a
first
calculation in Pionless EFT at NLO. This calculation was later extended to
N$^4$LO by \textcite{Butler:2001jj}.  An NLO calculation using a dibaryon field
to resum effective-range corrections was presented by \textcite{Ando:2008va}.
\textcite{Chen:2012hm} extended the calculation of the astrophysical $pp$
S-factor to also include its energy derivatives.

The inverse process, neutrino-deuteron breakup scattering, was considered by
\textcite{Butler:2000zp} to N$^2$LO, along the lines of an earlier NLO
perturbative-pion calculation \cite{Butler:1999sv}.  At NLO, the axial-vector
counterparts of Eq.~\eqref{eq:L-mag-L1L2} appear, with two analogous LECs
usually denoted $L_{1,A}$ and $L_{2,A}$. However, because of the quantum numbers
of initial and final states, only the isovector $L_{1,A}$, which contributes to
$pp \to d e^+ \nu_e$ as well, is significant.  Various constraints on $L_{1,A}$
have been discussed by \textcite{Butler:2002cw}, \textcite{Chen:2002pv},
\textcite{Balantekin:2003ep} and \textcite{Chen:2005ak}, confirming SNO's
conclusions about neutrino oscillations.

Additionally, single-nucleon properties can be inferred from nuclear data.
Compton scattering is influenced by the nucleon polarizabilities, which are
response functions that carry much information about hadron dynamics and thus
QCD.  While proton polarizabilities can be extracted directly, neutron
polarizabilities can only be probed in nuclear Compton scattering.  Compton
scattering on the deuteron was studied to N$^2$LO by
\textcite{Griesshammer:2000mi}, where effective ranges were resummed and $Z_d$
fitted.  Values for the isoscalar, scalar electric and magnetic
polarizabilities were extracted by \textcite{Griesshammer:2000mi}.
Additional features of the
cross section were considered by \textcite{Chen:2004wv} and
\textcite{Chen:2004wwa}.  Sum rules for vector and tensor polarizabilities
were given by \textcite{Ji:2003ia}, while a low-energy theorem for the
spin-dependent Compton amplitude was obtained by~\textcite{Chen:2004fg}.

All in all, these calculations support the convergence of Pionless EFT for
momenta below the pion mass, with the power counting discussed in
Sec.~\ref{sec:PowerCounting}. They provide theoretically-controlled cross
sections that impact astrophysics and particle physics.  Heavier probes, such
as pions~\cite{Beane:2002aw}, can be considered as well through a heavy-field
treatment.  Most interesting for nuclear physics are processes with additional
nucleons, which we consider next.

\subsection{Light nuclei: bound and scattered}
\label{sec:pionlesslightnuclei}

Pionless EFT extends effective-range theory into the nuclear realm, where it
leads to a striking emergence of structure related to the Efimov
phenomenon~\cite{Efimov:1970zz,Efimov:1981aa}, which we discuss in more detail,
in the context of Halo/Cluster EFT, in Sec.~\ref{sec:efimovhalos}.

\subsubsection{Extension to three particles}

\begin{figure}[tb]
\centering
\includegraphics[clip,width=0.75\columnwidth]{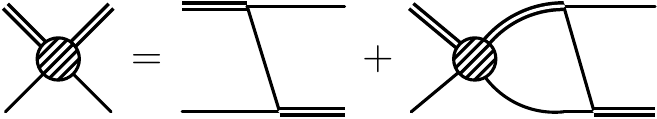}
\caption{Integral equation for $nd$ scattering in the spin-quartet channel.
 A pair of similar, but coupled, equations describes the doublet channel.
}
\label{fig:nd-IntEq-Q}
\end{figure}

The simplest three-body system that can be studied in Pionless EFT is
neutron-deuteron ($nd$) scattering in the quartet $S$-wave channel (total spin
$\nicefrac32$ and zero orbital angular momentum).  The Pauli principle dictates
that only the same configuration can appear in the intermediate state.
\textcite{Bedaque:1997qi} calculated the $nd$ quartet scattering length in a
framework using a deuteron dibaryon field (see Sec.~\ref{sec:Dibaryons}).
The driving mechanism is the exchange of a nucleon (neutron) between in and
outgoing deuterons.  The EFT power counting gives that all diagrams with an
arbitrary number of such exchanges are of the same order.  Quite analogous to
the two-body bubble chain they can be conveniently resummed into an integral
equation for the scattering amplitude, shown diagrammatically in
Fig.~\ref{fig:nd-IntEq-Q}.  The loop integrals are convergent, but for a
numerical treatment it is still convenient to introduce a momentum cutoff.
Resumming effective range corrections to all orders in the deuteron sector,
\textcite{Bedaque:1997qi} calculated the scattering length to be $6.33~\fm$, in
very good agreement with the experimental value
{6.35(2)~\fm~\cite{Dilg:1971ab}.}
{A perturbative treatment of effective-range corrections according to
Eq.~\eqref{eq:ERE-d} gives $(5.09 + 0.98 + 0.21 = 6.28)~\fm$ to \NNLO, with an
estimated 3\% uncertainty.  The LO result of $5.09~\fm$ agrees} with the much
older result
of \textcite{Skorniakov:1957aa} who used a zero-range model that is equivalent
to Pionless EFT at LO.  \textcite{Bedaque:1998mb} and \textcite{Bedaque:1999vb}
extended the EFT calculation to $nd$ scattering at finite energy.

\subsubsection{The triton as a near-Efimov state}
\label{sec:Triton}

Three nucleons can also couple to an $S$-wave state with total spin
$\nicefrac12$, which is the channel of the trinucleon bound states: triton
($^3$H) and helion (\ThreeHe).  The formalism used to calculate quartet-channel
scattering can be extended directly to the doublet channel, where now also
\OneSNot intermediate states are allowed.  The result can be written as an
integral equation for the $nd$ $T$ matrix with the same structure as in
Fig.~\ref{fig:nd-IntEq-Q}, but for the two coupled channels $n+np$(\ThreeSOne)
and $n+np$(\OneSNot)~\cite{Skorniakov:1957aa}.  The triton should show up as a
pole in this amplitude at a negative energy $E = {-}B(\ThreeH) =
{-}8.4818~\MeV$.  Since its relevant momentum scale is given by $\gamma_T =
\sqrt{2\MN B(\ThreeH)/3} \sim 80~\MeV$, it is within the expected range of
validity of the EFT.

However, it has been known for a long time that the three-nucleon system
is unstable when described solely with nonderivative two-body short-range
interactions: as the range of such a potential is sent to zero, one encounters
the ``Thomas collapse,'' \ie, the binding energy diverges~\cite{Thomas:1935zz}.
\textcite{Bedaque:1999ve}, generalizing their previous work on the three-boson
system~\cite{Bedaque:1998kg,Bedaque:1998km}, showed that the same happens in
Pionless EFT: as the cutoff $\Lambda\gg \Mlo$ {is increased},
the ground-state energy grows as $\Lambda^2/m_N$, and excited states appear
{repeatedly.  Since the $N\!N$ scattering lengths are large, one encounters
an approximate realization of the Efimov
effect~\cite{Efimov:1970zz,Efimov:1981aa}, \ie,
a tower of three-body
states with the ratio of neighboring binding energies approaching a universal
constant.}

\paragraph{The three-body force}

\begin{figure}[tb]
\centering
\includegraphics[clip,width=0.7\columnwidth]{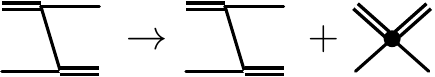}
\caption{Modification of the driving mechanism in Fig. \ref{fig:nd-IntEq-Q},
 with the LO three-body force (solid circled) in the $nd$ doublet channel.
}
\label{fig:ThreeBodyForce}
\end{figure}

The scattering amplitude in the doublet channel, obtained from the integral
equations analogous to Fig.~\ref{fig:nd-IntEq-Q}, does not approach a stable
limit as the cutoff is increased.  This lack of renormalization is a genuine
nonperturbative effect since every diagram generated by iterations is finite by
itself.  \textcite{Bedaque:1999ve} showed that the system can be stabilized by
adding a nonderivative three-body contact interaction.  Fierz rearrangements
show that there is only one such interaction, which can be written in any one
of various equivalents forms, for example
\begin{equation}
 \LL_{\text{3b}}
 = {-}4h_0 C_{0{\rm t}}^2\big(N^T (P_{\rm t})^k N\big)^\dagger
 \big(N^\dagger \sigma^k\sigma^l N\big)\big(N^T (P_{\rm t})^l N\big) \,,
\label{eq:L-NN-3}
\end{equation}
where $h_0$ is a new LEC to be determined.  In the formalism with dibaryon
fields, every nucleon-exchange diagram has to be accompanied by a
dibaryon-nucleon interaction with strength $h_0$, as shown in
Fig.~\ref{fig:ThreeBodyForce}. Attaching the two-nucleon-dibaryon
{vertex $g$}
from Eq.~\eqref{eq:L-NN-simple-d} on both dibaryon ends recovers the six-nucleon
operator \eqref{eq:L-NN-3} in the theory without dibaryon field.

\begin{figure}[tb]
\centering
\includegraphics[width=0.9\columnwidth]{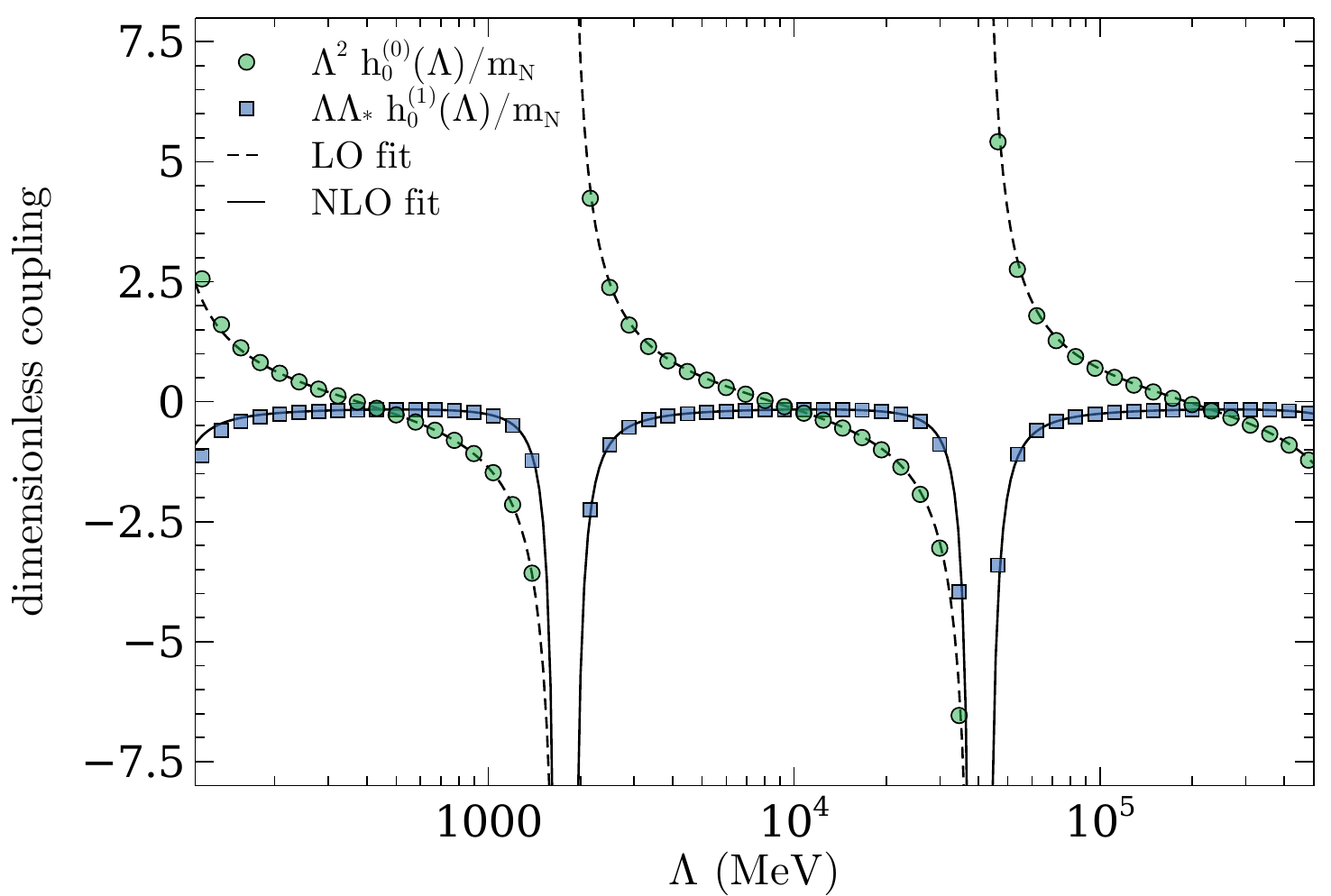}
\caption{RG running of the three-body coupling $h_0$ at LO and NLO. Numerical
 data points are fitted using Eq.~\eqref{eq:H0-0} at LO and an analogous
 expression~\cite{Ji:2011qg} at NLO.
}
\label{fig:ThreeBodyRunning}
\end{figure}

The $3N$ force is symmetric~\cite{Bedaque:1999ve} under the group of combined
spin and isospin transformations, Wigner's
$SU(4)$~\cite{Wigner:1936dx,Wigner:1937zz}.  Because the two-body amplitude is
also $SU(4)$~symmetric for momenta
$a_{\mathrm{s}}^{-1}<Q<\Mhi$~\cite{Mehen:1999qs}, the coupled integral equation
illustrated in Fig.~\ref{fig:nd-IntEq-Q} is symmetric in the limit where all
momenta are large {compared to the inverse scattering lengths}.  This
allowed \textcite{Bedaque:1999ve} to study the UV behavior of the amplitude
{based on decoupling the two integral equations,} with one of the
rotated amplitudes behaving exactly like the amplitude for the three-boson
system with two-body scattering length $a_2$.  This in turn leads to the
{analytical} result~\cite{Bedaque:1998kg,Bedaque:1998km}
\begin{equation}
 {\frac{\Lambda^2 h_0(\Lambda)}{\MN}} \equiv
 H(\Lambda)
 \approx {-}
 \frac{\sin\left(s_0\log(\Lambda/\Lambda_*)-\arctan(s_0^{-1})\right)}
 {\sin\left(s_0\log(\Lambda/\Lambda_*)+\arctan(s_0^{-1})\right)} \,,
\label{eq:H0-0}
\end{equation}
conveniently written as a dimensionless function.  Here, $s_0\simeq 1.0064$ is
a {universal constant~\cite{Danilov:1961aa}} and $\Lambda_*$ is a parameter
that has to be fixed to a three-body datum.  The striking log-periodic
dependence on the cutoff is shown in Fig.~\ref{fig:ThreeBodyRunning}, where the
overall prefactor in Eq.~\eqref{eq:H0-0} depends on the details of the
regularization scheme employed in a given
calculation~\cite{Platter:2004he,Braaten:2011sz}. \textcite{Hammer:2000nf}
studied this ``ultraviolet limit cycle'' and derived
the RG equation of which Eq.~\eqref{eq:H0-0} is a solution.  They realized that
the explicit three-body force can be set to zero by working at a set of
log-periodically spaced cutoffs {$\Lambda_n =
\Lambda_*\exp[s_0^{-1}(n\pi+\arctan s_0^{-1})]$} where $n$ is an
integer.  \textcite{Braaten:2003eu} have argued that the UV limit cycle
observed in Pionless EFT hints at an underlying \emph{infrared} cycle in QCD.

Such a $3N$ force would be of higher order according to na\"ive dimensional
analysis.  The fact that it has to be included already at LO to renormalize the
three-body system is another consequence of the fine tuning encountered in the
two-body sector.  After renormalization the Efimov tower of states is cutoff
independent, its position determined by $\Lambda_*$.  If the $N\!N$ scattering
lengths were in fact infinite, one would have a tower of shallow three-body
states accumulating at zero energy.  The large but finite physical scattering
lengths cut off this spectrum in the IR, whereas the breakdown scale $\Mhi$ of
the EFT sets a limit for the deepest state.  In nuclear physics at physical
quark masses, $a_{\rm s}$ and $m_\pi$ are not large enough for the appearance of
an excited $3N$ state.  However, \textcite{Rupak:2018gnc} show, in agreement
with earlier model calculations~\cite{Adhikari:1982zz}, {that a shallow
virtual state in $nd$ scattering, known to exist for a long
time~\cite{vanOers:1967lny,Girard:1979zza},} becomes the first excited
bound state as $a_{\rm s}$ increases.  Other situations are discussed by
\textcite{Braaten:2003eu}.

\paragraph{The Phillips line}

\begin{figure}[tb]
\centering
\includegraphics[width=0.9\columnwidth]{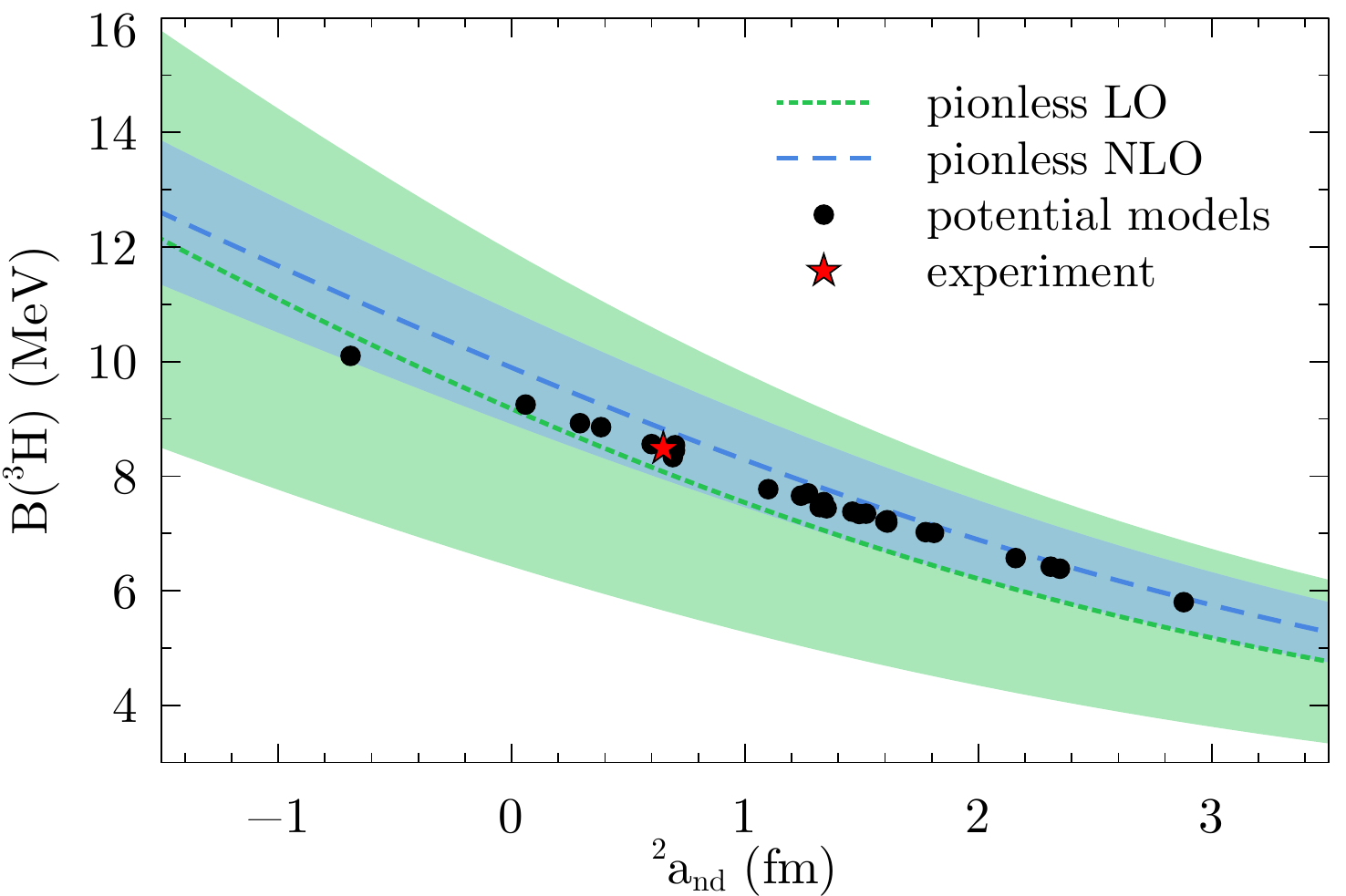}
\caption{Correlation between the triton binding energy $B(^3{\mathrm{H}})$ and
the doublet $nd$ scattering length $^2a_{nd}$ (Phillips line)
at LO and NLO, compared to results from various potential models
and experiment.
Bands indicate estimates of higher-order corrections:
the larger band around the dotted line is LO, the smaller band
around the dashed line, NLO.}
\label{fig:PhillipsLine}
\end{figure}

Pionless EFT at LO offers a striking but simple explanation of the well-known
``Phillips line'', \ie, the fact that different model potentials for the
nuclear interaction tuned to the same $N\!N$ scattering data give different but
highly correlated results for the triton binding energy and the doublet $nd$
scattering length~\cite{Phillips:1968zze}.  Pionless EFT allows one to
understand this and other correlations among three-body observables as a
consequence of the RG, \ie, as a correlation originating in the variation of
$\Lambda_*$~\cite{Bedaque:1999ve}.
{This is shown in Fig.~\ref{fig:PhillipsLine}.}
The proximity of the LO EFT line to the experimental point means that, whichever
observable is used as input, the other comes out correct.

\subsubsection{More neutron-deuteron scattering}
\label{sec:NdScattering}

\paragraph{Range corrections, partial resummation,
and two-body parametrizations}
\label{sec:NdScattering-PR}

\begin{figure}[tb]
\centering
 \vspace*{0.9em}
 \vcenteredhbox{\includegraphics[width=11em]{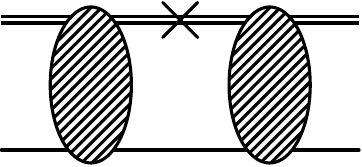}} $\,,\,$
 \vcenteredhbox{\includegraphics[width=5em,clip]{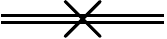}}
 \vcenteredhbox{\scalebox{1}{$\;\sim\;
 \left(p_0-\frac{\vecp^2}{4\MN}\right)$}}
\caption{Dibaryon kinetic-energy corrections for the $nd$ quartet-channel
 scattering amplitude at NLO. In the doublet channel, there are analogous
 diagrams with \OneSNot dibaryons.
}
\label{fig:nd-NLO-Corr}
\end{figure}

At NLO one needs to account for the two-body ranges. In the dibaryon framework
that means one insertion of each dibaryon kinetic-energy operator between LO
amplitudes, as shown in Fig.~\ref{fig:nd-NLO-Corr}.  At \NNLO, the procedure of
perturbative range insertions becomes tedious, and a direct calculation of the
corrections requires fully off-shell LO amplitudes.  To avoid this, range
corrections can be resummed with Eq.~\eqref{eq:Prop-d-resum}.   Already
\textcite{Bedaque:1997qi} noted that this resummation introduces an artificial
deep pole in the deuteron propagator.  Located at a momentum scale of roughly
$200~\MeV$, it is outside the range of validity of the EFT and thus in
principle
an irrelevant UV artifact, although it limits the range of cutoffs that can be
used in the numerical solution of the scattering equations.  This is especially
true in the doublet $S$ channel unless measures are taken to remove the pole.
In the quartet channel, due to the Pauli principle, the solution is not
sensitive to this deep pole and the cutoff can be made arbitrarily large.
Considering effective ranges as LO as proposed by \textcite{Beane:2000fi}
effectively cuts off the integral of the three-body equation at $\sim
r_0^{-1}$, eliminating the UV limit cycle and leaving only the IR limit cycle
manifest in the Efimov effect.  However, in general there is no guarantee that
the Efimov tower is at the correct location without a three-body force.
Similar
results are expected from any selective resummation of higher-order effects,
such as relativistic corrections~\cite{Epelbaum:2016ffd}.

\textcite{Bedaque:2002yg} proposed a middle ground that partially re-expands
the resummed propagators and uses terms up to order $n$ for a calculation at
N$^n$LO.  Using these ``partially resummed'' propagators generates all desired
terms at a given order, but still retains some higher-order corrections, which
have to be assumed to be negligible.  {We note that for such an approach to
be valid it is important to keep the cutoff at or below the breakdown scale of
the theory.}  $^3S_1$-$^3D_1$ mixing as well as
relativistic corrections formally enter at \NNLO but were not included by
\textcite{Bedaque:2002yg}.  \textcite{Griesshammer:2004pe} implemented the
two-body parametrization~\eqref{eq:Z-d} and found a
substantially better description of data, particularly in the doublet $S$ wave.

The first strictly perturbative \NLO calculation of $nd$ scattering in the
doublet $S$ channel was carried out by~\textcite{Hammer:2001gh}, implementing
the procedure suggested by \textcite{Bedaque:1999ve}.
\textcite{Vanasse:2013sda} developed a scheme that avoids the numerically
expensive determination of full off-shell amplitudes made in earlier
perturbative calculations~\cite{Ji:2011qg,Ji:2012nj}, requiring even slightly
less effort than using the partial resummation of \textcite{Bedaque:2002yg}.
Beyond the practical benefit, \textcite{Vanasse:2013sda}'s \NNLO calculation
also showed that an anomalous (unitarity-violating) behavior of the quartet
$S$-wave $nd$ phase shift above the deuteron breakup threshold, {known to
practitioners as a consequence of the partial-resummation scheme, does not
occur with a fully perturbative treatment of range corrections.  The lesson is
that individually small, undesirable effects can be generated by their
\emph{infinite} resummation.}  {Overall, \textcite{Vanasse:2013sda} obtains
$nd$ phase shifts at \NNLO which are in good agreement with the empirical
behavior up to laboratory energies of $\simeq 24~\MeV$.}

\paragraph{Higher partial waves}

Even with only $S$-wave interactions in the two-body sector, the
nucleon-exchange diagram driving the $nd$ scattering equations
(Fig.~\ref{fig:nd-IntEq-Q}) generates contributions in all partial waves.
\textcite{Gabbiani:1999yv} calculated the scattering phase shifts up to $G$
waves ($\ell=4$) to \NNLO using the full resummation~\eqref{eq:Prop-d-resum}
but omitting $^3S_1$-$^3D_1$ mixing.  They found good agreement with both
potential-model calculations and available experimental data up to about
$140~\MeV$ center-of-mass momentum, indicating that the breakdown scale of
Pionless EFT is indeed close to being set by the pion mass, at least for these
particular observables.  Better results still are obtained with the two-body
parametrization \eqref{eq:Z-d}~\cite{Griesshammer:2004pe}.  The fully
perturbative \NNLO calculation by \textcite{Vanasse:2013sda} did include
$^3S_1$-$^3D_1$ mixing and found reasonable agreement with potential-model
results.

The vector analyzing power $A_y$ has defied explanation with potential models
even at energies as low as a few \MeV---the ``$A_y$ puzzle''.
\textcite{Margaryan:2015rzg} employed the fully perturbative approach to
calculate  $nd$ polarization observables at \NNNLO.  They found that varying the
${}^3P_J$ LECs (first entering at \NNNLO) with the expected error (a $15\%$
band around their central values) covers a range for $A_y$ that is consistent
with experimental data.

\paragraph{Ordering of three-body forces}

\textcite{Bedaque:1999ve} argued that at \NLO range corrections force a shift
in the LEC of the $3N$ force~\eqref{eq:L-NN-3}, which was confirmed by
\textcite{Hammer:2001gh} in an explicit \NLO calculation.
Thus,
\begin{equation}
 h(\Lambda) = h_{0}^{(0)}(\Lambda) + h_{0}^{(1)}(\Lambda) + \cdots \,,
\end{equation}
where $h_{0}^{(0)}(\Lambda)$ is given in Eq.~\eqref{eq:H0-0} and
$h_{0}^{(1)}(\Lambda)$, determined in a fully perturbative calculation with
physical  $N\!N$ effective-range parameters, is shown in
Fig.~\ref{fig:ThreeBodyRunning}.  At NLO, the numerical data has been fit with
the analytical result found by \textcite{Ji:2011qg}.
{The existence of the correction $h_{0}^{(1)}$  does
\emph{not} mean} that there is a new three-body force at \NLO---it
is merely an
adjustment of the LO coefficient carried out by demanding that the observable
used to fixed $h_{0}^{(0)}(\Lambda)$ stays invariant after the inclusion of
range corrections.  Because there is no new three-body parameter at \NLO, simple
correlations through $\Lambda_\star$ survive with small shifts, as can be seen
in the \NLO line in Fig.~\ref{fig:PhillipsLine}.
{Since $h_{0}^{(1)}(\Lambda)$ depends on the two-body scattering
lengths, if the latter are changed, further experimental input
is needed to determine the NLO LEC \cite{Ji:2010su}.}

The conclusion that, despite having canonical dimension 9, the nonderivative
three-body force appears at LO together with the nonderivative two-body force
of dimension 6 raised the questions: Do other three-body interactions have to
be promoted?  More generally, what is the ordering of three-body forces in the
pionless power counting?  Since according to \textcite{Bedaque:1999ve} a new
$3N$ force enters at \NNLO, \textcite{Bedaque:2002yg} used a Lepage-plot
analysis~\cite{Lepage:1997cs} to show that its inclusion reduces the errors in
the calculation.  A general and comprehensive analysis of
pionless three-body forces using the asymptotic techniques of
\textcite{Bedaque:1998kg,Bedaque:1998km,Bedaque:1999ve} was carried out by
\textcite{Griesshammer:2005ga}, who identified a  logarithmic divergence at
\NNLO that mandates the inclusion of a new three-body force at this order.
\textcite{Griesshammer:2005ga} also cataloged the minimal orders at which $3N$
forces must first appear in various channels for proper renormalization.
\textcite{Platter:2006ev}, using a subtractive renormalization scheme, argued
that the LO three-body force is sufficient to achieve cutoff independence up to
\NNLO, contradicting the findings of \textcite{Bedaque:2002yg}.
\textcite{Ji:2011qg} and \textcite{Ji:2012nj}, studying the three-boson system,
later explained this discrepancy by noting that the conclusion of
\textcite{Platter:2006ev} only holds in the limit where the three-body UV
cutoff is taken to infinity, with the partial resummation of range corrections
affecting the perturbative expansion at smaller cutoffs.  Using a fully
perturbative inclusion of range corrections, \textcite{Ji:2012nj} concluded
that a new three-body force indeed enters at \NNLO.  This term can be
implemented using the same $SU(4)$-symmetric spin-isospin structure as the \LO
three-body force, with appropriate time derivatives included to give a linear
dependence on the energy~\cite{Ji:2012nj,Vanasse:2013sda}.

Generalizing these results, the $3N$ potential takes the form~\eqref{Vexp} with
(\cf~\cite{Griesshammer:2005ga}) $\mu= d +2 -s_3$, where $s_3=0,1,2$ is the
number of nucleon-deuteron ($N\!d$) Wigner-symmetric $S$ waves connected by
the operator.  The first orders are represented in Fig.~\ref{fig:pionlesspot}.
Amplitudes have the form~\eqref{Texp} with the
normalization~\eqref{pionlesspotnorm} for $A=3$.

\subsubsection{Proton-deuteron scattering and helion}

\paragraph{Nonperturbative Coulomb effects}

As discussed in Sec.~\ref{sec:pionlessCoulomb}, at the low energies potentially
reached in scattering Coulomb-photon exchange needs to be treated
nonperturbatively, which poses additional technical challenges.  The first
attempt to study Coulomb effects was made in the simpler quartet $S$-wave $pd$
scattering by \textcite{Rupak:2001ci}.  They developed a power counting
that, with some approximations, amounts to iterating a Coulomb potential between
proton and deuteron to all orders, along with the one-nucleon exchange diagram
that also enters in $nd$ scattering.  \textcite{Rupak:2001ci} calculated the
Coulomb-subtracted $S$-wave phase shift in the quartet channel, but could not
reach convergence below $pd$ center-of-mass momenta of $20~\MeV$ (the regime
where Coulomb effects really are nonperturbative).  Convergence down to $3~\MeV$
was later achieved by \textcite{Konig:2011yq} owing to an improved numerical
procedure.  \textcite{Konig:2011yq} also extended the analysis to the doublet
$S$-wave channel, including helion, and applied the partial-resummation approach
of \textcite{Bedaque:2002yg} to calculate higher orders.

\textcite{Ando:2010wq} carried out a direct momentum-space calculation of helion
based on a generalization of the $nd$ integral equation discussed in
Sec.~\ref{sec:Triton}.  Recasting methods developed by
\textcite{Kok:1979aa,Kok:1981aa} into EFT, \textcite{Ando:2010wq} included
Coulomb effects via a fully off-shell Coulomb $T$ matrix and obtained, at a
single momentum cutoff $\Lambda_0=380.689~\MeV$ (the first cutoff value where
the experimental triton binding energy is reproduced without a $3N$ force,
$H(\Lambda_0) = 0$) a \ThreeHe binding energy $B(\ThreeHe) \simeq 7.66~\MeV$,
close to the experimental value of about $7.72~\MeV$.  \textcite{Ando:2010wq}
treated all Coulomb effects nonperturbatively, {but considered the strong
sector only at LO.}  \textcite{Kirscher:2009aj} obtained similar numerical
results in a calculation that included \NLO and selected higher-order
interactions as part of an effective potential which was treated exactly.

The pionless calculation of $pd$ scattering and helion was revisited and
extended by \textcite{Koenig:2013} and \textcite{Konig:2014ufa}, who argued that
the \NLO $pd$ system, within the partial-resummation approach, is not properly
renormalized by the isospin-symmetric $3N$ force alone.  The same conclusion was
reached in a parallel analysis by \textcite{Vanasse:2014kxa}, who calculated
$pd$ scattering and \ThreeHe at \NLO in strict perturbation theory.  Using an
asymptotic analysis, it was shown that Coulomb at LO requires a nonderivative
$3N$ interaction at \NLO (but not \LO) to properly renormalize the $pd$ system.
{At this level, one can no longer predict low-energy
$pd$ scattering from $nd$ scattering without further input.}
Fixing the corresponding LEC to the \ThreeHe binding energy gives good
agreement with an analytically-derived expression, and it also provides
cutoff-stable
results for the phase shift.  In contrast, \textcite{Kirscher:2015zoa} recently
argued that \ThreeHe is renormalized at NLO without an additional counterterm
($pd$ scattering was not investigated).  While both calculations use Pionless
EFT, they differ in the numerical implementation and regularization scheme.

A comparison of the different schemes for the counting of Coulomb effects
in $pd$ scattering up to \NLO was provided by \textcite{Konig:2013cia}.
These authors argued that for the scattering of composite charged particles
there is a certain arbitrariness in the definition of Coulomb-subtracted
quantities (phase shifts and modified ERE parameters), namely whether or not
information about the deuteron substructure is included in the definition of
the pure Coulomb phase shift.

\paragraph{Perturbative Coulomb effects}

Renewed interest and developments in the strict application of perturbation
theory also motivated a new look into the counting of Coulomb effects.  The
characteristic trinucleon momentum scale $\gamma_T\sim 80~\MeV \gg \alpha m_N$
suggests that Coulomb effects should be a perturbative correction to the
\ThreeHe binding energy (compared to the triton as its isospin mirror state).
Already \textcite{Konig:2014ufa} showed that the calculation of
\textcite{Ando:2010wq} can be reproduced essentially unchanged when the fully
off-shell Coulomb $T$ matrix is replaced by one-photon exchange diagrams.
However, the calculation of \textcite{Konig:2014ufa} is still nonperturbative
because $B(\ThreeHe)$ is extracted from the pole in the off-shell $pd$ amplitude
obtained from an integral equation that resums both one-nucleon exchange as well
as $\OO(\alpha)$ Coulomb diagrams.  \textcite{Konig:2015aka} instead calculated
the binding-energy difference $B(\ThreeH)-B(\ThreeHe)$ as a perturbation
around an isospin-symmetric LO including a contribution missed in the earlier
calculation of \textcite{Konig:2014ufa}.  Once the logarithmic divergence
generated by $pp$ Coulomb effects is isolated and properly renormalized, the
\NLO \ThreeHe binding energy converges as the cutoff increases without an
isospin-breaking $3N$ force.  \textcite{Konig:2015aka} find $B(\ThreeHe) = (7.62
\pm 0.17)~\MeV$, supporting the perturbative nature of Coulomb in this bound
state.  The same conclusion was reached independently by
\textcite{Kirscher:2015zoa}.  These results suggest that for nuclear ground
states Coulomb is an \NLO effect and isospin-breaking $3N$ forces do not enter
up to this order.
{The same holds for $pd$ scattering for center-of-mass momenta
$k\gsim20~\MeV$, which \textcite{Konig:2016iny} showed to be
predicted from $nd$ scattering up to NLO.}

\paragraph{Dineutron constraints}

\textcite{Kirscher:2011zn} used Pionless EFT to constrain the neutron-neutron
($nn$) scattering length, for which there exist conflicting experimental
determinations of ${-}16.1\pm0.4~\fm$~\cite{Huhn:2000zz,Huhn:2001yk} and
${-}18.7\pm0.7~\fm$~\cite{GonzalezTrotter:1999zz,GonzalezTrotter:2006wz}.
Using a model-independent correlation between the difference of the $nn$ and
(Coulomb-modified) $pp$ scattering lengths on the one hand and the
\Triton-\ThreeHe binding-energy difference on the other,
\textcite{Kirscher:2011zn} extracted $\ann=(-22.9\pm 4.1)~\fm$ from an \LO
calculation where isospin-breaking, nonderivative $N\!N$ contact interactions
are included.

\textcite{Kirscher:2011zn} only considered negative values for $\ann$, thus
excluding the possibility of a bound shallow state, the existence of which would
correspond to $\ann$ large and positive.  Motivated by renewed experimental
interest in the existence of such a state, \textcite{Hammer:2014rba} revisited
the calculation and argued that the relevant parameter that enters the pionless
calculation is not $\ann$ directly, but rather its inverse, such that going from
large negative to large positive $\ann$ is only a small change.  Extending the
calculation of \textcite{Kirscher:2011zn} to \NLO, and taking into account the
new $pd$ counterterm identified by \textcite{Vanasse:2014kxa},
\textcite{Hammer:2014rba} concluded that Pionless EFT currently does not
exclude a bound dineutron state.

\subsubsection{Infrared regulators}
\label{infrareg}

Solving the EFT beyond the three-nucleon system poses significant technical
challenges.  All calculations so far have relied on the transition to the
Hamiltonian and a solution of the Schr\"odinger equation or one of its many-body
variants.  One way to mitigate difficulties is to introduce an IR regulator in
the form of a confining potential that produces discrete energy levels and,
together with the UV regulator, reduces the solution of the Schr\"odinger
equation to matrix inversion.

A simple choice is to confine the system to a periodic cubic box, as first
considered for EFT by \textcite{Muller:1999cp} and \textcite{Abe:2003fz}.
The case of $N\!N$ in Pionless EFT was dealt with by \textcite{Beane:2003da},
where a relation between phase shifts and energy levels within the box,
originally obtained by \textcite{Luscher:1986pf,Luscher:1990ux}, was rederived.
The relations between $N\!N$ LECs and ERE parameters for a large lattice
were found by \textcite{Seki:2005ns}.  Several papers have studied the
limit-cycle of the three-body system and the finite-volume corrections to
three-body binding energies in periodic cubic boxes
numerically~\cite{Kreuzer:2008bi,Kreuzer:2009jp,Kreuzer:2010ti,Kreuzer:2012sr}.
An analytical expression for the volume dependence of the three-body binding
energy in the unitary limit was obtained by \textcite{Meissner:2014dea} and
\textcite{Hansen:2016ync}.  \textcite{Konig:2017krd} have studied the volume
dependence of arbitrary $N$-body bound states, providing a more general
perspective that reproduces the leading exponential dependence of the explicit
three-body results just mentioned.  The formulation of the three-particle
quantization condition in a finite volume using the dibaryon formalism, which is
required for the extraction of scattering phase shifts from lattice
calculations, was considered by \textcite{Briceno:2012rv},
\textcite{Hammer:2017kms}, and \textcite{Hammer:2017uqm}.  {Alternative
approaches~\cite{Hansen:2014eka,Mai:2017vot} are reviewed by
\textcite{Hansen:2019nir}.}  $N\!d$ scattering in the
quartet $S$- and $P$-wave channels was calculated on a lattice by
\textcite{Elhatisari:2016hby}, and found to be in good agreement with continuum
results.  Other reactions, such as $np\to d\gamma$~\cite{Rupak:2013aue}
and $pp$ fusion~\cite{Rupak:2014xza}, are also accessible with this method.

Another widely employed confining potential is the harmonic oscillator, which
can also be deployed to EFT~\cite{Stetcu:2009ic}.  The analog of the L\"uscher
formula, due to \textcite{Busch:1998}, also follows from Pionless
EFT~\cite{Stetcu:2010xq,Luu:2010hw}.  Using this relation to determine the two-
and three-nucleon LECs from, respectively, $N\!N$ and $nd$ phase shifts,
\textcite{Rotureau:2011vf} generalized an earlier calculation for spin
$\nicefrac12$ fermions~\cite{Rotureau:2010uz} and reproduced previous \NLO
results for two- and three-nucleons in the limit of a wide harmonic oscillator.
{\textcite{Tolle:2010bq,Tolle:2012cx}} investigated the related
problem of up to 6 spinless bosons in a harmonic trap and provided explicit
expressions for the running coupling constants.  Using smeared contact
interactions, they improved the convergence of the energy levels considerably.
{A recent refinement of the Busch formula has been worked out by
\textcite{Zhang:2019cai}.}

\subsubsection{Four nucleons}
\label{sec:Helion}

One of the virtues of encoding the Efimov effect in the three-body force
is that its consequences for systems with more nucleons can be assessed in a
model-independent way.  A potential obstacle is the {relatively strong
binding} of the $A=4$ ground state, the alpha particle (\FourHe), whose binding
energy $B(\FourHe) = 28.296~\MeV$ can be associated with a momentum scale
$\gamma_\alpha = \sqrt{\MN B(\FourHe)/2} \sim 110~\MeV$ that is not necessarily
within the realm of Pionless EFT.

\begin{figure}[tb]
\centering
\includegraphics[width=0.9\columnwidth]{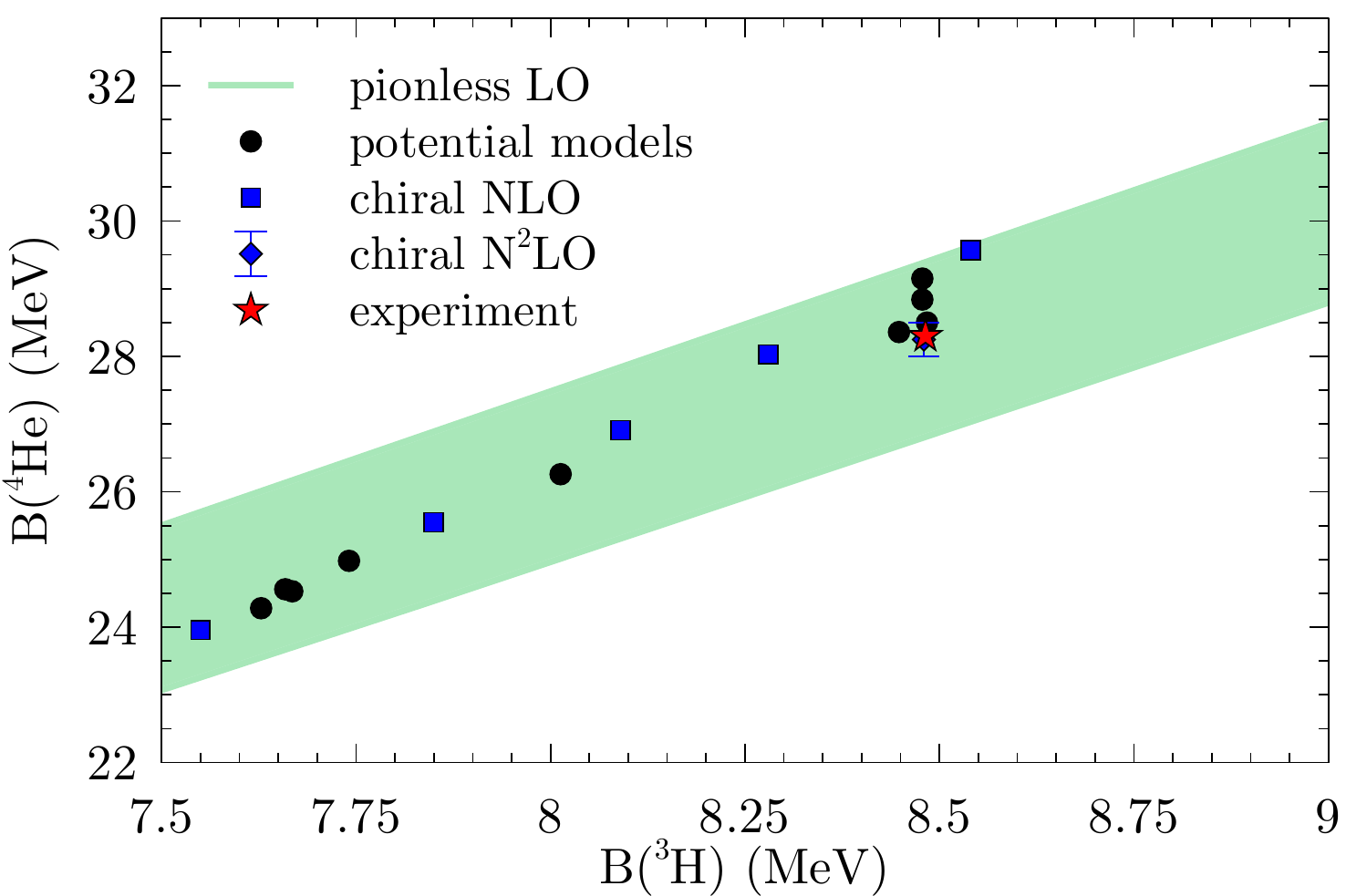}
\caption{Correlation between alpha-particle and triton binding energies,
 respectively $B(\FourHe)$ and $B(\Triton)$ (Tjon line){,
at LO in Pionless EFT, compared to results from various
chiral and phenomenological potentials, and experiment.
The band indicates an estimate of higher-order corrections.}
}
\label{fig:TjonLine}
\end{figure}

The application of Pionless EFT to the four-nucleon system was initiated
by \textcite{Platter:2004zs}, extending their previous work on the four-boson
system with large two-body scattering length~\cite{Platter:2004he}.  It was
found that {\it no} $4N$ force is required for renormalization at LO, \ie, the
alpha-particle binding energy converges as a function of increasing UV cutoff.
Low-energy four-nucleon observables are then determined at LO only by two- and
three-body input parameters.  This means that, as for the Phillips line
(Fig. \ref{fig:PhillipsLine}), Pionless EFT provides a natural explanation also
for the phenomenological Tjon line~\cite{Tjon:1975zz}, an empirical correlation
between $B(\FourHe)$ and $B(\Triton)$.  The surprising success of this LO
calculation~\cite{Platter:2004zs} is apparent
{in Fig.~\ref{fig:TjonLine}.}
The renormalizability and good description of the alpha-particle binding at LO
found by \textcite{Platter:2004zs} have been confirmed in other calculations,
for example using the resonating-group~\cite{Kirscher:2009aj,Kirscher:2015yda}
and auxiliary-field diffusion Monte Carlo \cite{Contessi:2017rww} methods.

\textcite{Hammer:2006ct} later studied a four-body generalization of the Efimov
effect in the four-boson system, demonstrating that the three-body ground state
is associated with two four-body states, one very near the particle-trimer
threshold, another deeper by a factor $\simeq 4$.  The alpha-particle ground
state is 3.7 times more bound than helion, and it has an excited, $0^+$ state
just below the neutron-helion threshold.  The $0^+$ excited state was obtained
at LO by \textcite{Stetcu:2006ey}, who solved the Schr\"odinger equation in a
harmonic-oscillator basis, as done in the no-core shell model approach.  The
three LECs were fitted to the deuteron, triton and alpha ground-state energies,
and the binding energy of the excited state, extrapolated in both UV and IR
regulators, was found to be within 10\% of the experimental value.

These calculations did not include the Coulomb interaction, which is consistent
with the power counting developed later by \textcite{Konig:2014ufa}.
{Coulomb interactions} have so
far been included only in calculations based on effective potentials {solved}
nonperturbatively, as suggested originally by Weinberg in the context of Chiral
EFT~\cite{Weinberg:1991um}.  {A variety} of methods was used to solve the
Schr\"odinger equation: resonating group~\cite{Kirscher:2009aj}, stochastic
variational~\cite{Lensky:2016djr}, and no-core shell
model~\cite{Bansal:2017pwn}.  {While results improve within a range of
cutoff values, the resummation of subleading interactions
(Sec.~\ref{subsubsec:regren}) limits this range to small values and prevents
conclusions about the RG beyond LO based on these calculations.}
{In the first study of perturbative range corrections in $A\geq4$ systems,
\textcite{Bazak:2018qnu} found that a four-body forces is required to
renormalize the universal four-boson system at \NLO.
This result directly carries over to {Pionless} EFT and implies that an
additional
observable (most conveniently taken to be the \FourHe binding energy) is
required as input at \NLO to set the scale of the four-body force.}

\textcite{Kirscher:2009aj} and \textcite{Kirscher:2011uc} have pioneered the
calculation of nucleon-trinucleon scattering in Pionless EFT.  New correlations
between the neutron-helion and neutron-triton scattering lengths, and the triton
binding energy were identified.  Proton-helion scattering was found in good
agreement with an existing {phase-shift analysis}.  These successes have
been
tempered by an LO calculation~\cite{Deltuva:2011djw} of the lowest, $0^-$
resonance in neutron-helion scattering---without Coulomb and an explicit $3N$
force, but with specific cutoff values for which the helion energy is correct.
However, it is difficult to check the absence of regularization artifacts due
to the absence of an explicit $3N$ interaction.  Except for this one
calculation, all evidence so far points to an unexpected triumph of the
theory for $A=4$.

\subsubsection{Beyond four nucleons}

Nuclear binding momenta generally increase with the number of nucleons, but
it is not clear what momentum scale (\eg, total binding energy or binding energy
per nucleon) is most relevant for the EFT power counting.  Consequently, it is
an open question up to which number of nucleons Pionless EFT should work.
Successful applications to \FourHe, which is already significantly more deeply
bound than \ThreeHe and \ThreeH, indicate that the binding energy per nucleon
might be the relevant scale to estimate the binding momentum, but this remains
to be firmly tested.

The first Pionless EFT calculation beyond four nucleons was carried out by
\textcite{Stetcu:2006ey} with the no-core shell model.  With the same parameters
that led to an excellent postdiction of the \FourHe excited state, the
ground-state energy of \SixLi came out at 70\% of the experimental value, which
is consistent with the \apriori \LO uncertainty estimate from the pionless power
counting.  With the resonating-group method, \textcite{Kirscher:2015ana} found
that Pionless EFT does not predict a bound five-nucleon state, and carried out
an exploratory study for the \SixHe system.  It was concluded that Pionless EFT
appears to support a shallow \SixHe bound state, but a lack of numerical
convergence prevented a strong assertion.

More recently, \textcite{Contessi:2017rww} used the auxiliary-field diffusion
Monte Carlo method to study \FourHe and \SixteenO at LO.  No evidence was found
for a \SixteenO that is stable with respect to breakup into four alpha
particles.  LO Pionless EFT does not fail to provide sufficient saturation, but
a small effect such as the \SixteenO energy relative to four alphas ($\sim 15\%$
of the \SixteenO binding energy) requires a higher-order calculation.
\textcite{Bansal:2017pwn} used the coupled-cluster method to study the same
systems and also $^{40}$Ca, with qualitatively similar results at \LO. With \NLO
interactions treated exactly, they found \SixteenO and ${}^{40}$Ca to be stable,
and in reasonable agreement with experiment.

{Leading-order} calculations all show convergence with increasing cutoff in
the absence of $4N$ and higher-body forces, as is the case for bosons
\cite{Bazak:2016wxm}.
{\textcite{Bazak:2018qnu} show that the five- and six-boson
systems are renormalized by the four-body force at \NLO and conjecture that
yet-higher-body forces enter at subsequent orders.  For nucleons, the Pauli
principle is expected to suppress these forces relative to the bosonic case
because at least two derivatives are required.}
{Together, \NLO results suggest that Pionless EFT
\emph{might} work over a much wider range of $A$ than originally anticipated.}

There have been ambitious attempts to consider even larger systems.
\textcite{Kirscher:2017xpj} has recently investigated the possibility that a
sufficiently large number of neutrons might bind.  The properties of dilute,
low-temperature neutron matter on a spacetime
lattice~\cite{Lee:2004qd,Lee:2005is,Lee:2005it,Abe:2007fe,Abe:2007ff}
were found in qualitative agreement with potential-model calculations and
expectations from other fermionic systems.  While Pionless EFT reproduces the
longstanding results for the low-density expansion in a uniform Fermi
system~\cite{Hammer:2000xg}, various resummations relevant for nuclear matter
have been discussed by \textcite{Schafer:2005kg}, \textcite{Kaiser:2011cg}, and
\textcite{Kaiser:2012sr}.  Convergence of Pionless EFT at saturation density is,
however, not obvious.

\subsubsection{External currents and reactions}
\label{sec:Currents-ThreeBody}

The coupling to external currents works exactly as in the two-body sector (see
Sec.~\ref{sec:Currents-TwoBody}).  Given the increased technical challenges,
there are fewer {calculations}, and they mostly tackle triton and helion
properties.  Within this limited scope, they confirm the convergence of Pionless
EFT at low energies.

The simplest observable is the triton charge form factor, in particular the
charge radius, which determines the leading momentum dependence.
\textcite{Platter:2005sj} used an effective quantum-mechanical framework to
obtain the form factor in the impulse approximation, \ie, considering the
electric charge operator between triton wavefunctions obtained by solving a
Faddeev equation for the effective potential derived from the pionless
Lagrangian at LO.  Similar to the Phillips line (see Sec.~\ref{sec:Triton}), the
existence of a single LO three-body parameter in Pionless EFT explains a
correlation between the triton binding energy and charge radius that had
previously been observed with different potential models~\cite{Friar:1985aa}.
An N$^2$LO calculation was reported by \textcite{Sadeghi:2009dm}, but few
details are given.  Recently, the fully perturbative treatment of higher-order
corrections was extended to the triton and helion charge radii at
\NNLO~\cite{Vanasse:2015fph,Vanasse:2017kgh}, and magnetic moments and radii at
\NLO~\cite{Vanasse:2017kgh}.  Even though Coulomb interactions were neglected
for \ThreeHe, good agreement with experiment was found.  Analogous results were
obtained with a resummation of higher-order effects by
\textcite{Lensky:2016djr}, who also showed the correlation between the \FourHe
charge radius and other three- and four-nucleon observables.

\textcite{Sadeghi:2004es}, \textcite{Sadeghi:2006fc}, and
\textcite{Sadeghi:2007qy} calculated the $nd \to \ThreeH\,\gamma$ capture cross
section, finding good agreement with available experimental data at \NNLO.
\textcite{Sadeghi:2009rf} calculated the inverse process of triton
photodisintegration, which features the same amplitudes due to time-reversal
symmetry.  Several significant flaws in the calculation of
\textcite{Sadeghi:2006fc} were identified by \textcite{Arani:2014qsa}, who
presented an updated calculation.  At \NNLO, they still find reasonable
agreement of the thermal $nd$ capture cross section with experiment.  The total
cross section for the related reaction $pd \to \ThreeHe\,\gamma$ was calculated
to \NLO in a range of energies amenable to perturbative Coulomb interactions by
\textcite{Nematollahi:2016zjg}.  The $3N$ force was fixed by the helion binding
energy and data were reasonably described considering the uncertainty of the
calculation.  The even more challenging process of deuteron-deuteron radiative
capture ($dd\to \FourHe\,\gamma$) was calculated at \LO by
\textcite{Sadeghi:2014yia}, with the $3N$ force fitted to the alpha-particle
energy.  While available data for the astrophysical $S$-factor are apparently
well described, a lack of technical details makes it difficult to assess
the validity of the calculation.

New ground was broken with the extension to electroweak processes made by
\textcite{De-Leon:2016wyu}, who studied triton $\beta$ decay to \NLO.  This
work establishes a new way of fixing the LEC $L_{1,A}$ of the axial-vector
counterpart of Eq.~\eqref{eq:L-mag-L1L2}, which is relevant for other
electroweak processes ($pp$ fusion, in particular) as well.  Calculations like
this reveal the potential of Pionless EFT to tackle interesting reactions
involving more than two nucleons{, based for example on the general framework
developed by \textcite{De-Leon:2019dqq}.}

\subsection{Outstanding issues and current trends}
\label{sec:pionlesssummary}

Pionless EFT has fulfilled the longstanding goal of a renormalizable quantum
field theory for nuclear physics.  Although it has a narrow regime of strict
validity, it seems to apply to, at least, $A\le 3$ bound states and possibly
to extend to $A=4$ and beyond.  RG invariance, combined with the fine tuning
that places two-body bound states near zero energy, has led to a power counting
that flies in the face of NDA, as summarized for the potential in
Fig.~\ref{fig:pionlesspot}.  Yet, unresolved questions remain, such as:
\begin{itemize}
\item
How far in $A$ can we describe nuclei within this framework?  So far all LO
calculations ($A\le 40$) give binding energies in agreement with experiment
within the expected theoretical uncertainty, but finer details such as relative
energies and thresholds are not reproduced.  Calculations for $A=4,16,40$ where
subleading interactions are resummed reinforce the surprising success of LO.
However, at the moment, no calculation exists {for more than four} nucleons
where \NLO and higher orders are treated perturbatively.  There also remain
issues about the power counting of Coulomb interactions and other
isospin-breaking interactions.  Higher-order calculations for {$A > 4$} are
sorely needed.
\item
{\textcite{Wigner:1936dx,Wigner:1937zz}} proposed an $SU(4)$
spin-isospin symmetry to explain the strong binding of nuclei containing integer
numbers of alpha particles.  Since the $3N$ force~\eqref{eq:L-NN-3} is $SU(4)$
symmetric~\cite{Bedaque:1999ve}, one cannot but wonder whether there are signs
of $SU(4)$ symmetry also in light nuclei.  It was shown by
\textcite{Chen:2004rq} that binding energies of $A\le 4$ nuclei satisfy
inequalities obtained from $SU(4)$ symmetry.  Accordingly,
\textcite{Vanasse:2016umz} developed an expansion around an $SU(4)$-symmetric LO
based on average \OneSNot and \ThreeSOne scattering lengths.  They showed that
this expansion is promising also for observables other than binding energies,
since it converges well for the triton charge radius up to \NLO in the
symmetry-breaking parameter (and including range corrections as well).
\item
In the same spirit, how far can we push the expansion around the nontrivial
fixed point of the $N\!N$ amplitude, \ie, the unitary limit where both the
deuteron and the \OneSNot virtual state have zero energy?  In this limit the LO
EFT has not only exact $SU(4)$ symmetry but also discrete scale invariance:
while the two-body amplitude is invariant under continuous scale
transformations~\cite{Mehen:1999nd}, the three-body force \eqref{eq:H0-0} is
symmetric only under discrete scale
changes~\cite{Bedaque:1998kg,Bedaque:1998km}.  This remaining symmetry leads to
Efimov states in the three-body system and its descendants in the
four-~\cite{Hammer:2006ct} and higher-body systems.  \textcite{Konig:2016utl}
and \textcite{Konig:2016iny}, generalizing an earlier approach to the \OneSNot
channel~\cite{Konig:2015aka}, have proposed an expansion around the
unitary (or unitarity) limit also in the \ThreeSOne channel: expansions in both
$1/(Q a_{\rm t})$ and $1/(Q a_{\mathrm{s}})$ are added to the standard Pionless
EFT expansion.  A \emph{single} LO parameter $\Lambda_*$ provides the
nonperturbative scaffolding on top of which more quantitative results are built
in by perturbation theory.  This quite radical expansion appears to converge
remarkably well for three- and four-nucleon binding
energies~\cite{Konig:2016utl,Konig:2016iny}.  At LO all binding energies are
functions of $\Lambda_*$, and for bosons~\cite{Carlson:2017txq} they saturate
according to the liquid-drop formula.  The correlation between nuclear-matter
saturation energy and density expressed in the Coester
line~\cite{Coester:1970ai} would emerge from variation of
$\Lambda_*$~\cite{Kolck:2017zzf} just as the Tjon line---\emph{if}, that
is, Pionless EFT holds all the way to heavy nuclei.  {Related work that
aims to simplify nuclear physics based on the closeness of the real world to
the unitarity limit and/or Wigner $SU(4)$ limit has been carried out by
\textcite{Kievsky:2015dtk,Kievsky:2018xsl,Gattobigio:2019omi,Lu:2018bat}.}
\end{itemize}

\section{Halo/Cluster EFT} \label{sec:halo}
\subsection{Motivation}

In this section we discuss efforts to go one step further in the application
of low-energy universality by including tightly bound clusters of nucleons as
explicit fields in the effective Lagrangian.  This Halo/Cluster EFT
framework is appropriate for halo nuclei and nuclei with a cluster structure.
In both cases, the energy required to remove clusters or halo nucleons,
characterized by a momentum scale $\Mlo$, is much smaller than the energy
required to break clusters apart, associated with a momentum scale $\Mhi$.
The classic example is $^6$He, where the energy to separate two neutrons
from an alpha-particle core is $S_{2n}\simeq 0.975~\MeV \ll B(\FourHe)$.  This
class of systems can be thought of as nucleons orbiting one or more clusters,
all separated by distances much larger than the cluster sizes.  They typically
lie at the limits of nuclear stability represented in the nuclear chart by the
so-called driplines, and are target of a vigorous experimental program at
rare-isotope facilities worldwide.  As we discuss below, they can display more
than one low-momentum scale, \eg{,} when Coulomb interactions are present or
when a cluster has an isolated low-energy excited state.

As in Eq.~\eqref{Texp}, observables are expanded in powers of $\Mlo/\Mhi$ and
$Q/\Mhi$, where $Q$ is a typical momentum.  While Halo/Cluster EFT is
mathematically similar to the Pionless EFT for nucleons discussed in the
previous section---and in fact is a theory without explicit pions by itself,
becoming Pionless EFT for light nuclei when the cores are nucleons---there
are a number of new aspects. First, higher partial waves between clusters,
or between clusters and nucleons, are often enhanced, as for the $n\alpha$
scattering relevant for $^6$He.  This causes a richer structure already in the
two-body sector and requires modified power counting schemes.  Second,
the antisymmetrization between nucleons in a cluster (which are not active
degrees of freedom) and halo nucleons is not explicit.

{One} might ask what kind of error is introduced by using explicit fields
for tightly bound clusters.  The effect on observables of exchanging nucleons
in the core with halo nucleons is governed by the overlap of the wavefunctions
of the halo nucleons with the wavefunctions of the core nucleons.  Since the
range of the former is $\Mlo^{-1}$ while the range of the latter is
$\Mhi^{-1}$, this overlap is suppressed by $\Mlo/\Mhi$ compared to the overlap
of two halo nucleons.  Therefore, these effects are controlled by the EFT
expansion in $\Mlo/\Mhi$ and are encoded in the LECs of Halo/Cluster EFT.  The
same argument applies for nucleons in different, widely separated clusters.
For momenta $Q$ of the order of the breakdown scale $\Mhi$ or above, when
distances compared to the core size are probed, full antisymmetrization and
other short-range physics have to be included explicitly.

Halo/Cluster EFT exploits the scale separation between $\Mlo$ and $\Mhi$
independently of the mechanism creating it.  Thus it complements \abinitio
approaches to halo nuclei by zooming out to large distances and providing
universal relations between different few-body observables.  These relations can
be combined with input from an underlying EFT or experiment to predict halo
properties.  Moreover, they allow us to test the consistency of different
approaches and/or experiments.  A particular strength lies in the possibility to
describe the electroweak structure and reactions of halo nuclei in a
model-independent way with controlled error estimates.

Halo/Cluster EFT can be viewed as a generalization of nuclear cluster models
and is usually referred to simply as Halo
EFT~\cite{Bertulani:2002sz,Bedaque:2003wa}.  We give a brief overview here,
starting with $S$-wave neutron halos and Efimov states in Secs.~\ref{sec:Shalo}
and \ref{sec:efimovhalos}, respectively.  The complementarity with \abinitio
methods, useful to explore heavy halos, is discussed in
Sec.~\ref{sec:halo+abinitio}, before higher partial waves are tackled in
Sec.~\ref{sec:highpar}.  We show how Halo EFT connects with electromagnetic
processes in Sec.~\ref{sec:haloelectric}, before sketching the changes needed
for proton halos (Sec.~\ref{sec:phalos}) and multi-cluster systems
(Sec.~\ref{sec:clusters}), and ending with an outlook (Sec.~\ref{sec:haloook}).
Our emphasis is complementary to Sec.~\ref{sec:pionless}.  A more in-depth
discussion can be found in the recent review of Halo EFT by
\textcite{Hammer:2017tjm}.

\subsection{$S$-wave neutron halos}
\label{sec:Shalo}

We start with the case of $S$-wave halo nuclei or cluster states without
Coulomb interactions.  For definiteness, we consider one- and
two-neutron halo nuclei using the formalism of \textcite{Hagen:2013xga}.  The
extension to other cases is straightforward.  We also restrict our analysis to
LO in $\Mlo/\Mhi$.  Higher-order effects can be included as discussed in the
previous section for Pionless EFT.

The effective Lagrangian for neutrons ($n$) and {a spinless core} ($c$)
can be  written as the sum of one-, two- and three-body contributions, $\LL =
\LL_{1b} + \LL_{2b} + \LL_{3b} + \cdots$, where the ellipses denote higher-order
terms and
\begin{align}
 \LL_{1b} &= \psi_0^\dagger
  \left(\ii\partial_0 + \frac{\Laplace}{2m_0}\right)\psi_0
  +
  \psi_1^{\dagger}
  \left(i\partial_0 + \frac{\Laplace}{2m_1}\right) \psi_1 \,,
  \nonumber\\
 \LL_{2b} &= \Delta_1 \,
  d_1^{\dagger}
  d_1
  - g_1\left[
  d_1^{\dagger} \,
  (\psi_1 \, \psi_0) + \hc\right]
  \nonumber\\
  &\hspace{1em}\null+ \Delta_0 \, d_0^\dagger d_0
  - \frac{g_0}{2}\left[d_0^\dagger \,
  (
  \psi_1^\text{\;T} P \,
  \psi_1) +  \hc \right] \, ,
  \nonumber\\
 \LL_{3b} &= \Omega \, t^\dagger \, t
  - h\left[t^\dagger (\psi_0\,d_0) + (\psi_0\,d_0)^\dagger t\right] \, .
\label{eq:lag_05}
\end{align}
The notation is slightly changed compared to the previous section
(\cf~Eq.~(\ref{eq:L-NN-simple-d})) in order to efficiently account for neutron
and core fields, the Pauli spinor $\psi_1$ and the scalar $\psi_0$,
respectively.  The two-body part $\LL_{2b}$ includes two dimerons, the scalar
$d_0$ corresponding to an $\OneSNot$ $nn$ pair and the Pauli spinor $d_1$ for a
$cn$ pair.  $P=\ii\sigma_2/\sqrt{2}$ projects the two neutrons on the spin
singlet.  Finally, $\LL_{3b}$ represents the three-body interaction written in
terms of a  trimeron auxiliary field~\cite{Bedaque:2002yg}, which is
particularly
useful for form{-}factor calculations~\cite{Hagen:2013xga,Vanasse:2015fph}.
It includes the bare trimeron residual mass $\Omega$ and the bare coupling $h$
of the trimeron $t$ to the $d_0$ dimeron ($nn$) and the $c$ field $\psi_0$.
Only the parameter combination $h^2/\Omega$ contributes to observables at LO.
As in Pionless EFT, there exists a whole class of equivalent theories with
three-body forces acting in different channels.  Integrating out the auxiliary
fields, different choices of $\LL_{2b}$ and $\LL_{3b}$ can be transformed into
the same theory without dimeron and trimeron fields up to four- and higher-body
interactions.

In the following, we focus on the properties of the $cn$, $nn$, and $cnn$
systems.  For compact notation, we define the mass parameters:
\begin{eqnarray}
 &&M_\text{tot} = m_0+2m_1\, , \;
  M_i = M_\text{tot}-m_i \, ,\;
  m_{ij}= M_i-m_j\,,
  \nonumber\\
 &&\mu_i = \frac{m_0m_1^2}{m_iM_i}\, ,  \;
  \tilde \mu_i = \frac{m_i M_i}{M_\text{tot}} \,.
\label{eq:lag_40}
\end{eqnarray}

The diagrams for the dressed dimeron propagator, Fig.~\ref{fig:DibaryonProp},
are completely analogous to that for the dibaryon field discussed in
Sec.~\ref{sec:Dibaryons}.  At LO, the full dimeron propagator for the dimeron
$d_i$
is
\begin{equation}
 \ii D_i(p_0, \textbf{p} ) = \frac{2\pi\ii}{s_i \,g_i^2\mu_i} \left[
 1/a_i - \sqrt{ -2\mu_i \tilde{p}_0 - \ii\varepsilon} \,\right]^{-1} \, ,
\label{eq:2ps_40}
\end{equation}
where $\tilde{p}_0=p_0-\textbf{p}^2/(2M_i)$ and $s_i=\delta_{i0}/2 +
\delta_{i1}$ is a symmetry factor.  As before, $a_i$ stands for the respective
scattering length.  For positive $a_i$, the propagator has a bound-state pole on
the first Riemann sheet with residue $Z_i = 2\pi/(s_i \, g_i^2\mu_i^2 a_i)$.
For negative $a_i$, there is a pole on the second sheet corresponding to a
virtual state.

The leading correction to the propagator~\eqref{eq:2ps_40} is due to the
effective range.  It can be included by making the dimeron fields dynamical as
discussed in Sec.~\ref{sec:Dibaryons}.  Here, we stay at LO in the EFT expansion
and neglect effective-range corrections.  The pole momentum $\gamma_i$ is then
given by $\gamma_i=1/a_i$.

Observables in the $cnn$ system can be obtained from the $T$-matrix for the
scattering process of a dimeron and a particle.   The  universal properties and
structure of two-neutron halos were also explored in an effective
quantum-mechanical framework~\cite{Canham:2008jd,Canham:2009xg,Acharya:2013aea}
by solving the Faddeev equations for an effective potential reflecting the
expansion in $\Mlo/\Mhi$ and for the renormalized zero-range
model~\cite{Amorim:1997mq,Delfino:2007zu}.  For a review of the latter work
see~\cite{Frederico:2012xh}.

We consider the center-of-mass frame, in which the on-shell $T$-matrix only
depends on the total energy $E$ and the relative momenta in the ingoing and
outgoing channels $\textbf{p}$ and $\textbf{k}$, respectively. External dimeron
legs are renormalized with the wavefunction renormalization factors
$\sqrt{|Z_i|}$.  The absolute value is only required for $i=0$ because $Z_0<0$,
corresponding to the unbound $nn$ pair.  Here, the factor provides a convenient
redefinition of the amplitude but has no physical significance.

There are two possibilities for the initial or final state, depending on the
identity of the particle and dimeron.  Here we label the $T$-matrix element
$T_{ij}$ by the index $i$ ($j$) of the dimeron and particle in the incoming
(outgoing) channel.  Keeping the matrix structure of $T_{ij}$ implicit, the
integral equation for the $T$ {} matrix is given by
Fig.~\ref{fig:nd-IntEq-Q}
with the substitution in Fig.~\ref{fig:ThreeBodyForce}, where the contact
three-body coupling is $h^2/\Omega$.
The $T$ {} matrix can be decomposed into
partial-wave contributions $T_{l m,l'm'}=\delta_{ll'}\delta_{mm'}T_{l}$.
The resulting $2\times2$-matrix integral equation for angular momentum $l$ is
a generalization of the Skorniakov-Ter-Martyrosian
equation~\cite{Skorniakov:1957aa} and reads
\begin{multline}
 T_{l}(E,p,k) = \int_0^\Lambda \! \dd q \, R_{l}(E,p,q) \,
 \bar{D}(E, q) \, T_{l}(E,q,k) \\
 \null + R_{l}(E,p,k)
\label{eq:3ps_30}
\end{multline}
when a sharp momentum-space cutoff $\Lambda$ is imposed on the loop momentum in
the three-body sector.  For simplicity we focus on the $S$ wave, $l=0$, and
drop the subscript $l$ on $R$ and $T$.  The components of the interaction matrix
$R$ are given by
\begin{align}
 R_{ij}(E,p,k) &= \frac{ 2\pi \, \chi_{ij} \,m_{ij}}{
 \left|a_i a_j s_i s_j\right|^{\frac{1}{2}} \mu_i\mu_j}
 \, \frac{1}{pk} \ Q_0\big(c_{ij}\big)
 - \delta_{i0} \delta_{j0} H \,, \nonumber \\
 c_{ij} &= \frac{m_{ij}}{pk} \left(\frac{p^2}{2\mu_j} +
 \frac{k^2}{2\mu_{i}} -E - \ii\varepsilon\right) \,,
\label{eq:3ps_40}
\end{align}
where $\chi_{ij}=1-\delta_{i0}\delta_{j0}$ and $Q_0$ is a Legendre function of
the second kind.  Moreover, $H=|Z_0|h^2/\Omega$ is the dimensionless three-body
coupling defined in Eq.~(\ref{eq:H0-0}) which depends log-periodically on the
cutoff $\Lambda$.  It only contributes for angular momentum $l=0$.  The dimeron
matrix is diagonal in the channel indices: $\bar{D} =
\text{diag}(\bar{D}_0,\bar{D}_1)$ with
\begin{equation}
 \bar{D}_i(E,q) = \frac{\mu_i |a_i|q^ 2}{2\pi^2} \ \left[-1/a_i
 + \sqrt{-2\mu_i\tilde{E}- \ii\varepsilon}\right]^{-1} \, ,
\label{eq:3ps_50}
\end{equation}
and $\tilde{E}=E-q^2/(2\tilde\mu_i)$.

{The transition amplitude near the energy of an $S$-wave three-body bound
state can be decomposed into a regular and an irregular part.}
This yields the homogeneous bound-state equation
\begin{equation}
 \mathcal B (p)= \int_0^\Lambda \dd q \, R(E,p,q) \, \bar{D}(E, q) \,
 \mathcal B(q) \,,
\label{eq:3ps_70}
\end{equation}
which has nontrivial solutions only at the bound-state energy $E=-B_{cnn}$.

For a given cutoff $\Lambda$, we can fix the unknown three-body parameter $H$
such that Eq.~\eqref{eq:3ps_70} has a solution at the desired value
$E=-B_{cnn}$.  However, any other three-body observable can be used as well.  In
this way, the three-body coupling is renormalized and other three-body
observables (including other three-body bound states) can be predicted.  In
particular, Eq.~\eqref{eq:3ps_30} can be solved numerically in order to
determine the $T$ matrix for three-body scattering observables.
Since the two-neutron system is not bound, only the element $T_{11}$ describes a
particle-dimeron scattering process, namely the scattering of a neutron from a
$cn$ bound state at energy $E=p^2/(2\tilde\mu_1) - 1/(2\mu_1 a_1^2)$:
\begin{equation}
 T_{11}\!\left(E,\,p,\,p\right)
 = \frac{2\pi}{\tilde \mu_1} \, \frac{1}{p\cot\delta_{cn-n}(p)- \ii p} {\,.}
\label{eq:3ps_75}
\end{equation}
{The other elements contribute to three-body scattering and
breakup.}

A fully perturbative extension of this formalism to NLO was recently presented
by \textcite{Vanasse:2016hgn}. NLO equations with resummed range corrections
were previously given by~\textcite{Canham:2009xg}.

\subsection{Excited Efimov states in halo nuclei}
\label{sec:efimovhalos}

The bound-state solutions of Eq.~(\ref{eq:3ps_70}) are a specific variant of
Efimov states~\cite{Efimov:1970zz,Efimov:1973awb}.  Thus the Efimov effect
provides a natural binding mechanism for two-neutron halos with dominantly
$S$-wave interactions.  However, the contributions of higher partial waves and
partial-wave mixing complicate the situation.  While Halo EFT naturally
accommodates resonant interactions in higher partial waves as discussed in
subsection~\ref{sec:highpar}, there is no Efimov effect in this
case{,
see, \eg, \cite{PhysRevA.77.043611,Nishida:2011np,Braaten:2011vf}.}
A general overview of Efimov states
in nuclear and particle physics was given by \textcite{Hammer:2010kp}.
Here, we review the possibility of identifying Efimov states in halo nuclei.

\begin{figure}[tb]
 \centering
 \includegraphics[width=6cm,clip=true]{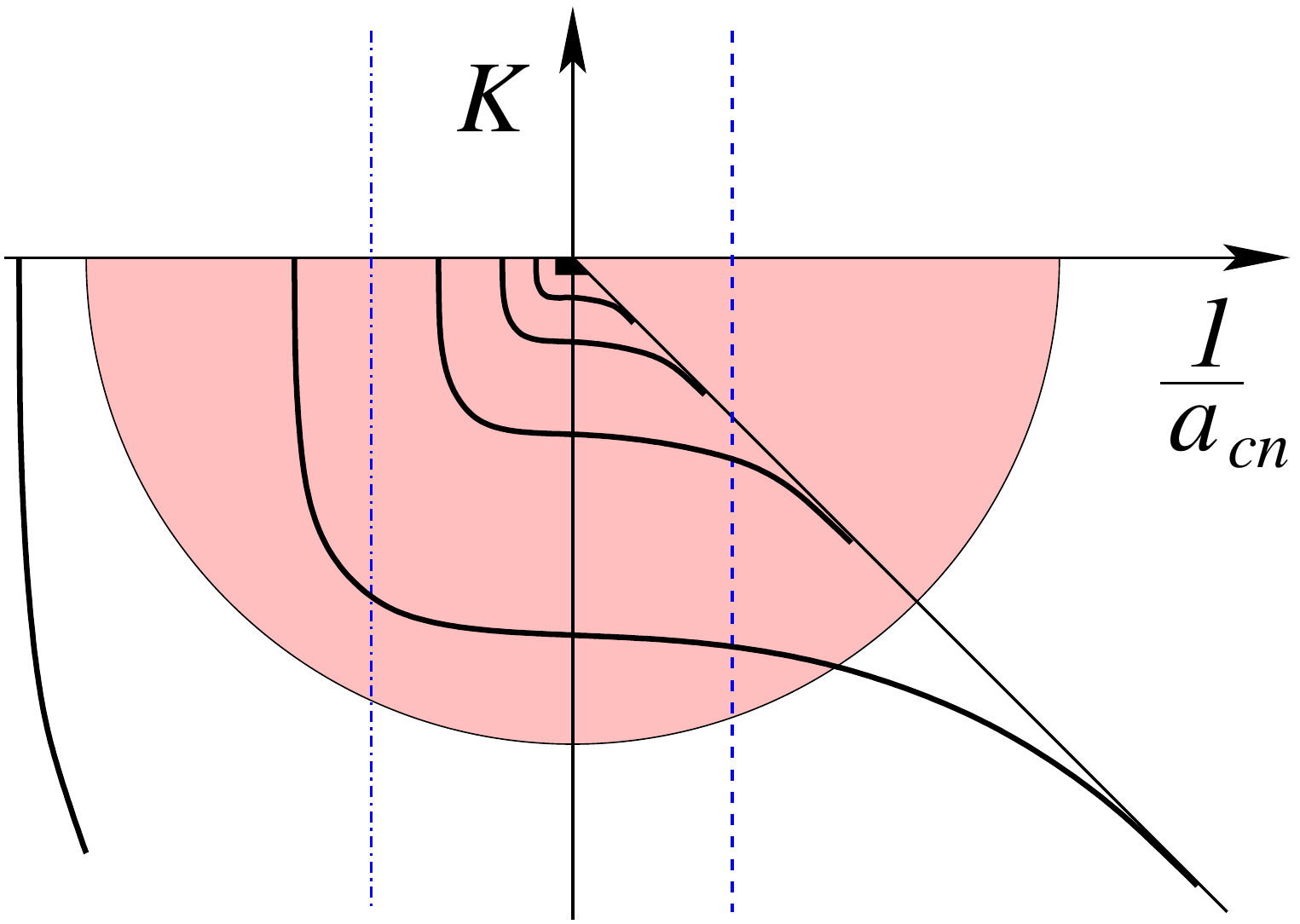}
 \caption{Illustration of the Efimov spectrum as a function of the three-body
 momentum $K$ (\cf~Eq.~\eqref{eq:3bbm}) and the inverse neutron-core
 scattering length $a_{cn}^{-1}$.  The diagonal line in the fourth quadrant
 represents the neutron-core threshold.  The solid lines indicate the Efimov
 states.  The window of universality for bound states is represented by the
 hashed half circle, while the dashed and dot-dashed lines indicate a typical
 system with $a_{cn}>0$ and $a_{cn}<0$, respectively.
}
\label{fig:efiplot}
\end{figure}

Since the strength of the interaction between the neutrons and the core is
fixed, the identification of Efimov physics is more delicate than for
ultracold atoms, where the effective scattering length can be dialed through an
external magnetic field.  In particular, the log-periodic dependence of
observables on the scattering length cannot be used to identify Efimov physics.
Instead one may look for excited states which (approximately) satisfy the
universal scaling relation for Efimov
states~\cite{Efimov:1970zz,Efimov:1973awb}.  Note that there are two relevant
scattering lengths for a $cnn$ system, $a_0\equiv a_{nn}$ and $a_1\equiv
a_{cn}$.  Since $a_{nn}$ is the same for all halo nuclei and negative, we focus
only on the dependence on $a_{cn}$.  We define the three-body momentum as
\begin{equation}
  K = \mathrm{sgn}(E)\sqrt{2\tilde{\mu}_1 |E|}\,,
  \label{eq:3bbm}
\end{equation}
where the sign of the square root is taken as the sign of the energy $E$.
The schematic dependence of the Efimov spectrum on $K$ and the inverse
neutron-core scattering length $a_{cn}^{-1}$ is illustrated in
Fig.~\ref{fig:efiplot}.  The breakdown scale $\Mhi$ defines a region outside of
which details of short-range physics matter and the bound states cease to be
universal.  Two typical situations are shown, with two universal states (at
$a_{cn}>0$) and one universal state (at $a_{cn}<0$).

In the hypothetical unitary limit $a_{cn}^{-1}=a_{nn}^{-1}=0$, the Efimov
spectrum becomes geometric,
\begin{equation}
 K^{(n)} = -\lambda_0^{-n} \kappa_*\,,
\label{B3-Efimov-uni}
\end{equation}
where $\lambda_0 = e^{\pi/ s_0}$ is the discrete scaling factor and $\kappa_*$
is the binding momentum of the state with label $n=0$. In general, $s_0$ and
$\lambda_0$ depend on the number of interacting pairs and the masses and
symmetry properties of the particles, {and} $\lambda_0\approx 22.7$ for the
equal-mass nucleons discussed in Sec.~\ref{sec:pionless}.  The value of
$\kappa_*$ is related by a {(regulator-dependent)} constant factor
to the three-body parameter
$\Lambda_\star$ that determines the three-body force $H$ in Eq.~\eqref{eq:H0-0}.
{An explicit value for the case of identical bosons is given}
{by \textcite{Braaten:2004rn}.}
The spectrum shown in Fig.~\ref{fig:efiplot} is invariant under discrete
scaling transformations with $\lambda_0$:
\begin{equation}
 \kappa_* \longrightarrow \kappa_* \,,
 \quad
 a_{cn}  \longrightarrow  \lambda_0^{m} a_{cn} \,,
 \quad
 K  \longrightarrow  \lambda_0^{-m} K \,,
\label{dssym}
\end{equation}
where $m$ is any integer.  This discrete scale invariance holds for all
few-body observables and is a clear signature of an RG limit cycle in the
three-body system~\cite{Bedaque:1998kg}.

If more particles are added, no new parameters are needed for renormalization
at LO~\cite{Platter:2004he,Platter:2004zs}.  As a consequence, all four-body
observables in the universal regime are governed by the same limit cycle and
can be characterized by $a$ and $\kappa_*$.  This leads to the universal
correlations between three- and four-body observables already discussed
in Secs.~\ref{sec:Triton} and~\ref{sec:Helion}.  A similar behavior is expected
for higher-body observables.

\begin{figure}[tb]
\centering
\includegraphics*[width=7.cm, angle=0,clip=true]{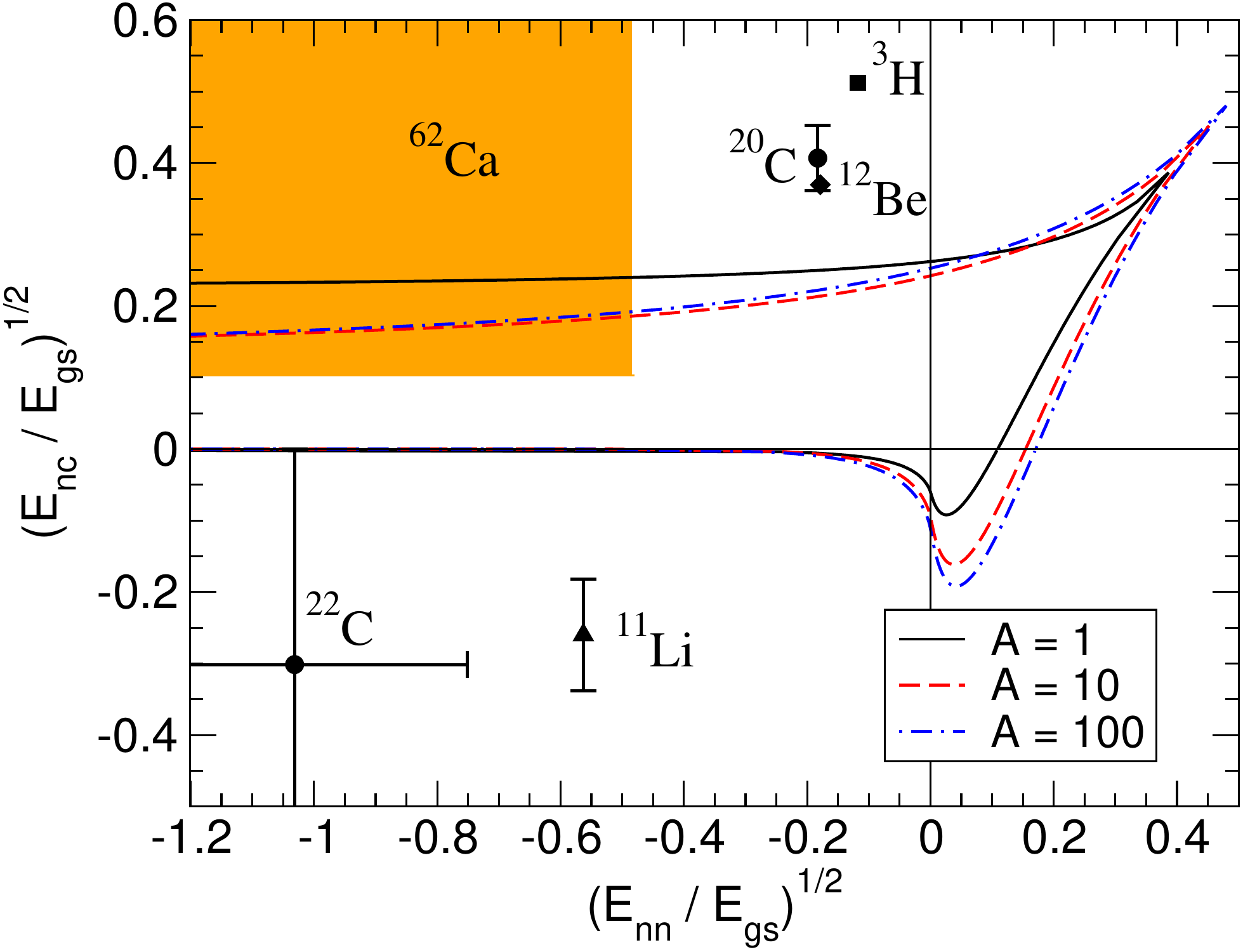}
\caption{Boundary curves for the existence of {excited Efimov
  states in two-neutron halo nuclei with different core masses $A$} as
  function of the {square roots of the} neutron-core energy $E_{nc}$ and
  neutron-neutron energy
  $E_{nn}$ in units of the three-body ground-state energy $E_{gs}$.
  The sign of the square root is taken positive (negative)
  if the corresponding two-body system, $nc$ or $nn$,
  has a bound state (virtual state).  Established halo nuclei are
  shown by the data points while the shaded area gives the parameter
  range for the predicted halo nucleus $^{62}$Ca.
  Reprinted by permission from Springer Nature:
  \cite{Hammer:2018fbe}, copyright (2018).
}
\label{fig:efihalo}
\end{figure}

Halo nuclei have been discussed as possible candidates for Efimov states for
more than 30 years~\cite{Fedorov:1994zz}. As the full Efimov plot for $cnn$
systems is three-dimensional and depends on the two scattering lengths $a_{nn}$
and $a_{cn}$, it is more instructive to plot candidate nuclei in a
two-dimensional plane characterized by the neutron-core energy $E_{nc}$ and
the neutron-neutron energy $E_{nn}$, in units of the three-body ground-state
energy $E_{gs}$, as introduced by \textcite{Amorim:1997mq}.  If a given nucleus
lies within a certain boundary curve that weakly depends on the mass number $A$
of the core, it should display an excited Efimov state.  The candidate nuclei
$^{11}$Li, $^{12}$Be, $^{14}$Be, $^{18}$C, and $^{20}$C were investigated in
LO Halo EFT assuming only resonant $S$-wave interactions by
\textcite{Canham:2008jd}.  An update of this analysis with current halo
candidates and established halo nuclei is shown in Fig.~\ref{fig:efihalo}.  The
triton has also been added, since it can be interpreted as an two-neutron halo
with a proton core.

In 2010, $^{22}$C was established as the {then}
heaviest halo nucleus.
In particular, $^{22}$C was found to display a large matter
radius~\cite{Tanaka:2010zza} and a large $S$-wave component {in its}
$n$-$^{20}$C
{subsystem}~\cite{Horiuchi:2006ds}.  Since the information on the
neutron-core energy
$E_{nc}$ was ambiguous, \textcite{Acharya:2013aea} used Halo EFT to explore the
correlation between the $n$-$^{20}$C energy and the two-neutron separation
energy of $^{22}$C.  Combining this correlation with the matter-radius
measurement, they demonstrated that an excited Efimov state in $^{22}$C
is unlikely.  A recent update of this analysis by \textcite{Hammer:2017tjm},
using the more precise matter radius from \textcite{Togano:2016wyx} as input,
reached the same conclusion.

Whether heavier neutron halos than $^{22}$C exist is still an open question,
although there is some experimental evidence that the ground states of $^{31}$Ne
and $^{37}$Mg have a low one-neutron separation energy and can be considered
deformed $P$-wave halos~\cite{Nakamura:2014hxa,Kobayashi:2014owa}.
This makes it
worthwhile to investigate the possibility for Efimov states in heavier nuclei.

\subsection{\Abinitio methods and Efimov states in heavier nuclei}
\label{sec:halo+abinitio}

Halo EFT can be used in conjunction with \abinitio calculations
to extend the reach of the latter or to test the consistency of different
approaches.  Here, we discuss an example of the former in the context of Efimov
physics.  Further examples regarding electromagnetic reactions will be given
below.

Coupled-cluster calculations by \textcite{Hagen:2012fb} of neutron-rich
calcium isotopes---which used a chiral potential with schematic $3N$ forces and
included coupling to the scattering continuum---suggested that a large $S$-wave
scattering length might occur in the $^{61}$Ca system, with interesting
implications for $^{62}$Ca.  Subsequently, \textcite{Hagen:2013jqa} computed the
elastic scattering of neutrons on $^{60}$Ca obtaining quantitative estimates for
the scattering length and the effective range, and confirming that a large
scattering length can be expected.  These results were then used as input for
Halo EFT in the study of the $^{60}$Ca-$n$-$n$ system.

\begin{figure}[tb]
 \centering
 \includegraphics*[width=7cm, angle=0,clip=true]{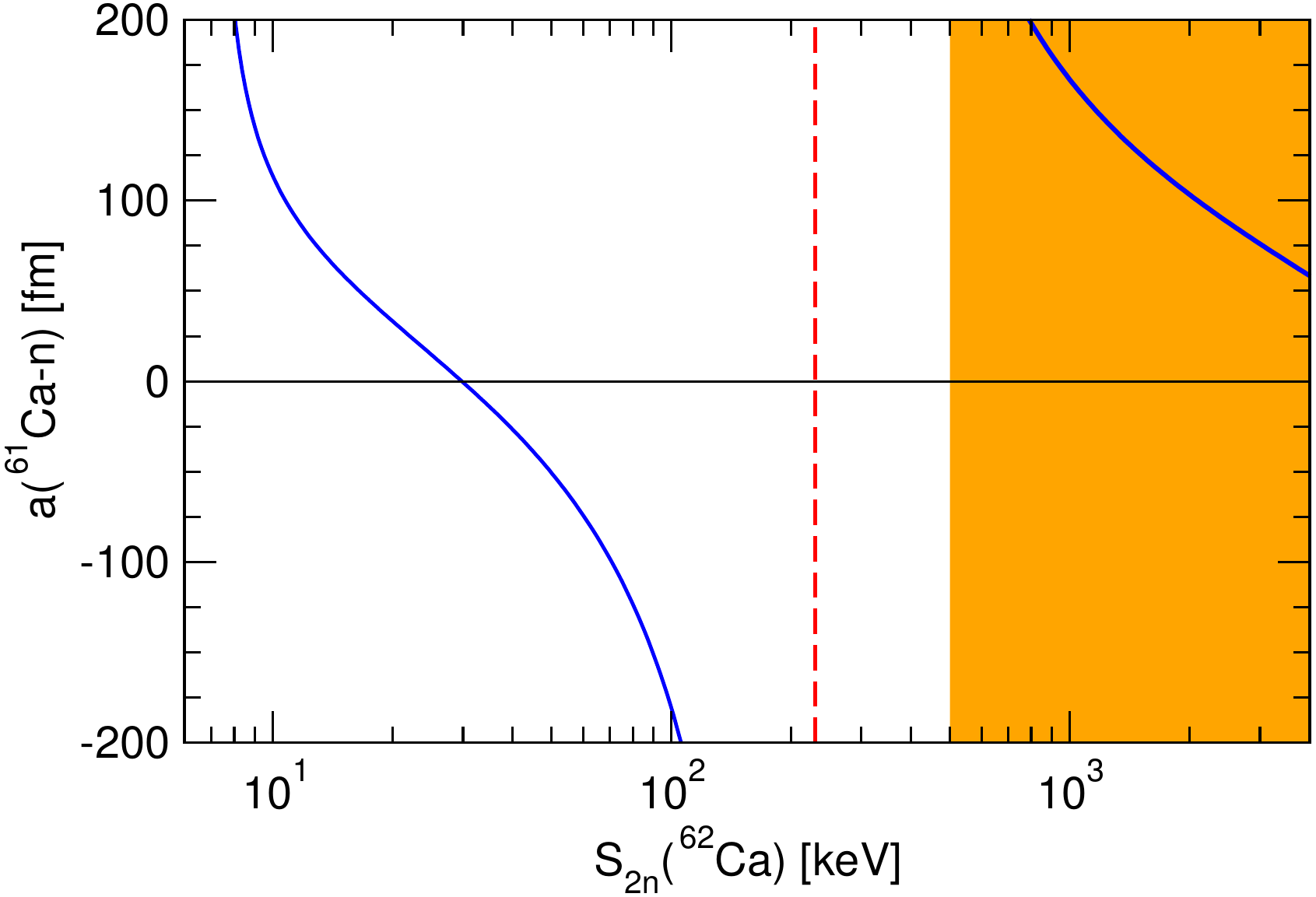}
 \caption{
  Correlation between the $^{61}$Ca-$n$ scattering length and the
  two-neutron separation energy $S_{2n}$ of $^{62}$Ca.  The emergence of an
  excited Efimov state around $S_{2n}=230$ keV is indicated by the vertical
  dashed line.  The shaded area indicates the region where Halo EFT
  breaks down.
  Figure taken from \cite{Hammer:2016exd}
  under the terms of the Creative Commons Attribution License,
  https://creativecommons.org/licenses/by/4.0/.
}
\label{fig:Ca62plot}
\end{figure}

Specifically, the focus was on signals of Efimov physics that are a consequence
of the large scattering lengths in the $^{60}$Ca-$n$ and $n$-$n$ systems.
This is illustrated in Fig.~\ref{fig:Ca62plot}, where the universal correlation
between the  $^{61}$Ca-$n$ scattering length and the two-neutron separation
energy of $^{62}$Ca is shown.  For $^{62}$Ca with $m_0 = 60 m_1$,
the discrete scaling factor governing the energy spectrum is approximately
$16^2=256$~\cite{Braaten:2004rn}, which is slightly more favorable than in the
case of equal mass particles.  The asymptotic scaling ratio applies only for
deep states or in the unitary limit of infinite scattering length.  Away from
the unitary limit, however, the ratio of energies near threshold can be
significantly smaller---see Fig.~\ref{fig:efiplot} and the corresponding
discussion {by \textcite{Braaten:2004rn}}.  In the case of $^{62}$Ca,
the whole energy region between $S_{2n}\approx 5-8~\keV$ and the breakdown
scale $S_{\mathrm{hi}}\approx 500~\keV$ is available for Efimov states.
At $S_{2n}\approx 230~\keV$, the $^{60}$Ca-$n$ scattering length jumps from
$+\infty$ to $-\infty$ and an excited Efimov state appears.  It is thus
conceivable that $^{62}$Ca would display an excited Efimov state and unlikely
that it would not display any Efimov states at all.  The matter radius of
$^{61}$Ca relative to the $^{60}$Ca core was found to be 4.9(4)~\fm, while the
matter radius of $^{62}$Ca could be even larger, depending on the precise value
of $S_{2n}$.

One can summarize the situation on (excited) Efimov states in halo nuclei as
follows.  While the ground {states} of many $S$-wave halo nuclei {are}
close to the
Efimov limit, there is currently no observed halo nucleus that displays an
excited Efimov state or is likely to display such a state.  There is some
theoretical evidence that the situation could be different for $^{62}$Ca.
The corresponding parameter range is indicated by the shaded square in
Fig.~\ref{fig:efihalo}.   The results of \textcite{Hagen:2013jqa} imply that
$^{62}$Ca is possibly the largest and heaviest halo nucleus in the chart of
nuclei and demonstrated that a large number of observables would display
characteristic features of Efimov physics.  {Measurement} of these
observables
clearly poses a significant challenge for experiment.  For example, $^{58}$Ca is
the heaviest Calcium isotope that has been observed
experimentally~\cite{Tarasov:2009hb}.  However, future radioactive-beam
facilities might provide access to calcium isotopes as heavy as $^{68}$Ca.

\subsection{Higher partial waves and resonances}
\label{sec:highpar}

Next we discuss systems with resonant interactions in higher partial waves.
Such interactions are ubiquitous in halo and cluster nuclei and lead to a
richer power counting structure.

Consider two-body scattering with reduced mass $\mu$ and energy $E=k^2/2\mu$ in
the center-of-mass frame.  Resonance behavior arises when the $S$-matrix has a
pair of poles in the two lower quadrants of the complex $k$ plane.  The
projection of $S$ into the resonant partial wave $l$ can be written
\begin{equation}
 \frac{S_l}{s_l(k)} = \frac{k+k_+}{k-k_+}\, \frac{k+k_-}{k-k_-}
 = \frac{E-E_0-\ii\Gamma(E)/2}{E-E_0+\ii\Gamma(E)/2} \,.
\label{smat}
\end{equation}
{Here $k_\pm= \pm k_R - \ii k_I$ with $k_{R,I} >0$ are the pole positions,
$s_l(k)$ is a smooth function in the energy region under consideration, $E_0
=(k_R^2 + k_I^2)/2\mu$ is the position of the resonance (where the corresponding
phase shift crosses $\pi/2$), and $\Gamma(E)/2= k k_I/\mu$ is referred to as
the half-width of the resonance.  A \emph{narrow} resonance is one for which
$\Gamma(E_0)/{(2E_0)}\ll 1$, that is, for which the poles are near the real
axis,
{$k_I/k_R\ll 1$}.  We call the resonance \emph{shallow} if $|k_\pm| \equiv
\Mlo
\ll \Mhi$.  An example of a shallow, narrow resonance, is given by the
$^2P_{3/2}$ resonance in $n\alpha$ scattering (the ground state of $^5$He),
which has~\cite{Bertulani:2002sz} $\Gamma(E_0)/2\simeq 0.3~\MeV \ll E_0 \simeq
0.8~\MeV \ll E_\alpha \simeq 20\, \MeV$, or $k_I \simeq 6 \, \MeV \ll k_R
\simeq 34~\MeV \ll \sqrt{m_N E_\alpha} \simeq 140~\MeV$, where $E_\alpha$ is
the excitation energy of the $\alpha$ core and $m_N$ is the nucleon mass.}

Shallow resonances can be described in Halo EFT, just as bound states.  For
notational simplicity we take the two scattering particles to be identical, with
mass $m=2\mu$ and no spin.  Generalization to other situations is
straightforward.  Like for bound states, it is convenient to introduce a
dimeron field with the quantum numbers of the resonance.
{Note that the formulation with dimeron fields is equivalent to a
formulation with particle contact interactions.}
{(For details see the discussion by~\textcite{Bertulani:2002sz}
and, for $S$-wave states, Sec.~\ref{sec:pionless}.)}
In the following, we focus on the case where the resonance is in the $l=1$
state, {in which case the dimeron field $d$ has three components
corresponding to spin 1.}  In a notation
similar to Eq.~\eqref{eq:L-NN-simple-d},
\begin{multline}
 {\LL = \psi^\dagger
  \left(\ii\partial_0
  + \frac{\Laplace}{2m}\right)
  \psi
  + d^\dagger_{l}
   \bigg[\eta_1 \bigg(\ii\partial_0+\frac{\Laplace}{4m}\bigg)
   -\Delta_1 \bigg] d_{l}}
   \\
  {\null + \frac{g_1}{4} \left[d^\dagger_{l} (\psi \galnab_{l} \psi)
  +\hc\right]
  + \cdots \,,}
\label{lagdPwave}
\end{multline}
where the Galilean combination $\vNablaLR*=\vNablaL-\vNablaR$ places the two
particles in a $P$ wave.

The full dimeron propagator, depicted in Fig.~\ref{fig:DibaryonProp}, is the
bare propagator given by the inverse of the dimeron kinetic term and dressed by
the bubbles generated by the $d\psi\psi$ interactions in Eq.~\eqref{lagdPwave}.
The two-particle $T$ matrix is obtained by attaching external particle
legs.
The result reproduces the ERE for $P$ waves, Eq.~\eqref{eq:T-kcot}, with
\begin{equation}
 \frac{1}{a_1}
 = \theta_3\Lambda^3-\frac{12\pi\Delta_1}{mg_1^2}
 \mathtext{,}
 {-}\frac{r_1}{2} = \theta_1\Lambda+ \eta_1\frac{12\pi}{m^2g_1^2}\,,
\label{matchingPwave}
\end{equation}
where, as in Eq. \eqref{eq:I0-cutoff}, $\theta_{1,3}$ are numbers that depend on
the chosen regularization.  We see that both the scattering length and effective
range need to be included at LO to absorb all
divergences~\cite{Bertulani:2002sz}.  This requirement has to be reflected by
the power counting for $P$ waves.  We note that the need to include additional
interactions for renormalization at LO will become more severe in higher partial
waves.

For the scaling of the parameters $a_1, r_1, \ldots$, different scenarios can
be envisioned:
\begin{itemize}
\item[(i)]
Na\"ive dimensional analysis suggests that the typical size for the ERE parameters
$a_1, r_1, \ldots$ is given by the appropriate power of the momentum scale
$\Mhi$ where the EFT breaks down.
\end{itemize}
For instance, if the interaction between the particles is described by a
potential of depth $\sim \Mhi$ and range $\sim 1/\Mhi$, one would expect
$a_1\sim 1/\Mhi^3$ and $r_1\sim \Mhi$.  In particular, a resonance or bound
state, if present, generally occurs at the momentum scale $\Mhi$.

Scenario (i) is clearly not appropriate for halo nuclei with shallow resonances
or bound states.  In such systems, the interactions are finely tuned in such a
way as to produce a resonance or bound state close to threshold, at a scale
$\Mlo$ much smaller than $\Mhi$, violating the NDA estimate.  This situation
can occur when one or more of the ERE parameters have unnatural sizes
related to $\Mlo$.
\begin{itemize}
\item[(ii)]
  \textcite{Bertulani:2002sz} proposed a different power counting
  assuming $a_1\sim 1/\Mlo^3$ and $r_1\sim \Mlo$, while all  higher
  ERE parameters scale with $\Mhi$.
\end{itemize}
With this scaling, all three terms of the ERE shown explicitly in
Eqs.~\eqref{eq:T-kcot-a} and~\eqref{eq:T-kcot-b} are of the same order for
momenta $k\sim \Mlo$ and must be retained at LO.  Higher ERE parameters are
suppressed by powers of $\Mlo/\Mhi$ and thus are subleading.

Scenario (ii) requires that \emph{two} combinations of constants,
$\Delta_1/g_1^2$ and $1/g_1^2$, be fine tuned against the large values
$\Lambda \simge \Mhi$ in Eq.~(\ref{matchingPwave}) in order to produce a result
containing powers of the small scale $\Mlo$.  From a naturalness perspective,
this makes it less likely to occur in nature than a scenario with one fine
tuning like the one for an $S$-wave bound state.
\begin{itemize}
\item[(iii)]
  An alternative scaling was suggested by \textcite{Bedaque:2003wa}, where
  $a_1\sim 1/(\Mlo^2\Mhi)$, $r_1\sim \Mhi$, and all other
  ERE parameters again scale with appropriate powers of $\Mhi$.\footnote{A
  similar scheme was applied to the $\Delta(1232)$ resonance
  in Chiral EFT by ~\textcite{Pascalutsa:2002pi} and \textcite{Long:2009wq}.}
\end{itemize}
This scenario requires only \emph{one} combination of constants, namely
$\Delta_1/g_1^2$, to be fine tuned.

With option (iii), the terms proportional to $1/a_1$ and $r_1 k^2$
in the dimeron propagator are of the same order for momenta $k\sim\Mlo$.
The term stemming from the unitarity cut, $\ii k^3$, is suppressed by one power
of $\Mlo/\Mhi$ and is, therefore, subleading.  The remaining terms in the
ERE are even more suppressed.  Thus, LO corresponds to taking the bare dimeron
propagator while the effects of loops and higher-derivative interactions enter
as higher-order corrections.

At LO the difference between the scalings~(ii) and~(iii) is the presence of the
unitarity-cut term $\sim \ii k^3$.  This difference disappears if instead of
considering generic momenta $k$ of order $\Mlo$ we focus onto a narrow region
around the position of the resonance at $k=\sqrt{2/a_1r_1}$.  Due to the near
cancellation
{within a window of size  $\Delta k =
2/a_1r_1^2$ around the pole
between the two
terms that are leading in scenario (iii),}
the unitarity-cut term has to be resummed to all
orders, and provides a width to the resonance.  In this kinematic range there
are two fine tunings: one implicit in the short-distance physics leading to the
unnatural value of $a_1$, and another one explicitly caused by the choice of
kinematics close to the pole.
{%
Power counting for resonances is further discussed by \textcite{Gelman:2009be},
\textcite{Alhakami:2017ntb},} {and \textcite{Schmidt:2018vvl}.}

If the underlying theory cannot be solved, the appropriate scaling for a
specific physical system can be
{inferred} from the data, \ie, from the
numerical values of the ERE parameters.  However, such a determination is not
always unique and/or different scalings might apply in different kinematic
regions.
In the first papers on Halo EFT, \textcite{Bertulani:2002sz} and
\textcite{Bedaque:2003wa} applied the scalings~(ii) and~(iii),
respectively, to the lowest resonance in $n\alpha$ scattering.  The experimental
ERE parameters can be accommodated in both scalings such that both appear
viable.  Although the unitarization implicitly carried out in~(iii) is not
necessary except near the resonance, it improves the description throughout the
low-energy region.  In either case, scattering data determine the $n\alpha$
interaction parameters.

The two-neutron halo nucleus $^6$He offers a further testing ground for
Halo EFT with resonant $P$ waves.  The $n\alpha$ interaction in that nucleus is
dominated by the $^2P_{3/2}$ resonance.  The structure and renormalization
of $^6$He were investigated by \textcite{Rotureau:2012yu} and
\textcite{Ji:2014wta}.  \textcite{Rotureau:2012yu} calculated $^6$He at LO in
the Gamow shell model using scenario~(ii) and found that a three-body
force---the analog of $\LL_{3b}$ in Eq. \eqref{eq:lag_05}---is required to
stabilize the system. \textcite{Ji:2014wta} solved the Faddeev equations in
scenario~(iii), but demoted the $^2S_{1/2}$ $n\alpha$ interaction to NLO.
They also found that a three-body force is required for renormalization at LO
and determined its running over a wide range of cutoffs.  The observed behavior
is not log-periodic, although some periodicity is observed.  Alternative
formulations at LO were investigated by \textcite{Ryberg:2017tpv} and shown to
be equivalent{,} {while momentum-space probability densities of ${}^6$He
were calculated by \textcite{Gobel:2019jba}.}

The power counting for resonant partial waves with $l\geq 2$ was also discussed
by \textcite{Bertulani:2002sz} and \textcite{Bedaque:2003wa}.  Their analysis of
the power divergences of the one-loop self-energy showed that
the first $l+1$ ERE parameters are required to absorb all divergences.  This
was confirmed by the Wilsonian RG analysis of \textcite{Harada:2007ua}, which
considered the cases $l=1,2$ explicitly.  An alternative power counting for
bound states with $l=2$ was proposed by \textcite{Braun:2018hug} and applied to
{the description of $D$-wave states in $^{15}$C
and $^{17}$C~\cite{Braun:2018vez}.}

\subsection{Electromagnetic properties and reactions}
\label{sec:haloelectric}

For one-neutron halos{,} Halo EFT essentially reproduces the ERE, but their
electromagnetic structure and reactions can be predicted.  The formalism is
similar to that of Pionless EFT (Sec.~\ref{sec:Currents-ThreeBody}) and serves
to illustrate it.  Moreover, the accuracy limits of cluster models can be
estimated from the order at which gauge-invariant couplings to currents appear.

In the following, we exemplify the power of Halo EFT in the electromagnetic
sector using the example of $^{11}$Be~\cite{Hammer:2011ye} and give a brief
overview of results in other systems.  The first excitation of
$^{10}$Be is $3.4~\MeV$ above the ground state, which has $J^P=0^+$.
Meanwhile, $^{11}$Be has a $1/2^+$ state with neutron separation energy
$B_0=500~\keV$, and a $1/2^-$ state with $2n$ separation energy
$B_1=180~\keV$~\cite{AjzenbergSelove:1990zh}, which we denote $^{11}$Be$^*$.
The shallowness of these two states of ${}^{11}$Be compared to the bound states
of ${}^{10}$Be suggests that they have significant components in which a loosely
bound neutron orbits a ${}^{10}$Be core.  In Halo EFT, the $1/2^+$
state is described as an $S$-wave bound state of the neutron and the core,
while $1/2^-$ is a $P$-wave bound state governed by scenario~(iii) from the
previous section.

The effective Lagrangian for the system can be obtained by combining
Eq.~\eqref{eq:lag_05} for the $1/2^+$ ground state and Eq.~\eqref{lagdPwave}
(generalized to unequal masses) for the $1/2^-$ excited state.  Photons are
included via the minimal substitution, Eq.~\eqref{eq:minimal}, and through the
field strength.  (See \cite{Hammer:2011ye} for explicit expressions.)

Here our focus is on electric properties, and the dominant pieces of the
electric response follow from the minimal substitution~\eqref{eq:minimal}.
But, at higher orders in the computation of these properties, gauge-invariant
operators (counterterms) appear involving the electric field $\mathbf{E}$ and
the fields $c$ for the $^{10}$Be core, $n$ for the halo neutron, {$d$ for
the $^{11}$Be} {ground-state}
{dimeron, and $d^*$ for the $^{11}$Be$^*$} {excited-state} {dimeron.}
Possible one- and
two-derivative operators with one power of the photon field are
\begin{spliteq}
 {\LL_\text{EM} = \;}& {L_{C0} \, d^\dagger
 (
 \vNabla\cdot\mathbf{E})\,d
  + L_{C0}^{(*)} \,d^{*\dagger} (
 \vNabla\cdot \mathbf{E}) \,d^*}
  \\
  & {\null + \ii L_{E1}^{(1/2)} \,\left( [d d^{*\dagger}]_l \,\mathbf{E}_l
  + \hc\right) + \cdots \,,}
\label{eq:nonminimalEM}
\end{spliteq}
where {$[\cdots]_l$} indicates the projection on $l=1$.  If magnetic
properties are to be considered, we have to include operators involving the
magnetic field
${\bf B}$ as well.

The electric interactions in Eq.~\eqref{eq:nonminimalEM} are gauge invariant
by themselves, and so we must determine the order at which they occur.
Rescaling the fields to absorb all powers of $\Mlo$ as
{
done by \textcite{Beane:2000fi},}
the scaling of the coupling constants with $\Mlo$ can be obtained from
NDA~\cite{Hammer:2011ye}.  As a consequence, the leading effects in the
charge-radius-squared of the $1/2^-$ state in ${}^{11}$Be are
$\sim (r_1 \Mlo)^{-1} \sim (\Mlo\Mhi)^{-1}$.  The operator proportional to
{$L_{C0}^{(*)}$} produces effects of order $(r_1 \Mhi)^{-1} \sim
(\Mhi)^{-2}$, and thus affects the prediction for the charge radius at NLO.
Similarly, the E1($1/2^+ \rightarrow 1/2^-$) matrix element has parametric
dependence $\Mlo^{-1} (\Mlo\Mhi)^{-1/2}$.  Including the proper wavefunction
renormalization factors, the operator with LEC $L_{E1}^{(1/2)}$
yields an effect $\sim \Mhi^{-1} (\Mlo\Mhi)^{-1/2}$, and so also  occurs already
at NLO.  {Thus for electric quantities involving the shallow $1/2^-$
excited state of $^{11}$Be there are two
parameters in the Halo EFT description
at NLO which cannot be fixed with ${}^{10}$Be-$n$
scattering information
alone. There are none at LO and presumably more at N$^2$LO.}

\subsubsection{Form Factors}

\begin{figure}[tb]
 \centerline{\includegraphics[width=3cm]{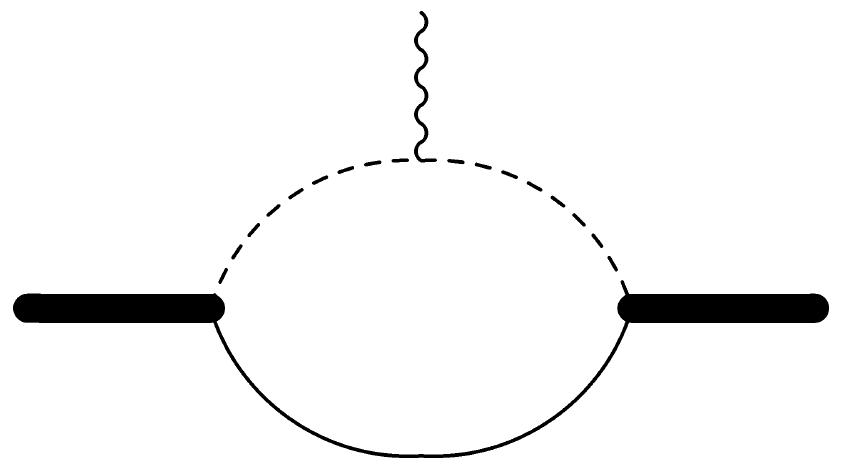}}
 \caption{The LO contribution to the irreducible vertex for an $A_0$ photon
  coupling to the ${}^{10}$Be-neutron $S$-wave bound state.
  The thick solid line
  indicates the ${}^{11}$Be ground state, while the solid, dashed, and curly
  lines represent the neutron, ${}^{10}$Be core, and photon, respectively.
 }
\label{fig:formfactor-11Be}
\end{figure}

The form factor of the $^{11}$Be ground state is computed by calculating the
contribution to the irreducible vertex for {$A_0 dd$} interactions
shown in Fig.~\ref{fig:formfactor-11Be}.  There is no diagram coupling the
photon to the neutron at this order since $Q_n=0$.  In the Breit frame, where
the four-momentum of the virtual photon is $q=(0,\mathbf{q})$, the
irreducible vertex for the $A_0$ photon coupling to the {$d$} field is
${-}\ii e Q_c G_E(|\mathbf{q}|)$ where $Q_c$ is the charge of the core.
A straightforward calculation yields
\begin{equation}
 G_E^{}(|\mathbf{q}|)= \frac{2 \gamma_0}{f|\mathbf{q}|}
 \arctan\!\left(\frac{f|\mathbf{q}|}{2 \gamma_0}\right)\,,
\label{eq:Gc}
\end{equation}
with $\gamma_0 =\sqrt{\mathstrut 2\mu_1 B_0}$ and $f=(1+m_0/m_1)^{-1}$, in the
notation of Eq.~\eqref{eq:lag_40}.  For the deuteron $m_0=m_N$ and $f=1/2$,
thus Eq.~\eqref{eq:Gc} reduces to the LO Pionless EFT result of
\textcite{Chen:1999tn}.

The form factor is a function of $\mathbf{q}^2$ only and the charge radius is
defined as $\langle r_E^2 \rangle^{} =-6 (\dd/\dd\mathbf{q}^2)
G_E|^{}_{\mathbf{q}^2=0}$.  Applying this expression to
Eq.~\eqref{eq:Gc} yields
\begin{equation}
 \langle r_E^2 \rangle^{}=\frac{f^2}{2 \gamma_0^2} \,,
\label{rcqs}
\end{equation}
which gives the charge radius of the $^{11}$Be ground state  relative to the
charge radius of $^{10}$Be.  Thus we have
$ \langle r_E^2 \rangle_{^{11}{\rm Be}}-\langle r_E^2 \rangle_{^{10}{\rm Be}}
= f^2/2 \gamma_0^2$.  This relation can be understood by writing the charge
distribution of $^{11}$Be as a convolution of the charge distribution of
$^{10}$Be with that of the $^{10}$Be-$n$ halo system.  Using the convolution
theorem for the Fourier transform, one finds that the total mean-square radius
is the sum of the squared radii for $^{10}$Be and the $^{10}$Be-$n$ halo system.

The latter effect can be calculated in Halo EFT.  We note that the finite size
of the core will also appear in Halo EFT at higher orders~\cite{Chen:1999tn}.
{An extended power-counting scheme that explicitly takes into account the
scaling of the mass ratio $f$ to move these contributions to lower orders was
given by \textcite{Ryberg:2019cvj}.}

Inserting $\gamma_0=0.15~\fm^{-1}$, the relative radius becomes
$\langle r_E^2\rangle_{^{11}{\rm Be}}-\langle
r_E^2\rangle_{^{10}{\rm Be}}=0.19~\fm^2$.  This is consistent with the
experimental result, $0.51(17)~\fm^2$~\cite{Nortershauser:2008vp}, within the
$40\%$ uncertainty from NLO effects in this system.  Using the experimental
result for the $^{10}$Be charge radius as further input, we find $\langle r_E^2
\rangle^{1/2}_{^{11}{\rm Be}}=2.40~\fm$ at LO.  This is 2--3\% smaller than the
atomic-physics measurement which yields $\langle r_E^2 \rangle^{1/2}_{^{11}{\rm
Be}}=2.463(16)~\fm $~\cite{Nortershauser:2008vp}.

At NLO a new operator associated with gauging the term {$\sim d^\dagger
\partial_0 d$} in the effective Lagrangian contributes.  The calculation
produces an increased charge radius, as long as the $S$-wave $n$-$^{10}$Be
effective range $r_0$ is positive (\cf~\cite{Beane:2000fi}),
\begin{equation}
 \langle r_E^2 \rangle_{^{11}{\rm Be}}-\langle r_E^2 \rangle_{^{10}{\rm Be}}
 =\frac{f^2}{2 (1- r_0 \gamma_0) \gamma_0^2}\,.
\end{equation}
{Using the value $r_0=2.7~\fm$ determined from Coulomb dissociation
of $^{11}$Be (see below), the relative radius becomes
$\langle r_E^2\rangle_{^{11}{\rm Be}}-\langle
r_E^2\rangle_{^{10}{\rm Be}}=0.31(5)~\fm^2$ at NLO, which improves the
agreement with the atomic-physics measurement. The change is
of order $40$\%, in agreement with the \apriori  expectation.
As a consequence,
the result for the full
charge radius of the ${}^{11}$Be ground state increases to
$\langle r_E^2 \rangle^{1/2}_{{}^{11}{\rm Be}}=2.42~\fm$.}
In contrast to observables involving the $1/2^-$ state, the radius
of the ${}^{11}$Be ground state does not receive any corrections from
short-distance physics until N$^3$LO{~\cite{Chen:1999tn}.}
The remaining difference between
NLO and experimental values is consistent with the presence of the
short-distance operator {$\sim L_{C0}^{}$} from
Eq.~\eqref{eq:nonminimalEM} at N$^3$LO in the expansion for the radius.

For the charge form factor of the $1/2^-$ excited state, NLO corrections might
be expected to be smaller since its typical momentum is lower.  However, a
counterterm enters already at NLO for this observable.  The form factor is given
by the contribution to the irreducible vertex for {$A_0 d^* d^*$}
interactions.
There are two diagrams at LO, the first of which is analogous to that for the
$1/2^+$ state shown in Fig.~\ref{fig:formfactor-11Be}, while the second diagram
represents a direct coupling of the photon from gauging the {$d^{*\dagger}
\partial_0 d^*$} term in the effective Lagrangian.  The latter contributes at
LO because the effective range $r_1$ corresponds to an LO operator for the
$1/2^-$ state.  The charge form factor of the $1/2^-$ state at LO is obtained
as~\cite{Hammer:2011ye}
\begin{equation}
 {G_E^{(*)}(|\mathbf{q}|)
 = 1 - \frac{\gamma_1}{r_1} +\frac{f^2
  \mathbf{q}^2
 + 2 \gamma_1^2}{|\mathbf{q}| f r_1}\,
 \arctan\!\left(\frac{f|\mathbf{q}|}{2 \gamma_1}\right)
 \,,}
\label{eq:Gc-p}
\end{equation}
where $\gamma_1 =\sqrt{2\mu_1 B_1}$ and $r_1$ is the $P$-wave effective range
for $n$-$^{10}$Be scattering.
Note that {$G_E^{(*)}(0)=1$}, as required by charge conservation,
while the charge radius of the $1/2^+$ state relative to the $^{10}$Be
ground state is
\begin{equation}
 {\langle r_E^2 \rangle^{(*)}  = {-}\frac{5 f^2}{2 \gamma_1 r_1} \,.}
\label{eq:LOpradresult}
\end{equation}
This scales as $1/(\Mlo\Mhi)$ as expected.
It seems counterintuitive that there is a short-distance contribution
to {$\langle r_E^2 \rangle^{(*)}$} already at NLO---especially when the
corresponding effect does not occur in $\langle r_E^2 \rangle^{}$
until N$^3$LO \cite{Chen:1999tn}.  The reason for this enhanced sensitivity is
that the probability distribution of $P$-wave states is drawn in to shorter
distances than the one of $S$-wave states, as it gets caught between the
attractive potential that produces the $P$-wave state and the centrifugal
barrier.  Observables associated with a shallow $P$-wave bound state will,
therefore, generically exhibit counterterms at lower order than those of their
$S$-wave counterparts.

Numerical evaluation of the LO expression~\eqref{eq:LOpradresult} leads to
the prediction $\langle r_E^2 \rangle_{^{11}{\rm Be}^*}-\langle r_E^2
\rangle_{^{10}{\rm Be}} = 0.36~\fm^{2}$, where we have used the value
$r_1=-0.66~\fm$ from the $B(E1)$ value as input (see below).  The NLO radius
includes contributions from the counterterm {$L_{C0}^{(*)}$}
in Eq.~\eqref{eq:nonminimalEM}, the coefficient of which is unknown.
\textcite{Hammer:2011ye} estimated the NLO contributions to be of order
20\%, assuming that the short-distance effects in $\langle r_E^2
\rangle^{1/2}_{^{11}{\rm Be}^*}$ scale with $f$. This {assumption}
is in agreement with the expectation from the power counting.  Using again the
experimental result for the $^{10}$Be charge radius~\cite{Nortershauser:2008vp},
the prediction for the charge radius of $^{11}$Be$^*$ at LO is $\langle r_E^2
\rangle^{1/2}_{^{11}{\rm Be}^*} = (2.43 \pm 0.1)~\fm$.  To date there is no
experimental determination of this charge radius.

Halo EFT calculations for the charge and magnetic form factors of $^{11}$Be and
$^{19}$C were performed to NLO by~\textcite{Fernando:2015jyd}.  They considered
$^{15}$C as well and suggested the inclusion of the effective range as an LO
effect in this particular case.

\subsubsection{E1 transition and photodisintegration}

\begin{figure}[tb]
 \centerline{\includegraphics[width=6cm]{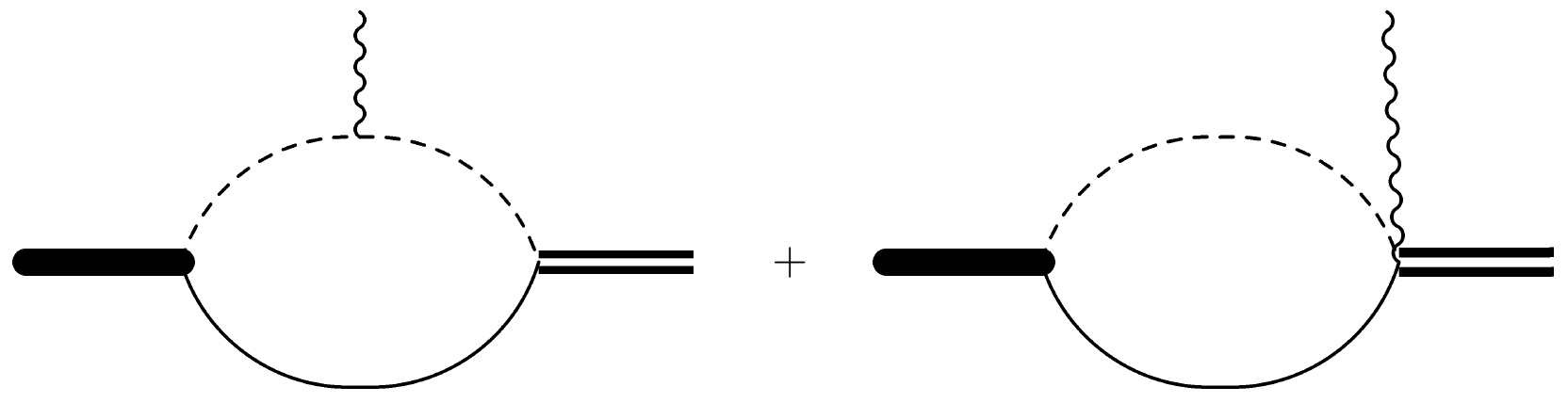}}
 \caption{The two diagrams contributing to the irreducible vertex for the
   $S$-to-$P$-state transition, $\Gamma_{j \mu}$, at LO.
   The double line indicates the ${}^{11}$Be excited state; the other
   lines are as in Fig.~\ref{fig:formfactor-11Be}.
 }
\label{fig:Gammajmu}
\end{figure}

Next we discuss the E1 transition from the $1/2^+$ state to the $1/2^-$ state.
The irreducible vertex for this transition is depicted in
Fig.~\ref{fig:Gammajmu}.  We compute the transition for a photon of arbitrary
four momentum $k=(\omega,{\bf k})$, and the sum of diagrams yields
${-}\ii\Gamma_{j\mu}$, where $j$ is the angular momentum index of the {$d^*$}
field and $\mu$ is the polarization index of the photon.
The diagrams depicted in Fig.~\ref{fig:Gammajmu} are divergent, but their
divergences cancel---providing a nontrivial check on the calculation.  As long
as both diagrams are considered current conservation is also
satisfied~\cite{Hammer:2011ye}, $k^\mu \Gamma_{j \mu} = 0$.  Note that if only
the long-distance E1 mechanism on the left-hand side of Fig.~\ref{fig:Gammajmu}
is considered---as was done, for example, by \textcite{Typel:2008bw}---then
current conservation is not satisfied and it appears that some input from
short-distance physics is needed in order to define the prediction for this
observable.

Evaluating the diagrams in
Fig.~\ref{fig:Gammajmu}, we obtain the LO Halo EFT result for ${\rm B(E1)}$,
\begin{equation}
 {\rm B(E1)} =
 -\frac{Z_\text{eff}^2 e^2}{3 \pi} \frac{\gamma_0}{r_1}
 \left[\frac{2 \gamma_1 + \gamma_0}{(\gamma_0 + \gamma_1)^2}\right]^2 \,,
\label{eq:BE1}
\end{equation}
with $Z_\text{eff} = f Q_c \approx 0.366$ the effective charge.
No regularization is needed in order to get a finite result.  We note that the
result~\eqref{eq:BE1} is ``universal'' in the sense that it applies to any E1
$S$-to-$P$-wave transition in a one-neutron halo nucleus.  Once $r_1$,
$\gamma_1$, and $\gamma_0$ are known for a given one-neutron halo,
the prediction (\ref{eq:BE1}) is accurate up to corrections of
{${\mathcal O}(\Mlo/\Mhi)$}.

Since there is no experimental value for the $P$-wave effective range $r_1$,
\textcite{Hammer:2011ye} extracted it from the experimental number ${\rm
B(E1)}(1/2^+ \rightarrow 1/2^-) = 0.105(12)~e^2$ \fm$^2$ from
\textcite{Summers:2007du}, yielding $r_1^{\rm LO} = {-}0.66~\fm^{-1}$.
Short-distance effects enter B(E1) through a counterterm
in the NLO corrections.  The B(E1) ($1/2^+ \rightarrow 1/2^-$) transition
therefore cannot be predicted at NLO, which can be seen from the presence of the
operator with LEC $L^{(1/2)}_{E1}$ in Eq.~\eqref{eq:nonminimalEM}.

Comparing this calculation with a shell-model treatment of ${}^{11}$Be, it is
clear that one effect which is subsumed into the NLO counterterm
$L^{(1/2)}_{E1}$ is the transition of a neutron from a $d_{5/2}$ to a $p_{3/2}$
orbital, with that neutron coupled to the $2^+$ state of ${}^{10}$Be.  This
$2^+$ state is $3.4~\MeV$ above the ${}^{10}$Be ground state, so the dynamics
associated with it takes place at distances $\sim \Mhi^{-1}$.  Hence in Halo EFT
it can only appear in short-distance operators such as that multiplying
$L^{(1/2)}_{E1}$.  The computation of
\textcite{Millener:1983zz} suggests that such a contribution reduces the E1
matrix element by $\sim 30$\%, which is the anticipated size of an NLO effect
when the $\Mlo/\Mhi$ expansion is employed in the ${}^{11}$Be system.
There are other effects of a similar size that will affect B(E1) at NLO.
Specifically, there are NLO corrections from the wavefunction renormalization
factors associated with the $S$- and $P$-wave fields.  Both tend to increase
B(E1) over the LO prediction.

\begin{figure}[tb]
 \centerline{\includegraphics[width=18em]{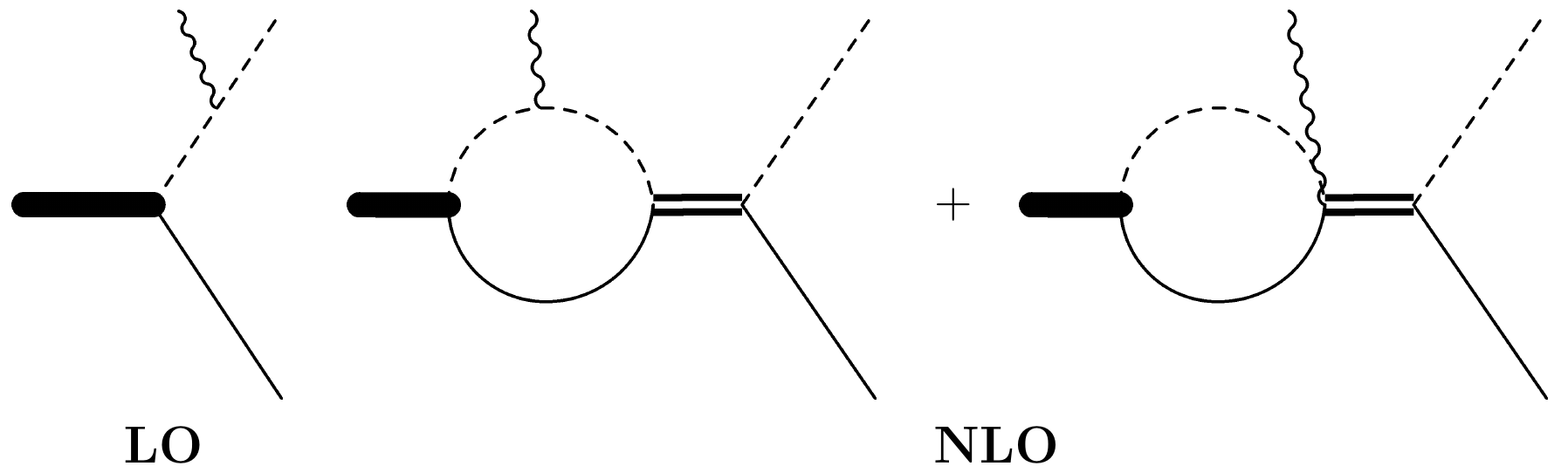}}
 \caption{
  Diagrams contributing to photodissociation of the ${}^{11}$Be
  ground ($S$-wave) state. The notation is as in Fig~\ref{fig:Gammajmu}.}
\label{fig:photodis-11Be}
\end{figure}

We move on to the photodisintegration of ${}^{11}$Be into ${}^{10}$Be
plus a neutron.  In practice this process is measured using Coulomb
excitation of the ${}^{11}$Be nucleus, with the two reactions connected within
the equivalent-photon approximation.
There are three contributions to this process, as depicted in
Fig.~\ref{fig:photodis-11Be}.  The first diagram, denoted ``LO'' in the figure,
corresponds to the contribution from the plane-wave impulse approximation.  The
second and third diagrams, denoted ``NLO'', include the final-state interactions
between the neutron and the core in the $J=1/2$ channel.  As we will show below,
the first diagram is dominant over diagrams involving $P$-wave final-state
interactions.  From these diagrams, we obtain the differential B(E1) strength
distribution at NLO~\cite{Hammer:2011ye},
\begin{multline}
 {\frac{\dd{\rm B(E1)}}{\dd E}
 = \frac{e^2 Z_\text{eff}^2}{4\pi}
 \frac{12\mu_1 \gamma_0 {|\mathbf p'}|^3}{\pi^2({\mathbf p'}^2
 + \gamma_0^2)^4}} \\
 {\null \times \left(1+r_0 \gamma_0 + \frac{2 \gamma_0}{3 r_1}
 \frac{3 {\mathbf p'}^2+ \gamma_0^2}{{\mathbf p'}^2 + \gamma_1^2}\right) \,.}
\label{eq:dBdE1NLO}
\end{multline}
where ${\mathbf p'}$ is is the relative momentum of the outgoing $^{10}$Be-$n$
pair and $E={\mathbf p'}^2/(2\mu_1)$ is the kinetic energy of the $^{10}$Be-$n$
pair in the center-of-mass frame.

The LO result corresponds to taking only the 1 in the {parenthesis}
of Eq.~\eqref{eq:dBdE1NLO}.  The NLO correction comes from two sources.  The
first is the shift of of the wavefunction renormalization to larger values due
to $r_0 > 0$, which tends to increase the B(E1) strength.  Second, final-state
interactions between the neutron and the core in the $J=1/2$ channel enter at
this order.  Accurate measurements of the Coulomb dissociation spectrum
therefore provide information on the $S$-wave $n$-${}^{10}$Be effective range,
if the $P$-wave effective range is already fixed from another observable.

Up to LO accuracy for bound-to-bound state transition and NLO for
bound-to-continuum, there are four LECs: $\gamma_0$ and $\gamma_1$ (which are
known from separation energies) and the $S$- and $P$-wave effective ranges $r_0$
and $r_1$.  At the next order, the counterterm $L_{E1}^{(1/2)}$ from
Eq.~\eqref{eq:nonminimalEM} enters as well.

\begin{figure}[tb]
 \centerline{\includegraphics*[width=7cm,clip=true]{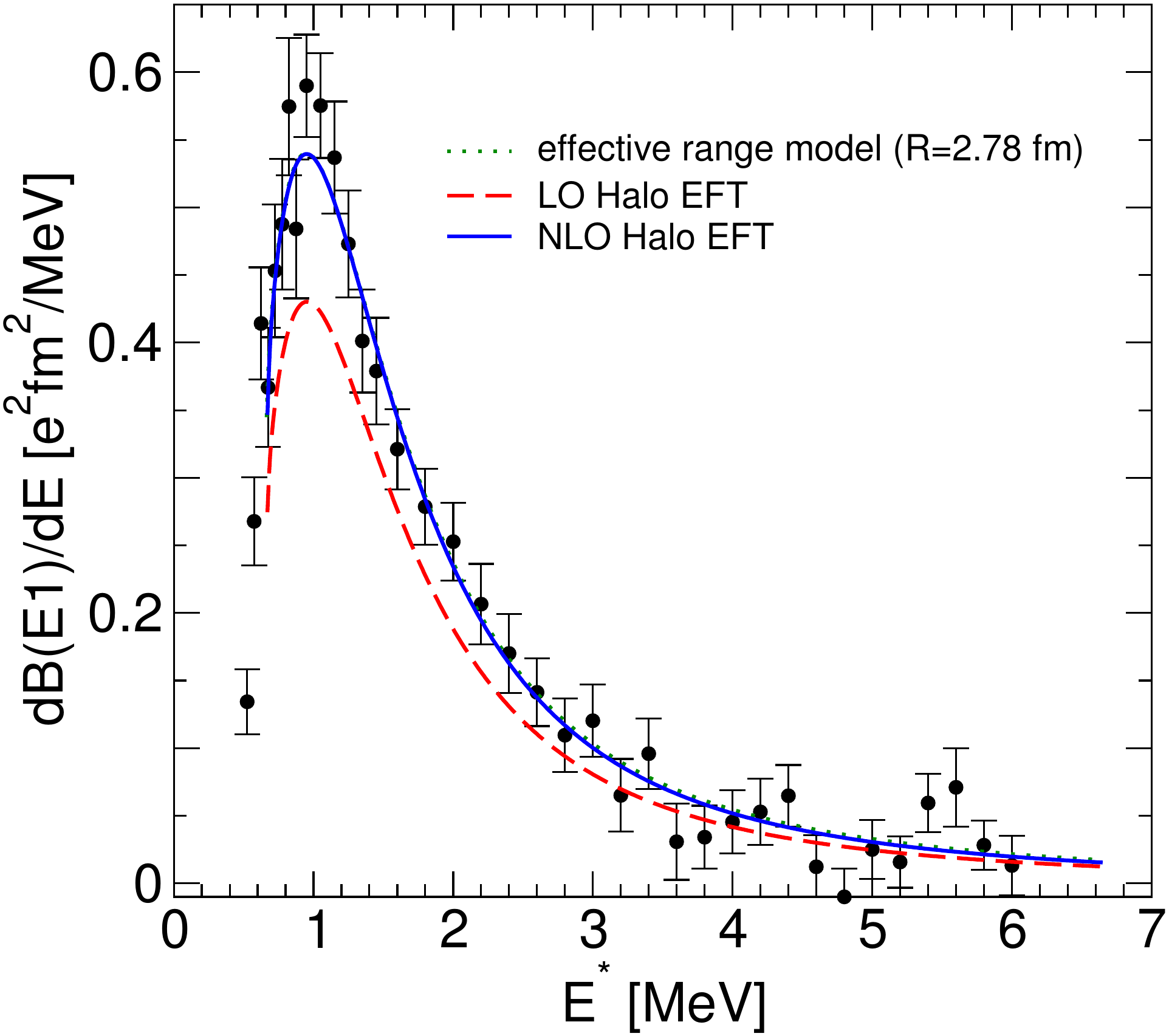}}
 \caption{Differential B(E1) strength for Coulomb dissociation of $^{11}$Be
  into $^{10}$Be + $n$ as a function of excess energy of the
  neutron, $E^*$.  The dashed (solid) lines show the Halo EFT result of
  \textcite{Hammer:2011ye} at LO (NLO) folded with the detector resolution.
  The experimental data are from \textcite{Palit:2003av}, while the
  dotted line {(almost on top of the solid line)} gives the effective range
  model of \textcite{Typel:2004us}.
  Reprinted from \cite{Hammer:2011ye}, with permission from Elsevier.
 }
\label{fig:results}
\end{figure}

Folding the Halo EFT result~\eqref{eq:dBdE1NLO} with the neutron detector
resolution and the spectrum of E1 photons, the experimental data of
\textcite{Palit:2003av} are well described, as shown in Fig.~\ref{fig:results}.
At NLO, if we take the value of $r_1$ fixed above, we have one free parameter,
the value of the $S$-wave effective range $r_0$.  A reasonable fit is found for
$r_0=2.7~\fm$, very close to the effective-range result of
\textcite{Typel:2004us} with all integrals cut off at $R=2.78~\fm$.
This choice of the cutoff corresponds to specific assumptions about the
counterterms.  Another experiment by~\textcite{Fukuda:2004ct} can be described
equally well but suggests a 3-4\% larger value for $r_0$.

The Coulomb dissociation of the one-neutron halo nucleus $^{19}$C was studied
by \textcite{Acharya:2013nia} using the $^{18}$C core and the neutron as
effective degrees of freedom.  In this case, there is no excited state present.
The authors demonstrated the power of Halo EFT by calculating various
observables and extracted the ERE parameters and the separation energy of the
halo neutron from the Coulomb dissociation data of \textcite{Nakamura:1999rp}.
In particular, they obtained the values $(575 \pm 55 \pm 20)~\keV$ for the
one-neutron separation energy of $^{19}$C, and $(7.75 \pm 0.35 \pm 0.3)~\fm$ for
the $^{18}$C-neutron scattering length, where the first error is statistical and
the second error is an estimate of the EFT uncertainty.  Their prediction for
the longitudinal-momentum distribution is in good agreement with the data of
\textcite{Bazin:1995zz} and confirms the $S$-wave dominance for $^{19}$C.

The charge form factor and the Coulomb breakup of two-neutron halo nuclei
were first calculated by \textcite{Hagen:2013xga}, \textcite{Hagen:2014tt}, and
\textcite{Acharya:2015tt} {at} LO.  The calculation of the charge form
factor was recently extended to NLO by \textcite{Vanasse:2016hgn}. (Vanasse also
corrected an error in the  prefactor of one term in the form factor calculation
of \textcite{Hagen:2013xga}.)  Since the value of the neutron-core
effective range is unknown and can only be estimated, we only quote
in Table~\ref{tab:electricradii} the LO charge radii for
${}^{11}$Li, ${}^{12}$Be, and ${}^{22}$C from~\cite{Vanasse:2016hgn},
together with the input value for the two-neutron separation energy.
The charge radius for ${}^{11}$Li has been measured by
\textcite{Puchalski:2006zz} and \textcite{Sanchez:2006zz}.
The result $\langle r_E^2 \rangle=1.104(85)$ fm$^2$ is consistent with the LO
result within the estimated $40$\% uncertainty due to range effects.
The charge radii of $^{14}$Be and $^{22}$C have not yet been measured.

\begin{table}
   \centering
   \begin{tabular}{|c| c c|} \hline
     Nucleus & $S_{2n}$ [MeV] & $\langle r^2_{E}\rangle$ [fm$^2$] \\
     \hline\hline
     $^{11}$Li & 0.3693(6) & 0.744 \\
     $^{14}$Be & 1.27(13)  & 0.126 \\
     $^{22}$C  & 0.11(6)   & 0.519 $^{+\infty}_{{-}0.274}$ \\
     \hline
 \end{tabular}
 \caption{Two-neutron separation energies and LO charge radii squared
  for four different two-neutron halos.  Adapted from \cite{Vanasse:2016hgn}.
 }
\label{tab:electricradii}
\end{table}

Halo EFT has also been used to calculate the matter radii of the two-neutron
halos {listed} in Table~\ref{tab:electricradii} up to NLO using (i) dimeron
propagators with resummed range effects~\cite{Canham:2008jd,Canham:2009xg} and
(ii) with a fully perturbative treatment of range
corrections~\cite{Vanasse:2016hgn}.  Both methods lead to consistent results.

\subsubsection{Correlations}

EFTs in general, and Halo EFT in particular, provide model-independent
correlations between different observables.  In Pionless EFT, the most prominent
universal correlations were discussed in Secs.~\ref{sec:Triton}
and~\ref{sec:Helion}. Such correlations have also proven useful in the analysis
of universal properties of ultracold atoms~\cite{Braaten:2004rn}.

In the previous sections, we have expressed the electromagnetic properties of
the ${}^{11}$Be system through the ERE parameters for $n$-${}^{10}$Be
scattering: $\gamma_0$, $\gamma_1$, $r_0$, and $r_1$.  These expressions can
be interpreted as correlations between scattering observables and
electromagnetic properties.  Analogously, there are correlations between
different electromagnetic observables.

As a specific example, we consider the correlation between the B(E1) strength
and the radius of the $1/2^+$ state in ${}^{11}$Be at LO.  Using
Eqs.~\eqref{eq:LOpradresult} and \eqref{eq:BE1} we obtain
\begin{equation}
 {\rm B(E1)} = \frac{2 e^2 Q_c^2}{15 \pi}
 \left(\langle r_E^2 \rangle_{^{11}{\rm Be}^*}
 - \langle r_E^2 \rangle_{^{10}{\rm Be}}
 \right)
 \, x \left[\frac{1 + 2x}{(1 + x)^2}\right]^2 \,,
\label{eq:BE1corr}
\end{equation}
where $Q_c$ is the charge of the core and $x=\sqrt{B_1/B_0}$ is the
{square root of the} ratio of
the neutron separation energies for the $1/2^-$ and $1/2^+$ states.  The B(E1)
strength is thus proportional to the mean-square radius of the {$1/2^-$}
state.
In the limit of vanishing neutron separation energy for the {$1/2^-$}
state, the B(E1) strength vanishes linearly with $x$.
Equation~\eqref{eq:BE1corr} can also be
used to obtain the charge radius of the  {$1/2^-$} state
{$\langle r_E^2 \rangle_{^{11}{\rm Be}^*}$} directly from
the measured value  of ${\rm B(E1)}$ and the neutron separation energies $B_1$
and $B_0$.  This gives
$\langle r_E^2 \rangle_{^{11}{\rm Be}^*} - \langle r_E^2 \rangle_{^{10}{\rm
Be}} = 0.35\ldots0.39~\fm^2$, depending on which experimental value for B(E1) is
used.  Similar correlations can be derived for other observables.

These correlations make Halo EFT a powerful tool to test the consistency of
experimental data and/or \abinitio calculations based on general
assumptions about the scaling of observables with $\Mlo$ and $\Mhi$.  They can
be combined with \abinitio results to obtain predictions for low-energy
observables as discussed in \ref{sec:halo+abinitio} and below. In this spirit,
\textcite{Braun:2018hug} used a correlation between the B(E2) value
for the transition $5/2^+ \to 1/2^+$ and the quadrupole moment of the
$5/2^+$-state in $^{15}$C to predict the quadrupole moment from \abinitio
calculations of the  B(E2) value. {\textcite{Lei:2018toi} used a
correlation between the $d\alpha$ $S$-wave scattering length and the
amount by which $^6$Li is bound with respect to the $np\alpha$ threshold
to argue that $^6$Li is a two-nucleon halo nucleus.}

\subsubsection{Neutron capture}

The inverse reaction of the photodissociation of one-neutron halo nuclei is
radiative neutron capture on the core nucleus, which can be relevant for a
variety of astrophysical processes.  The corresponding efforts in
Halo EFT have been reviewed by \textcite{Higa:2015yvf} and
\textcite{Rupak:2016mmz}.

One example is the radiative neutron capture on $^7$Li.  This reaction was
investigated in Halo EFT by \textcite{Rupak:2011nk}. They expressed the cross
section in terms of $n$-$^7$Li scattering parameters and showed that the LO
uncertainty comes from the poorly known $P$-wave effective range $r_1$.  The
low-energy data for this reaction can be described well by a one-parameter fit
yielding $r_1 =-1.47~\fm^{-1}$.  In subsequent work, \textcite{Fernando:2011ts}
extended this calculation to higher energies, {where} the $3^+$ resonance
becomes important.  Their results suggest a resonance width about three times
larger than the experimental value.  They also presented  power-counting
arguments that establish a hierarchy for electromagnetic one- and two-body
currents.

The radiative neutron capture on $^7$Li was refined by \textcite{Zhang:2013kja}
in an approach combining Halo EFT and \abinitio calculations.  They presented a
Halo EFT calculation that describes neutron capture to both the ground and
first excited states of $^8$Li.  Each of the possible final states were treated
as halo bound-state configurations of $^7$Li plus a neutron,
including low-lying excited states of the $^7$Li core.  The asymptotic
normalization coefficients of these bound states were taken from an \abinitio
calculation using a phenomenological potential.  In contrast to
\textcite{Rupak:2011nk}, they found good agreement with the ratio of partial
cross sections for different initial spin states.  Moreover, they obtained
excellent agreement with the measured branching rations between the two final
states.

\textcite{Rupak:2012cr} applied Halo EFT to the dominant E1 contribution to
radiative neutron capture on $^{14}$C including contributions from both
resonant and non-resonant interactions.  They found that significant
interference between these two mechanisms leads to a capture contribution that
deviates from simple Breit-Wigner resonance form.

\subsection{Proton halos}
\label{sec:phalos}

Proton halos are less common due to the delicate interplay between attraction
from the strong interaction and the Coulomb repulsion.  The presence of the
Coulomb barrier introduces the Coulomb momentum,
\begin{equation}
    \label{eq:Cb-mom}
 {k_C=Z_1 Z_2 \alpha\mu_{12}\,,}
\end{equation}
with {$Z_{1,2}$} {the particle charges and $\mu_{12}$ their reduced
mass,  as a new scale corresponding to the inverse of the Bohr radius of the
system.}  This scale is
independent of the hadronic scales and complicates the power counting
{(cf. the discussion for protons in Sec.~\ref{sec:pionless}).}
In general, the correct scaling of the Coulomb momentum with respect to
strong-interaction scales strongly depends on the system considered.
One focus of recent studies in Halo EFT has been, therefore, on the underlying
scaling relations in systems and reactions with Coulomb forces.

An EFT for $S$-wave proton halo nuclei was developed
by \textcite{Ryberg:2013iga}.  They analyzed the universal features of proton
{halos} bound due to a large $S$-wave scattering length and derived
LO expressions for the charge form factor and the radiative proton-capture
cross section.  In subsequent work~\textcite{Ryberg:2015lea} extended
the calculation to higher orders and analyzed the effect of finite-range
corrections.  They calculated the charge radius to NLO and the
astrophysical $S$-factor for low-energy proton capture to fifth order in the
low-energy expansion. Higher-order ERE parameters cannot contribute
to the E1 capture reaction, and thus the accuracy is only limited by
gauge-invariant counterterms.  As an application, \textcite{Ryberg:2015lea}
considered the $S$-factor for proton capture on $^{16}$O into the excited
$1/2^+$ state of $^{17}$F and quantified an energy-dependent model error to be
utilized in data fitting.  They also provided a general discussion of the
suppression of proton halos compared to neutron halos by the need for two fine
tunings in the underlying theory.
{\textcite{Schmickler:2019dcy,Schmickler:2019ewl} investigated
universal binding in few-body systems of up to four charged particles.
They showed that range corrections are generically enhanced in the
strong Coulomb case relevant for most nuclei.
}

\begin{figure}[tb]
 \centerline{\includegraphics*[width=7cm,clip=true]{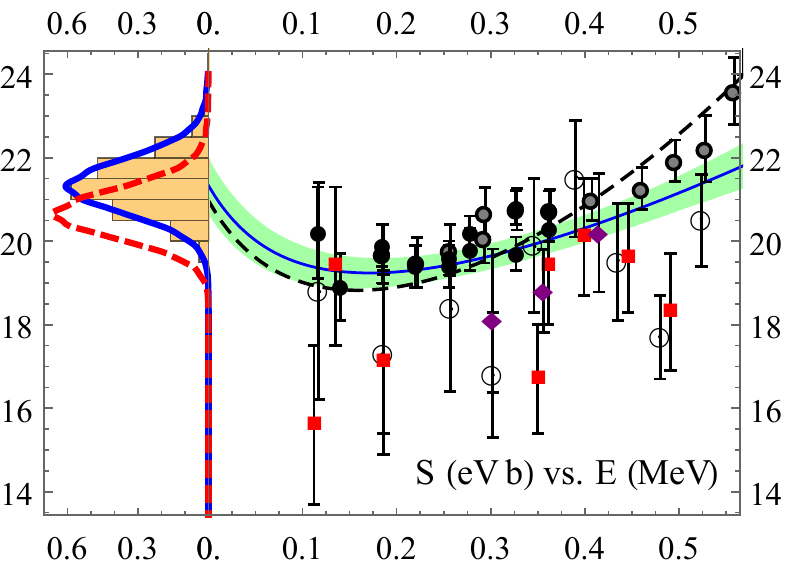}}
 \caption{Right panel: NLO $S$-factor as function of energy (solid blue
  curve). Shading indicates 68\% interval.  The dashed line gives
  the LO result.
  Left panel: 1d probability distribution functions for $S(0)$
  (blue line and histogram) and $S(20~\mathrm{keV})$ (red-dashed
  line).
  Figure taken from~\cite{Zhang:2015ajn},
  under the terms of the Creative
  Commons Attribution License,
  \url{https://creativecommons.org/licenses/by/4.0/}.
 }
 \label{fig:results3}
\end{figure}

The inclusion of Coulomb effects in $P$-wave halos was pioneered by
\textcite{Higa:2010zi}, who looked at low-energy $p\alpha$ scattering.
More extensive calculations were carried out later
by \textcite{Zhang:2014zsa}---extending their previous work for neutron
capture to proton halos---for $^7\mathrm{Be}(p,\gamma)^8\mathrm{B}$.
This reaction is important for analyzing solar neutrino
experiments~\cite{Adelberger:2010qa,Robertson:2012ib}.  However,
due to the Coulomb barrier it cannot be measured at the very low energies
required for this purpose and the data must be extrapolated.
\textcite{Zhang:2014zsa} demonstrated that Halo EFT together with
input from \abinitio calculations constitutes a powerful tool
to carry out this extrapolation.  They treated $^8\mathrm{B}$ as a shallow
$P$-wave bound state of a proton and a $^7\mathrm{Be}$ core and
included the first core excitation explicitly.  The couplings were fixed using
measured binding energies and $p$-$^7\mathrm{Be}$ $S$-wave scattering lengths,
together with $^8\mathrm{B}$ asymptotic normalization coefficients from
\abinitio calculations. They  emphasized the important role of
$p$-$^7\mathrm{Be}$ scattering parameters in determining the energy
dependence of $S(E)$ and demonstrated that their present uncertainties
significantly limit attempts to extrapolate the data to stellar energies.
\textcite{Zhang:2015ajn} extended this  calculation to NLO and used
Bayesian methods to determine the EFT parameters and the low-energy $S$-factor,
using measured cross sections and scattering lengths as inputs.
The results of their analysis, which reduced the uncertainty of $S(0)$ by a
factor of two, are shown in Fig.~\ref{fig:results3}.
Further details are given by \textcite{Zhang:2017yqc}.

In related work, \textcite{Ryberg:2014exa} pointed out that the charge
radius of $^8$B and the $S$-factor for $^7\mathrm{Be}(p,\gamma)^8\mathrm{B}$
are correlated at LO in Halo EFT.  This correlation thus provides indirect
access to the $S$-factor at low energies and serves as a consistency check.

\subsection{Cluster systems}
\label{sec:clusters}

Many nuclear states are close to a threshold for break-up into smaller
clusters, and are therefore amenable to an EFT approach where
these smaller clusters are the relevant degrees of freedom. For example,
several states of nuclei with $A=4 (n+1)$, $n\ge 1$ an integer,
and equal numbers of proton and neutrons are thought to be made of
alpha-particle clusters~\cite{Ikeda:1968}.  The most famous example is the Hoyle
state, the first $0^+$ excited state of $^{12}$C, which {owing} to its
position
near the $3\alpha$ threshold plays an important role in the creation of
$^{12}$C and $^{16}$O---and thus our type of life---in the universe.
Traditionally, these states have been investigated with a variety
of phenomenological approaches~\cite{Freer:2017gip}.

\begin{figure}[tb]
 \centerline{\includegraphics*[width=7.0cm,clip=true]{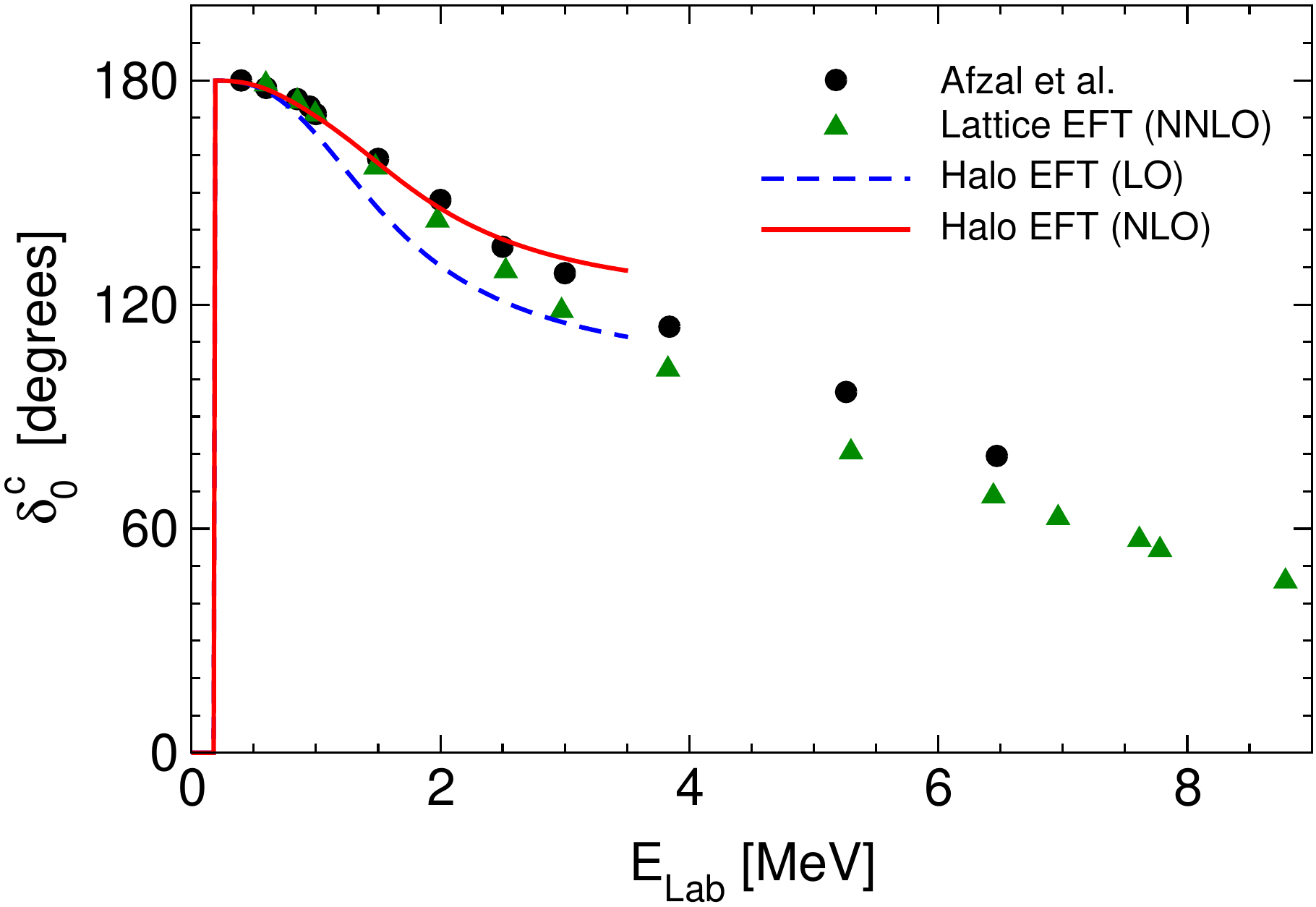}}
 \vspace*{-0pt}
 \caption{Halo EFT results for the $\alpha\alpha$ $S$-wave phase shift
  by \textcite{Higa:2008dn} as a function of the laboratory energy $E_{Lab}$.
  LO and NLO phase shifts are given by the (blue) dashed and (red) solid lines,
  respectively. The experimental data are from~\cite{Afzal:1969zz}, while the
  lattice EFT results are from \cite{Elhatisari:2015iga}.
 }
\label{fig:alal_pshift}
\end{figure}

The first step to study these systems in Halo EFT is $\alpha\alpha$
scattering.  \textcite{Higa:2008dn} developed a power counting for
this system, which is highly fine tuned.  Due to the subtle interplay of
strong and electromagnetic forces there is a narrow resonance at an energy of
about $0.1~\MeV$, the $^8$Be ground state.  The scenario
explored by \textcite{Higa:2008dn} can be viewed as an expansion around the
limit where, when electromagnetic interactions are turned off,
the $^8$Be ground state is at threshold and exhibits conformal invariance.
This implies treating the Coulomb momentum
{$k_C=2\alpha m_\alpha\simeq 60$ MeV},
where $m_\alpha$ is the alpha particle mass,
as a high-momentum scale and expanding observables in powers of
{$Q/(3k_C)$}, where $Q$ is a typical external momentum, in addition to the
standard expansion in the strong interactions.
{The Coulomb-modified scattering length is very large, and the
corresponding effective range almost saturates the Wigner bound for charged
systems~\cite{Koenig:2012bv}.}
The corresponding phase shifts are shown
in Fig.~\ref{fig:alal_pshift} together with the experimental data
from~\cite{Afzal:1969zz} and an \abinitio lattice EFT calculation
from~\cite{Elhatisari:2015iga}.
{Agreement with data seems to extend
somewhat beyond the laboratory energy $E_{\rm Lab} = 2$ MeV corresponding
to $k_C$.}
The sharp rise in the phase shift at low
energies is {a fine-tuned effect that is}
very difficult to describe in the \abinitio calculation,
{which displays a bound state instead.  In contrast,
the \abinitio calculation extends to much higher energies than Halo EFT.}

An RG analysis of the coupled channels $p+^7$Li and $n+^7$Be
which couple to a $2^-$ state of $^8$Be very close to the
$n+^7$Be  threshold was carried out {by \textcite{Lensky:2011he}.}
A more recent study involving $^8$Be concerned a reported anomaly in the
$e^+e^-$ production from the decay of one of the $1^+$ resonances to the ground
state. A careful {analysis inspired by Halo EFT} was carried by
\textcite{Zhang:2017zap}, who concluded that nuclear physics is unlikely to
explain the experimental result.

C and O production in stars also depends on the radiative capture of alpha
particles by $^{12}$C, $^{12}$C$(\alpha,\gamma)^{16}$O, at low energies.
As in Sec. \ref{sec:haloelectric}, parameters from
elastic $\alpha$-$^{12}$C scattering enter in a Halo EFT approach
to the capture process.  {\textcite{Ando:2016ban,Ando:2017uqc}}
developed a description of the elastic reaction taking the $^{12}$C ground state
as pointlike, and obtained asymptotic normalization coefficients
for some of the $^{16}$O states from a fit to phase shifts.

Very little has been done using Halo EFT for other cluster systems.
{An early application of Halo EFT for $l=2$ to the reaction $d + t \to n +
\alpha$ was carried out by \textcite{Brown:2013zla}.  However, they used
dimensional regularization with minimal subtraction and thus missed relevant
parameters.}  \textcite{Higa:2016igc} recently investigated another
important  process in the Sun, namely the radiative capture of an $\alpha$
particle on $^3$He.  They extracted an $S$-factor slightly above the average in
the literature, but consistent within error bars. {\textcite{Zhang:2018qhm}
recently performed a Bayesian analysis of this reaction without relying on
existing phase shift analyses {as a} constraint.}

\subsection{Outlook}
\label{sec:haloook}

In this section, we have reviewed the progress in Halo/Cluster EFT, a
short-range EFT with explicit fields for nucleon and cluster
degrees of freedom, designed for the description of halos and cluster nuclei.
Such systems have a very rich structure due to the emergence of new scales from
the Coulomb interaction between the clusters.  The Efimov effect plays an
important role for neutron halos and its application in the presence of Coulomb
interactions presents an exciting opportunity for the discovery of new
phenomena.

Halo EFT is conceptually very similar to the Pionless EFT discussed
in Sec.~\ref{sec:pionless}.  As a description of nuclei not limited to the
few-nucleon sector, it complements \abinitio approaches by parametrizing
universal relations between low-energy observables in systems dominated by
shallow bound states and low-lying resonances, and it quantifies the corrections
to these relations.  While such \abinitio calculations can be based on
interactions from Chiral EFT (which is reviewed in the following section), for
systems within its reach of applicability Halo EFT sets up a more
effective/efficient expansion.  Halo EFT promises a quantitative description
on the same footing of both nuclear structure and reactions of clusterized
systems and exotic isotopes, which is is a major challenge of contemporary
nuclear theory. {Nuclear reactions have been investigated both
in strict Halo EFT \cite{Schmidt:2018doj} as well as in accurate
phenomenological models of reactions with a Halo EFT motivated description
of the projectile \cite{Yang:2018nzr,Capel:2018kss}.}
{Many} of these reactions have impact on
astrophysical processes and even on the quantification of nuclear uncertainties
in experimental anomalies.

Future challenges include a better integration of \abinitio methods and
Halo EFT in order to maximally benefit from the strengths of both approaches.
The use of Bayesian statistics for the estimation of higher-order corrections
provides a method to account for the different sources of theory errors beyond
simple scaling arguments.  Finally, hypernuclei are a new and almost unexplored
arena for Halo EFT and universality.  {As we discuss in
Sec.~\ref{subsection:hyperchiral},} experimental data are not abundant
and a combination of Halo EFT and \abinitio methods {therefore} appears to
be especially
promising.

\section{Chiral EFT} \label{sec:chiral}
\subsection{Motivation}
\label{subsec:chiralintro}

As the typical momentum in a nuclear process increases beyond the pion mass,
pion effects can no longer be approximated by an expansion around the
zero-range limit.  As a response to the failure of early attempts to achieve
RG invariance in pion theories, an approach gradually emerged in the 1950s
where the nuclear potential and currents took purely phenomenological forms or,
at best, came from
the single (and, very occasionally, double) exchange of an arbitrary selection
of mesons.  The potential was almost always constructed so as to be regular
(\ie, not a singular potential~\cite{Frank:1971xx}), which in the case of meson
exchange was ensured by including physical form factors.  The relative ease of
solving the two-nucleon Schr\"odinger equation numerically made it possible to
produce exquisite fits to a large amount of two-nucleon data, frequently of the
same quality even with very different physical input into the potential.  In
contrast, a comparable description of $A>2$ systems seems to require
three-nucleon forces and two-nucleon currents, but the large variety of
possible structures posed a significant obstacle to this purely
phenomenological approach.  Moreover, the connection to QCD and the assignment
of systematic errors are not addressed.

Chiral EFT attempts to overcome these shortcomings by solving the RG problems
of earlier pion theories.  In hindsight, the latter were both too restrictive,
in the sense of not including all interactions consistent with symmetries,
and not restrictive enough, in the sense of not incorporating the
constraints of chiral symmetry.  Early forms of mesonic Chiral Perturbation
Theory (ChPT) date back to the 1960s and were extremely important in the
development of the EFT paradigm.  The mature version of mesonic ChPT took shape
in the 1980s~\cite{Weinberg:1978kz,Gasser:1983yg,Gasser:1984gg}, and processes
for $A=1$~\cite{Gasser:1987rb,Jenkins:1990jv,Jenkins:1991es,%
Bernard:1991rq,Bernard:1991rt} and $A\ge2$~\cite{Weinberg:1990rz,Rho:1990cf,%
Weinberg:1991um,Ordonez:1992xp,Weinberg:1992yk,VanKolck:1993ee} started
receiving significant attention in the late 1980s and early 1990s.  As this
section reviews, substantial progress has been made in understanding the
structure of the nuclear potential and currents, but Chiral EFT has not yet
produced a complete solution to the RG problems that plagued earlier pion
theories.

Extensive reviews exist of Chiral EFT applications to nuclear phenomenology
{by, for example, \textcite{vanKolck:1999mw,Beane:2000fx,Bedaque:2002mn,
Epelbaum:2008ga,Machleidt:2011zz,Epelbaum:2012vx}.}
We focus here on some of the conceptual
issues, which parallel those of Pionless EFT (Sec.~\ref{sec:pionless}) and
Halo/Cluster EFT (Sec.~\ref{sec:halo}).  Chiral EFT extends Pionless EFT
to processes with characteristic momentum $Q\sim \Mlo$, where $\Mlo \sim
m_\pi\ll \MQCD$.  As discussed in Sec.~\ref{subsec:chiralbasis}, the
breakdown scale $\Mhi \simle \MQCD$ depends in part on the degrees of freedom
kept explicit.  Section~\ref{subsec:chiralbasis} also discusses the pertinent
symmetries and Lagrangian.  The nuclear potential and currents, defined in
Sec.~\ref{subsec:NuclearEFTS}, are free of the IR enhancement that leads to
nuclear bound states and resonances.  As a consequence, contributions to the
potential can be treated similarly to contributions to amplitudes in ChPT, as
discussed in Sec.~\ref{subsec:chiralpot}.  The relation to experiment
\textit{via} amplitudes
{
and the more complex issue of their renormalization are}
reviewed in Sec.~\ref{subsec:chiralamps}.  Analogous considerations afflict
reactions with external light probes such as photons and pions, which are
sketched in Sec.~\ref{subsec:chiralreact}.  Section~\ref{subsec:Outissues}
lists some of the outstanding issues facing Chiral EFT.

\subsection{Basic elements}
\label{subsec:chiralbasis}

\subsubsection{Degrees of freedom and symmetries}
\label{subsubsec:Dofssymms}

By extending the Pionless EFT of Sec.~\ref{sec:pionless} to include an
isovector field $\vec\pi$ that collects the three charged pion states one
develops a
representation of QCD for $Q\sim m_\pi$, with $m_\pi$ now among the low-energy
scales collectively denoted by $\Mlo$.  In this EFT, pion exchange among
nucleons generates amplitudes that are no longer given by the ERE or a simple
generalization thereof.  Instead, there appear non-analytic functions of
$Q/m_\pi$ in all amplitudes.

The lightness of the pions relative to other hadrons can be explained naturally
if they are identified with the pseudo-Goldstone bosons of the spontaneous
breaking of (approximate) chiral symmetry, $SU(2)_{\rm L}\times
SU(2)_{\rm R}$ for two flavors.  In the chiral limit ($\bar m=0$,
$\varepsilon=0$, $e=0$), the QCD Lagrangian~\eqref{QCDL} has an exact chiral
symmetry.  In contrast, the spectrum shows only an approximate isospin
symmetry, $SU(2)_{\rm V}$.  {Close to} the chiral limit
$SU(2)_{\rm L}\times SU(2)_{\rm R}$ is an approximate symmetry of Eq.~\eqref{QCDL}.
Pions, which have vanishing mass in the chiral limit, acquire a relatively
small but
non-zero common squared mass $m_\pi^2=\OO(\MQCD \bar m)$ and a square-mass
splitting $\delta m_\pi^2=\OO(\alpha \MQCD^2/4\pi, \varepsilon^2 m_\pi^4/\MQCD^2)$
between charged and neutral states.

Pions should play a special role as long as $\bar m\ll \MQCD$.  In this section
we consider this regime, where we integrate out all other mesons because they
are expected to have masses $\OO(\MQCD)$.  The lightest of these is the
$\sigma$ with spin $S=0$ and isospin $I=0$, and a mass (half-width)
$m_\sigma=441~\MeV$ ($\Gamma_\sigma/2 =2 71~\MeV$)~\cite{Caprini:2005zr}.  This
suggests that the EFT radius of convergence is no larger than $\Mhi\sim
\sqrt{m_\sigma^2+ \Gamma_\sigma^2/4} \simeq 500~\MeV$.  The hadronic
EFT of QCD beyond this scale is not known.  The problem is power counting: by
NDA, interactions with derivatives will produce powers of $Q/\MQCD$ in
amplitudes; thus, as $Q$ approaches $\MQCD$ all interactions are equally
important.  To incorporate mesons in an EFT we need an argument that, at least
at a formal level, justifies treating their masses and interactions as small
with respect to $\MQCD$.  Typically this is accomplished by assuming QCD to
have further approximate symmetries.  For example, scale symmetry has been
invoked in the context of three flavors to justify the inclusion of a
scalar-isoscalar meson~\cite{Crewther:2013vea} and a dynamical ``vector''
symmetry~\cite{Georgi:1989gp} postulated for the $\rho$ meson.  Although very
interesting, such schemes have so far met with limited success, if any, away
from the mesonic sector.

One must, however, consider the effects of nucleon excitations. As mentioned in
Sec.~\ref{sec:introduction}, for $Q\ll m_N \sim \MQCD$, the nucleon mass is
inert.  For baryon-number conserving processes, the relevant mass scale
for other baryons is their mass splitting from the nucleon.  The Delta isobar
with $S=3/2$ and $I=3/2$ lies at $m_\Delta - m_N -i \Gamma_\Delta/2 \simeq (270
- 50\ii)~\MeV$~\cite{Arndt:2006bf}.  Although the mass difference $m_\Delta -
m_N$ does not vanish in the chiral limit, it is relatively small, in line with
arguments based on a large number of colors $N_{\rm c}$: when QCD is generalized
to an $SU(N_{\rm c})$ gauge theory, $m_\Delta - m_N=\OO(\MQCD/N_{\rm c})$.
A Deltaless version of Chiral EFT exists where the Delta isobar is integrated
out, but it fails before one reaches the Delta region in $A=1$ processes, which
leads to relatively large errors in $A\ge 2$
systems~\cite{Pandharipande:2005sx}.  To increase the radius of convergence
beyond $\sim \sqrt{(m_\Delta - m_N)^2 + \Gamma_\Delta^2/4} \simeq 275~\MeV$,
one {introduces~\cite{Jenkins:1991es,Hemmert:1997ye}}
a heavy field $\Delta$---a
four-component
object in spin and isospin space with the nucleon mass $m_N$ removed from its
rest energy.  As a consequence, the Delta kinetic energy and interactions are
also expanded around the non-relativistic limit.  Apart from the spin/isospin
structure, the main difference with respect to the nucleon is that a term
linear in the mass difference $m_\Delta - m_N$ (included in $\Mlo$) remains in
the Lagrangian.
{%
Explicit Delta propagation improves the description of data beyond threshold
\cite{Fettes:2000bb} and} enlarges the realm
of Chiral EFT beyond the Delta region once the power counting is properly
reformulated~\cite{Pascalutsa:2002pi,Long:2009wq}.

Whether other nucleon excitations should be introduced in Chiral EFT is less
clear.  The Roper resonance~\cite{Roper:1964zza} is special for several
reasons~\cite{Long:2011rt}.  First, its pole appears at an energy not much
above the Delta, $m_R - m_N - \ii\Gamma_R/2 \simeq (420 -
80\ii)~\MeV$~\cite{Arndt:2006bf}.   Other resonances lie at least $\Mhi\simeq
500~\MeV$ above threshold---the next resonance ($S_{11}$) has a mass $m_{S_{11}}
- m_N\simeq 500~\MeV$~\cite{Arndt:2006bf}---and it is difficult to see why they
should be incorporated in the EFT without the concomitant inclusion of meson
resonances.  Second, the Roper width is, numerically, $\Gamma_R\sim
\Gamma_\Delta (m_R - m_N)^3/(2(m_\Delta - m_N)^3)$, as expected from ChPT widths
scaling as $Q^3/\MQCD^2$.  This is not true for higher resonances, which
typically have relatively smaller widths.  As a consequence, the Delta and the
Roper nearly saturate the Adler-Weisberger sum rule, a result which suggests
that, together with the nucleon{,} they fall into a simple reducible
representation of the chiral group~\cite{Weinberg:1969hw,Beane:2002ud}.
Inclusion of an explicit Roper
{field~\cite{Banerjee:1995wz,Gegelia:2016xcw}}
improves the convergence of Chiral EFT around the Delta
resonance~\cite{Long:2011rt}, but it has not been systematically investigated.
In the following we consider Chiral EFT with nucleon and Delta fields only.

\subsubsection{Chiral Lagrangian}
\label{subsubsec:chiLag}

The construction of the most general chiral Lagrangian is based on the
theory of the non-linear realization of a
symmetry~\cite{Weinberg:1968de,Coleman:1969sm,Callan:1969sn}.  Different
parametrizations of the three-dimensional sphere $SU(2)_{\rm L}\times SU(2)_{\rm
R}/SU(2)_{\rm V} = SO(4)/SO(3)\sim S^3$ correspond to different choices of
pion fields.  Observables are of course independent of this choice.  Pions
appear in the chiral Lagrangian always as $\vec{\pi}/f_\pi$, where the
pion decay constant $f_\pi\simeq 92~\MeV$ is determined by the radius of $S^3$.
Because the three pions cannot provide a linear realization of $SO(4)$, they
transform non-linearly under chiral symmetry, so each term in the chiral
Lagrangian is associated with an infinite tower of interactions in powers of
$(\vec{\pi}/f_\pi)^2$.  Nucleon
{%
$N=( p \; n)^T$ and Delta $\Delta =(\Delta^{++}
\; \Delta^{+} \; \Delta^{0} \; \Delta^{-})^T$ fields}
can be chosen to transform under
chiral symmetry just as under an isospin rotation, but with an angle linear in
the pion field.  Covariant derivatives of the pion and baryon fields can be
defined so that they transform in the same way.  They are
$D_\mu=(1-\vec{\pi}^2/4f_\pi^2 +\cdots)\partial_\mu$ and
{$\mathcal{D}_\mu=\partial_\mu  + \ii \vec{\tau} \cdot (\vec{\pi}\times
\partial_\mu\vec{\pi})/4f_\pi^2+\cdots$}
for the pion and nucleon, respectively, where $\vec{\tau}$ are the Pauli
matrices in isospin space.  For the Delta, the form is the same as for the
nucleon with $\vec{\tau}$ replaced by the $I=3/2$ representation of $SO(3)$.
Delta-nucleon transition operators involve a set of $2\times 4$ isospin
matrices $\vec{T}$.
{%
Details
are given, \eg, by \textcite{Ordonez:1995rz}.}

The chiral Lagrangian is automatically chiral-invariant if it is built from
isospin-symmetric operators involving the baryon fields, their covariant
derivatives, and the pion covariant derivative.  Chiral-symmetric interactions
of the pions are thus proportional to the momentum.  Away from the chiral limit,
quark masses and electromagnetic interactions break chiral symmetry and even
the isospin subgroup.  The symmetry breaking pattern is known from
Eq.~\eqref{QCDL}, and interactions in the EFT are constructed to behave the
same way.  Thus, although chiral symmetry is not exact, information about QCD
is contained also in the {chiral-symmetry breaking} interactions.
These interactions {do not necessarily involve derivatives, but must be}
proportional to powers of the small parameters
$\bar m/\MQCD$, $\varepsilon$ and $e$, as well as the coefficients of
higher-dimensional operators (including violations of parity, time-reversal,
and possibly baryon-number and Lorentz invariance{)}.
The parameter $\bar m/\MQCD$ can be traded for
{$m_\pi^2/\MQCD^2$,}
while $\varepsilon$ and $e$ govern isospin-breaking quantities.
{
Electromagnetic
interactions are constrained by $U(1)_{\rm em}$ gauge invariance and appear in
two ways:
\textit{i)}
between low-energy photons and other fields \via chiral-covariant derivatives
enlarged to be gauge-covariant as well, and \via the electromagnetic field
strength;
\textit{ii)} among hadronic fields that originate in integrating
out energetic photons.
}

{Overall, chiral symmetry and its known breaking pattern}
{lead to a} {low-energy expansion because all interactions of pions
among} {themselves} {or with nucleons involve} {%
derivatives (which bring powers of $Q\sim \Mlo$ to amplitudes),
powers of $m_\pi^2\sim \Mlo^2$, or powers of smaller parameters.}
Choices of fields with
different chiral-transformation properties do not change this feature, but will
in general require delicate cancelations among different interactions.

As in Pionless EFT, it is most convenient to choose a heavy nucleon for which
the Dirac matrices reduce to the Pauli spin matrices $\boldsigma$.
Analogously, one can employ a heavy Delta field using the corresponding $S=3/2$
matrices.  Nucleon-Delta bilinears can be constructed with $2\times 4$ spin
transition matrices $\boldS$ analogous to the isospin
transition matrices $\vec{T}$.  Incorporating Lorentz invariance---in an
expansion in $Q/m_N$---is thus no more difficult in Chiral EFT than in Pionless
EFT.  In recent years it has become popular to use ``covariant'' baryon fields
from which the nucleon mass is not subtracted.  As any field redefinition, such
choices cannot affect observables in an essential way: amplitudes obtained from
different fields but the same power counting can only differ by higher-order
terms.  Although these differences are sometimes interpreted as an
indication of the ``best'' field choice, they merely reflect the error of
the truncation.

The baryon-number-conserving chiral Lagrangian can be split into pieces with
even numbers of fermion fields, $\LL= \LL_{f=0} + \LL_{f=2} + \LL_{f\ge 4}$,
where
\begin{subequations}
\label{chiLag}
\begin{multline}
 \LL_{f=0} =
 \frac{1}{2}
 \left[ \left(D_0 \vec\pi\right)^2 -(\boldD\vec\pi)^2
 - m_\pi^2\vec\pi^2 \left(1-\frac{\vec{\pi}^2}{4f_\pi^2}+\cdots\right)\right] \\
 \null + \cdots \,,
\label{chiLagpi}
\end{multline}
\begin{multline}
 \LL_{f=2} =
 N^\dagger \left(\ii \mathcal{D}_0 + \frac{\boldcalD^2}{2m_N} \right) N
 + \frac{g_A}{2f_\pi} N^\dagger\vec\tau \boldsigma N \cdot \boldD \vec{\pi} \\
 \null + \Delta^\dagger  \left(\ii \mathcal{D}_0 +m_N - m_\Delta \right)\Delta
 \\
 + \frac{h_A}{2f_\pi} \left(N^\dagger \vec{T} \boldS \Delta + \hc\right)
 \cdot \boldD \vec{\pi}
 + \cdots \,,
\label{chiLag1bar}
\end{multline}
\begin{widetext}
\begin{multline}
 \LL_{f\ge 4} =
 {-}\frac{C_{0\textrm{t}}}{2}
 \left(N^T\pPt N \right)^\dagger \left(N^T \pPt N \right)
 - \frac{1}{2}\left[C_{0\textrm{s}}
 + D_{2\textrm{s}}m_\pi^2\left(1-\frac{\vec\pi^2}{2f_\pi^2}+\ldots\right)\right]
 \left(N^T\pPs N \right)^\dagger \left(N^T \pPs N \right) \\
 \null - \frac{C_{2\textrm{s}}}{8}\Big\{\left(N^T\pPs N \right)^\dagger
 \big[N^T \pPs \boldcalD^2 N  + (\boldcalD^2 N)^T \pPs N\big] + \hc\Big\}
 - \frac{C_{2\textrm{t}}'}{4}
 \Big[
 \left(N^T \pPt \boldcalD N \right)^\dagger \cdot
 \left(N^T \pPt \boldcalD N \right) \\
 \null + \left((\boldcalD N)^T \pPt N \right)^\dagger \cdot
 \left((\boldcalD N)^T \pPt N \right)
 \Big]
 + \frac{G_A}{2f_\pi}  N^\dagger\!N \, N^\dagger \boldsigma\vec\tau N
 \cdot \boldD\vec\pi
 -H_0 \, N^\dagger\!N \, N^\dagger\!N \, N^\dagger\!N
 + \cdots \,,
 \label{chiLag2bar}
\end{multline}
\end{widetext}
\end{subequations}
with LECs $g_A$, $h_A$, {$C_{0\textrm{s,t}}$, $D_{2\textrm{s}}$,
$C_{2\textrm{s}}$, $C_{2\textrm{t}}'$, $G_A$}, and $H_0$, and where we used a
notation similar to Eq.~\eqref{eq:L-NN}.  Only a few representative interactions
are shown explicitly here, others (including more fields, derivatives, powers of
$m_\pi^2$, isospin breaking, \etc) being relegated to the ``$\cdots$''.  Note
that many terms can be written in different forms with Fierz reordering and/or
field redefinitions.  One can also introduce dibaryon fields as described in
Sec.~\ref{sec:Dibaryons}
{and done,
for example, by~\textcite{Soto:2011tb} and \textcite{Long:2013cya}.}

Particularly convenient for nuclear processes, where nucleon energies and
momenta are of very different magnitudes, is to use field redefinitions to
eliminate time derivatives of the nucleon field in favor of spatial derivatives.
When interaction terms appear in the classical Lagrangian which depend on time
derivatives, the effective Lagrangian obtained \via the path integral of the
Hamiltonian contains additional
terms~\cite{Charap:1970xj,Salam:1971sp,Honerkamp:1996va,Charap:1971bn,
Gerstein:1971fm}.  These do not vanish in general if a momentum cutoff is
used.  Generally, the easiest way to respect symmetries is to implement
regulators as operators in the chiral Lagrangian constructed from
chiral-covariant objects
{\cite{Slavnov:1971aw,Djukanovic:2004px,Long:2016vnq}.}

If $m_\Delta-m_N$ is considered a large scale, the Delta is integrated out
and appears only through LECs starting at one order higher than in the Deltaful
EFT.  If $m_\pi$ is also considered a large scale, pions are integrated
out as well.  Although the chiral Lagrangian formally reduces to the pionless
form~\eqref{eq:L-NN} when terms with pions and Deltas are omitted from
{%
Eq.~\eqref{chiLag}}, one should keep in mind that the remaining LECs depend on
what degrees of freedom appear in the EFT.

\subsection{Chiral Perturbation Theory and the nuclear potential}
\label{subsec:chiralpot}

A great advantage of EFT over earlier attempts to describe nuclear physics from
field theory is its explicit focus on the regime of momenta well below the
nucleon mass, where the theory splits into sectors of fixed nucleon number $A$.
As pointed out in Sec.~\ref{sec:introduction}, there are significant
differences between $A\le 1$ and $A\ge2$ processes.

\subsubsection{Power counting}
\label{subsubsec:PC}

To express amplitudes in an expansion in powers of $Q/\MQCD$, as in
Eq.~\eqref{Texp}, one needs to count powers of both $Q\sim \Mlo$ and $\MQCD$.
For $Q$ one first relates nucleon energies and momenta, and this relation in
general depends on the sector of the theory.  For $A\le 1$, typically (but not
always) $E=\OO(Q)$, while $A\ge 2$ processes with only nucleons in external
legs involve energies $E=\OO(Q^2/m_N)$.  For pions, since we count $m_\pi$ as
$\Mlo$, $E=\OO(Q)$.  The crucial assumption in counting powers of $\MQCD$ is
naturalness, namely that an LEC needed to eliminate cutoff dependence of a loop
at a certain order has finite pieces of the same order.

For an $A\le 1$ Feynman diagram, the various elements scale (after
renormalization) as:
\begin{subalign}[eq:ChPTPC]
 \text{derivative} &\sim  Q \,,
\label{derivative} \\
 \text{baryon, pion propagator} &\sim Q^{-1}, Q^{-2} \,,
\label{props} \\
 \text{(pion) loop integral} &\sim (4\pi)^{-2}Q^4 \,,
\label{loopintegral}
\end{subalign}
where the factor of {$(4\pi)^{{-}2}$} is typical of relativistic loops.  The
sizes of LECs
can be estimated \via NDA, Eq.~\eqref{NDA}.  Chiral-symmetric operators
depend on arbitrary powers of the reduced strong-coupling constant $g_{\rm
red}=g/(4\pi)$, which for consistency should be taken as 1.  NDA applied to
Eq.~\eqref{chiLagpi} gives $f_\pi = \OO(\MQCD/4\pi)$, and for a generic
LEC~\cite{Manohar:1983md,Georgi:1986kr}
\begin{equation}
 c_i=\OO\left(\frac{c_{i,\text{red}}}{f_\pi^{f_i+p_i-2} \MQCD^{\Delta_i}}\right)
 \,,\quad \Delta_i\equiv d_i+f_i/2 -2 \,,
\end{equation}
where $d_i$, $f_i$ and $p_i$ the number of, respectively, derivatives, baryon
fields, and pion fields of the corresponding operator.  The reduced LEC
$c_{i,\text{red}} = \OO(1)$ for a chiral-symmetric operator.  NDA is consistent
with the non-relativistic expansion since applied to Eq.~\eqref{chiLag1bar} it
gives $m_N=\OO(\MQCD)$.  Keeping explicit Deltas means, however, that we are
taking $(m_\Delta -m_N)_{\rm red}=\OO(\Mlo/\MQCD)\ll 1$, as suggested by
large-$N_{\rm c}$ arguments.  A {chiral-symmetry breaking} operator stemming from
the quark masses will have a reduced LEC proportional to powers of
${\bar m}_{\rm red}={\bar m}/\MQCD=m_\pi^2/\MQCD^2=\OO(\Mlo^2/\MQCD^2)$ and
$\varepsilon\simle \OO(1)$, using that $m_\pi^2=\OO(\MQCD \bar{m})$ when NDA
is again applied to Eq.~\eqref{chiLagpi}.
The effect of integrating out hard photons is given by powers of
$e_{\rm red}^2 = (e/(4\pi))^2\simle \OO(\Mlo^3/\MQCD^3)$~\cite{VanKolck:1993ee,%
vanKolck:1995cb} in the corresponding
reduced LEC.\footnote{How one accounts for $e_{\rm red}$ relative to other
parameters is somewhat ambiguous, and to some extent a matter of convenience.
Sometimes the choice $e_{\rm red}^2=\OO(\Mlo^2/\MQCD^2)$ is made in the
literature.  This choice leads a pion mass splitting
$\delta m_\pi^2=\OO(\alpha \MQCD^2/4\pi)=\OO(m_\pi^2)$
and to a Coulomb potential comparable to OPE for momenta $Q\sim m_\pi$.
That means electromagnetic effects at LO, an overestimate.  A similar
ambiguity affects $\varepsilon\simeq 1/3$, which
can be counted as $\OO(1)$ or as $\OO(\Mlo/\MQCD)$.}
If we take  $\varepsilon=\OO(1)$, then
$c_{i,\text{red}} = \OO(\Mlo^{n_i}/\Mhi^{n_i})$ where $n_i$ counts the powers of the
low-energy scales $m_\pi$, $m_\Delta -m_N$ and $(e/(4\pi))^{2/3}m_N$.
It is convenient to enlarge the definition of $d_i$ to include $n_i$ as well.
The interactions displayed in Eqs.~\eqref{chiLagpi} and~\eqref{chiLag1bar}
then have $\Delta_i=0$, except for the nucleon recoil term $\boldcalD^2/2m_N$
with $\Delta_i =1$.  Chiral symmetry guarantees $\Delta_i\ge 0$ for all
interactions stemming from the terms shown explicitly in
Eq.~\eqref{QCDL}.\footnote{The choice of heavy baryon fields makes this evident
by removing positive powers of the large nucleon mass from the Lagrangian.}

Using standard identities for connected graphs, a diagram with $L$ loops and
$V_i$ vertices with chiral index $\Delta_i$ contributes to the
amplitude~\eqref{Texp} a term with~\cite{Weinberg:1978kz}\footnote{Note that
$\nu$ can be written in various ways that differ by an additive factor (and by
the overall normalization).  In writing Eq.~\eqref{chiralnu}---as well as
Eq.~\eqref{chiralnupot} below---we chose a form where LO corresponds to
$\nu=0$.}
\begin{equation}
 \nu = 2L + \sum_i V_i \Delta_i \,,\quad \mathcal{N} = f_\pi^{4-3A-E_b} \,,
\label{chiralnu}
\end{equation}
where $E_b$ is the number of external bosons.  The factor $2L$ implies that
ChPT amplitudes are in general perturbative, \ie, the non-analytic functions
$F^{(\nu)}$ in Eq.~\eqref{Texp} can be obtained from a finite number of Feynman
diagrams.  Because of the way NDA was inferred, these loop diagrams are
accompanied by higher-index interactions that provide the necessary
counterterms for RG invariance in the sense of
Eq.~\eqref{TRGtrunc}.  Because of chiral symmetry, $\nu \ge 0$.\footnote{If
interactions in the ``$\cdots$'' of Eq.~\eqref{QCDL} are considered, $\Delta$
and $\nu$ can be negative.  However, these interactions are small due to
strengths that are much smaller than our expansion parameter $Q/\MQCD$. Such
interactions can still be included perturbatively.}
LO ($\OO({\mathcal{N}})$) and NLO (relative $\OO(Q/\MQCD)$) consist of
tree-level ($L=0$) diagrams made out of interactions with chiral index
$\Delta=0$ and, respectively, no or one interaction with $\Delta=1$.  They are
equivalent to ancient current algebra.  Baryons are not only nonrelativistic,
but also approximately static.  Quantum-mechanical corrections ($L\ge 1$) start
at N$^2$LO (relative $\OO(Q^2/\MQCD^2)$).  As $\nu$ increases, progressively
more short-range physics is included, which account for details of hadron
structure.  Many good reviews of ChPT exist, see
for example~\cite{Bernard:1995dp,Bernard:2007zu}.

That is not to say that within certain regions of phase space perturbation
theory does not break down.  The power counting~\eqref{chiralnu} is only meant
as a general rule, which is bound to fail in specific situations.  For example,
within a momentum window of size $\OO(Q^3/\MQCD^2)$ around the Delta
pole---where $E\simeq m_\Delta -m_N$---the one-loop diagrams that make for most
of the Delta width become important and a resummation is necessary at
LO~\cite{Pascalutsa:2002pi,Long:2009wq}.  Similarly, around certain points
below threshold where energies are $\OO(Q^2/\MQCD)$, nucleon recoil needs to be
resummed and elevated to LO \cite{Lv:2016slh}.  The latter resummation is
naturally incorporated by the use of non-heavy baryon (``covariant'')
{fields~\cite{Becher:1999he,Fuchs:2003qc}}, but in the literature it is
often wrongly implied that such a choice is necessary. In general, the choice
of fields is unimportant, but one should always ensure that the power
counting~\eqref{chiralnu} applies to the kinematic region of interest.  Any
resummation needs to be done carefully so as not to break RG invariance.

Nucleon-only $A\ge 2$ processes have $E=\OO(Q^2/m_N)$ and require a
resummation as {well~\cite{Weinberg:1991um}}.  We will return to this in
Sec.~\ref{subsec:chiralamps}, focusing for now on the sum of ``irreducible''
diagrams involving $A\ge 2$ nucleons (and $E_b=0$), which is defined (see
Sec.~\ref{subsec:NuclearEFTS}) as the ``full'' nuclear
potential.\footnote{{For a recent attempt to treat pions dynamically
instead through quantum Monte Carlo methods, see \cite{Madeira:2018ykd}.}}
The analogous currents are briefly discussed in Sec.~\ref{subsec:chiralreact}.

By construction, the potential is free of IR enhancement, and we expect a power
counting similar to ChPT's to apply as long as interactions with $f\ge 4$ also
obey NDA.  A complication is that the full potential {introduced in
Sec. \ref{subsec:NuclearEFTS}} includes disconnected
diagrams.  Each disconnected piece scales as $(4\pi)^n Q^{-4}$, where $n$ is an
integer, coming from the fact that the extra four-dimensional delta function,
which enforces momentum conservation, also eliminates a loop integral. Weinberg
and others~\cite{Weinberg:1991um,Ordonez:1992xp,Weinberg:1992yk,VanKolck:1993ee}
assumed $n=2$ on the basis of Eq.~\eqref{loopintegral}, while
\textcite{Friar:1996zw} took $n=1$, which is consistent with the
nonrelativistic nature of reducible loops as discussed in
Sec.~\ref{subsubsection:conpiless}.  As a consequence, a diagram with
$1\le C\le A-1$ separately connected pieces contributes to the
potential~\eqref{Vexp} with~\cite{Weinberg:1991um,Friar:1996zw}
\begin{spliteq}
 \mu &= n (A-1-C) + 2L + \sum_i V_i \Delta_i \,, \\
 {\tilde{\mathcal{N}}} &= (4\pi)^{(2-n)A}f_\pi^{4-3A} \,.
\label{chiralnupot}
\end{spliteq}
This power counting (with $n=2$) has been used in most studies of
chiral potentials to date.

\subsubsection{Nuclear potential}
\label{subsubsec:nukepot}

In Chiral EFT, Fig.~\ref{fig:PionsOut} is undone: pion exchange appears
explicitly in the potential, with the remaining contact interactions accounting
for higher-momentum physics.  In contrast to Pionless EFT, the potential itself
involves (irreducible) loops, where energies are comparable to momenta and
nucleons are approximately static.  The long-range pion-exchange contributions
appear in all partial waves and yield many-body forces consistent with
$2N$ forces and  the hadronic physics described by ChPT.  They are not known to
violate the estimate~\eqref{chiralnupot}.

Pion loops also generate short-range contributions that cannot be separated
from contact interactions in the potential.  The piece of a LEC that removes
the cutoff dependence in irreducible loops, or more generally the piece that
obeys NDA, is sometimes referred to as a {``primordial
counterterm''}~\cite{Long:2011xw,Long:2012ve}.  This is to distinguish it from
another piece that renormalizes the reducible loops of the full amplitude.
This
{additional piece}
may violate NDA and be present at a lower order than the primordial
piece, as discussed in Sec.~\ref{subsec:chiralamps}.
As in Pionless EFT, the potential is \emph{not} cutoff independent.

\paragraph{Leading order}

The full LO potential has maximum $C$ ($C_{\rm max}=A-1$, from $A-2$
disconnected lines): it consists of the sum over pairs of the $2N$ potential at
tree level ($L=0$) constructed entirely from $\Delta=0$ interactions.  The
long-range $2N$ potential consists of static one-pion exchange (OPE) and the
primordial counterterms are the two LECs of the non-derivative chiral-symmetric
contact interactions in Eq.~\eqref{chiLag2bar}:
\begin{multline}
 V^{(0)} = -\frac{4\pi}{m_N\MNN}
 \frac{\vec{\tau}_1\cdot\vec{\tau}_2}{\boldq^2+m_\pi^2}
 \left(S_{12}(\boldq)-\frac{m_\pi^2}{3}\boldsigma_1\cdot\boldsigma_2\right) \\
 + {C_{0\mathrm{s}}P_{\mathrm{s}}  + C_{0\mathrm{t}}P_{\mathrm{t}}} \,,
\label{OPE}
\end{multline}
where the indices 1 and 2 label the two nucleons, $\boldq$ is the transferred
momentum{, and
$S_{12}(\boldq)=(\boldsigma_1\cdot\boldq)(\boldsigma_2\cdot\boldq)
-\boldq^2(\boldsigma_1\cdot\boldsigma_2)/3$ is the tensor operator.}
OPE is static because the transferred energy, related to nucleon recoil, is
small (relative $\OO(Q/\MQCD)$) compared to $\abs{\boldq}$.  The scale
that controls the OPE strength, in a form we can compare with short-range
interactions in Pionless EFT, is given by~\cite{Kaplan:1998tg,Kaplan:1998we}:
\begin{equation}
 \MNN = \frac{16\pi f_\pi^2}{g_A^2 m_N} = \OO(f_\pi) {\,,}
\label{OPEscale}
\end{equation}
{using NDA.}  OPE gives rise in coordinate space to
\textit{i)} a tensor potential that is as singular $\sim1/r^3$ as $r\to0$, and
\textit{ii)} the regular Yukawa potential.
The tensor potential is non-vanishing only for total spin $s=1$ and can mix
waves with $l=j\pm 1$.  It is attractive in some uncoupled waves like $^3P_0$
and $^3D_2$, and in one of the eigenchannels of each coupled wave.  The Yukawa
potential is attractive in isovector (isoscalar) channels for $s=0$ ($s=1$).
The other two terms in Eq.~\eqref{OPE} are contact interactions, which for
large cutoffs contribute only to the $^3S_1$ and $^1S_0$ channels.  A contact
interaction from OPE has been eliminated through the redefinition
\begin{equation}
 C_{0\mathrm{s}} + \frac{4\pi}{m_N\MNN} \to C_{0\mathrm{s}}
 = \OO\left(\frac{4\pi}{m_N \MNN}\right) \,.
\label{C0redef}
\end{equation}

\paragraph{Subleading orders}

\begin{figure}[tb]
\begin{center}
\includegraphics[width=\columnwidth]{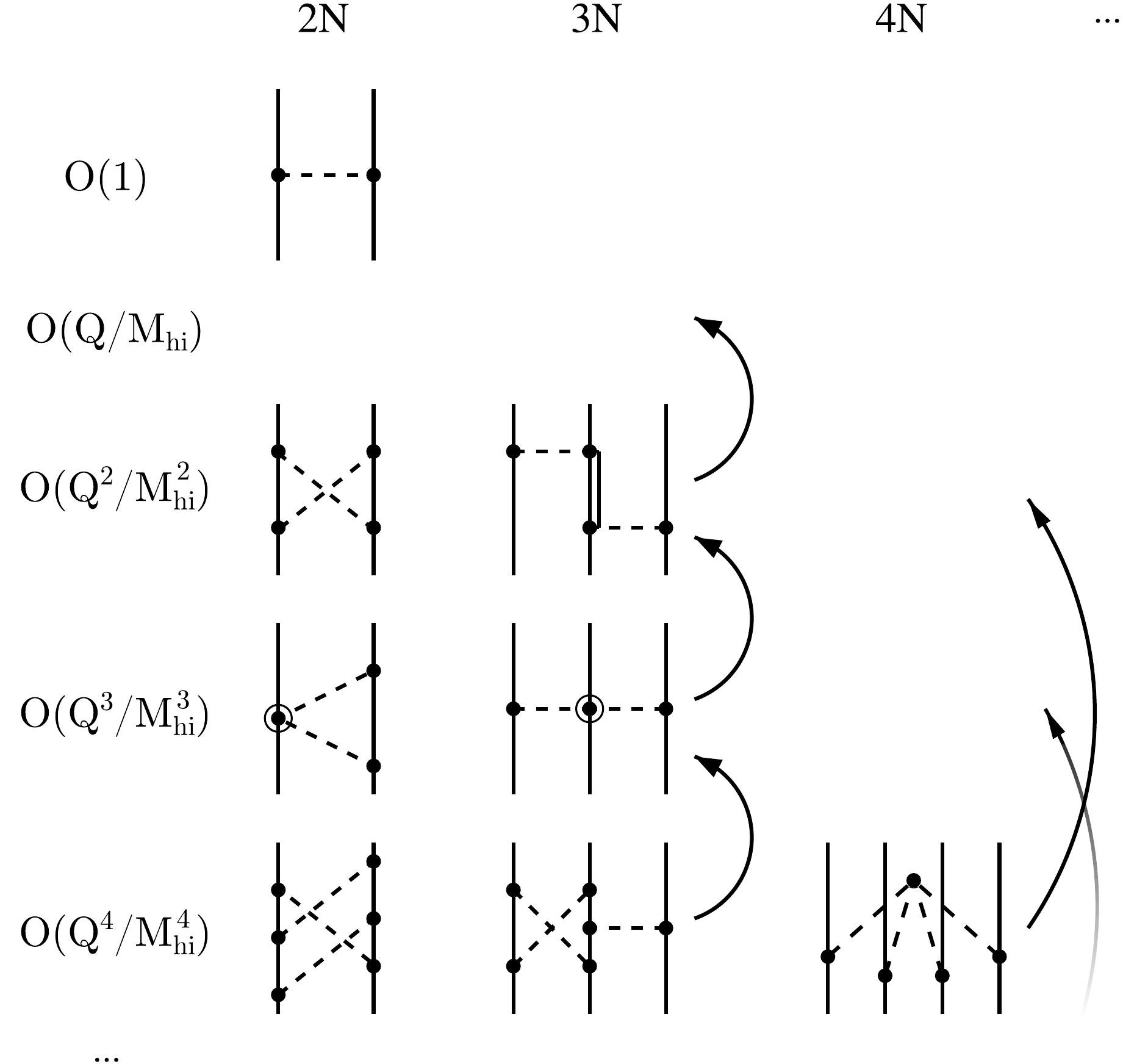}
\end{center}
\caption{Sample of diagrams representing the pion-range components of
 the $AN$ nuclear potential in Chiral EFT, according to
 Eq.~\eqref{chiralnupot} with $n=2$.  The order of the contributions is
 indicated as $\OO(Q^\mu/\Mhi^\mu)$, $\mu\geq0$, where $Q\sim\mpi$ and
 $\Mhi\sim\MQCD$.  Arrows in the $3N$ and $4N$ columns indicate the
 changes for $n=1$.  A solid (double) line stands for a nucleon (nucleon
 excitation), while a dashed line, for a pion. A circle around the central
 solid circle denotes an inverse power of $\Mhi$.
}
\label{fig:chiralpot}
\end{figure}

The order increases as the chiral index $\Delta$, the number of loops $L$, and
the number of nucleons in connected pieces increase.  In much of the literature
the potential at relative $\OO(Q^{\mu}/\MQCD^{\mu})$ is referred to as
N$^{\mu-1}$LO, but this notation is not flexible enough to accommodate changes
in the power counting described in Set.~\ref{subsec:chiralamps}, which suggest
$n=1$ in Eq.~\eqref{chiralnupot} and also
{departures} from NDA.  For clarity, we
denote the order of contributions using their explicit scaling throughout the
rest of this section.  The structure of the long-range nuclear potential
is shown schematically in Fig.~\ref{fig:chiralpot}.

The first few-body forces arise~\cite{VanKolck:1993ee,vanKolck:1994yi} at
$\OO(Q^n/\MQCD^{n})$ compared to LO, from $\Delta=0$ and $L=0$ with $C=A-2$:
\begin{itemize}
\item
a $3N$ TPE force via an intermediate Delta---the Fujita-Miyazawa
force~\cite{Fujita:1957zz}, shown in Fig.~\ref{fig:chiralpot};
\item
nucleon-only $3N$ and ``double-pair'' forces (for $A\ge 4$) when, in a
time-ordered diagram, a $2N$ interaction occurs while a pion is flying between
two nucleons.
\end{itemize}
Forces of the second type exactly cancel against nucleon recoil in the $2N$ OPE,
once the latter is inserted in the Lippmann-Schwinger
equation~\cite{Weinberg:1991um,Ordonez:1992xp,VanKolck:1993ee,vanKolck:1994yi}.
But recoil is an $\OO(Q/m_N)$ effect compared to LO: the cancellation
implies that $(4\pi)^{2-n}Q^2/\MQCD^2\sim Q/m_N$.  For $n=2$ one obtains $Q/m_N
\sim Q^2/\MQCD^2$ at variance with the NDA that underlies the power counting.
In contrast, if one takes NDA seriously, $m_N=\OO(\MQCD)=\OO(4\pi f_\pi)$, then
$n=1$ and $Q\sim f_{\pi}=\OO(M_{NN})$.  As we discuss in
Sec.~\ref{subsec:chiralamps}, this is consistent with the counting of factors
of $4\pi$ suggested by Pionless EFT.
{%
Not all authors count $m_N$, and thus implicitly choose $n$, in the same
way. However, regardless of how $m_N$ is counted, Chiral EFT, just as
Pionless EFT, implements the constraints of Lorentz invariance in a $Q/m_N$
expansion.}

After this cancellation, the $2N$ potential vanishes at relative $\OO(Q/\MQCD)$
if we neglect parity
violation~\cite{Weinberg:1991um,Ordonez:1992xp,VanKolck:1993ee,vanKolck:1994yi}.
For $n=1$, the Fujita-Miyazawa $3N$ force survives at this order.
It is demoted
to relative $\OO(Q^2/\MQCD^2)$ if $n=2$, in which case the full potential
vanishes at $\OO(Q/\MQCD)$.  More generally, the first $aN$ force ($L=0$, all
interactions with $\Delta=0$) is expected to appear at relative
$\OO(Q^{n(a-2)}/\MQCD^{n(a-2)})$.  The relative suppression of few-body forces
\cite{Weinberg:1991um,Ordonez:1992xp,VanKolck:1993ee,vanKolck:1994yi} is in
agreement with the experience drawn from phenomenological potentials containing
explicit pion exchange, for which $3N$ forces are usually found necessary at
the 10\% level---for example, to provide $\sim 1$ MeV to the triton binding
energy $\simeq 8.5~\MeV$.  The explanation for the smallness, but
non-negligibility, of {phenomenological} few-body forces was an early
success of Chiral EFT.

\paragraph{$2N$ potential}

At relative $\OO(Q^2/\MQCD^2)$, corrections to OPE merely shift existing
couplings---for example, $g_A$ in Eq.~\eqref{OPEscale} receives a
contribution proportional to $m_\pi^2$, the so-called Goldberger-Treiman
discrepancy.  The long-range $2N$ potential consists of two-pion exchange (TPE),
the so-called box, crossed-box (shown in Fig.~\ref{fig:chiralpot}), triangle
and football diagrams built out of $\pi N^\dagger N$ and $2\pi N^\dagger N$
interactions with chiral index $\Delta=0$.  For the latter three types, all
combinations of nucleons and Deltas need to be considered in intermediate
states.  For the box diagram with nucleons only, once-iterated OPE needs to be
subtracted.  The primordial counterterms consist of all possible two-derivative
chiral-symmetric contact
interactions~\cite{Ordonez:1992xp,VanKolck:1993ee,Ordonez:1995rz}, such as the
$C_{2\mathrm{s}}$ and $C_{2\mathrm{t}}'$ terms in Eq.~\eqref{chiLag2bar}, and
no-derivative {chiral-symmetry breaking} terms linear in the quark masses,
such as $D_{2\mathrm{s}}$.  The constraints imposed by relativity on these
primordial counterterms have been discussed by \textcite{Girlanda:2010ya}.

At relative $\OO(Q^3/\MQCD^3)$, apart from further contributions to OPE
parameters, the $2N$ potential is made up of TPE with one $\Delta=1$ vertex,
such as the triangle diagram shown in Fig.~\ref{fig:chiralpot}.  For $n=1$,
Galilean corrections ($\propto m_N^{-1}$) should be kept, while for $n=2$ they
contribute only at next order.  There are no new contact interactions at this
order.

The isospin-symmetric $2N$ potential up to $\OO(Q^3/\MQCD^3)$ was derived early
on~{\cite{Ordonez:1992xp,VanKolck:1993ee,Ordonez:1995rz,Kaiser:1997mw,%
Kaiser:1998wa}} and rederived many times since.
{\textcite{Epelbaum:1998ka} introduced the unitary transformation
method which allows for the separation of the iterated OPE with
a consistent set of $1/m_N$ corrections.}
{\textcite{Friar:1999sj} discusses the various
forms, including pre-EFT results, and the issues involved
in the separation of iterated OPE. The potential at this order resembles}
phenomenological potentials where pion exchange is supplemented by
a short-range structure.  {The TPE part, which carries information about
the chiral symmetry of QCD, involves LECs that can be determined
from pion-nucleon scattering.  (For recent work, see
\cite{Hoferichter:2015tha,Siemens:2016jwj}.)
It is a chiral analog of the van der Waals potential}
and behaves at short distances as $1/r^5$, $1/r^6$, or $1/r^7$ depending on
order and number of intermediate Deltas, and has the qualitative features of
heavier-meson exchange potentials~\cite{Kaiser:1997mw,Kaiser:1998wa}.  The TPE
from Chiral EFT without explicit Deltas successfully replaces heavier-meson
exchange in  the Nijmegen partial-wave analysis of $2N$
data~\cite{Rentmeester:1999vw,Rentmeester:2003mf};
for a modern version, see \cite{Perez:2013oba,Perez:2014bua}.

The $2N$ potential has now been extended to higher orders.  One- and
two-loop TPE and two-loop three-pion exchange (see Fig.~\ref{fig:chiralpot})
diagrams at $\OO(Q^4/\MQCD^4)$ were calculated by \textcite{Kaiser:1999ff,%
Kaiser:1999jg,Kaiser:2001pc,Kaiser:2001at,Kaiser:2015yca}.  More recently the
long-range Deltaless potential has been constructed at
$\OO(Q^5/\MQCD^5)$~\cite{Kaiser:2001dm,Epelbaum:2014sza,Entem:2015xwa}, and even
$\OO(Q^6/\MQCD^6)$~\cite{Entem:2015xwa}.  By parity conservation, primordial
counterterms only appear at even orders.

\paragraph{$3N$ potential}

Beyond the Fujita-Miyazawa term, $3N$ forces have a similar hierarchy.  At
$\OO(Q^{n+1}/\MQCD^{n+1})$, the $3N$ potential contains TPE diagrams where one
interaction has $\Delta=1$ (see Fig.~\ref{fig:chiralpot}).  Again, the form of
TPE is constrained by chiral symmetry, and provides a chiral-corrected
version of the earlier Tucson-Melbourne (TM) potential~\cite{Coon:1978gr},
sometimes called the TM$'$
potential~\cite{Friar:1998zt,Huber:1999bi,Coon:2001pv}, and close in form to
the Brazil potential~\cite{Coelho:1984hk}.  There are no additional
isospin-symmetric contributions from Deltas~\cite{Epelbaum:2007sq}, but there
are mixed one-pion/short-range and purely short-range components originating in
the interactions with LECs $G_A$ and $H_0$, respectively, in
Eq.~\eqref{chiLag2bar}~\cite{vanKolck:1994yi,Epelbaum:2002vt}.  Again, parity
conservation implies primordial counterterms only at every second order.

The primordial counterterms at $\OO(Q^{n+3}/\MQCD^{n+3})$ have been listed
by \textcite{Girlanda:2011fh}.  Relativistic corrections, which appear at this
order for $n=1$, have been calculated by \textcite{Bernard:2011zr}. At
$\OO(Q^{n+2}/\MQCD^{n+2})$ the first loops in the $3N$ force appear as indicated
in Fig.~\ref{fig:chiralpot}, and have been derived without Deltas by
\textcite{Ishikawa:2007zz} {and \textcite{Bernard:2007sp,Bernard:2011zr}.}
The long-range {Deltaless and Deltaful
potentials at one order higher ($\OO(Q^{n+3}/\MQCD^{n+3})$) are found
in~\cite{Krebs:2012yv,Krebs:2013kha,Krebs:2018jkc}.}

One must as well look into higher-order double-pair or other disconnected
diagrams where more than two clusters of nucleons interact at the same time.
\textcite{Epelbaum:2006eu,Epelbaum:2007us} finds that double-pair diagrams with
a recoil correction ($\OO(Q^{2n}/\MQCD^{2n})$), with one insertion of a
$\Delta=2$ interaction ($\OO(Q^{n+2}/\MQCD^{n+2})$), or with $L=1$
($\OO(Q^{n+2}/\MQCD^{n+2})$) all add to nothing without Deltas.

\paragraph{$4N$ potential}

Four-body forces first appear at relative $\OO(Q^{2n}/\MQCD^{2n})$, among them
the one from a
{four-pion} interaction displayed in Fig.~\ref{fig:chiralpot}.  They
are all of long range  and contain no free parameters.  The components without
Deltas can be found in~\cite{Epelbaum:2006eu,Epelbaum:2007us}.  A first
estimate~\cite{Rozpedzik:2006yi} of the effect of {these} components in
$^4$He gives an additional binding of a few hundred keV.  The first contact
$4N$ force is of $\OO(Q^{2(n+1)}/\MQCD^{2(n+1)})$; since it has no derivatives,
the exclusion principle allows only one such interaction, as has been verified
explicitly by \textcite{Girlanda:2011fh}.

\paragraph{Isospin violation}

As discussed in Sec.~\ref{sec:pionless}, Coulomb exchange is
{nonperturbative}
only at small energies; in the region Chiral EFT power counting is designed for,
the Coulomb potential can be treated in perturbation theory.  The way
$e_{\text{red}}$ is counted above ensures that the Coulomb potential  appears
at $\OO(\Mlo/\MQCD)$, not LO.  Other purely electromagnetic components are even
smaller and can be incorporated as in ChPT.  More interesting is the isospin
breaking coming from interactions in Eq.~\eqref{chiLag} where hard photons have
been integrated out and/or the quark mass difference $\bar{m}\varepsilon$ (see
Eq.~\eqref{QCDL}) appears.  These interactions lead to the charge-neutral pion
mass splitting $\delta m_\pi^2=\OO(\Mlo^3/\MQCD)>0$ and the neutron-proton mass
difference $\delta m_N=\OO(\Mlo^2/\MQCD)>0$.  Other isospin-violating effects
are, likewise, suppressed by at least one power of
$\MQCD^{-1}$~\cite{VanKolck:1993ee,vanKolck:1995cb}, which means isospin is an
accidental symmetry: although broken in QCD, it is a symmetry of the LO EFT.

In contrast to many models, Chiral EFT produces relatively simple
isospin-violating forces that are invariant under both gauge
{transformations and pion-field redefinitions.}  The isospin-violating $2N$
potential has been calculated up to relative $\OO(Q^3/\MQCD^3)$ by
\textcite{VanKolck:1993ee,vanKolck:1995cb,vanKolck:1996rm,vanKolck:1997fu,%
Friar:1999zr,Niskanen:2001aj,Friar:2003yv,Friar:2004ca,Epelbaum:2005fd,%
Epelbaum:2007sq,Epelbaum:2008td}, including the pion mass splitting in OPE and
TPE, the most important isospin-breaking pion-nucleon coupling in OPE,
simultaneous photon-pion exchange, the nucleon mass difference in TPE, and
primordial counterterms. In standard terminology~\cite{Miller:1990iz}, ``class
I'' forces refer to isospin symmetry, ``class II'' to forces that break charge
independence but not charge symmetry---defined as a rotation of $\pi$ around
the 2-axis in isospin space---, ``class III'' to forces that break charge
symmetry but vanish in the $np$ system, and ``class IV'' to those that break
charge symmetry but cause isospin mixing in the $np$ system.  Class M forces
are first found at $\OO(Q^{{\rm M}-1}/\MQCD^{{\rm M}-1})$, which provides a
justification for the pre-EFT phenomenology, where this hierarchy was
observed~\cite{Miller:1990iz}.

Isospin violation first appears in the $3N$ potential at relative
$\OO(Q^{n+1}/\MQCD^{n+1})$ where it breaks charge symmetry through the nucleon
mass difference in TPE~\cite{Friar:2004ca,Epelbaum:2004xf,Friar:2004rg}.
However, since $\delta\MN$ is the result of a partial cancellation between
quark-mass and electromagnetic effects, isospin breaking in the $3N$ potential
is relatively small.

\paragraph{Summary}

The nuclear potential---its long-range form and its primordial
counterterms---has been derived in Chiral EFT to a considerably high order.
Although some of its elements had been anticipated using phenomenological
methods, new forces have also been found, particularly those carrying the
hallmark of QCD via chiral symmetry.  Small differences of implementation
remain regarding the related
assignments of order to few-body forces and to the inverse nucleon mass.
A more detailed exposition of the chiral potential can be found in
{\cite{Epelbaum:2005pn,Machleidt:2011zz}}.
We turn now to some of the important issues that
arise in connecting these forces to data.

\subsection{Nuclear amplitudes and observables}
\label{subsec:chiralamps}

Observables are determined by the $T$ matrix, which in turn is obtained by
using the potential with the appropriate dynamical framework---the
Lippmann-Schwinger or Schr\"odinger equation, or one of its many-body variants.
This process involves reducible diagrams for which the power counting of the
previous section does not apply.  Weinberg's original
prescription~\cite{Weinberg:1990rz,Weinberg:1991um} was to truncate the
potential and solve the corresponding equation exactly.  The hope,
based on experience with regular potentials, was that, if corrections are small
in the potential, they would only generate small corrections at the amplitude
level even if treated {nonperturbatively}.  However, chiral potentials are
only
regular due to the regularization procedure, which means that reducible
diagrams generate further regulator dependence.  As in Pionless EFT
(Sec.~\ref{sec:pionless}),
{non-negative} powers of $\Lambda$ are generated this way
which, if not compensated by the LECs, lead not only to potentially large
corrections from subleading orders, but also to model dependence through the
regulator choice.  The relevant question is to which extent
Eq.~\eqref{TRGtrunc} affects the ordering of the short-range interactions in
the potential.

\subsubsection{Weinberg's prescription}
\label{sebsubsec:weinbergprescription}

The first numerical study of chiral potentials with Weinberg's
prescription by \textcite{Ordonez:1993tn,Ordonez:1995rz} yielded a reasonable
description of $2N$ data at $\OO(Q^3/\MQCD^3)$ with explicit Deltas for a
Gaussian regulator on the transferred momentum with cutoff values $\Lambda=500,
800, 1000~\MeV$, {but used an over-complete set of interactions}.
A drawback of such a local (but non-separable) regulator is
that it allows a contact interaction to contribute to all partial waves
consistent with the exclusion principle.  In the large-$\Lambda$ limit the
contribution of a contact interaction to all but one wave disappears, but at
any finite cutoff data fitting is highly coupled and complicated.
{\textcite{Epelbaum:1999dj} carried out the first fit with
the minimum number of} {seven LECs} {at $\OO(Q^2/\MQCD^2)$.}
{That work as well as subsequent} fits
have employed different regulators
for the potential and for the dynamical equation, with a separable, non-local
regulator for the latter.  Fits of higher quality were achieved, and it
eventually emerged that they do depend on the choice of regulator; only for a
limited range of cutoff values $\Lambda\simle \MQCD$ have good fits been
obtained~\cite{Epelbaum:2006pt,Marji:2013uia}.  A milestone was a fit
\cite{Entem:2003ft} to $2N$ data at $\OO(Q^4/\MQCD^4)$ without explicit
Deltas for a non-local, super-Gaussian exponential regulator with
$\Lambda=500~\MeV$.  This achieved accuracy comparable to that of
phenomenological potentials ({for the $^3S_1$ phase shifts,}
see curve labeled ``EM500'' in
Fig.~\ref{fig:ThreeSOnePhase}).  Since then other high-quality fits have been
achieved at this or even lower and/or incomplete
orders~\cite{Epelbaum:2003xx,Epelbaum:2004fk,Ekstrom:2013kea,Ekstrom:2014iya,%
Epelbaum:2014efa,Piarulli:2014bda}.  The state-of-the-art
{are the $\OO(Q^5/\MQCD^5)$}
{fits of \textcite{Entem:2017gor} and \textcite{Reinert:2017usi}}.\footnote{
{The relatively large size of TPE for
the cutoff values employed by \textcite{Reinert:2017usi}
might call into question
the expansion of the potential shown in Fig. \ref{fig:chiralpot}.
However, the potential is not directly observable and its expansion
must be judged according to its effects on renormalized amplitudes.}}
We might expect increasingly accurate results as newly developed
fitting-optimization procedures are applied to higher-order potentials.

To the orders where good fits to $2N$ data have been achieved, the chiral
potential is expected, as discussed in the previous section, to include $3N$
forces whichever value one takes for $n$.  In most calculations, where $n=2$ is
assumed and the Delta is integrated out, the leading $3N$ forces appear at
$\OO(Q^3/\MQCD^3)$ and its two parameters $G_A$ and $H_0$
(Eq.~\eqref{chiLag2bar}) are fitted to few-nucleon data.  An obvious observable
to fit with $H_0$ is the triton binding energy, as frequently done in Pionless
EFT.  Possible ways to determine $G_A$ include a $2N$ process such as $N\! N\to
N\! N\pi$~\cite{Hanhart:2000gp,Baru:2009fm}, another $3N$ quantity, such as the
doublet neutron-deuteron ($nd$) scattering length~\cite{Epelbaum:2002vt} or the
triton{%
half-life~\cite{Gardestig:2006hj,Nakamura:2007vi,Gazit:2008ma,Ekstrom:2014iya,%
Baroni:2016xll}}, and a $4N$ quantity, such as the \FourHe binding
energy~\cite{Nogga:2005hp,Ekstrom:2013kea}.
{\textcite{KalantarNayestanaki:2011wz} review chiral $3N$ forces
in light nuclei.}  Recently a simultaneous
fit to $A\le 4$ properties~\cite{Carlsson:2015vda} has been performed up to
$\OO(Q^3/\MQCD^3)$ without explicit Deltas.  Overall, a good description of
$A\le 4$ systems, including scattering, can be achieved at this order and
higher, as long as a ``good'' regulator with a cutoff parameter $\Lambda \simle
\MQCD$ is employed.

Owing to their symmetry connection with QCD, chiral potentials have become
increasingly popular within the nuclear structure/reaction community,
particularly after the milestone fit of \textcite{Entem:2003ft}.  Remarkable
progress has been achieved in the development of ``\abinitio'' many-body
methods for the solution of the Schr\"odinger equation starting from a given
potential.
{Typically additional UV and IR regulators (see Sec.~\ref{infrareg}) are
introduced, and EFT suggests extrapolations to mitigate their effects
\cite{Coon:2012ab,Furnstahl:2012qg,Tolle:2012cx,Kruse:2013qaj,More:2013rma,%
Furnstahl:2013vda,Furnstahl:2014hca,Coon:2014nja,Konig:2014hma,Wendt:2015nba}.}
Extensive benchmarking---for {examples}, see
\cite{Kamada:2001tv,Hagen:2007hi,Abe:2012wp}---has ensured that, while not
entirely controlled, results are found in very satisfactory agreement with each
other.  \Abinitio methods are now at a stage where they can contribute
substantially to the understanding of the input interactions, by relating
parameters of these interactions to $A>4$ data.  The majority of today's
\abinitio calculations uses chiral potentials as input, for many of the new
methods are flexible enough to accommodate the non-localities of both the
interactions themselves and the chosen regulators.

Weinberg's prescription
is simple to implement because it is the same as that used for a
phenomenological potential: the various components are treated equally
in the solution of the Schr\"odinger equation.  The availability of many-body
calculations has led to an increased use of $A>4$ data to constrain the
potential parameters---particularly those of the $3N$ force, which proves
important in describing some nuclear quantities such as the ground-state spin of
$^{10}$Be~\cite{Navratil:2007we}, the dripline in oxygen
isotopes~\cite{Otsuka:2009cs,Hagen:2012sh}, and the evolution of shell structure
in calcium isopes~\cite{Holt:2010yb,Hagen:2012fb}.  These achievements have been
reviewed by \textcite{Hammer:2012id}{, and more recently in a compilation
of articles~\cite{Dudek:2016phscr} celebrating the 40-year anniversary of the
1975 Nobel Prize.}
While a large number of nuclear data have been well
{described---for example an excellent reproduction of
$A\le 12$ spectra \cite{Piarulli:2017dwd}---there}
are also challenges in reproducing bulk properties
of both {nuclei~\cite{Soma:2013xha,Binder:2018pgl} and
nuclear matter~\cite{Hagen:2013yba,Drischler:2017wtt},
as well as some $A=3$ scattering observables, including
the recalcitrant $A_y$ puzzle \cite{Piarulli:2017dwd,Binder:2018pgl}.}

\begin{figure}[tbp]
\centering
\includegraphics[width=0.9\columnwidth]{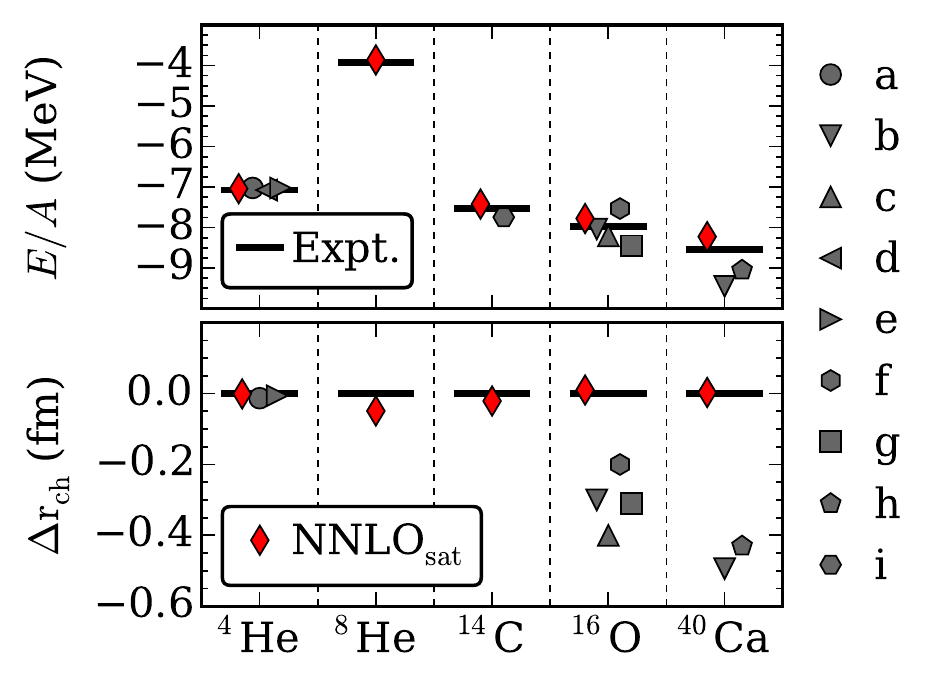}
\caption{Ground-state energy (negative of binding energy) per nucleon (top),
 and residuals (differences between computed and experimental values) of charge
 radii (bottom) for selected nuclei computed with
 a variety of chiral potentials (labeled a---i and
 $\text{NNLO}_{\text{sat}}$).
 Figure and description taken from \cite{Ekstrom:2015rta}.
 Courtesy of A.~Ekstr\"om.
}
\label{fig:1502_04682v2_fig1}
\end{figure}

The issue of the optimal set of data to fit has come to the fore.  In a
controlled EFT, a change in input data at a given order is not, ordinarily, a
systematic improvement because it represents a change that can be compensated by
higher orders.  However this is not necessarily true when correlated data
are employed~\cite{Lupu:2015pba}.  If only data at $Q\sim \Mlo$ are used as
input, we expect no particular correlations.  Guaranteeing that this is
the case is made difficult by the relative closeness between $\Mlo$ and $\Mhi$
in Chiral EFT, aggravated by the use of cutoff parameters $\Lambda \simle \Mhi$.
If one employs data characterized by $Q< \Mlo$, which are better described by a
lower-energy EFT, correlations might appear.  Examples of correlations among
few-nucleon observables were
given in Secs.~\ref{sec:pionless}
and~\ref{sec:halo}.  If one attempts to use, say, the triton binding energy
and the doublet $nd$ scattering length to fix two parameters of the $3N$ force,
one might expect one parameter combination to be relatively poorly determined.
These are low-energy data within the realm of Pionless EFT, which reorganizes
interactions into the appropriate
low-energy combinations.  Since one expects
larger systems to be increasingly within the regime of Chiral EFT, it is
possible that using properties of heavier nuclei provides real improvement.
A concrete example is the so-called NNLO$_{\text{sat}}$
potential~\cite{Ekstrom:2015rta}, a Deltaless chiral potential at
$\OO(Q^3/\MQCD^3)$ where the LECs are simultaneously adjusted not only for
$A\le 4$, but also to binding energies and radii of carbon and oxygen isotopes.
Examples of the corresponding predictions for other nuclei are shown in
Fig.~\ref{fig:1502_04682v2_fig1}.  In a similar spirit,
\textcite{Elhatisari:2015iga,Elhatisari:2016owd}, implementing Chiral EFT in a
lattice framework, have shown that the alpha-alpha interaction can be used as a
sensitive handle to determine
{internucleon interactions}.

The role of the regulator has also been under increasing scrutiny in \abinitio
calculations.  A powerful tool for many-body calculations is quantum Monte Carlo
(QMC), which pioneered the modern solution of many-nucleon systems but is
best suited for local potentials.  In Chiral EFT, it favors local interactions
and regulators.  Antisymmetrization among nucleons can be used to eliminate
some non-local contact
interactions~\cite{Gezerlis:2013ipa,Gezerlis:2014zia,Lynn:2014zia,%
Piarulli:2016vel,Logoteta:2016nzc}, enabling various new calculations.  For
example, the effects of the  $\OO(Q^2/\MQCD^2)$ potential in neutron matter
have been
studied~\cite{Gezerlis:2013ipa,Gezerlis:2014zia,Tews:2015ufa,Lynn:2015jua}.
Because of antisymmetrization, different spin-isospin forms of the $3N$ contact
interaction can be written, which are all equivalent and consistent with zero
within the EFT truncation error for $\Lambda\simge \MQCD$.  In contrast, the
limitation to $\Lambda\simle \MQCD$ leads to relatively large regulator
artifacts above saturation density~\cite{Lynn:2015jua}.  This indicated that a
lack of RG invariance, besides destroying model-independence, also has
undesirable phenomenological consequences.

\subsubsection{Renormalization of singular potentials}
\label{subsubsection:singrenorm}

The regulator dependence of Weinberg's prescription has been understood for a
long time.  Historically the first observation was that the
potential~\eqref{OPE} can be solved semi-analytically in the $^1S_0$ channel:
if $\boldp$ ($\boldp'$) denotes the relative incoming (outgoing) momentum in
the center-of-mass frame and $k^2/m_N$ the energy, the LO amplitude can be
written~{\cite{Kaplan:1996xu,Gegelia:1999ja,Eiras:2001hu,Long:2012ve}}
\begin{equation}
 T^{(0)}(\boldp',\boldp;k)= T_{\mathrm{Y}}(\boldp',\boldp;k)
 + \frac{\chi(\boldp';k)\chi(\boldp;k)}
 {C_{0\mathrm{s}}^{-1}(\Lambda) - I(\Lambda;k)} \,,
\label{T01S0}
\end{equation}
where $T_{\mathrm{Y}}$ is the amplitude for a pure Yukawa potential and
\begin{align}
 \chi(\boldp;k) &=
 1+m_N\int \frac{\dd^3l}{(2\pi)^3}
 \frac{T_{\mathrm{Y}}(\boldl,\boldp;k)}{k^2-\boldl^2 + \ii\epsilon} \,,
 \\
 I(\Lambda;k) &= m_N\int \frac{\dd^3l}{(2\pi)^3}
 \frac{\chi^{{2}}(\boldl;k)}{k^2-\boldl^2 + \ii\epsilon} {\,,}
\end{align}
{where we keep the regularization implicit.}
When $T_{\mathrm{Y}}=0$, $I(\Lambda;k)$ reduces to the $I_0(\Lambda;k)$
of Eq.~\eqref{eq:I0-cutoff}{,} and Eq.~\eqref{T01S0} to Eq.~\eqref{eq:T-C0}.
While $T_{\mathrm{Y}}$ and $\chi$ have no
{non-negative-power} dependence on
$\Lambda$, $I(\Lambda;k)$ has two types of inadmissible cutoff dependence:
$\propto \Lambda$ and $\propto (m_\pi^2/\MNN)\ln (\Lambda/\MNN)$.  The former is
the same cutoff dependence one sees in Pionless EFT (Sec.~\ref{sec:pionless}),
and can be absorbed in $C_{0\mathrm{s}}$.  The latter cutoff dependence, which
appears despite the fact that the Yukawa interaction is regular by itself
{and is formally the same as the logarithmic divergence generated by the
Coulomb interaction (see Sec.~\ref{sec:pionlessCoulomb})}, comes
from the interference between contact interaction and OPE.  It does not appear
in other singlet channels~\cite{Eiras:2001hu,Nogga:2005hy}, but it cannot be
absorbed in $C_{0\mathrm{s}}$, which is the LEC of a chiral-invariant
interaction and thus is not linear in $m_\pi^2$.  Instead, one has to modify the
LO potential \eqref{OPE} by~{\cite{Kaplan:1996xu}}
\begin{equation}
 C_{0\mathrm{s}} \to C_{0\mathrm{s}} + m_\pi^2 D_{2\mathrm{s}}
 = \OO\left(\frac{4\pi}{m_N \MNN}\right) \,.
\label{C0D0}
\end{equation}
Even though NDA estimates $D_{2\mathrm{s}} = \OO(C_{0\mathrm{s}}/\MQCD^2)$,
renormalization requires $D_{2\mathrm{s}} = \OO(C_{0\mathrm{s}}/\Mlo^2)$.
Because of the different transformation properties under chiral symmetry, the
two operators with LECS $C_{0\mathrm{s}}$ and $D_{2\mathrm{s}}$ differ in their
pion interactions, see Eq.~\eqref{chiLag2bar}.  Of course, one cannot see this
problem numerically in the $2N$ system unless one varies $m_\pi$, which is not
done in most nuclear work, and even then the divergence is only logarithmic.
The additional pion interactions from $D_{2\mathrm{s}}$ should have effects on
other processes, such as pion-nucleus scattering, but only when there is a
significant contribution from the $^1S_0$ $2N$ partial wave.  Regardless of its
phenomenological (ir)relevance, this is the simplest example where the
renormalization of observables in Chiral EFT is not guaranteed by NDA.  This
result has been confirmed in many other
studies~\cite{Beane:2001bc,PavonValderrama:2003np,PavonValderrama:2005wv}.

A similar but more dramatic renormalization effect concerns the momentum
dependence of OPE.  The tensor potential is singular, behaving at short
distances as $-\alpha/r^n$ with $n=3$ and, in some channels, $\alpha>0$.
It is well known~\cite{Frank:1971xx} that such potentials need to be treated
carefully because both solutions of the radial Schr\"odinger equation are
irregular at $r=0$.  For two particles of reduced mass $\mu$, the zero-energy
$S$-wave radial wavefunction behaves at short distances as\footnote{Particularly
interesting is $n=2$, which is equivalent~\cite{Efimov:1971zz} to the
three-boson system at unitarity (Sec.~\ref{sec:pionless}).  In this case
$\sqrt{2\mu \alpha} \, r^{1-n/2}/(n/2-1)\to \sqrt{2\mu \alpha -1/4} \, \ln
(r/r_0)$, with $r_0$ an arbitrary dimensionful parameter, and
$\phi_2=\phi_2(r_0)$.  This is an example of an
anomaly~\cite{Camblong:2003mz,Camblong:2003mb}
where the scale invariance of the
classical system is broken by renormalization to discrete scale invariance.}
\begin{equation}
 \psi(r)=r^{n/4-1} \cos\left(\frac{\sqrt{2\mu \alpha}\, r^{1-n/2}}{n/2-1}
 + \phi_n\right) + {\cdots} \,,
\end{equation}
where $\phi_n$ is a phase related to the scattering length and, more generally,
the phase shifts.  In EFT, the phase is determined by a contact interaction,
the LEC of which displays {an} oscillatory dependence on the
{cutoff~\cite{Beane:2000wh,Bawin:2003dm,Braaten:2004pg,PhysRevA.71.022108,%
Hammer:2005sa,Bouaziz:2014wxa,Odell:2019wjq}},
characteristic of a limit cycle or more
complicated attractor.  For RG analyses and reviews of limit cycles,
see~\cite{Barford:2002je,Barford:2004fz,PavonValderrama:2007nu},
and~\cite{Hammer:2011kg,Bulycheva:2014twa}, respectively.
{Without the contact interaction, the increasing attraction of the
singular potential leads to the repeated appearance of low-energy bound states
as the momentum cutoff increases.
With the contact interaction, not only the two-,
but also the three-body system is renormalized properly~\cite{Odell:2019wjq},
at least for $n=3$.}

The argument can be generalized to the tensor force
{\cite{Beane:2001bc,Birse:2005um}}, which is attractive in some
uncoupled channels and has one attractive eigenvalue in coupled channels.  In
$^3S_1$-$^3D_1$, the $C_{0\mathrm{t}}$ interaction in Eq.~\eqref{OPE} is
sufficient to absorb the cutoff dependence and fix the low-energy phase
shifts~\cite{Frederico:1999ps,Beane:2001bc,PavonValderrama:2005gu,%
PavonValderrama:2005wv,Yang:2007hb}, as suggested by NDA.  However, in higher
partial waves the effects of $C_{0\mathrm{t}}$ are cutoff artifacts that
disappear at large cutoffs.  As the cutoff increases, bound states accrue in the
higher partial waves where the tensor OPE is attractive ($^3P_0$,
$^3P_2$-$^3F_2$, $^3D_2$, $^3D_3$-$^3G_3$, \etc), leading to wild variation in
the corresponding low-energy phase
shifts~\cite{Nogga:2005hy,PavonValderrama:2005uj}.  This
problem can be cured~\cite{Nogga:2005hy} by a short-range interaction in each
such wave, \eg, including a term {$C'_{2\mathrm{t}}(\vecp'\cdot\vecp)
\pPt$} in Eq.~\eqref{OPE}.  As primordial counterterm,
$C'_{2\mathrm{t}}=\OO(C_{0\mathrm{s}}/\MQCD^2)$ appears only
at $\OO(Q^2/\MQCD^2)$, and similarly for the counterterms in other attractive,
singular waves.  The absence in Weinberg's prescription of the appropriate
counterterms explains the need for a ``physical cutoff''
$\Lambda_{\text{phys}}\simle 1~\GeV$, where $^3P_0$ would develop a bound
state~\cite{Nogga:2005hy}. {In other waves, bound states cross threshold
at higher cutoffs.} In triplet waves where OPE is repulsive there is no
need for such counterterms at LO~\cite{Eiras:2001hu,Nogga:2005hy}.

Renormalization problems have been
{reported~\cite{PavonValderrama:2005wv,PavonValderrama:2005uj,Entem:2007jg,%
Yang:2009kx,Yang:2009pn,Zeoli:2012bi}}
within Weinberg's prescription also for higher-order potentials, which are
increasingly singular and attractive in other waves as well.  In contrast,
a perturbative treatment of more-singular
corrections to singular potentials can be properly
renormalized~\cite{Long:2007vp} with counterterms containing more derivatives,
as expected from NDA.  Renormalization of chiral potentials seems to demand
that at least some parts of the potential be treated in perturbation theory,
just like in Pionless EFT.\footnote{
{%
The simple toy model of a regular long-range potential
plus a short-range interaction that yields a natural two-body
scattering length nicely illustrates how treating the subleading
EFT contact interaction nonperturbatively,
similar to the ``peratization'' of Fermi
theory~\cite{Feinberg:1963zz,Feinberg:1964zza},
prevents a large cutoff~\cite{Epelbaum:2009sd}.}}

\subsubsection{Connection with Pionless EFT}
\label{subsubsection:conpiless}

Experience with Pionless
EFT~\cite{Bedaque:1997qi,vanKolck:1997ut,Kaplan:1998tg,Kaplan:1998we,%
Bedaque:1998mb,vanKolck:1998bw} shows that the factors associated with reducible
loops are
\begin{subalign}[eq:ConnectionPionlessPC]
 \text{potential}
 &\sim 4\pi m_N^{-1} \Mlo^{-1}\left(Q \MQCD^{-1}\right)^{\mu} \,,
 \label{potsize} \\
 \text{nucleon prop.} &\sim m_N Q^{-2} \,, \\
 \text{reducible loop int.} &\sim (4\pi m_N)^{-1} Q^5 \,,
\end{subalign}
where the factor of {$(4\pi)^{{-}1}$} is typical of integrals involving
Schr\"odinger
propagators.  One iteration of the order-$\mu$ potential adds a
reducible loop and two nucleon propagators, or $(Q/\Mlo)(Q/\MQCD)^{\mu}$.  This
is an IR enhancement of $m_N/(4\pi Q)$ over the factor that arises from
Eqs.~\eqref{props} and~\eqref{loopintegral}.  As a consequence, the
{perturbative} series in
the LO potential ($\mu=0$) fails to converge for $Q\sim \Mlo$, while subleading
potentials ($\mu\ge 1$) should be amenable to perturbation theory.

The LO chiral potential~\eqref{OPE} has the form~\eqref{potsize} if $\MNN =
\OO(\Mlo)$.  Since bound states indicate a breakdown of perturbation theory,
one expects binding energies per nucleon
\begin{equation}
 \frac{B_A}{A}\sim \frac{\MNN^2}{\MQCD}\sim \frac{f_\pi}{4\pi} \sim 10~\MeV \,,
\label{B/A}
\end{equation}
which is in the right ballpark for heavy nuclei.  Thus chiral symmetry together
with this IR enhancement provides a natural explanation~\cite{Bedaque:2002mn}
for the shallowness of nuclei compared to $\MQCD$, $B_A/A\ll \MQCD$, long
considered a mystery.

The factor of $4\pi$ in the IR enhancement was not recognized before Pionless
EFT was developed, but it has implications for the natural size of few-body
forces.  Connecting an $aN$ potential to another nucleon to make it an $(a+1)N$
potential without changing $L$ or $\Delta$ involves an additional factor $4\pi
m_N^{-1} \Mlo^{-2}$ from the extra $2N$ interaction and the extra nucleon
propagator inside the $aN$ potential.  At the same time, it adds a reducible
loop and one nucleon propagator at the amplitude level, resulting in an overall
suppression by $Q/m_N$.  For $m_N=\OO(\MQCD)$---as dictated by NDA for
{$A=1$}, where it works well---this is the $n=1$ suppression of
\textcite{Friar:1996zw}.  In contrast, missing the $4\pi$ in the IR enhancement
would require $Q/m_N\sim Q^2/\MQCD^2$, as found in Sec.~\ref{subsubsec:nukepot}
for $n=2$~\cite{Weinberg:1991um,Ordonez:1992xp,Weinberg:1992yk,VanKolck:1993ee}.
Thus, counting factors of $4\pi$ in reducible loops leads to Friar's power
counting{, which, however, has not been widely tested so far.}

\subsubsection{Perturbative pions}
\label{subsubsection:pertpions}

A radical solution to the renormalization problems of Weinberg's prescription
was proposed by \textcite{Kaplan:1998tg,Kaplan:1998we}: assume the contact
interactions carry a low-energy scale
{characteristic of the binding momenta of light nuclei},
$\Mlo\ll \MNN$, and treat $\MNN$ as a
high-energy scale.  Pion exchange in nuclear amplitudes appears in two
expansions:
\begin{enumerate}
\item
The expansion in $Q/(4\pi f_\pi)$ of the nuclear potential, which, as discussed
in Sec.~\ref{subsubsec:nukepot}, is similar to the ChPT expansion for $A\le 1$.
\item
An expansion in $Q/\MNN$ in the solution of the dynamical equation, which is
similar to the Pionless EFT expansion for $A\ge 2$.
\end{enumerate}
Thus, if $Q \sim m_\pi\ll \MNN\simle \MQCD$ one can treat all pion exchanges
in perturbation theory.  Numerically $\MNN$ could be larger than the NDA
estimate $\MNN\sim f_\pi$.

This version of Chiral EFT closely resembles Pionless EFT
(Sec.~\ref{sec:pionless}),
{with similar $\Mlo$ scaling of the LECs
but different values, and additional pion exchanges.}
The range of validity of the EFT is enlarged, at least
near the chiral limit where integrating out pions becomes a very restrictive
condition.  At LO the two EFTs are formally the same, so
{the corresponding} results from
Pionless EFT carry over (there are two non-derivative $2N$ contact interactions
and one non-derivative three-nucleon force), but $m_\pi$ is now counted as
$\Mlo$ together with the $2N$ binding momenta.

At relative $\OO(Q/\MNN)$, however, there are not only two-derivative
two-nucleon contact terms ($\propto Q^2$) but also OPE, which provides a shape
function that goes beyond the first two terms in the effective-range expansion.
In perturbation theory, the $m_\pi^2\ln \Lambda$ cutoff dependence in the
$^1S_0$ channel (Sec.~\ref{subsubsection:singrenorm}) comes from a diagram
where OPE appears in-between two LO interactions.
The corresponding {chiral-symmetry breaking}
counterterm ($\propto m_\pi^2$) must be NLO as well. Both $Q^2$ and $m_\pi^2$
corrections appear at the same order, as in ChPT, but they are suppressed by
only one power of $Q/\MNN$.  The two-nucleon amplitude is well
reproduced~\cite{Kaplan:1998tg,Kaplan:1998we,Soto:2007pg}.  (For the
renormalization issues associated with a resummation of effective-range effects
with and without dibaryon fields, see~\cite{Ando:2011aa}
and~\cite{Nieves:2003uu}, respectively.)  There is only one calculation of the
effects of perturbative OPE in the three-nucleon system---quartet $nd$
scattering below and above break-up~\cite{Bedaque:1999vb}---and it gives
results very similar to those from Pionless EFT (Sec.~\ref{sec:NdScattering}).

N$^2$LO~{{\cite{Fleming:1999bs,Cohen:1999ds,Fleming:1999ee,Soto:2009xy}},
\ie, {relative} $\OO(Q^2/\MNN^2)$ is a crucial test of this expansion,
since it is the first manifestation of iterated OPE.
It was demonstrated~\cite{Cohen:1999ds,Fleming:1999ee} that, while the
expansion works
well at small momenta, in the low, spin-triplet partial waves where the OPE
tensor force is attractive, it fails for momenta $Q\sim 100~\MeV$.
{\textcite{Fleming:1999ee} employed} dimensional regularization with a
power-divergence subtraction (PDS)~\cite{Kaplan:1998tg,Kaplan:1998we} designed
to facilitate power counting, but of course other regularization/subtraction
schemes give equivalent results~\cite{Cohen:1998bv,Mehen:1998zz,Mehen:1998tp}.
(For an RG discussion, see \cite{Harada:2010ba}.)  A calculation employing a
procedure similar to Pauli-Villars regularization gave better
results~\cite{Beane:2008bt} in channels with LECs, but not in the spin-triplet
channels lacking LECs at that order.  These signs of the breakdown of
perturbative pions are consistent with an expansion in $Q/\MNN$ where
$\MNN\sim f_\pi$ as indicated by NDA.

There has also been criticism of the perturbative-pion expansion based on the
poor convergence of threshold observables~\cite{Cohen:1998jr,Cohen:1999iaa}.
This suggests that the expansion in $m_\pi/\MNN$ is not great for the real
world, again pointing to the low value of $\MNN$.  However, it is the
reorganization of interactions in Pionless EFT that is
optimized for momenta
$Q\ll m_\pi$, where the ERE holds.  The effectiveness of a power counting in
Chiral EFT should be judged from the convergence of observables at $Q\sim
m_\pi$.  At such momenta, for example, the scattering length contribution is
small, and one might start from the unitary limit
instead~\cite{Soto:2007pg,Soto:2009xy}, as discussed in
Sec.~\ref{sec:pionlesssummary}.
{In the $^1S_0$ channel, the perturbative-pion expansion does
converge~\cite{Beane:2001bc} despite claims to the contrary based on an NLO
calculation~\cite{Gegelia:1998ee}.
The slow convergence can be attributed to the short-range interactions.}
For analyses of perturbative pions in the
better-controlled context of toy models, see
\cite{Steele:1998zc,Rupak:1999aa,Kaplan:1999qa}.

In the real world, this version of Chiral EFT does not seem to work much beyond
the regime of validity of Pionless EFT. Because the latter is simpler and holds
for larger pion masses, it has been preferred in most low-energy applications.
However, Chiral EFT retains the constraints of chiral symmetry that are lost in
Pionless EFT; when such constraints are useful, Chiral EFT with perturbative
pions can be deployed.  Moreover, at smaller pion masses Chiral EFT with
perturbative pions is expected to have a considerably larger range of
applicability than Pionless EFT.

\subsubsection{Partly perturbative pions}
\label{subsubsection:partpertpions}

For a couple of years the choice facing the field was between a power counting
that lacks counterterms in the sense discussed in
Sec.~\ref{subsubsection:singrenorm} (\ie, ensuring that all divergences at a
given order can be absorbed) but worked well phenomenologically
(Sec.~\ref{sebsubsec:weinbergprescription}), and another one that has all
counterterms but failed to converge even at relatively small energies
(Sec.~\ref{subsubsection:pertpions}).  A way out was suggested by
\textcite{Nogga:2005hy} and carried out by
\textcite{Valderrama:2009ei,Valderrama:2011mv,Long:2011qx,Long:2011xw,%
Long:2012ve}.  Perhaps not surprising in hindsight, this solution is a middle
ground between Weinberg's prescription and fully perturbative pions.  It is
based on two observations:
\begin{enumerate}
\item
Pions are perturbative in sufficiently high two-nucleon partial waves.  For an
orbital angular momentum $l> l_{\text{cr}}$, where
$l_{\text{cr}}(l_{\text{cr}}+1)\sim \Mhi/\MNN$, the centrifugal barrier
dominates over OPE at all distances $r \simge 1/\Mhi$ relevant when $Q\lesssim
\Mhi$.  In these waves OPE should be perturbative for the external momenta where
Chiral EFT holds~\cite{Nogga:2005hy}.  In other words, OPE in the radial
Schr\"odinger equation is an expansion in $Q/\MNN^{(l)}$, where $\MNN^{(0)} =
\MNN$ but $\MNN^{(l)}$ increases with $l$.  For $l \le l_{\text{cr}}$, OPE is
{nonperturbative}.  In these waves, and in these waves only, OPE needs to be
iterated at LO, as observed by \textcite{Fleming:1999ee}.  The LECs needed for
renormalization (Sec.~\ref{subsubsection:singrenorm}) should, of course, be
iterated as well.  All subleading interactions are to be treated in
distorted-wave perturbation theory for $l \le l_{\text{cr}}${, and in
ordinary perturbation theory for $l > l_{\text{cr}}$.}
\item
Multiple pion exchange, being suppressed by powers of $Q/(4\pi f_\pi)$, should
be small after renormalization, and thus amenable to perturbation theory in all
waves.  It is more singular than OPE, but can be renormalized perturbatively
with a finite number of LECs~\cite{Long:2007vp}.
\end{enumerate}
Since the OPE tensor force survives in the chiral limit, for $m_\pi \simle
\MNN$ one can perform an additional expansion around the chiral
limit~\cite{Beane:2001bc}, but, as discussed in
Sec.~\ref{subsubsection:pertpions}, this expansion is not likely to be useful
much beyond the physical pion mass.

For $\Mhi >\MNN$, \ie, $\MNN$ counted as a low-energy scale, one expects
$l_{\text{cr}}\ge 1$.  Of course, other dimensionless factors stemming from
spin and isospin make the transition from {nonperturbative} to perturbative
OPE somewhat fuzzy, which however does not mean that such a transition does not
exist.  Early studies of perturbative pions with and without
Deltas~\cite{Kaiser:1997mw,Ballot:1997ht,Kaiser:1998wa}, which did not
discriminate between iterated pion exchange and multiple pion exchange in the
potential, indicated that pion exchange might be perturbative for $l\simge 3$.
This interpretation is also consistent with subsequent investigations of
peripheral waves with chiral potentials up to
$\OO(Q^6/\MQCD^6)$~{\cite{Entem:2002sf,Epelbaum:2003gr,Krebs:2007rh,%
Entem:2014msa,Entem:2015xwa}}.  Qualitatively, this result has been confirmed
for OPE~\cite{Nogga:2005hy}.  A semi-analytical estimate~\cite{Birse:2005um} of
the momenta where the tensor part of OPE needs to be treated
{nonperturbatively}
in the lower triplet waves is given in Table~\ref{tab:OPEtensorcrit}.
{Some evidence thus points to $l_{\text{cr}}\approx 3$.
More detailed, recent analyses suggest, however, that pions are perturbative
up to a relatively high scale in all waves other than $^3S_1$-$^3D_1$ and
$^3P_0$ \cite{Wu:2018lai,Kaplan:2019znu}.}

\begin{table}[tb]
\begin{center}
\begin{tabular}{|c|c||c|c||c|c|}
\hline
Channel & $p_{\text{cr}}$/MeV
& Channel & $p_{\text{cr}}$/MeV
& Channel & $p_{\text{cr}}$/MeV \\
\hline
\hline
$^3S_1$-$^3D_1$ & 66 & $^3P_0$ & 182 & $^3P_1$ & 365 \\
$^3P_2$-$^3F_2$ & 470 & $^3D_2$ & 403 & $^3D_3$-$^3G_3$ & 382 \\
$^3F_3$ & 2860 & $^3F_4$-$^3H_4$ & 2330 & $^3G_4$ & 1870 \\
\hline
\end{tabular}
\end{center}
\caption{{Estimate of the critical values} $p_{\text{cr}}$ of the
relative momentum in the lowest
two-nucleon triplet channels above which the OPE tensor force cannot be treated
perturbatively~\cite{Birse:2005um}.
\label{tab:OPEtensorcrit}}
\end{table}

In the low $2N$ waves where $l \le l_{\text{cr}}$ and $\MNN^{(l)}\approx
\MNN$, the situation at LO is similar to Weinberg's prescription, except that
more short-range interactions are needed for
renormalization~\cite{Nogga:2005hy} than implied by NDA.
{For example, in the $^1S_0$ channel OPE at LO solves the problem of the
slow convergence of perturbative pions~\cite{Beane:2001bc}
at the cost of the additional $D_{2\rm s}$ LEC in Eq.~\eqref{C0D0}.
The residual $1/\Lambda$ dependence then
means~\cite{Long:2012ve} that a correction appears at $\OO(Q/\MQCD)$
from the two-derivative contact interaction responsible for the
short-range contribution to the effective range,
similarly to Pionless EFT (Sec. \ref{subsubsec:regren}).}
At higher orders in the lower partial waves,
multiple-pion exchanges appear and require at $\OO(Q^\mu/\MQCD^\mu)$ LECs with
up to $\mu$ derivatives more than the LECs appearing at LO~\cite{Long:2007vp}.

{%
This approach was confronted with empirical phase shifts for the lower $2N$
partial waves by~\textcite{Nogga:2005hy,Epelbaum:2006pt,Valderrama:2009ei,%
Valderrama:2011mv,Long:2011qx,Long:2011xw,Long:2012ve,Yang:2016brl}.
}
The results of~\textcite{Long:2011xw} are included in
Fig.~\ref{fig:ThreeSOnePhase}, whereas Fig.~\ref{fig:Valderrama_2011mv}
shows the $\ThreePNot$ results of~\textcite{Valderrama:2011mv} as
a further example.
{While in both cases $\OO(Q^2/\MQCD^2)$ improves on
$\OO(1)$, $\OO(Q^3/\MQCD^3)$ goes in the wrong
direction---perhaps an indication that a better description of the
pion-nucleon subamplitude with an explicit Delta isobar is needed.}

\begin{figure}[tbp]
\centering
\includegraphics[width=0.95\columnwidth]{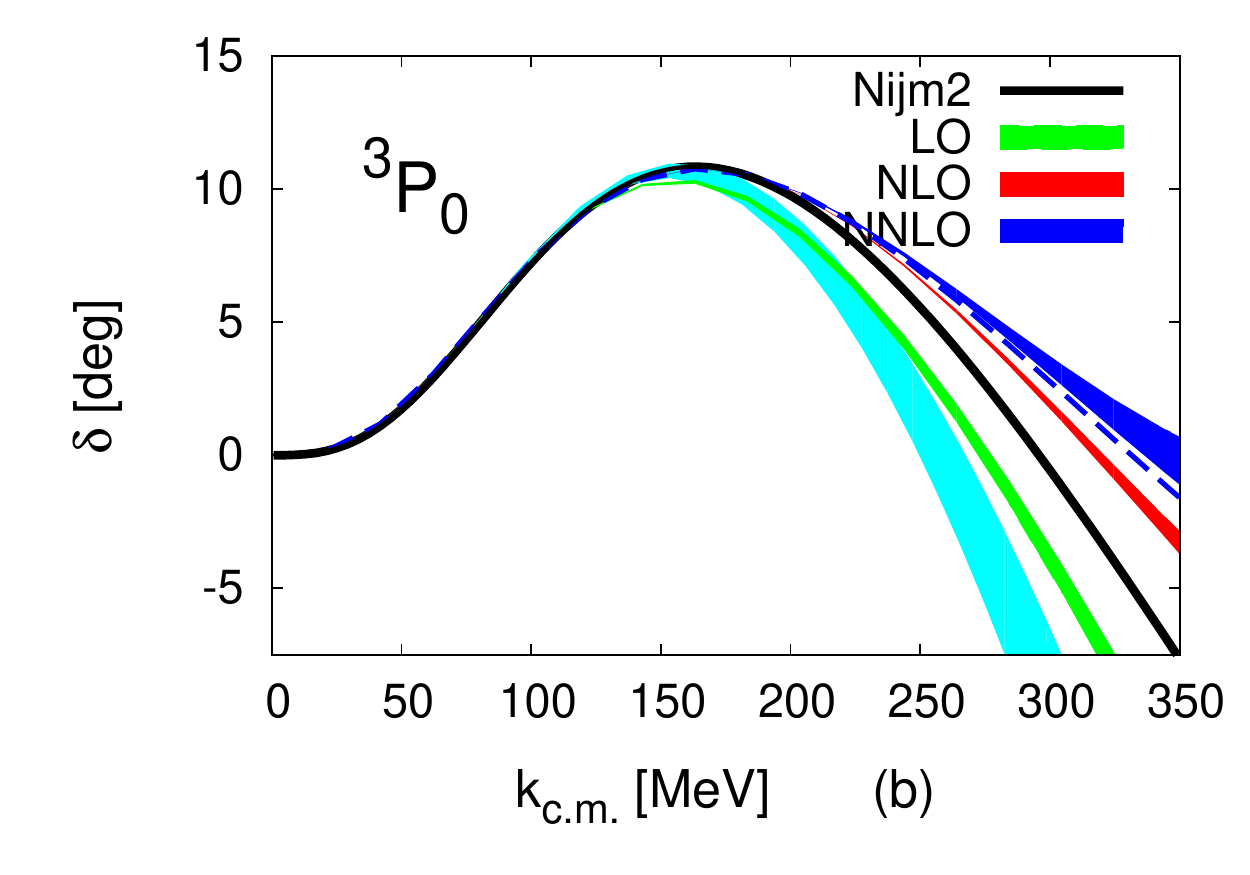}
\caption{{The $N\!N$ \ThreePNot phase shift $\delta$ as a function
 of the center-of-mass momentum $k_{c.m.}$
 at different chiral orders:
 ${\mathcal O}(1)$ (green),
 ${\mathcal O}(Q^2/\MQCD^2)$ (red), and
 ${\mathcal O}(Q^3/\MQCD^3)$ (blue),
 according to \textcite{Valderrama:2011mv}.
 Bands represent the variation of a
 coordinate-space cutoff in the range 0.6--0.9 fm, with the dashed
 line showing the ${\mathcal O}(Q^3/\MQCD^3)$ results with a 0.3 fm cutoff.
 The cyan band shows the ${\mathcal O}(Q^3/\MQCD^3)$
 potential in Weinberg's prescription with the regularization procedure
 of \textcite{Epelbaum:2003xx}, where the momentum-space cutoff in
 pion loops (Lippmann-Schwinger equation)
 is varied between 500 and 700 (450 and 650) MeV.
 The solid black line is the Nijmegen phase-shift analysis~\cite{Stoks:1993tb}.
 Figure from \textcite{Valderrama:2011mv}.}
 Courtesy of M.~Pav\'on Valderrama.
}
\label{fig:Valderrama_2011mv}
\end{figure}

Little is known quantitatively about partly perturbative pions beyond the
two-nucleon system. The three-nucleon system is renormalized properly without a
three-nucleon force at LO~\cite{Nogga:2005hy,Song:2016ale} and, without
explicit Deltas, also at NLO~\cite{Song:2016ale}.  The truncation in $l$ needed
at LO is reminiscent of the truncation in total {two-nucleon}
angular momentum typically
invoked in solutions of the Faddeev and Faddeev-Yakubovski equations for three-
and four-nucleon systems with phenomenological potentials. However, to go to
higher orders the $l$ dependence of $\MNN^{(l)}$ must be quantified.  So far,
this has been done only in singlet waves~\cite{PavonValderrama:2016lqn}.

{At LO, symmetric nuclear matter was found to saturate,
but with significant underbinding,
in a cutoff-converged Brueckner pair approximation
\cite{Machleidt:2009bh}. This is in contrast to Weinberg's prescription,
where Deltaless \cite{Sammarruca:2018bqh} or Deltaful \cite{Ekstrom:2017koy}
potentials of $\OO(1)$ and $\OO(Q^2/\MQCD^2)$
do not saturate within the EFT domain.
The fact that {higher-order} potentials with this
prescription do
saturate~\cite{Ekstrom:2017koy,Drischler:2017wtt,Sammarruca:2018bqh}
suggests that,
if nuclear matter is within the regime of Chiral EFT, the LO potential
requires more interactions than prescribed by NDA.}

Although they differ in detail from the
{field-theoretical} renormalization
outlined above, RG analyses of the Schr\"odinger
equation~\cite{Birse:2005um,Birse:2010fj,Valderrama:2016koj} support the
conclusion that counterterms in two-nucleon attractive singular waves appear at
lower order than expected on the basis of NDA.  The case is further
strengthened by removing the effect of OPE and (perturbative) TPE from
empirical two-nucleon phase shifts~\cite{Birse:2007sx,Ipson:2010ah}.
The Schr\"odinger RG analysis also
predicts the enhancement of some three-body forces~\cite{Birse:2010fj}.

\subsubsection{Other approaches}
\label{subsubsection:other}

{%
The renormalization of Chiral EFT described in the previous section was
criticized by~\textcite{Epelbaum:2018zli}, who provided examples where the
nonperturbatively renormalized amplitude exhibits positive powers of the cutoff
$\Lambda$ when expanded in Planck's constant $\hbar$.
{Since no observable {considered} in that work is affected, however,
the significance of this claim for an EFT is unclear.}
Moreover, \textcite{Valderrama:2019yiv} argued that these powers of $\Lambda$
can be eliminated by changing the $\Lambda^{-1}$ running of the
LECs~\cite{Valderrama:2016koj}.  (See also the response
by~\textcite{Epelbaum:2019msl}.)
}

{%
An alternative approach to the renormalization woes of Weinberg's
prescription was articulated by~\textcite{Epelbaum:2018zli}, building
on earlier work~{\cite{Gegelia:1998gn,Gegelia:1998iu,Gegelia:1998ee,%
Gegelia:1999ja,Gegelia:2001ev,Gegelia:2004pz,Epelbaum:2009sd}}.
{It consists of renormalizing the perturbative series and subsequently
resumming the renormalized contributions, that is,
it includes} at each order the infinite number of LECs needed to
eliminate the cutoff dependence of all diagrams to be resummed.  These LECs
exist because an EFT contains all interactions allowed by symmetry, but even
without pions they are difficult or impossible to write down explicitly.
In a stark departure from naturalness, only the LECs prescribed by NDA are
assumed to contribute finite parameters, which amounts to an infinite number of
fine tunings.  The resummation of an infinite number of derivative interactions
introduces an intrinsic nonlocality in all channels at every order---not only
LO in, say, the $N\!N$ $^1S_0$~\cite{Beane:2000fi,SanchezSanchez:2017tws} or
the $N\alpha$ $^{2}P_{3/2}$~\cite{Bertulani:2002sz} channels, where there are
shallow poles.
}

{
In this approach, there are no constraints on the EFT from renormalization,
for example no Wigner bound on the effective range
(\textcite{Wigner:1955zz}; see
Sec.~\ref{subsubsec:regren}) when $C_2$ or higher-order interactions are
resummed~\cite{Gegelia:1998gn}} and no
explanation~\cite{Epelbaum:2016ffd,Epelbaum:2019msl} for the emergence of a
single, independent three-body scale in the three-body system at LO, which
determines the position of the Efimov tower of states
{(\textcite{Efimov:1970ab}; see Sec.~\ref{sec:Triton})}.  Thus, the
justification for power counting from the combination of renormalization and
naturalness, which in the perturbative context gives
NDA~\cite{Manohar:1983md,Georgi:1986kr}, is absent.  NDA becomes an \textit{ad
hoc} rule{. It does not, \eg, reproduce the established scaling of range
corrections in amplitudes resulting from short-range potentials, which is
represented in Pionless EFT through Eq.~\eqref{eq:scaling-re}.}

{%
So far this approach has been implemented only in the $^1S_0$ channel, where
significant dependence on the choice of (low-energy) subtraction points
is seen~\cite{Gegelia:1998iu,Gegelia:1998ee,Gegelia:1999ja,Gegelia:2001ev}.
On the basis of a toy model, \textcite{Epelbaum:2017byx} conclude that
Weinberg's prescription is satisfactory as long as the renormalization scale
$\mu=\OO(\Mhi)$.  Chiral EFT's overlapping integrals in other channels
prevent the explicit resummation of ``renormalized diagrams'' and it is not
known whether this procedure, if it can be carried out at all, {reproduces
the nonperturbative} solution of the Lippmann-Schwinger equation.
Further discussion of renormalization from the perspective of subtraction
schemes can be found in~\cite{Timoteo:2010mm,Szpigel:2011bc,Batista:2017vao}.
}

{%
\textcite{Gegelia:2001ev} offer the solution that the cutoff should not be
varied significantly around the breakdown scale in Chiral EFT.  In the absence
of renormalization, non-negative powers of the cutoff should appear in the
truncated amplitude {given by Eq.}~\eqref{Ttrunc} as corrections of
$\OO(Q^{\mathcal{V}+1-i} \Lambda^i/(\Mhi^{\mathcal{V}+1-j} \Mlo^{j}))$
{with nonnegative integers $i,j$.}  If $j=0$ the corrections should be small
for $\Lambda\ll \Mhi$~\cite{Gegelia:2004pz}, but $j> 0$ arises when the LO
potential, which does not involve $\Mhi$, is singular.
}

{%
An attempt to mitigate cutoff artifacts was made
by~\textcite{Djukanovic:2007zz}, \textcite{Epelbaum:2012ua},
and \textcite{Epelbaum:2013naa,Epelbaum:2015sha}} with the most recent
formulations developed
{by~\textcite{Behrendt:2016nql} and \textcite{Baru:2019ndr}}.
A nucleon propagator with faster large-momentum
fall-off is constructed by demanding that states satisfy a relativistic
(Lorentz-invariant) normalization condition, while overall the treatment is
still nonrelativistic.  This softer UV behavior helps to obtain LO amplitudes
with well-defined {large-cutoff} limits.  While \textcite{Behrendt:2016nql}
state
that higher orders should be treated in perturbation theory if this feature is
to be maintained beyond LO, from a practical point of view they still advocate a
nonperturbative treatment (where the cutoff is then limited again to a finite
range, argued to be larger than what is typically used with standard Weinberg
counting). Moreover, \textcite{Behrendt:2016nql} find that with their
approach a $^3P_0$ LEC has to be promoted compared to NDA, {as in the purely
nonrelativistic context~\cite{Nogga:2005hy}.}
{There is, nevertheless, growing interest in the development of
a covariant version of Chiral EFT, which could perhaps be used as input
to relativistic formulations of nuclear physics
\cite{Petschauer:2013uua,Ren:2016jna,Ren:2017yvw}.}

\subsection{{Pion and electroweak reactions}}
\label{subsec:chiralreact}

One of the great advantages of a quantum-field-theoretical foundation of
nuclear physics is that not only many-body forces can be constructed
consistently with two-body forces, but also many-body currents can be derived
consistently with inter-nucleon interactions.  This virtue was realized early
on~\cite{Rho:1990cf,Weinberg:1991um}, and some of the pioneering papers on
reactions have been dedicated to electroweak
currents~\cite{Park:1993jf,Park:1995pn,Phillips:1999am}, neutron radiative
capture on the proton~\cite{Park:1994sr,Park:1997kp}, proton-proton
fusion~\cite{Park:1998wq}, Compton scattering on the
deuteron~\cite{Beane:1999uq}, pion photo-~\cite{Beane:1995cb,Beane:1997iv} and
electro-~\cite{Bernard:1999ff} production off the deuteron,
pion photoproduction off the
trinucleon~\cite{Lenkewitz:2011jd,Lenkewitz:2013hka},
pion scattering on the deuteron~\cite{Beane:1997yg}
and
{helion}~\cite{Liebig:2010ki}, as well as pion production in
$2N$ collisions~\cite{Park:1995ku,Cohen:1995cc,vanKolck:1996dp,Sato:1997ps}.
The goal is not only to supply information to other areas of physics where
these reactions play a role, but also to extract nucleon properties (in
particular for the neutron, for which good targets do not exist) to infer
properties of the QCD dynamics.

The early work, {reviewed by \textcite{vanKolck:1999mw} and
\textcite{Bedaque:2002mn},} was based on
Weinberg's prescription, where, in addition to the potential, also the kernel
of the reaction process is expanded according to NDA.  For probes with energies
$E\sim Q \sim m_\pi$, the kernel is defined as the subdiagrams to which the
external probes are attached, energies are comparable to momenta, and nucleons
are approximately static.  Like the potential, kernels can be multiply
connected, and power counting is similar to Eqs.~\eqref{chiralnu}
and~\eqref{chiralnupot}.  For any kernel, a figure like
Fig.~\ref{fig:chiralpot} can be drawn.  However, some subtleties need to
be taken into account when probes have other typical {energy or momentum}:
\begin{itemize}
\item
$E\sim \MNN^2/\MQCD$ (\eg~Compton scattering): a resummation is needed
between kernels because infrared enhancements appear in intermediate states,
where nucleons are not static~\cite{Beane:1999uq}.
\item
$Q\sim \sqrt{m_\pi m_N}$ (\eg~pion production):\footnote{{Although
parametrically $\sqrt{m_\pi m_N}= {\mathcal O}(\sqrt{m_\pi \MQCD})<\MQCD$,
one should keep in mind that the breakdown scale $\MQCD$ of Chiral EFT is
not known precisely.}} intermediate states containing
only nucleons can be part of the kernel, but these nucleons are not
static~\cite{Cohen:1995cc}.
\end{itemize}

The full amplitude is given by the matrix element of the kernel with
wavefunctions obtained from the potential.  In Weinberg's prescription, these
wavefunctions are exact solutions of a truncated potential.  Much of the work
predating phenomenologically successful chiral potentials employed a ``hybrid''
approach where the kernel was calculated in Chiral EFT but wavefunctions from
``realistic'' phenomenological potentials were used.  Emphasis has since been
shifting towards increased consistency between wavefunctions and kernels.
Reactions are an area of renewed interest in Chiral EFT, in consonance with the
revival of development in the broader area of nuclear reactions, including
\abinitio approaches.  Of particular recent interest have been electroweak
currents, where earlier work was revisited and significantly improved
upon---see the excellent reviews by
{\textcite{Phillips:2016NN} and \textcite{Riska:2016cud}}.  There has
also been substantial work on reactions involving pions, particularly pion
production in two-nucleon collisions, which has now been calculated up to three
orders in the chiral expansion~\cite{Baru:2013zpa}.

The process that has been most thoroughly examined in Chiral EFT is Compton
scattering (CS).  It gives access to nucleon polarizabilities---response
functions that carry much information about hadron dynamics.  While proton
polarizabilities can be extracted directly, neutron polarizabilities can only
be probed in nuclear CS.  Chiral EFT allows for a consistent treatment of both
these cases, and at the same time enables a connection with lattice QCD through
variation in the pion mass.  Recent work has capitalized on all advances
in EFT and provides an analysis of CS that is a model for future work on
nuclear reactions.  At the one-nucleon level, CS was calculated in ChPT with
an explicit Delta isobar~\cite{McGovern:2012ew}, including the
resummation~\cite{Pascalutsa:2002pi} needed to go through the Delta peak.  At
the nuclear level, the kernel was calculated consistent with Weinberg's
prescription~\cite{Griesshammer:2012we}.  Proton~\cite{McGovern:2012ew} and
deuteron~\cite{Myers:2014ace,Myers:2015aba} data were fitted and
polarizabilities were extracted, see Fig.~\ref{fig:compton}.  The average quark
mass was then varied to produce predictions{, with uncertainties determined
\via Bayesian
techniques~\cite{Furnstahl:2014xsa,Furnstahl:2015rha,Wesolowski:2015fqa},} for
the polarizabilities at unphysical pion masses~\cite{Griesshammer:2015ahu}, to
which lattice QCD results can be compared.  Analyses of this type for other
processes should increasingly become standard in this field, allowing to bridge
from QCD to nuclear reactions just as to nuclear structure.

\begin{figure}[tb]
\begin{center}
\includegraphics[width=\columnwidth]{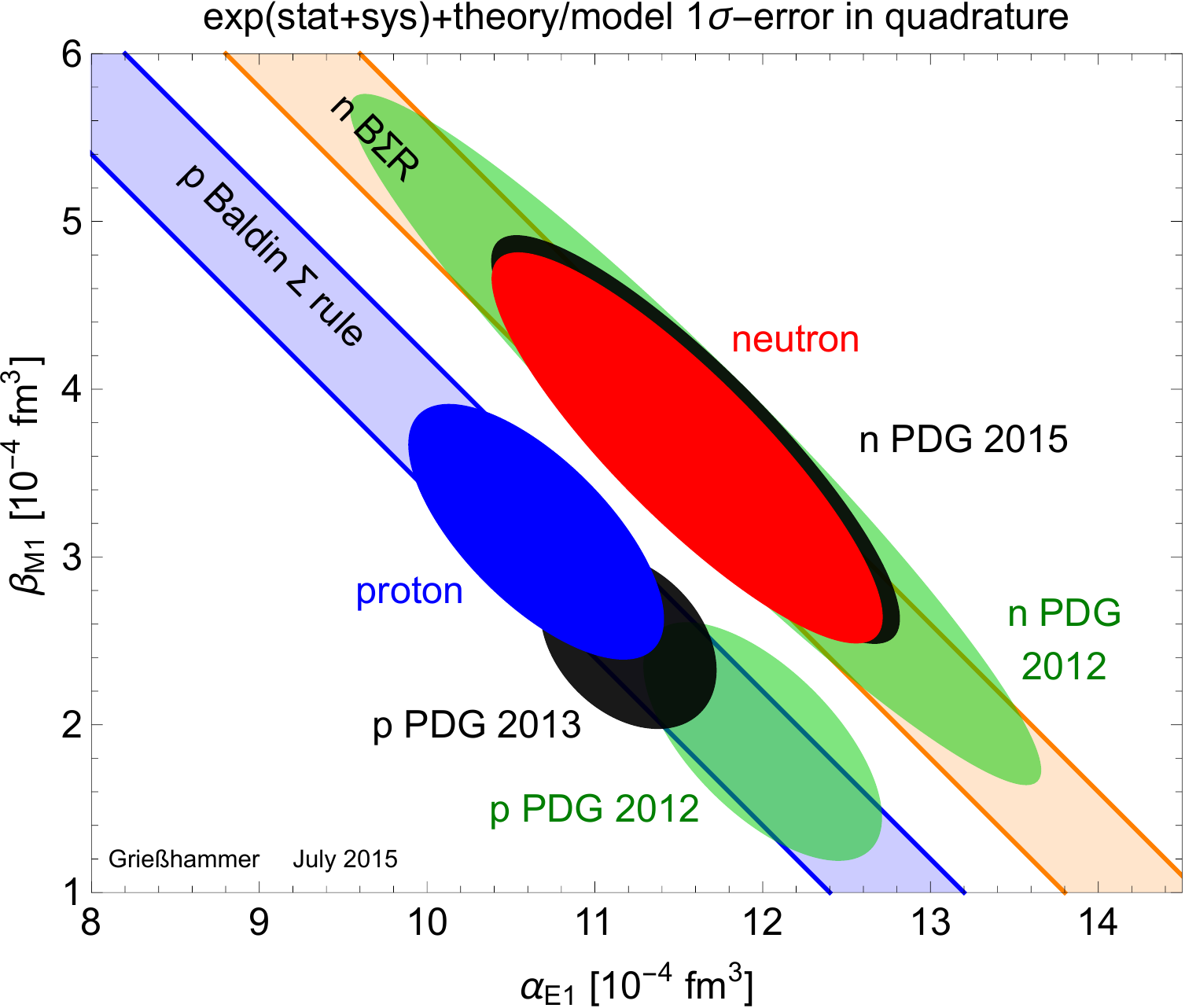}
\end{center}
\caption{
Nucleon electric ($\alpha_{{\rm E}1}$) and magnetic ($\beta_{{\rm M}1}$)
polarizabilities extracted with Chiral EFT.
Ellipses show the 1$\sigma$-error fits of \textcite{McGovern:2012ew}
for proton (blue, on lower band) and neutron (red, on upper band),
where statistic, systematic, and theory errors were added in quadrature.
For comparison, Particle Data Group averages
before (large, green ellipses) and after (small, black ellipses)
these fits are also given.
The bands are the constraints from the Baldin sum rule.
Reprinted from \cite{Griesshammer:2015wha}, with the permission of AIP
Publishing.
}
\label{fig:compton}
\end{figure}

The renormalization problems of Weinberg's prescription demand scrutiny in the
treatment of reactions as well.  The emergence of a perturbative-pion
formulation of Chiral EFT has led to a re-examination of many of the
reactions that had been studied earlier with Weinberg's prescription.
With perturbative pions, not only the kernel but also the wavefunction is
expanded in perturbation theory.  For the deuteron---target of most, if not
all, perturbative-pion studies---the {analytical} nature of the
calculations makes it easier to establish proper renormalization.
In most cases, the LO calculation is the same as in
Pionless EFT (Secs.~\ref{sec:Currents-TwoBody}
and~\ref{sec:Currents-ThreeBody}), and pion effects enter explicitly at
subleading orders.  For example, the charge, magnetic dipole and electric
quadrupole form factors of the deuteron have been calculated to NLO by
\textcite{Kaplan:1998sz}.  Other reactions include neutron radiative capture on
the proton~\cite{Savage:1998ae}, deuteron Compton scattering and
polarizability~\cite{Chen:1998vi,Chen:1998ie,Chen:1998rz}, and
neutrino-deuteron scattering~\cite{Butler:1999sv}.  Generally, this approach
has been successful for the low-energy properties of the deuteron, both in
comparison with data and with
{nonperturbative-pion} calculations.

\subsection{Outstanding issues and current trends}
\label{subsec:Outissues}

Within Weinberg's prescription, work continues in pushing potential and kernels
to higher orders, as well as developing {better} fitting strategies.
Meanwhile, an
RG-invariant formulation of Chiral EFT has not yet achieved the level of
phenomenological impact of Weinberg's prescription.  Some of the outstanding
questions are:
\begin{itemize}
\item
What is the role of fine-tuning in Chiral EFT?  The scattering length is
particularly large in the $^1S_0$ two-nucleon channel, where short-range
interactions show strong energy dependence~\cite{Birse:2010jr}.  In particular,
short-range contributions to the effective range are anomalously large, which
has led to the suggestion that the two-derivative contact interaction
($C_{2\mathrm{s}}$ in Eq.~\eqref{chiLag2bar}) should be treated as
LO~{\cite{Beane:2001bc,Long:2013cya}}, following a similar proposal in
Pionless EFT~\cite{Beane:2000fi} (Sec.~\ref{sec:Dibaryons}) and Chiral EFT with
perturbative pions~\cite{Ando:2011aa} (Sec.~\ref{subsubsection:pertpions}).
Although an improved description is achieved, convergence deteriorates quickly
with momentum, at least in a calculation without Deltas~\cite{Long:2013cya}.
The role of the relatively low-energy zero of the amplitude remains to be fully
investigated~\cite{Lutz:1999yr,SanchezSanchez:2017tws}.
\item
What is the pion mass variation of nuclear amplitudes?  To determine this
variation, we need lattice data within the region of validity of Chiral EFT.
While lattice QCD data exist at relatively large $m_\pi$, not all competing
collaborations are in agreement.  Moreover, differences exist within Chiral EFT
depending on the approach used.  Among the issues for extrapolations is the
role of ``radiation'' pions, which are present only at relatively high order.
Real pions can be produced in $2N$ collisions for momenta
$Q\simge \sqrt{m_\pi m_N}$;
at lower $Q$ the effects of the corresponding virtual pions are
indistinguishable from the LECs \cite{Ordonez:1995rz,Epelbaum:1998ka}.
These contributions have been investigated in the
perturbative-pion context~\cite{Mehen:1999hz,Mondejar:2006yu,Soto:2011tb}
where they give rise to powers of $m_\pi^{1/2}$.
\item
The NDA-based organization of pion-exchange contributions to the potential is
not affected by the renormalization issues that plague Weinberg's prescription,
except for the possible enhancement factors of $4\pi$ in few-nucleon forces
(Sec.~\ref{subsubsec:nukepot}).  Once the $2N$ system has been properly
renormalized, one must ask whether short-range many-body forces are immune to
the enhancements seen in Pionless EFT (Sec.~\ref{sec:Triton}).
Most work on many-body forces in Chiral EFT takes Weinberg's power counting for
granted.  \textcite{Nogga:2005hy} and \textcite{Song:2016ale} have found no
renormalization evidence for a $3N$ force in LO or $\OO(Q/\MQCD)$, but they
obtain a triton binding energy which is only about half of the experimental
value, perhaps because this is a very-low energy observables in the sense of
being within the regime of Pionless EFT.  \textcite{Kievsky:2016kzb} argue from
continuity with Pionless EFT that the $H_0$ three-body force in
Eq.~\eqref{chiLag2bar} should be included at LO.
\item
The most important problem facing reaction theory in Chiral EFT echoes the
renormalization woes of Weinberg's prescription for nuclear structure.
In order to maintain model independence, one must ensure that the average of
the reaction kernel has a well-defined limit as the cutoff is increased.  Only
for electroweak reactions on the deuteron has this been
investigated~\cite{Valderrama:2014vra,Phillips:2016NN}, with the conclusion
that enhancements over NDA appear there as well.  The impact of this
observation on previously studied reactions and future reaction theory remains
to be investigated.
\item
To which extent can an RG-invariant formulation of Chiral EFT be incorporated
in calculations of larger nuclei?  Distorted-wave perturbation theory usually
becomes very demanding in second order where the fully off-shell $A$-body
propagation is needed in intermediate states.  In Pionless EFT, this problem
has been side-stepped by a reformulation in terms of the solution of further
integral equations~\cite{Vanasse:2013sda,Vanasse:2015fph}.  It is an open issue
whether this or another method can be  applied to the $A$-nucleon problem
{in Chiral EFT}.
\item
Although not RG invariant, Weinberg's prescription has several practical
advantages because it is most closely related to previous phenomenological
approaches. For example, it provides fits to data of comparable quality to
``realistic'' phenomenological potentials
(Sec.~\ref{sebsubsec:weinbergprescription}){,} it explains some of the
qualitative features of these potentials (Sec.~\ref{subsubsec:nukepot}), and it
can be employed in already existing \abinitio codes.
Its successes beg the question, is it possible that a
{particular choice of regulator allows for a small-cutoff
formulation of Chiral EFT that is equivalent to its RG-invariant form?}
It is conceivable that one can iterate subleading
terms of the latter, as it is done in the former, within a limited range of
cutoffs, just as iterating momentum-dependent contact interactions in Pionless
EFT requires cutoff values $\Lambda \simle 1/r_0$ (Wigner bound,
Sec.~\ref{subsubsec:regren}).  If achieved, such an understanding of Weinberg's
prescription might justify current uses of Chiral EFT without requiring further
development of \abinitio methods.  {Furthermore, it will likely make it
desirable to choose a regulator which optimizes convergence for the problem at
hand,} {for example neutron matter \cite{Lynn:2015jua}.
First steps in this direction were just made by
\textcite{Tews:2018sbi} and \textcite{Valderrama:2019lhj}.
}
\item
Irrespective of other issues discussed above and throughout this section, it is
a highly nontrivial task to determine the LECs of Chiral EFT---quite
substantial in number at higher orders---from fitting calculated observables to
data. {One wants to do this in such a way that the EFT can fulfill its
promises of providing systematic model independence and fully quantified
uncertainties.  For that, one should account for the truncation error of the
EFT expansion directly as part of the fitting procedure, and propagate
forward all this information to the final result of a calculation.}  Following
the initial suggestion of \textcite{Schindler:2008fh}, Bayesian methods have
emerged in recent
{years~\cite{Furnstahl:2014xsa,Furnstahl:2015rha,Wesolowski:2015fqa,
Melendez:2017phj,Wesolowski:2018lzj}}
as an important tool to address the issue in a robust and comprehensive way.
\end{itemize}

\section{Broader applications} \label{sec:other}
Many of the ideas originating in nuclear EFTs have found applications to other
systems.  Pionless EFT, while being clearly connected to QCD as a low-energy
limit, is driven largely by the universal features that arise from the large
$N\!N$ scattering lengths and the associated large sizes of light nuclei.
As a consequence, the EFT for nuclear halo states discussed in
Sec.~\ref{sec:halo} can be constructed as a generalization of Pionless EFT.
But universality goes beyond nuclear physics: it is relevant to any system
dominated by short-range interactions, when one is interested in distance scales
much larger than the range of the interactions.

As one probes distances comparable to the interaction range, issues similar to
the ones discussed in Chiral EFT emerge.  For example, can the long-range part
of the interaction be treated in perturbation theory?  In other hadronic
systems, the long-range interaction might still be one-pion exchange, and then
Chiral EFT applies except that other heavy particles are substituted for
nucleons.

In this section we briefly describe some of the systems where versions of
Pionless and Chiral EFTs have found application.  We start by discussing how
these EFTs arise from QCD.

\subsection{Connection with QCD}
\label{subsec:qcdconn}

The inclusion of all possible interactions consistent with QCD symmetries
ensures that nuclear EFTs capture the low-energy limit of QCD.  This means
that in principle one can follow a top-down approach and determine low-energy
constants that appear in an EFT from a direct solution of the more
general theory.  While such solutions of QCD (in the highly nonperturbative
low-energy regime) are elusive analytically, lattice calculations have made
significant progress towards nuclear systems.  Matching EFT to
LQCD serves to extend the predictions of QCD in essentially two directions:
\begin{enumerate}
\item
Larger distances: solution by \abinitio methods of an EFT with parameters fixed
by LQCD allows for predictions of properties of larger nuclei and their
reactions, which are difficult to simulate directly from QCD.
\item
Smaller pion masses: with the relative importance of chiral-symmetry-breaking
interactions understood, Chiral EFT can be used as an extrapolation tool
from larger quark---and thus pion---masses to the physical point.
\end{enumerate}

{%
Moreover, remnants of QCD's color gauge symmetry can be traced down to nuclear
EFTs.  In particular, it is possible to consider the inverse number of colors,}
{$1/N_{\rm c}$},
as an expansion parameter to constrain nuclear forces.  This has
been studied, for example, by
\textcite{Kaplan:1996rk,Phillips:2013rsa,Phillips:2014kna,Samart:2016ufg,%
Schindler:2018irz}.
}

\subsubsection{Nuclear physics at large quark masses}
\label{subsubsection:lattice nuclei}

The pion mass is a tunable parameter in LQCD.  With calculations getting more
expensive for lower pion masses, typically results are extracted for values of
$\Mpi$ well above the physical point.  Observables like hadron spectra can by
now be described with amazing accuracy~\cite{Kronfeld:2012uk},
and results for hadronic properties have become available even at or below the
physical pion masses.  Direct QCD calculations of few-nucleon systems, however,
still use relatively large pion masses
{(\textcite{Beane:2010em} review the framework)}, and even then significant
discrepancies exist among the outcomes from various groups.  EFT calculations
have used a subset of these results, and tested their consistency with
increasing nucleon number $A$.

\textcite{Barnea:2013uqa} have shown how LQCD input for few-nucleon systems at a
fixed pion mass can be used in conjunction with \abinitio solutions of EFT to
predict the properties of larger nuclei.  Using the LQCD results of
\textcite{Beane:2012vq} at $\Mpi=805~\MeV$ for $A=2,3$ to fix the pionless LECs
at LO, \textcite{Barnea:2013uqa} predicted binding energies for $A=4,5,6$.
The $A=4$ result was consistent with the direct LQCD prediction, lending
credibility to both approaches.  Within the large uncertainties of the LQCD
input and of LO Pionless EFT, the pattern of binding energies resembles that of
the physical world, but with larger binding momenta which nevertheless remain
below the pion mass.  These results were extended to $A=16$ and to lattice input
from \textcite{Yamazaki:2012hi} at $\Mpi=510~\MeV$ by
\textcite{Contessi:2017rww}, and further to resummed \NLO and $A=40$ by
\textcite{Bansal:2017pwn}.

Simple reactions can be calculated directly in LQCD.  \textcite{Beane:2015yha}
extracted the pionless LEC $L_1$ that appears in Eq.~\eqref{eq:L-mag-L1L2}, thus
allowing for a parameter-free calculation of the $np \to d\gamma$ capture
process (as well as the inverse photodisintegration process) in very good
agreement with the experimental capture cross section. {The nuclear matrix
element determining the $pp\to d e^+ \nu_e$ fusion cross section and the
Gamow-Teller matrix element contributing to tritium $\beta$-decay were
calculated in LQCD by \textcite{Savage:2016kon}, allowing for a direct
extraction of the leading two-nucleon axial counterterm
$L_{1,A}=3.9(0.1)(1.0)(0.3)(0.9)$~fm$^3$ in Pionless EFT.} For larger nuclei,
one can use Pionless EFT to turn LQCD bound-state input into predictions for
reactions, as shown by \textcite{Kirscher:2015yda}. {They calculated $nd$
scattering observables at LO and obtained the Phillips and Tjon line
correlations at unphysical pion masses.}  Other observables such as magnetic
moments and polarizabilities of light nuclei
\cite{Chang:2015qxa,Kirscher:2017fqc} have been studied as well.  Mapping the
patterns of nuclear properties at unphysical pion masses could shed light into
the nature of the fine tuning that {pervades} nuclear physics.

\subsubsection{Fine-tuning in Chiral EFT and infrared limit cycle}
\label{subsubsection:fineorigin}

Chiral EFT, constituting the many-nucleon generalization of ChPT, yields the
pion-mass dependence of nuclear observables.  If LQCD data are within the
limit of validity of the theory, the latter can be used to extrapolate towards
smaller values of the quark---and thus pion---masses.

If $\MNN$ is considered a low-energy scale, the pion-mass dependence arises at
LO from the explicit pion mass in OPE and from the
{chiral-symmetry-breaking $D_{2{\rm s}}$} interaction in the $^1S_0$
channel, Eq.~\eqref{C0D0}.  At subleading  orders, it enters not only through
the explicit pion mass in multiple-pion  exchange, but also through the
quark-mass dependence of other LECs.  There have  been various calculations of
the pion-mass dependence of
two-~\cite{Beane:2001bc,Beane:2002vs,Epelbaum:2002gb,Beane:2002xf,%
Epelbaum:2002gk,Chen:2010yt,Soto:2011tb,Berengut:2013nh} and
three-~\cite{Hammer:2007kq} nucleon observables, which differ in the power
counting used and related issues (order, range of cutoffs, \etc), and
assumptions about presently unknown LECs.  Qualitatively, the observation of
\textcite{Beane:2001bc} has been confirmed: the deuteron ($^1S_0$ virtual
state) becomes unbound (bound) at a pion mass close (very close) to physical.

Although we do not know why these critical values of $m_\pi$ are close to
physical, the pion-mass dependence offers a plausible mechanism for the
fine-tuning observed in the real world, where the $N\!N$ binding energies are
small compared to the scale set by Eq.~\eqref{B/A}, and scattering lengths
large with respect to $\MNN^{-1}$.  Except in the vicinity of an $m_\pi$
critical value, these quantities attain values more in line with expectation.
In order to produce shallow poles, short-range physics does not need to be
particularly strong, but must be fine-tuned to negatively interfere with OPE, or
\viceversa.  {(For example, in the $^1S_0$ channel with nonperturbative OPE
the finite, energy-independent terms of {$I(\Lambda;k)$} in
Eq.~\eqref{T01S0} must partially cancel the short-range term.)}

\begin{figure}[tbp]
\centering
\includegraphics[clip,width=0.75\columnwidth]{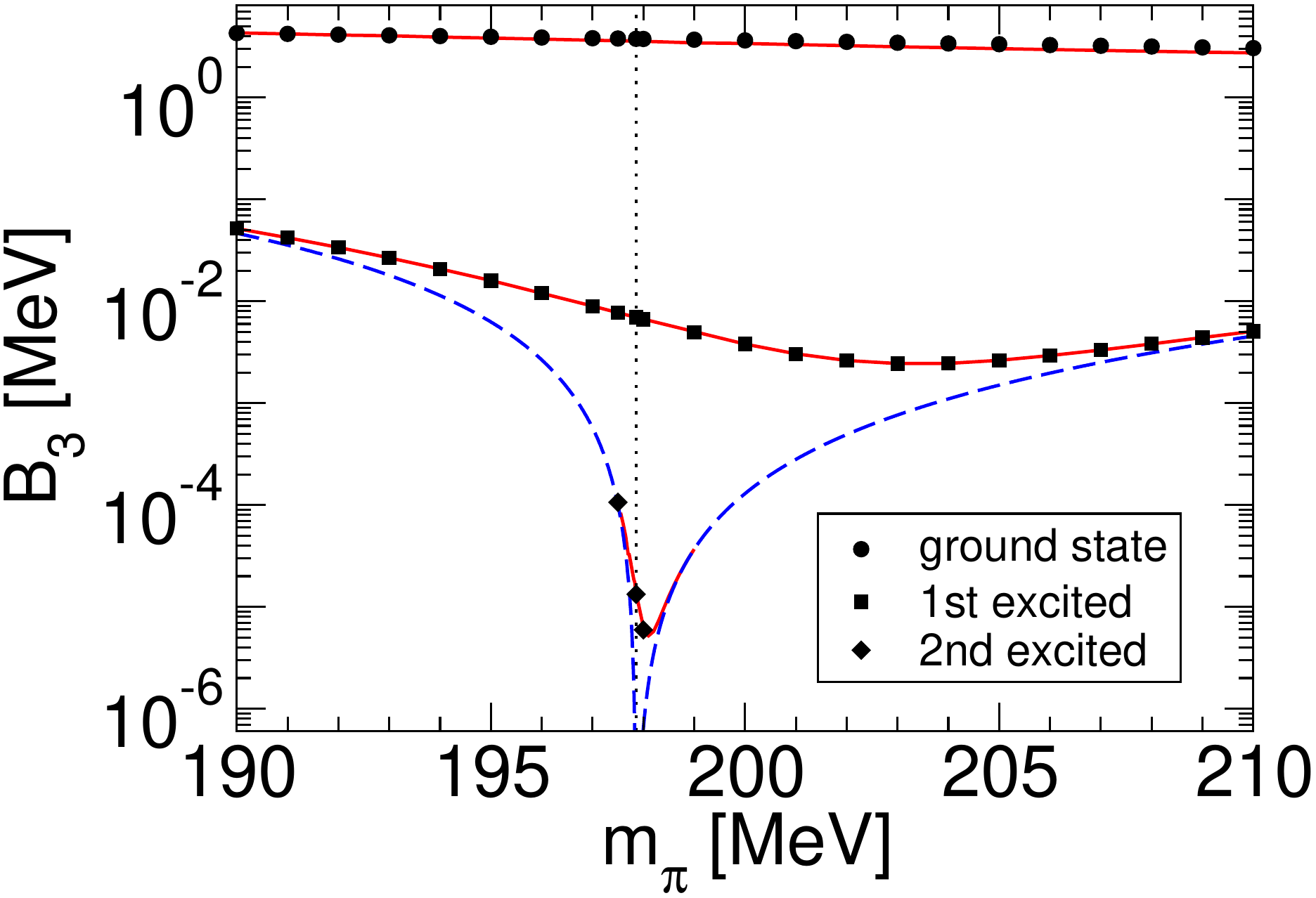}
\caption{%
Triton spectrum in the vicinity of a limit cycle as function of the pion mass.
The circles, squares, and diamonds give the {chiral-potential result of}
\textcite{Epelbaum:2006jc}, while the solid lines are N$^2$LO calculations in
{Pionless} EFT~\cite{Hammer:2007kq}.
The vertical dotted line indicates the critical pion mass. The thresholds for
stable three-nucleon states are given by the dashed lines.
Reprinted by permission from Springer Nature:
\cite{Hammer:2007kq}, copyright (2007).
}
\label{fig:B3_spectrum2}
\end{figure}

A variation of the pion mass has an effect similar to the variation of an
external magnetic field near a Feshbach resonance.  \textcite{Braaten:2003eu},
using the pion-mass dependence of the $N\!N$ $S$-wave scattering lengths
calculated within Chiral EFT by \textcite{Epelbaum:2002gb}, studied the
consequences for the three-body spectrum as a function of the pion mass.  They
found that an excited state of the triton appears at $\Mpi\approx175~\MeV$,
indicating that slightly changing the parameters to increase the pion mass
brings QCD closer yet to an infrared limit cycle.  Based on this it is
conjectured that it should be possible to tune QCD exactly to the limit cycle by
changing the up and down quark masses separately.  In this case, the triton
would have infinitely many excited states.  \textcite{Epelbaum:2006jc} showed
that parameters sets exist which make both $N\!N$ $S$ waves diverge for critical
pion masses between $179$ and about $200~\MeV$.  \textcite{Hammer:2007kq}
extended this analysis to higher orders and also calculated three-nucleon
scattering observables as a function of the pion mass.  The triton spectrum in
the vicinity of a limit cycle found in that work is shown in
Fig.~\ref{fig:B3_spectrum2}.   {Some of the implications to primordial
nucleosynthesis are pointed out by \textcite{Bedaque:2010hr} and
\textcite{Berengut:2013nh}.}

\subsection{Antinucleon systems}
\label{subsection:antinuclei}

While nucleon-antinucleon pairs can be integrated out at low energies, nuclear
EFTs apply as well to systems where antinucleons, a consequence of the Lorentz
invariance of QCD, appear in initial states.  The simplest such system is
antinucleon-nucleon ($\bar{N}\!N$) scattering.  While the details of
annihilation involve short distances, low-energy antinucleon-nucleus scattering
can be described in ways similar to nucleon-nucleus scattering provided that
the loss of flux to annihilation is accounted for with complex LECs.  Other
ingredients are very similar and related by charge conjugation ($C$) to the
nuclear potential summarized in Fig.~\ref{fig:chiralpot}.

The use of Chiral EFT for $\bar{p}p$ was pioneered by \textcite{Chen:2010an} and
\textcite{Chen:2011yu}, along the lines of \textcite{Kaplan:1996xu} for the
spin-singlet $N\!N$ channel.  The $\bar{N}\!N$ potential was derived to
$\OO(Q^4/\MQCD^4)$ by \textcite{Kang:2013uia} and \textcite{Dai:2017ont}, who
also obtained a very successful description of $\bar{p}p$ observables using
Weinberg's prescription together with the promotion of some interactions and
the demotion of others.  The final-state interactions generated by these
potentials~\cite{Chen:2010an,Kang:2015yka} explain the near-threshold
enhancement in the $\bar{p}p$ invariant-mass spectrum seen in charmonium decays
and $e^+e^-$ annihilation.  Chiral two-pion exchange had already been
incorporated in the partial-wave analysis of elastic and charge-exchange
scattering $\bar{p}p$ data by \textcite{Zhou:2012ui}, following the earlier
Nijmegen approach to $N\!N$~\cite{Rentmeester:1999vw,Rentmeester:2003mf}.

\subsection{Hypernuclei}
\label{subsection:hyperchiral}

Among the ``$\cdots$'' in Eq.~\eqref{QCDL} we find the kinetic, mass, and strong
and electromagnetic interaction terms of the strange quarks.  Chiral EFT can be
extended to $SU(3)_{\rm L} \times SU(3)_{\rm R}$ in an attempt to incorporate
kaon and eta exchange to describe hypernuclei.  The difficulty is the
intermediate value of the strange quark mass: it prevents integrating the
strange quark out at a perturbative scale, as it is done for heavier quarks,
but leads to poor convergence of the ChPT expansion~\cite{Donoghue:1998bs}
because of the relatively large kaon and eta masses, $m_K$ and $m_\eta$.

Nevertheless, by counting $m_K$ and $m_\eta$ as low-energy scales, one can
formally apply the power counting of Sec.~\ref{subsubsec:PC} to organize
the inter-baryon potential along the lines of Sec.~\ref{subsubsec:nukepot}.
The two-baryon potential was derived up to $\OO(Q^2/\MQCD^2)$ by
\textcite{Polinder:2006zh}, \textcite{Haidenbauer:2009qn},
\textcite{Haidenbauer:2013oca}, and \textcite{Haidenbauer:2015zqb}.  With
Weinberg's prescription, a description of hyperon-nucleon data is obtained of
quality comparable to the most advanced phenomenological models.  The leading
three-baryon forces, which appear at $\OO(Q^{n+1}/\MQCD^{n+1})$, have also
been written down~\cite{Petschauer:2015elq}.
{A large-$N_{\rm c}$ analysis of hyperon-nucleon interactions
was carried out by \textcite{Liu:2017otd},
while a covariant formulation presented by \textcite{Li:2016paq,Li:2016mln}.}
If $m_K$ or $m_\eta$ are
considered large scales, the onset of $\eta$-nuclear binding can be considered
in a Pionless EFT approach in order to derive constraints on the $\eta N$
scattering length~\cite{Barnea:2017epo,Barnea:2017oyk}.

{Certain hypernuclei are also amenable to Pionless and Halo EFT.  The
process of $\Lambda d$ scattering and the properties of the hypertriton
${}_\Lambda^{3}$H in the $SU(3)$ limit were first studied
in Pionless EFT by \textcite{Hammer:2001ng}.
Since the hypertriton is extremely shallow, the low-energy
observables in this channel are insensitive to the exact values of the
$\Lambda N$ low-energy parameters, as any shift can be absorbed by changing
the three-body force.}
{By
constructing a system of coupled integral equations in the spin-isospin basis,
\textcite{Ando:2015fsa} investigated}
the viability of the $nn\Lambda$ bound
state suggested by the recent experiment of the HypHI collaboration at
GSI~\cite{Rappold:2013jta}.
{%
The three-body force present at LO prevented any definitive conclusions
} about the existence of the $nn\Lambda$ bound state.
{More recently, \textcite{Hildenbrand:2019sgp} calculated the structure of
$nn\Lambda$ and ${}_\Lambda^{3}$H{,}
and clarified the value of the corresponding scaling factors.
For physical
{hyperon} and nucleon masses, they obtained
the $\Lambda d$ scattering length
$a_{\Lambda d} = (13.8^{+3.8}_{-2.0})$~fm, where the error
is dominated by the uncertainty in the hypertriton binding energy.}
{Implications of three-body universality to systems with two neutrons
and a flavored meson (such as $K^-$ and $D^0$)
were considered by \textcite{Raha:2017ahu}.}

{Pionless EFT for states with strangeness ${-}1$
was extended up to $_{\Lambda}^{5}$He by \textcite{Contessi:2018qnz},
presenting a solution to the ``overbinding
problem'' observed with previous approaches based on nucleon-hyperon model
interactions.}
{Light nuclear states with strangeness ${-}2$ have also been
examined by \textcite{Contessi:2019csf}. With the $\Lambda\Lambda$
contact interaction estimated from correlations observed in
relativistic heavy ion collisions and the $\Lambda\Lambda N$ three-body force
constrained by the binding energy of $^{\ \ 6}_{\Lambda\Lambda}$He,
the conditions for $^{3,4}_{\Lambda\Lambda}n$, $^{4,5}_{\Lambda\Lambda}$He
and $^{\ \ 5}_{\Lambda\Lambda}$H binding were discussed.}

{In paralell, ${}_{\Lambda\Lambda}^{\ \ 4}$H and ${}_{\Lambda\Lambda}^{\ \
6}$He have been described in Halo EFT as three-body systems where the two
hyperons orbit around, respectively, deuteron \cite{Ando:2013kba} and
alpha-particle \cite{Ando:2014mqa} cores.  In the spin-singlet channel of
$S$-wave ${}_\Lambda^{\, 3}$H-$\Lambda$ scattering, there is no bound state and
no three-body force at LO.  In this case, the ${}_\Lambda^{ \, 3}$H-$\Lambda$
scattering length was found to be $a_0^{} = (16.0\pm 3.0)~\fm$. In the
spin-triplet channel, a $\Lambda\Lambda d$ contact interaction is required at LO
to obtain a cutoff-independent ${}_{\Lambda\Lambda}^{\ \ 4}$H binding energy.
Similarly, a $\Lambda\Lambda\alpha$ three-body force is needed for
${}_{\Lambda\Lambda}^{\ \ 6}$He renormalization already at LO, but the
correlation between the double-$\Lambda$ separation energy of
${}_{\Lambda\Lambda}^{\ \ 6}$He and the $S$-wave $\Lambda\Lambda$ scattering
length could be investigated.}

{The paucity of experimental information on hypernuclei represents
an important opportunity for Lattice QCD to impact nuclear physics
through EFT (see Sec. \ref{subsubsection:lattice nuclei}).}

\subsection{Hadronic molecules}
\label{subsec:hadromol}

Universality also bridges the gap between the seemingly unrelated domains of
atomic and hadronic physics.  In recent years, a large number of new
``quarkonium'' states in the charmonium and bottomonium region have been
identified by various experiments~\cite{Olive:2016xmw}.  Many of these states
are close to the thresholds for decays into charm and bottom mesons which
strongly influences their properties, see reviews by \textcite{Swanson:2006st}
and \textcite{Brambilla:2010cs}.
{A new ``flavored nuclear physics'' has emerged where nucleons are replaced
by hadrons containing heavy quarks, which is amenable to EFTs
that parallel those deployed in conventional nuclear physics.}

A prominent member of this family of so-called $XYZ$ states is the $X(3872)$,
where the number in parentheses refers to the center-of-mass energy (in \MeV) at
which the state was first observed.  The closeness of the $X(3872)$ to the
$\bar{D}^0D^{*0}$ threshold as well as its quantum numbers
$J^{PC}=1^{++}$~\cite{Aaij:2013zoa} quickly led to the conjecture that it can be
interpreted, at least in part, as a shallow bound or virtual state of these two
mesons.  \textcite{Braaten:2003he} first used an EFT assuming a large
$\bar{D}^0D^{*0}$ scattering length to describe the $X(3872)$, with a number of
further papers building upon this{, for example
\cite{Braaten:2004rw,Braaten:2004fk,Braaten:2004ai,Braaten:2005ai,%
Braaten:2007dw,Braaten:2007ft}}.  An extension of this EFT to the three-body
sector was given by \textcite{Canham:2009zq}, who studied $D$ and $D^*$
scattering off the $X(3872)$.
{Consequences of this pionless EFT for other states, including
the effects of heavy quark symmetry, are discussed by
\textcite{AlFiky:2005jd,Mehen:2011yh,Nieves:2012tt,Guo:2013sya,%
Wilbring:2013cha,Guo:2013xga,Albaladejo:2015dsa,Valderrama:2018sap,Liu:2018zzu}.
}

The corrections to universality can be calculated systematically using an
EFT for the $X$ with explicit pions, called XEFT, which was developed by
\textcite{Fleming:2007rp}.  The analog of the scale \eqref{OPEscale} is
\begin{equation}
 M_{DD^*}= \frac{8\pi f_\pi^2}{g^2\mu_{{D^0}D^{*0}}} \,,
 \label{OPEscaleDD*}
\end{equation}
where $\mu_{{D^0}D^{*0}}\simeq 967~\MeV$ is the reduced mass and $g\simeq
0.5-0.7$ is the transition coupling of the pion to $\bar{D}^0$-$D^{*0}$.
$M_{DD^*}$ is larger than $\MNN$, while the mass associated with OPE is
smaller, $[(m_{D^{*0}}-m_{D^0})^2-m_\pi^2]^{-1/2}\simeq 45$ MeV instead of
$m_\pi$.  As a consequence, one expects pions to be
perturbative~\cite{Fleming:2007rp,Baru:2011rs,Valderrama:2012jv,%
Alhakami:2015uea,Braaten:2015tga} in the region of energies where the bound
state might lie.  XEFT is analogous to the version of Chiral EFT discussed in
Sec.~\ref{subsubsection:pertpions}.

A number of aspects of exotic mesons were investigated in this approach,
{such as light-quark-mass~\cite{Jansen:2013cba} and
finite-volume~\cite{Jansen:2015lha} effects} on the $X(3872)$ binding energy,
various decays of the
$X(3872)$~\cite{Fleming:2007rp,Fleming:2008yn,Mehen:2011ds,Baru:2011rs,%
Fleming:2011xa,Mehen:2015efa}, {the decay $\psi(4160) \to X(3872) \gamma$ as
a probe the $X(3872)$'s molecular content \cite{Margaryan:2013tta}, the triangle
singularity in $e^+e^- \to X(3872) \gamma$ \cite{Braaten:2019gfj},} scattering
of low-energy pions on the $X(3872)$~\cite{Braaten:2010mg}, {the role of
exact Galilei invariance for the $X(3872)$ and its {line
shape}~\cite{Braaten:2015tga,Schmidt:2018vvl},} {$X(3872)$ production in
colliders~\cite{Braaten:2018eov,Braaten:2019yua},} and heavy and light-quark
symmetries~\cite{HidalgoDuque:2012pq}.

The role of pion exchange has been further discussed for the $X(3872)$ at
{physical~\cite{Nieves:2011vw,Kalashnikova:2012qf,Wang:2013kva,%
Baru:2015nea}} and unphysical~\cite{Baru:2013rta,Baru:2015tfa} quark masses,
as well as in the context of {other states and the implications of heavy
quark symmetry~\cite{Liu:2012vd,Cleven:2015era,Baru:2016iwj,Baru:2017gwo,%
Lu:2017dvm,Geng:2017hxc,Geng:2017jzr,Xu:2017tsr,Wang:2018jlv,Wang:2018atz,%
Baru:2019xnh}.}

\subsection{Fundamental symmetries}

According to QCD, nuclei ultimately emerge from the interaction between quarks
and gluons.  The quarks and gluons, however, are subject not only to the strong
and electromagnetic interactions displayed explicitly in Eq.~\eqref{QCDL},
but also to weak and possibly other interactions found in the ``$\cdots$''.
Allowing for violation of symmetries such as parity ($P$) and time reversal
($T$) from higher-dimensional interactions in Eq.~\eqref{QCDL} introduces other
components to the nuclear potential and currents.  Nuclear EFT enables us to
incorporate the effects of weak and beyond-the-Standard-Model interactions in
the description of low-energy hadronic and nuclear processes.
{Input from Lattice QCD is particularly desirable in this
context~\cite{Cirigliano:2019jig}.}

\subsubsection{Parity violation}

Besides being responsible for beta decay, weak interactions also imply that
there should be small $P$-violating operators in the nuclear force and currents.
Since they stem mostly from four-quark interactions proportional to the Fermi
constant $G_F \simeq 1.17 \cdot 10^{-5}~\GeV^{-2}$, NDA~\eqref{NDA} suggests
that for $T$-conserving $P$ violation the suppression factor is $G_Ff_\pi^2 \sim
10^{-7}$.  The framework for the incorporation of $P$-violating effects in
nuclear EFTs {was developed by \textcite{Zhu:2004vw}}.  A major motivation
for this program is to understand the tension that exists among different
experiments~\cite{Holstein:2010zza}, when analyzed with quark and meson-exchange
models~\cite{Desplanques:1979hn}.

In the pionless theory, $P$ violation in the nuclear force is manifest as
$S$-to-$P$-wave contact interactions, five of which are independent at
LO~\cite{Zhu:2004vw,Girlanda:2008ts}.  \textcite{Phillips:2008hn} pointed out
that Pionless EFT is well suited to describe a number of existing and planned
$N\!N$ scattering experiments and calculated the relevant relationships between
observables (typically spin-polarization asymmetries) at LO in the $P$-violating
sector.  \textcite{Schindler:2015nga} have argued that
{large-$N_{\rm c}$} arguments can
be used to reduce the number of LO $P$-violating operators from five to two.
\textcite{Griesshammer:2010nd} have shown that no $P$-violating $3N$
force occurs up to and including NLO, enabling predictions for $P$-violating
elastic neutron-deuteron scattering~\cite{Griesshammer:2011md,Vanasse:2011nd}
based on the two-nucleon LECs.  {However, \textcite{Vanasse:2018buq}
concludes that a $P$-violating $3N$ force is required at NLO, after all.}

First pionless calculations of the deuteron anapole (or toroidal dipole) moment
and $P$-violating effects in the $np \to d\gamma$ capture process were
presented by \textcite{Savage:2000iv}, building upon previous work in the theory
with explicit, perturbative
pions~\cite{Savage:1998rx,Kaplan:1998xi,Savage:1999cm}.
\textcite{Schindler:2009wd}, \textcite{Vanasse:2014sva}, and
\textcite{Shin:2009hi} looked at $P$-violating asymmetries in the $np \to
d\gamma$ process.  Spin polarization in the inverse process was studied by
\textcite{Ando:2011nv}.  \textcite{Arani:2011if} and \textcite{Arani:2014eua}
studied $P$ violation in the $nd \to \ThreeH\,\gamma$ radiative-capture
reaction, and \textcite{Mahboubi:2016sta} included electromagnetic effects to
calculate polarized $pd$ scattering.

Some of these processes have also been considered in Chiral EFT.  As far as
strong interactions are concerned, the power $\mu$ of a contribution to the
potential (see Eq.~\eqref{chiralnupot}) can now be negative, but of course the
corresponding terms are suppressed by small factors.  The lowest orders of the
$T$-conserving, $P$-violating potential and electromagnetic currents were
obtained by \textcite{Zhu:2004vw}, \textcite{Kaiser:2007zzb},
\textcite{Liu:2007fn}, \textcite{Girlanda:2008ts},
\textcite{Viviani:2014zha}, and \textcite{deVries:2014vqa}.  They display new
elements, such as TPE, compared to the one-meson-exchange potentials usually
employed to study $P$ violation~\cite{Desplanques:1979hn}.  Calculations of $P$
violation in few-nucleon systems have so far been based on Weinberg's
prescription, as reviewed by \textcite{deVries:2015gea}.

\subsubsection{Time-reversal violation}

In the case of $T$, there is potential violation from the QCD vacuum angle
$\bar\theta$ and from higher-dimensional operators{, all contained in the
``$\cdots$'' in Eq.~\eqref{QCDL}.}
While the former is
anomalously small {($\bar\theta \simle 10^{-10}$~\cite{Tanabashi:2018oca}),}
the latter are suppressed by at least two powers of a large scale.
All violation from operators of dimension
up to six is accompanied by $P$ violation.  These interactions induce
$T$-violating nuclear form factors, such as electric dipole and magnetic
quadrupole, which could be probed in proposed storage-ring
experiments~\cite{Pretz:2013us}.  In nuclear EFT they are calculated within the
same framework used for nucleon electric dipole moments.

The lowest-order $P,T$-violating potential calculated in Chiral EFT by
\textcite{Maekawa:2011vs} and \textcite{deVries:2012ab}
shows, as its $P$-conserving counterpart, new ingredients compared to
one-meson-exchange potentials~\cite{Liu:2004tq}.  The implications of an
additional {large-$N_{\rm c}$} expansion are discussed by
\textcite{Samart:2016ufg}.
Together with $P,T$-violating currents, form factors were calculated for the
deuteron in Chiral EFT with perturbative pions by \textcite{deVries:2011re} and
with Weinberg's prescription by \textcite{deVries:2011an},
\textcite{Bsaisou:2012rg}, \textcite{Liu:2012tra}, and
\textcite{Bsaisou:2014zwa}; good accord was found between these calculations.
\textcite{deVries:2011an} and \textcite{Bsaisou:2014zwa} have also calculated
the electric dipole moments of triton and helion with Weinberg's prescription.
The deuteron also possesses a $P$-conserving, $T$-violating form factor, the
toroidal quadrupole.  The contribution to this moment from the same
$P,T$-violating operators in conjunction with weak interactions was obtained
with perturbative pions by \textcite{Mereghetti:2013bta}.  $T$ violation in
few-nucleon systems is reviewed by \textcite{Mereghetti:2015rra}, where it is
shown how measurements of the electric dipole moments of nuclei, together with
further theoretical advances, could at least partially disentangle the
various possible sources of $T$ violation.

\subsubsection{Other symmetries}

Higher-dimensional operators in the Standard Model break also other symmetries
like lepton ($L$) and baryon ($B$) number, but less work exists in {the
context of} nuclear EFTs.  Of particular contemporary interest is $L$ violation,
especially through the only dimension-five operator, which leads to Majorana
neutrino masses. The most sensitive laboratory probe of $L$ violation (by two
units) is neutrinoless double-beta decay ($0\nu\beta\beta$), which is however
afflicted by severe nuclear-physics uncertainties.  Traditionally
$0\nu\beta\beta$ has been calculated from the exchange of an explicit Majorana
neutrino {together with phenomenological nuclear models, but more recently
nuclear EFTs have been deployed \cite{Cirigliano:2018yza}.
It has been uncovered} that a short-range LEC must enter at LO for proper
renormalization in both Pionless \cite{Cirigliano:2017tvr} and
Chiral~\cite{Cirigliano:2018hja} EFTs.  First \abinitio calculations of these
contributions in light nuclei are becoming
available~\cite{Pastore:2017ofx,Cirigliano:2018hja}.  Operators of higher
dimensions have been considered as
well~\cite{Prezeau:2003xn,Cirigliano:2017djv}.  Implementation of these
operators in the shell model are starting to
appear~\cite{Horoi:2017gmj,Neacsu:2018urf}.

Analogous to $L$ violation is $B$ violation by two units.  The
process of neutron-antineutron oscillation in a nucleus leads to decay after the
antineutron annihilates with a nucleon.
{An NLO calculation of this process in the deuteron~\cite{Oosterhof:2019dlo}
gives a lifetime in Pionless EFT that is comparable to earlier zero-range
models, while in Chiral EFT (with perturbative pions) it is a factor of
$\simeq 2.5$ smaller than existing potential-model calculations.
}

Extensions of the Standard Model can be constructed which account for possible
violation of Lorentz and (then unprotected) $CPT$ symmetries at high energies,
allowing for operators with low dimensions.  Most tests of these symmetries take
place at low energies where QCD is nonperturbative, impeding direct bounds on
operators involving strongly-interacting particles.  Among the
lowest-dimensional operators of this type is a Lorentz-violating (but
$CPT$-conserving) purely gluonic operator.  The nuclear potential induced by
this operator and its possible effects on atomic-clock comparisons and on the
spin precession of the deuteron and other light nuclei in storage-ring
experiments are discussed by \textcite{Noordmans:2017kji}.  Dimension-five
operators are similarly discussed by \textcite{Noordmans:2016pkr}.  A more
detailed analysis of the nuclear implications of these interactions is needed.

\subsection{Dark-matter detection}
\label{subsec:dark}

Nuclear EFTs have also been used to describe dark-matter scattering off heavy
nuclei in direct-detection experiments.  The dark-matter particles must be
nonrelativistic in order to be bound in the dark-matter halo by gravitation,
with typical velocities of order $0.001$ times the speed of light.  Since the
recoil momentum is comparable to the typical momentum of a nucleon in the
nucleus, it is crucial for the interpretation of current experimental limits
(\cf~\textcite{Liu:2017drf}) that nuclear-structure factors be properly
addressed.

The calculation of nuclear-structure factors in nuclear EFT has been organized
in two different {ways.  First,} a Pionless EFT for nucleon and dark-matter
fields~\cite{Fan:2010gt,Fitzpatrick:2012ix,Fitzpatrick:2012ib,Anand:2013yka}
allows a study of {nuclear response functions in terms of effective}
couplings, and {the extraction} of limits on the coefficients of the
operators.  This
approach reaches its limit at the largest momentum transfers for scattering off
heavy nuclei where the details of pion exchange are resolved.  Second, Chiral
EFT has been used to predict the nuclear-structure {factors} for
spin-independent
and spin-dependent scattering. The analysis within Chiral EFT establishes
relations between different operators in the pionless framework, and provides
a counting scheme that indicates at which order two-nucleon operators
contribute.  Recent work in this direction includes
{Chiral EFT-based}
structure factors for the spin-dependent
response~\cite{Menendez:2012tm,Klos:2013rwa},
aspects of spin-independent
scattering~\cite{Cirigliano:2012pq,Cirigliano:2013zta,Vietze:2014vsa},
{scattering off light nuclei \cite{Korber:2017ery},}
inelastic scattering~\cite{Baudis:2013bba}, as well as a general Chiral EFT
analysis of one- and two-body
currents~\cite{Hoferichter:2015ipa,Hoferichter:2016nvd,Hoferichter:2018acd}
and improved limits for dark-matter models from experimental
searches~\cite{Hoferichter:2017olk,Aprile:2018cxk}.

\subsection{Bosons with large scattering length}
\label{subsec:bosons}

Up to technical details that arise from spin and isospin degrees of freedom,
Pionless EFT is virtually identical to a theory that describes a system of
bosons with a large two-body scattering length.  Throughout the history of
Pionless EFT this fact has been used repeatedly, the few-boson system serving to
guide analogous analyses in the few-nucleon sector.

But such bosonic systems are relevant far beyond serving as a toy problem.
Experimentally, they are realized in cold atomic gases, where the two-body
interaction can in fact be tuned arbitrarily by varying an external magnetic
field---the Feshbach-resonance mechanism.  In particular, the Efimov effect has
been established experimentally by exploiting the fact that the occurrence of
three-body states close to points where the two-body scattering length is tuned
to infinity~\cite{Kraemer:2006nat}, with many more experiments since the first
observation.  A comprehensive discussion of the theoretical treatment of
universal few-body systems has been given by \textcite{Braaten:2004rn}.
The current status of the field was recently reviewed by
\textcite{Naidon:2016dpf}.

\section{Conclusion} \label{sec:conclusion}
As shown in Fig.~\ref{fig:EFTscape}, a significant portion of low-energy nuclear
physics is amenable to an EFT description, with different theories tailored
specifically for different regions.  With increasing energy, a tower of EFTs
starts from the simple pionless case, an expansion around the unitary limit 
of large $N\!N$ scattering lengths.  Its range of applicability can be extended
by the inclusion of pions---first perturbatively then nonperturbatively---in
Chiral EFT, constructed as an expansion around the eponymous chiral limit of
vanishing quark masses.  Although there is a fundamental difference regarding
how pions are treated---a heavy degree of freedom in Pionless EFT, but a light
one in Chiral EFT---these theories are low-energy limits of QCD.  They are both
formulated as theories of pointlike nucleons with interactions that give rise to
low-energy poles of the $S$ matrix.
 
The fact that nucleons, being composite hadrons, in reality have substructure is
encoded in the EFT  expansion, namely in local operators with an increasing
number of derivatives.  Establishing the ordering of such interactions is the
crucial element that enables a systematic description of observables.

The usefulness of EFT does not stop at this point because new scale separations
arise in nuclei.  A particular case, Halo/Cluster EFT, has been discussed in 
this review as {a promising} way to describe cluster-like nuclei.  On yet 
another
level, efforts are underway to construct EFTs for rotational and vibrational
modes in heavy nuclei~\cite{Papenbrock:2010yg}.  Moreover, applications to other
composite systems---from dark matter to cold atoms---show how nonperturbative
EFTs are a driving force behind many important developments in modern 
theoretical physics.

EFTs are ideally suited to root nuclear physics in the Standard Model EFT, 
elegantly exploiting its emergence from QCD as the underlying theory of the 
strong interaction---particularly through lattice simulations.  EFTs have become
the \textit{initio} of \abinitio methods for the solution of few- and 
many-nucleon dynamics and engendered such an explosion of activity that it is
difficult to draw a line to conclude this article.  We have already reached the
point where calculated nuclear properties are being used to identify 
deficiencies in the nuclear interactions used as input.

\Abinitio calculations now almost unanimously follow---mostly in Chiral EFT, 
but increasingly in Pionless EFT---Weinberg's original prescription, \ie, they
do not expand on the subleading components of the potential.  Our emphasis in
this review were approaches that pursue the longstanding goal of RG invariance
through the perturbative expansion of the $S$ matrix on the subleading
interactions.  This approach has led to a new and unified description of 
few-body ``Efimov physics'' under the umbrella of Pionless EFT.  It remains to 
establish, however, to which extent this framework can share the efficiency of
Weinberg's approach without dependence on the form of the regulator, or perhaps
explain its phenomenological success for larger nuclei within narrow cutoff
windows.

It is thus our hope that this review does not only provide a unified overview of
what has been done, but will also inspire future research towards a comprehensive 
and solid understanding of nuclear structure and reactions from an EFT, and 
ultimately QCD, perspective.

\begin{acknowledgments}
Many colleagues have sharpened our understanding of nuclear EFTs.
For discussions that directly affected this manuscript,
we thank S.~Fleming, R.~J.~Furnstahl, H.~W.~Grie{\ss}hammer, and S.~Meinel.
This work was supported in part by the U.S. Department of Energy, Office 
of Science, Office of Nuclear Physics, under award No.~DE-FG02-04ER41338,
by the European Union Research and Innovation program Horizon 2020 under grant 
agreement No.~654002, by the DOE-funded NUCLEI SciDAC Collaboration under award 
DE-SC0008533, by the NSF under award PHY--1306250, by the ERC Grant No.~307986 
STRONGINT, {by the German Federal Ministry of Education and Research
(BMBF) under contracts no.\ 05P15RDFN1 and 05P18RDFN1,
and by the Deutsche Forschungsgemeinschaft (DFG, German Research Foundation)
- Projektnummer 279384907 - SFB 1245.}
\end{acknowledgments}

\end{document}